\documentclass{aa}  
\usepackage{graphicx}
\usepackage{txfonts}
\usepackage{longtable}
\usepackage{xcolor}
\usepackage[section]{placeins}
\usepackage{float}
\usepackage{capt-of}
\usepackage{upgreek}
\usepackage{nccmath}
\usepackage{amsmath}
\usepackage{caption}
\usepackage{subcaption}
\usepackage{hyperref}
\hypersetup{
    colorlinks=true,
    linkcolor=blue,
    filecolor=blue,      
    urlcolor=blue,
    citecolor=blue,
}
\begin{document}

   \title{Interstellar anatomy of the TeV gamma-ray peak \\ in the IC443 supernova remnant}

   \author{P. Dell'Ova
          \inst{1,2}
          \and
          A. Gusdorf\inst{1,2}
          \and
          M. Gerin\inst{2}
          \and
          D. Riquelme\inst{3}
          \and 
          R. Güsten\inst{3}
          \and
          A. Noriega-Crespo\inst{4}
          \and
          L.N. Tram\inst{5,6}
          \and
          M. Houde\inst{7}
          \and
          \\
          P. Guillard\inst{8}
          \and
          A. Lehmann\inst{1,2}
          \and
          P. Lesaffre\inst{1,2}
          \and
          F. Louvet\inst{1,9}
          \and
          A. Marcowith\inst{10}
          \and
          M. Padovani\inst{11}
          }

   \institute{Laboratoire de Physique de l’École Normale Supérieure, ENS, Université PSL, CNRS, Sorbonne Université, Université de Paris, F-75005 Paris, France.\\
              \email{pierre.dellova@ens.fr}
         \and
             Observatoire de Paris, PSL University, Sorbonne Université, LERMA, 75014 Paris, France.
         \and
             Max-Planck-Institut für Radioastronomie (MPIfR), Auf dem Hügel 69, 53121 Bonn, Germany.
         \and
             Space Telescope Science Institute, 3700 San Martin Dr., Baltimore, MD 21218, USA
         \and 
             Stratospheric Observatory for Infrared Astronomy, Universities Space Research Association, NASA Ames Research Center, MS 232-11, Moffett Field, 94035 CA, USA
         \and
             University of Science and Technology of Hanoi, Vietnam Academy of Science and Technology, 18 Hoang Quoc Viet, Vietnam
         \and
             Department of Physics and Astronomy, The University of Western Ontario, London, Ontario, Canada, N6A 3K7.
         \and
             Institut d’Astrophysique de Paris, CNRS UMR 7095, Sorbonne Université, 75014 Paris, France
         \and
             AIM, CEA, CNRS, Universit\'e Paris-Saclay, Universit\'e Paris Diderot, Sorbonne Paris Cit\'e, 91191 Gif-sur-Yvette, France
         \and
             Laboratoire Univers et Particules de Montpellier (LUPM) Université Montpellier, CNRS/IN2P3, CC72, place Eugène Bataillon, 34095, Montpellier Cedex 5, France
         \and
             INAF-Osservatorio Astrofisico di Arcetri, Largo E. Fermi 5, 50125 Firenze, Italy
             }


 
  \abstract
   {Supernovae remnants (SNRs) represent a major feedback source from stars on the interstellar medium of galaxies. During the latest stage of supernovae explosions, shock waves produced by the initial blast modify the chemistry of gas and dust, inject kinetic energy in the surroundings, and may alter star formation characteristics. Simultaneously, $\gamma$-ray emission is generated by the interaction between the ambiant medium and the cosmic rays, including those locally accelerated in the early stages of the explosion.}
   {We study the stellar and interstellar contents of IC443, an evolved shell type SNR at a distance of 1.9 kpc, with an estimated age of 30 kyr. We aim to measure the mass of the gas and characterize the nature of infrared point sources within the extended G region, which corresponds to the peak of $\gamma$-ray emission detected by VERITAS and \textit{Fermi}.}
   {We performed 10$^\prime \times$ 10$^\prime$ mapped observations of $^{12}$CO and $^{13}$CO J=1--0, J=2--1 and J=3--2 pure rotational lines, as well as C$^{18}$O J=1--0 and J=2--1 obtained with the IRAM-30m and APEX telescopes over the extent of the $\gamma$-ray peak to reveal the molecular structure of the region. We first compared our data with local thermodynamic equilbrium (LTE) models. We estimated the optical depth of each line from the emission of the isotopologues $^{13}$CO and C$^{18}$O. We used the population diagram and large velocity gradient (LVG) assumption to measure the column density, mass, and kinetic temperature of the gas using $^{12}$CO and $^{13}$CO lines. We used complementary data (stars, gas, and dust at multiple wavelengths) and infrared point source catalogues to search for protostar candidates.} 
   {Our observations reveal four molecular structures: a shocked molecular clump associated with emission lines extending between -31 km s$^{-1}$ and 16 km s$^{-1}$, a quiescent, dark cloudlet associated with a linewidth of $\sim$2 km s$^{-1}$, a narrow ring-like structure associated with a linewidth of $\sim$1.5 km s$^{-1}$ and a shocked knot. We measured a total mass of $\sim$230 $\mathrm{M_{\odot}}$, $\sim$90 $\mathrm{M_{\odot}}$, $\sim$210 $\mathrm{M_{\odot}}$ and $\sim$4 $\mathrm{M_{\odot}}$ respectively for the cloudlet, the ring-like structure, the shocked clump and the shocked knot. We measured a mass of $\sim$1100 $\mathrm{M_{\odot}}$ throughout the rest of the field of observations where an ambient cloud is detected. We found 144 protostar candidates in the region.}
   {Our results emphasize how the mass associated with the ring-like structure and the cloudlet cannot be overlooked when quantifying the interaction of cosmic rays with the dense local medium. Additionally, the presence of numerous possible protostars in the region might represent a fresh source of CR, which must also be taken into account in the interpretation of $\gamma$-ray observations in this region.}

   \keywords{ISM: supernova remnants --
                ISM: individual object: IC443 --
                ISM: kinematics and dynamics --
                ISM: cosmic rays --
                Stars: formation --
                Physical data and processes: shock waves
               }

   \maketitle
%

\section{Introduction}
The violent end of some stellar objects, a supernova (SN) explosion, is the beginning of an incredible sequence of energy injection in the surrounding interstellar medium (ISM). A SN explosion ejects material with mass ranging from $\sim$1.4 to $\sim$20 $M_\odot$ and a typical energy of 10$^{50-52}$~ergs (see e.g. \citealt{Draine11}) and drives fast shock waves, about 10$^4$~km~s$^{-1}$ ahead of the ejecta (the expelled stellar material) through the interstellar medium. Already at early stages, SNe play a crucial, multifaceted role in the evolution of galaxies. The explosion and subsequent dispersion of matter is itself the most important source of elements heavier than nitrogen in the gas phase \citep{Francois04}. The fast shocks inject kinetic energy that gradually decays in turbulence, a key process in the regulation of star formation on galactic scales \citep{McLow04}. The fast shock fronts are also a recognized site for the production of the bulk of cosmic rays (CRs, at least at GeV energies; see e.g. \citealt{Bykov18, Tatischeff18} and references therein for recent reviews), although alternative origins draw more and more attention (such as superbubbles and Fermi bubbles, see e.g. \citealt{Grenier15, Gabici19} and references therein for recent reviews). Finally, the hot (10$^6 - 10^8$~K) and relatively dense ($1-10$~cm$^{-3}$) conditions in the ejecta could be favourable to the synthesis of cosmic dust up to hundred years after the explosion (see recent reviews by \citealt{Cherchneff14, Sarangi18, Micellota18}). After the first phases of expansion (free, then adiabatic), the temperature of the shock front drops to 10$^6$~K, allowing the gas to radiatively cool down. At this stage, the so-called supernova remnant (SNR) resembles a spherical shell of $10-20$~pc radius, delimitated by regions where shocks interact with the ambient medium. 

This evolved SNR stage is also of fundamental importance to many aspects of galactic evolution. The fastest remaining shocks (between $\sim$30 and a few hundreds of km s$^{-1}$) dissociate and ionize the pre- and post-shock medium, and generate far-ultraviolet photons (e.g. \citealt{Hollenbach89}). More generally, shocks heat, accelerate, and compress the ambient medium. They inject energy and trigger specific dust and gas-phase chemical processes (e.g. \citealt{Vandishoeck93}), hence significantly participating to the cycle of matter in galaxies. CRs accelerated in the earlier stages of the explosion and trapped in the shock fronts now interact with the dense medium, producing observable X-ray to $\gamma$-ray photons (\citealt{Gabici09, Celli19, Tang19} and references therein). Cosmic rays of Galactic origin can also be re-accelerated in the shock regions (like in W44, e.g. \citealt{Cardillo16}). Finally, evolved SNRs play a key role on star formation. Like all massive stars, the progenitor of a SN explosion has formed in a cluster, in a dense and inhomogeneous environment, where lower-mass stellar companions have a greater life expectancy \citep{Montmerle79}. During its life on the main sequence, the progenitor has driven stellar winds in the surrounding medium, possibly triggering a second generation of star formation in the neighbouring molecular clouds (e.g. \citealt{Koo08}). The compression and cooling caused by SNR shocks might also generate star formation (eg. \citealt{Herbst77}). In any case, the injection of energy exerted by SNRs in all possible forms (CRs, energetic photons, shocks) likely locally alters the characteristics of all possible star formation events over significant spatial scales and times.   

With the present study we aim to start characterizing as precisely as possible and on fields as large as possible the mechanisms of energy injection (shocks, photons, CRs) exerted by an evolved SNR, and its consequences on the local star formation. Such a study must be performed on an evolved object, since the energy injection effects can spread over the full duration of the SNR phase, and are all the more visible when the SNR is old. In particular, we want to provide support for the study of CRs properties (acceleration, composition, diffusion) in evolved SNRs. In these objects, CRs interact with the local medium through four processes that all generate $\gamma$-ray photons: pion decay from the collision of hadronic CRs with the dense material, Bremsstrahlung from the interaction of leptonic CRs with the local dense medium, inverse Compton scattering of leptonic CRs with the local radiation field, and synchrotron emission of leptonic CRs gyrating around the local magnetic fields. Our first aim is constraining the properties of the local medium that is the target of these interactions: mass and density of all observed components, magnetic field structure, and radiation field structure. 

Our second aim is to identify all possible sources of on-going acceleration of \lq fresh’ CRs additional to the \lq old’ injection of CRs previously accelerated by the SNR. These sources can be of two kinds: ionized regions where kinetic energy is deposited and where the magnetic field structure and the ionization fraction make the acceleration possible (like [\ion{H}{II}] regions, see \citealt{Padovani2019}), or protostellar jets and outflows (where these conditions can be naturally combined). Our work can thus provide support for further studies of CR-related questions only if the study of local star formation is performed simultaneously. Indeed, \citet{Padovani15,Padovani16} have shown that jets can accelerate low-energy CRs, that can be re-accelerated in the shock fronts of the remnant. Other studies have confirmed that supermassive star clusters neighbouring SNRs can be a source of CRs \citep{Hanabata14}. Conversely an optimal characterization of CR action on the local formation is mandatory to better understand star formation in older galaxies, up to $z\sim2$ for which the star formation efficiency is higher \citep{Madau14}. Indeed, such a star formation regime is reminiscent of starburst galaxies, where SNe from a given generation of stars affect the next one. 

\begin{figure*}[]
   \centering
   \includegraphics[width=0.75\hsize, trim={2cm 1cm 0cm 1.9cm},clip]{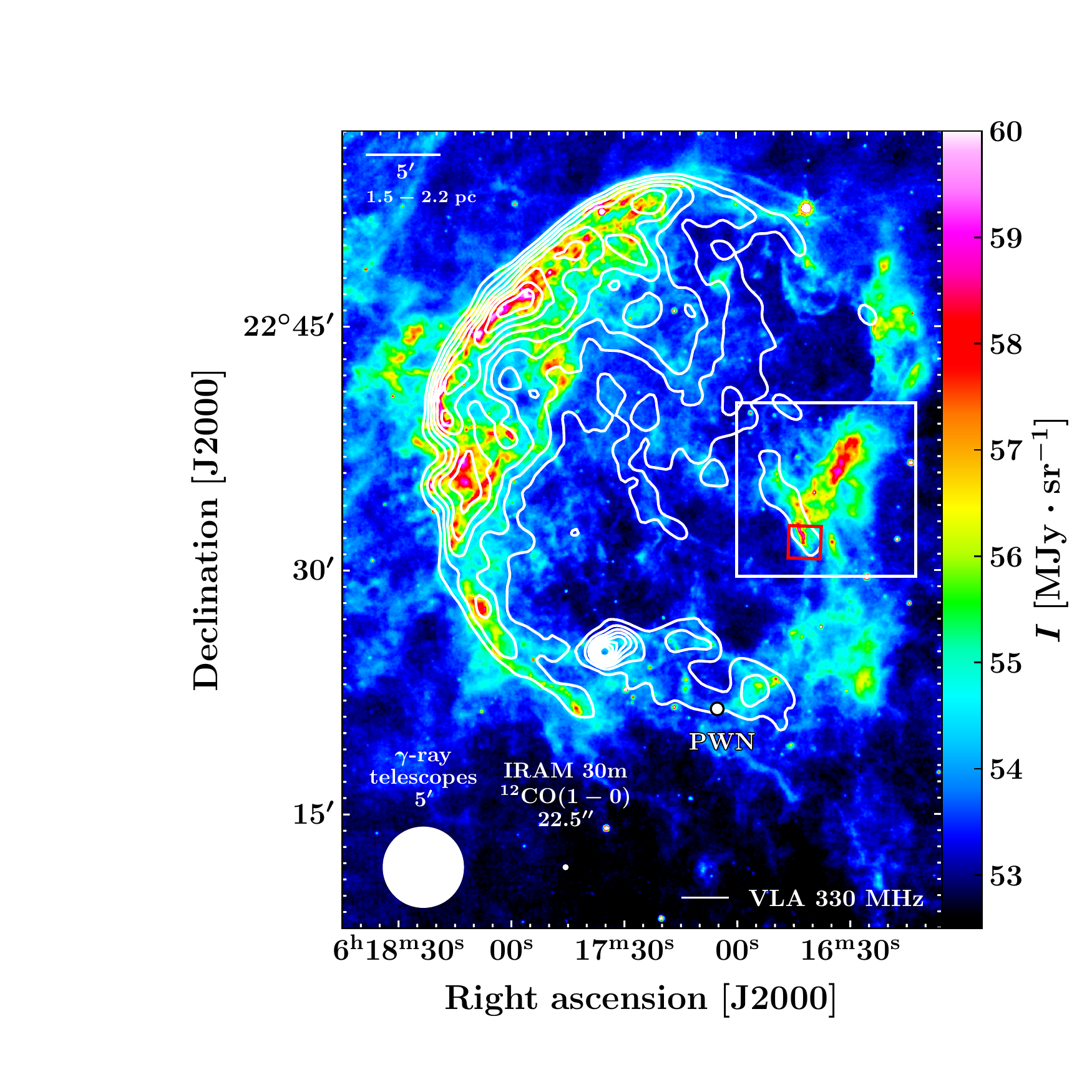}
      \caption{\textit{Spitzer}/MIPS map (colors) of IC443 at 24 $\upmu$m (\citealt{Pinheiro11}, \citealt{Noriega-crespo2009}). In contours, the Very Large Array (VLA) emission map displays the morphology of the synchrotron emission at 330 MHz \citep{Claussen1997}. The white dot marks the position of the pulsar wind nebula \citep{Olbert2001}. The red box represents one of the fields observed by \textit{Spitzer}-IRS \citep{Neufeld2004}, corresponding roughly to the G region defined by \citet{Huang86}. The white box represents the 10$^\prime \times$10$^\prime$ field of our observations, that we name 'the extended G region'. The IRAM-30m instrumental beam diameter corresponding to the $^{12}$CO(1--0) transition and the size of the typical PSF of $\gamma$-ray telescopes (5$^\prime$) are indicated by white disks in the lower left corner of the figure.} 
         \label{fig:finderchart}
   \end{figure*}

The threefold and intertwined goals of our study (interstellar medium, star formation and CR) make the study of large fields mandatory. Indeed, CR studies rely on $\gamma$-ray spectra obtained with telescopes that only provide a limited spatial resolution, typically a few arcminutes. This extent is the minimum field where we have to characterize the interstellar medium and the star formation as best as we can. This is why we have chosen to study a 10$^\prime \times$10$^\prime$ field in the relatively evolved IC443 SNR (see Fig. \ref{fig:finderchart}). More particularly, with the present paper we investigate the physical conditions and dynamical structure of the molecular gas and its association with protostars in such a field located at the peak of $\gamma$-ray emission detected in the IC443 SNR, based on observations of the CO emission lines and its isotopologues. In Sect.~\ref{section:source} we present a summarized review of the source. In Sect.~\ref{section:data} we present our observations and propose a description of the morphology and kinematics of the region, emphasising three distinct molecular components. Sect.~\ref{mass} focuses on the measure of the gas mass for these components. First we perform a local thermodynamical equilibrium (LTE) analysis of the $^{12}$CO, $^{13}$CO and C$^{18}$O emission lines and we build pixel-per-pixel, channel-per-channel population diagrams corrected for optical depth. Then we propose a second method using a radiative transfer code based on the Large Velocity Gradient (LVG) approximation. In Sect.~\ref{stars} we study the spectral energy distribution of point sources identified in infrared survey catalogs, as well as the spatial distribution of optical point sources detected with GAIA. Finally we summarize our findings in Sect.~\ref{conclusion}.

\section{The supernova remnant IC443}\label{section:source}
IC443 is a mixed-morphology supernova remnant, located at a distance of 1.5-2 kpc (\citealt{Denoyer78}, \citealt{Welsh03}), with recent measures suggesting a kinematic distance of 1.9~kpc \citep{Ambrociocruz17}. IC443 is an evolved SNR, yet its exact age is a matter of debate. The literature contains two kinds of value ($\sim$3 and $\sim$30 kyr), depending on the type of data that is analysed. A compelling finding concerning the age and origin of the IC443 SNR was the discovery of the CXOU J061705.3+222127 Pulsar Wind Nebula (PWN), based on \textit{Chandra} X-ray observatory and Very Large Array (VLA) images \citep{Olbert01}. The motion of the PWN is consistent with an age of 30~kyr for the SN event, and its detection strongly supports a core-collapse formation scenario for the SNR.

IC443 displays a shell morphology in radio, with two atomic sub-shells (shells A and B, \citealt{Braun&Strom1986}). It is one of the most striking example of a SNR interacting with neighbouring molecular clouds. The most up-to-date and complete description of the structure and kinematics of both the atomic and dense molecular environment of the SNR was offered by \citet{Lee2008, Lee2012}. Their $\sim 1^{\circ} \times 1^{\circ}$ map of the J=1-0 transition of $^{12}$CO allowed to characterize both the incomplete molecular shell interacting with shocks (toward the southern part of the SNR) and the molecular cloud that is associated with the remnant.
Continuum radio emission in IC443 is partly correlated with the molecular shell and the secondary [\ion{H}{I}] shell (see \citealt{Castelletti2011} for a detailed description of the low-frequency radio emission in IC443 at 74 and 330 MHz, and \citealt{Egron2017}, \citealt{Loru2019} for high and very-high frequency studies, respectively at 7 GHz and 21.4 GHz).

\begin{table}[]
\caption{Spectroscopic parameters corresponding to the observed lines. $J_{\mathrm{up}}$ is the rotational quantum number, $\nu_{ij}$ and $A_{ij}$ are respectively the frequency and the Einsten coefficient of the pure rotational transition ($\varv=0$ where $v$ is the vibrational quantum number). $E_{\mathrm{up}}$ and $g_{\mathrm{up}}$ are respectively the energy and degeneracy corresponding to the upper level. The values given are taken from the Cologne Database for Molecular Spectroscopy (\citealt{Muller2001}, \citealt{Muller2005}, \citealt{Endres2016}) and Jet Propulsion Laboratory database \citep{Pickett1998}, and are the numeric values used for all our measures.} 
\label{table:1}      
\centering                          
\begin{tabular}{c c c c c c}        
\hline\hline                 
specie & $J_{\mathrm{up}}$ & $\nu_{ij}~\mathrm{(GHz)}$ & $A_{ij}~\mathrm{(s^{-1})}$ & $g_{\mathrm{up}}$ & $E_{\mathrm{up}}~\mathrm{(K)}$ \\    
\hline \hline \\[-1.0em]
    $\mathrm{^{12}}$CO & 1 & 115.2712018 & $7.203 \times 10^{-8}$  & 3  & 5.53\\
    $\mathrm{^{12}}$CO & 2 & 230.5380000 & $6.910 \times 10^{-7}$  & 5  & 16.6\\
    $\mathrm{^{12}}$CO & 3 & 345.7959899 & $2.497 \times 10^{-6}$  & 7  & 33.19\\
\hline \\[-1.0em]
    $\mathrm{^{13}}$CO & 1 & 110.2013543 & $6.294 \times 10^{-8}$  & 3  & 5.29\\
    $\mathrm{^{13}}$CO & 2 & 220.3986841 & $6.038 \times 10^{-7}$  & 5  & 15.87\\
    $\mathrm{^{13}}$CO & 3 & 330.5879652 & $2.181 \times 10^{-6}$  & 7  & 31.73\\
\hline  \\[-1.0em]                             
    C$\mathrm{^{18}}$O & 1 & 109.7821734 & $6.266 \times 10^{-8}$  & 3  & 5.27\\
    C$\mathrm{^{18}}$O & 2 & 219.5603541  & $6.011 \times 10^{-7}$  & 5  & 15.81\\
\hline
\end{tabular}
\end{table}

\begin{table*}
\caption{Observed lines and corresponding telescope parameters for the observations of the extended G region of IC443: $\nu$ is the frequency of the transition, FWHM corresponds to the full width at half maximum of the instrumental beam, $F_{\mathrm{eff}}$ is the forward efficiency and $B_{\mathrm{eff}}$ the beam efficiency of the dish. $\overline{T_{\mathrm{sys}}}$ is the average system noise temperature and $\Delta \varv$ the nominal spectral resolution. PWV is the precipitable water vapor, and r.m.s is the standard deviation measured on the baseline resampled with a spectral resolution of 0.5 km s$^{-1}$ to allow direct comparison between the data from the IRAM-30m and APEX telescopes.}       
\label{table:2}      
\centering                          
\begin{tabular}{l  c  c  c  c c c c c c c}        
\hline \hline         \\[-1.0em]
species & $^{12}$CO & $^{12}$CO & $^{12}$CO & $^{12}$CO & $^{13}$CO & $^{13}$CO & $^{13}$CO & C$^{18}$O & C$^{18}$O \\
\hline \\[-1.0em]
line & (1--0) & (2--1) & (2--1) & (3--2) & (1--0) & (2--1) & (3--2) & (1--0) & (2--1)\\
telescope & \footnotesize{IRAM-30m} & \footnotesize{IRAM-30m} & \footnotesize{APEX} & \footnotesize{APEX} & \footnotesize{IRAM-30m} & \footnotesize{APEX} & \footnotesize{APEX} & \footnotesize{IRAM-30m} & \footnotesize{APEX}\\
\hline
\hline \\[-1.0em]
\footnotesize{$\nu$ (GHz)} & \footnotesize{115.271} & \footnotesize{230.538} & \footnotesize{230.538} & \footnotesize{345.796} & \footnotesize{110.201} & \footnotesize{220.399} & \footnotesize{330.588} & \footnotesize{109.782} & \footnotesize{219.560} \\
\footnotesize{FWHM ($^{\prime \prime}$)} & \footnotesize{22.5} & \footnotesize{11.2} & \footnotesize{28.7} & \footnotesize{19.2} & \footnotesize{23.5} & \footnotesize{30.1} & \footnotesize{20.0} & \footnotesize{23.6} & \footnotesize{30.2}  \\
\footnotesize{sampling ($^{\prime \prime}$)} & \footnotesize{3.5} & \footnotesize{3.5} & \footnotesize{6} & \footnotesize{6} & \footnotesize{3.5} & \footnotesize{6} & \footnotesize{6} & \footnotesize{3.5} & \footnotesize{6}\\
\hline
\footnotesize{receiver} & \footnotesize{EMIR} & \footnotesize{EMIR} & \footnotesize{PI230} & \footnotesize{FLASH345} & \footnotesize{EMIR} & \footnotesize{PI230} & \footnotesize{FLASH345} & \footnotesize{EMIR} & \footnotesize{PI230}\\
\footnotesize{obs. dates} & \footnotesize{19-02-20} & \footnotesize{19-02-20} &  \footnotesize{18-09-11} & \footnotesize{18-09-11} & \footnotesize{19-02-20} & \footnotesize{18-09-11} & \footnotesize{18-09-11} & \footnotesize{19-02-20} & \footnotesize{18-09-11} \\
 & \footnotesize{19-02-24} & \footnotesize{19-02-24} & \footnotesize{18-09-11} & \footnotesize{18-09-11} & \footnotesize{19-02-24} & \footnotesize{18-09-11} & \footnotesize{18-09-11} & \footnotesize{19-02-24} & \footnotesize{18-09-11}\\
\hline
\footnotesize{$F_{ \rm eff}$} & \footnotesize{0.94} & \footnotesize{0.92} & \footnotesize{0.95} & \footnotesize{0.95} & \footnotesize{0.91} & \footnotesize{0.95} & \footnotesize{0.95} & \footnotesize{0.94} & \footnotesize{0.95} \\
\footnotesize{$B_{ \rm eff}$} & \footnotesize{0.78} & \footnotesize{0.59}  & \footnotesize{0.73} & \footnotesize{0.63} & \footnotesize{0.51} & \footnotesize{0.73} & \footnotesize{0.63} & \footnotesize{0.78} & \footnotesize{0.73} \\
\hline \\[-1.0em]
\footnotesize{$\overline{T_{\rm sys}}$ (K)} & \footnotesize{172} & \footnotesize{321} & \footnotesize{146} & \footnotesize{291} & \footnotesize{118} & \footnotesize{129} & \footnotesize{330} & \footnotesize{118} & \footnotesize{129} \\
\footnotesize{$\Delta \varv$ (km s$^{-1}$)} & \footnotesize{0.5} & \footnotesize{0.5} & \footnotesize{0.1} & \footnotesize{0.1} & \footnotesize{0.5} & \footnotesize{0.1} & \footnotesize{0.1} &\footnotesize{0.5} & \footnotesize{0.1} \\    
\footnotesize{r.m.s (K)} & \footnotesize{0.035} & \footnotesize{0.039} & \footnotesize{0.069} & \footnotesize{0.066} & \footnotesize{0.025} & \footnotesize{0.084} & \footnotesize{0.080} &\footnotesize{0.022} & \footnotesize{0.130} \\   
\hline
\end{tabular}

\end{table*}

CO emission was observed by \citet{Denoyer79b} towards the SNR, revealing three shocked CO clumps along the southern molecular ridge (labelled A, B and C). Follow-up observations of CO J=1--0 over a 50$^\prime\times$50$^\prime$ field by \citet{Huang86} allowed to detect new areas of shock-cloud interaction and to identify 5 previously unknown CO clumps, extending the classification started by \citet{Denoyer79b} and providing the first mention of the G knot. The OH 1720~MHz line is a powerful diagnostic for the classification of SNRs interacting with molecular clouds \citep{Frail96}. 6 masing spots have been identified in IC443 by \citet{Claussen97}, all located in the G region delineated by \citet{Huang86}. He proposed that OH masers spots could be promising candidates for the sites of CR acceleration. \citet{Lockett99} improved the modelling of the shock origin for these maser lines, associated with moderate temperatures (50-125~K), local densities ($\sim 10^5$~cm$^{-3}$) and OH column densities of the order of 10$^{16}$~cm$^{-2}$, then \citet{Wardle99} added the effect of the dissociation of molecules by far ultraviolet (FUV) photons in molecular clouds subject to CR and X-ray ionization (see \citealt{Hoffman03}, \citealt{Hewitt06}, \citealt{Hewitt08}, \citealt{Hewitt09} for recent studies).
From J=1--0 $^{12}$CO and HCO$^+$ emission, \citet{Dickman1992} measured a mass of 41.6 $M_\odot$ for clump G, and estimated that a total molecular mass of 500-2000 M$_{\odot}$ is interacting with the SNR shocks, which corresponds to 5\%-10\% of the SN energy considering that the average velocity of the clumps is 25 km s$^{-1}$. \citet{Zhang2010} showed that two distinct structures are resolved in the region G, labelling G1 the strongest $^{13}$CO peak and G2 the previously mentionned shocked clump. \citet{Lee2012} measured a mass of 57.7$\pm$0.9 M$_\odot$ for the clump G2. Oddly, \citet{Xu11} measured a mass of 2.06$\times$10$^3$ M$_\odot$ for the 'cloud G', which is much higher than the previous estimates. Several molecular shocks were mapped within the SNR using $^{12}$CO lines (e.g. \citealt{White1987}, \citealt{Wang&Scoville1992} for clumps A, B and C). In particular, the kinematics of clump G were characterized in details by \citet{Vandishoeck93} who presented observations of the rotational transition J=3--2 of CO along the shocked molecular ring at a spatial resolution of 20$^{\prime \prime}$-30$^{\prime \prime}$. In the last ten years, the large scale molecular contents of IC443 have been scrutinized with increasing precision and completeness, since several authors have mapped the J=1--0 transition of the isotopologues $^{12}$CO, $^{13}$CO and C$^{18}$ over large fields (from 40$^\prime\times$45$^\prime$ to 1.5$^\circ \times$1.5$^\circ$, \citealt{Zhang2010}, \citealt{Lee2012}, \citealt{Su2014}). Several sub-millimeter observations of IC443 were performed to characterize the shocked molecular gas (e.g. \citealt{Vandishoeck93} for a study of the shock chemistry in the southern ridge, towards the clumps B, C and G). Notably, the ground state of shocked ortho-H$_2$O was detected towards the clumps B, C and G \citep{Snell2005}.

The incomplete shell-morphology is also observed in the J, H, K bands observed by 2MASS (Two-Micron All-Sky Survey, \citealt{Rho2001}), as well as in infrared and far infrared observations by \textit{Spitzer}-MIPS (\citealt{Pinheiro11}, \citealt{Noriega-crespo2009}) and WISE (Wide-field Infrared Survey Explorer, \citealt{Wright10}). Excitation by shocks was suggested as the most likely scenario for the emission lines detected in IR, instead of X-ray or FUV mechanisms (e.g. [\ion{O}{I}], \citealt{Burton1990}). \citet{Rho2001} analysed the emission of [\ion{O}{I}] with shock models, suggesting a fast J-type shock ($\sim$100 km s$^{-1}$) in the NE atomic shell, and a C shock (v$_\mathrm{s}$=$\sim$30 km s$^{-1}$) propagating in the southern molecular ridge. In the effort to study molecular shocks, H$_2$ pure rotational transitions were mapped towards the clump G by ISOCAM  (ISO, \citealt{Cesarsky1999}) and compared to non-stationary shock models, as well as towards the clumps C and G by \text{Spitzer}-IRS (Infrared Spectrograph) \citep{Neufeld2007}. The molecular clumps B, C and G were also observed by AKARI \citep{Shinn2011} and by the The Stratospheric Observatory for Infrared Astronomy (SOFIA) \citep{Reach2019}. All these studies were carried out in small fields ($\sim$ 1$^\prime\times$1$^\prime$) and allowed to put constraints on the shock velocity ($ \sim $ 30 km s $ ^{-1} $) and pre-shock density ($ \sim $ 10$^4$ cm$ ^{-3} $), and to outline similarities with protostellar shocks in the southern ridge where we aim to focus on the extended G region.

The optical emission is well correlated with radio and [\ion{H}{I}] features, reproducing shells A and B. In particular, the SNR displays bright, filamentary structures towards the northeastern part of the remnant (\citealt{Fesen80}, \citealt{Alarie&Drissen2019}) . IC443 was fully mapped by the Sloan Digital Sky Survey (SDSS, \citealt{York2000}). There are no bright features towards the extended G region, but optical studies offered constraints on the global characteristics of IC443. \citet{Ambrociocruz17} estimated an age of $\sim$30 kyr and an energy of 7.2$\times$10$^{51}$ erg injected in the environment by the SNR from the comparison of observations of [H$\alpha$] with SNR models \citep{Chevalier1974}. 

The shell-like structure of IC443 in radio, centrally filled in X-rays puts the remnant into the category of mixed-morphology SNRs (\citealt{Petre1988}, \citealt{Rho1998}). Observations of the hard X-ray contents of IC443 (up to 100 keV) by BeppoSAX show hints of shock-cloud interaction \citep{Bocchino00}. XMM-Newton mapped IC443 in the ranges 0.3-0.5 keV and 1.4-5.0 keV with an unprecedented field of view and spatial resolution \citep{Bocchino&Bykov2003}, and showed that the soft X-ray emission is partly absorbed by the nearby molecular cloud \citep{Troja2006}. \citet{Troja2008} reported the detection of a ring-shaped ejecta encircling the PWN, associated with a hot metal rich plasma which abundances are in agreement with a core-collapse scenario.

The CR content and $\gamma$-ray emission in IC443 have been scrutinized by multiple observatories. \textit{Fermi}-LAT (Large Area Telescope) detected an extended $\gamma$-ray source in the 200 MeV - 50 GeV energy band (\citealt{Abdo2010}, \citealt{Ackermann2013}). They showed that the spectrum is well reproduced by the decay of neutral pions and pinpointed the clouds B, C, D, F and G as targets of interaction. \citet{Tavani2010} reported the detection of $\gamma$-ray enhancement towards the NE shell in the 100 MeV - GeV range observed by AGILE (Astro-Rivelatore Gamma a Immagini Leggero). In their hadronic model, the cloud \lq E' is the suggested target for the interaction with CRs. Interestingly, the location of the $\gamma$-ray peak differs between the lower energy emission (\textit{Fermi} and EGRET detections, \citealt{Esposito1996}) and the VHE emission (VERITAS and MAGIC detections, \citealt{Albert2007}). This tendancy is also verified by \citet{Tavani2010} who located the 100 MeV $\gamma$-ray peak close to the \textit{Fermi}/EGRET position and further towards NE. Recently, \citet{Humensky15} presented an updated \textit{Fermi} map where the position of the peak was shifted and consistent with VERITAS, exposing the uncertainty on the localization of the peak from the analysis of $\gamma$-ray observations. The choice of the region studied in this paper is based on the TeV $\gamma$-ray significance map from \citet{Humensky15}, that emphasizes the extended G region as a favorable target of high-energy CR interaction with dense molecular gas.

Explicit magnetic field studies towards IC443 are scarce. \citet{Wood91} performed 6.1~cm polarimetric observations of the northeast rim of IC443. They found that the local magnetic field is rather correlated with the rim structure, but with no clear orientation (i.e. parallel or perpendicular) to it. Toward the clump G, only \citet{Hezareh13} conducted a polarization study. They observed circular and linear polarization of the CO (1--0) and (2--1) lines with the Institut de Radioastronomie Millimétrique 30m antenna (hereafter IRAM-30m), and linear polarization maps from the dust continuum with the Acatama Pathfinder EXperiment (hereafter APEX). Their study constitute a crucial step towards the characterization of the interaction of CRs with the magnetic field in the extended G region.

Star formation in IC443 was the focus of a few studies, but early investigations lacked sufficient data to make clear findings. Recently, \citet{Xu11} used color criteria for the point sources of IRAS and the Two Micron All Sky Survey (2MASS) to identify protostellar objects and Young Stellar Object (YSO) candidates and showed their association to several regions of the SNR interacting with neighbouring molecular structures, including clump G. \citet{Su2014} confirmed the shell structure of the distribution of YSO candidates using a selection method based on color-color diagrams inferred from WISE band 1, 2 and 3 and 2MASS band K. Both studies concluded that the formation of this YSO population is likely to have been triggered by the stellar winds of the progenitor.

\section{Observations, reduction, and dataset}\label{section:data}
\subsection{Observations}
\subsubsection{APEX}

A mosaic of the IC443G extended region was carried out with APEX\footnote{This publication is partly based on data acquired under project M9508A\_102 with the Atacama Pathfinder EXperiment (APEX). APEX is a collaboration between the Max-Planck-Institut f\"ur Radioastronomie (MPIfR), the European Southern Observatory, and the Onsala Space Observatory.}. APEX observations towards IC443G were conducted on September 11, 2018. The heterodyne receivers PI230 and FLASH345 (First Light APEX Submillimeter Heterodyne receiver, \citealt{Heyminck06}), operating at $230~\mathrm{GHz}$ and $345~\mathrm{GHz}$ respectively, were used in combination with the FFTS4G and the MPIfR fast Fourier transform spectrometer backend (XFFTS, \citealt{Klein12}). This setup allowed to cover a total field of $10^\prime \times 10^\prime$ towards the center of the molecular region G. The observations were performed in position-switching/on-the-fly mode using the APECS software \citep{Muders06}. Tab.~\ref{table:1} contains the spectroscopic parameters of the observed lines and Tab.~\ref{table:2} contains the corresponding observing setups\footnote{APEX telescope efficiencies are taken from \footnotesize{\url{http://www.apex-telescope.org/telescope/efficiency/index.php}}, and IRAM telescope efficiencies are taken from \footnotesize{\url{https://www.iram-institute.org/medias/uploads/eb2013-v8.2.pdf}}.}.

The central position of all observations was $\alpha_{[\rm{J}2000]}$=$6^{\mathrm{h}}16^{\mathrm{m}}37.5^{\mathrm{s}}$, $\delta_{[\rm{J}2000]}$=$+22^\circ35'00^{\prime \prime}$. The off-position used was $\alpha_{[\rm{J}2000]}$=$6^{\mathrm{h}}17^{\mathrm{m}}35.8^{\mathrm{s}}$, $\delta_{[\rm{J}2000]}$=$+22^\circ33'08^{\prime \prime}$, in the inner region of the SNR. We checked the focus during the observing session on the stars IK Tau and R Dor. We checked line and continuum pointing every hour locally on V370 Aur, Y Tau, IK Tau and R Dor. The pointing accuracy was better than $\sim$3$^{\prime \prime}$ rms, regardless of which receiver we used. The absolute flux density scale was also calibrated on these sources. The absolute flux calibration uncertainty was estimated to be $\approx 15 \%$ during our observations. We used the GILDAS\footnote{The Grenoble Image and Line Data Analysis Software is developed and maintained by IRAM to reduce and analyze data obtained with the 30m telescope and Plateau de Bure interferometer. See  \url{www.iram.fr/IRAMFR/GILDAS}} package to calibrate and merge the data of all sub-fields to produce a mosaic, and extract the spectral bands containing the signal corresponding to the (2-1) rotational transitions of $^{12}$CO, $^{13}$CO, C$^{18}$O, and the (3-2) rotational transitions of $^{12}$CO and $^{13}$CO. The reduction included 1$^\mathrm{st}$ order baseline subtraction, spatial, and spectral regridding. The final products of this reduction process are spectral cubes centered on the previously cited rotational lines with a resolution of $0.5~\mathrm{km ~ s^{-1}}$ to increase the signal-to-noise ratio, which is more than enough for the expected linewidths (nominal spectral resolutions are indicated in Tab. \ref{table:2}). The observed area is showed in the white box in Fig. \ref{fig:finderchart}.

\subsubsection{IRAM-30m}
The same mosaic of the IC443G extended region was carried out with the IRAM-30m\footnote{This work is based on observations out under project number 169-18 with the IRAM-30m telescope. IRAM is supported by INSU/CNRS (France), MPG (Germany) and IGN (Spain).}. IRAM observations towards IC443G were conducted during one week, from February 20, 2019 to February 24, 2019. The heterodyne receiver EMIR (Eight MIxer Receiver, \citealt{EMIR12}), operating at $115~\mathrm{GHz}$ and $230~\mathrm{GHz}$ simultaneously, was used in combination with the FTS200 and VESPA backends. This setup allowed to cover the same total field of $10^\prime \times 10^\prime$ towards IC443. The observations were performed in position-switching/on-the-fly mode. Tab.~\ref{table:2} contains the corresponding observing set-ups.

The central position of all observations was the same as the one used for APEX observations, $\alpha_{[\rm{J}2000]}$=$6^{\mathrm{h}}16^{\mathrm{m}}37.5^{\mathrm{s}}$, $\delta_{[\rm{J}2000]}$=$+22^\circ35'00^{\prime \prime}$. The off-position used was $\alpha_{[\rm{J}2000]}$=$6^{\mathrm{h}}17^{\mathrm{m}}54^{\mathrm{s}}$, $\delta_{[\rm{J}2000]}$=$+22^\circ47'40^{\prime \prime}$, in the northeastern ionized region of the SNR. We checked the continuum pointing and focus every hour during the observing sessions on several bright stars. The pointing accuracy was better than $\sim$3$^{\prime \prime}$ rms. The absolute flux calibration uncertainty was estimated to be $\approx 15 \%$. We also used the GILDAS package to calibrate and merge the data of all sub-fields to produce a mosaic, and extract the spectral bands containing the signal corresponding to the  (1-0) rotational transitions of $\mathrm{^{12}}$CO, $\mathrm{^{13}}$CO, C$^{18}$O and (2-1) rotational transition of $^{12}$CO. The reduction process and final products are identical to what was presented in the previous subsection. The observed area is showed in the white box in Fig. \ref{fig:finderchart}.

   \begin{figure}
   \centering
   \includegraphics[width=0.9\hsize, trim={0.4cm 0.6cm 1.5cm 1.5cm},clip]{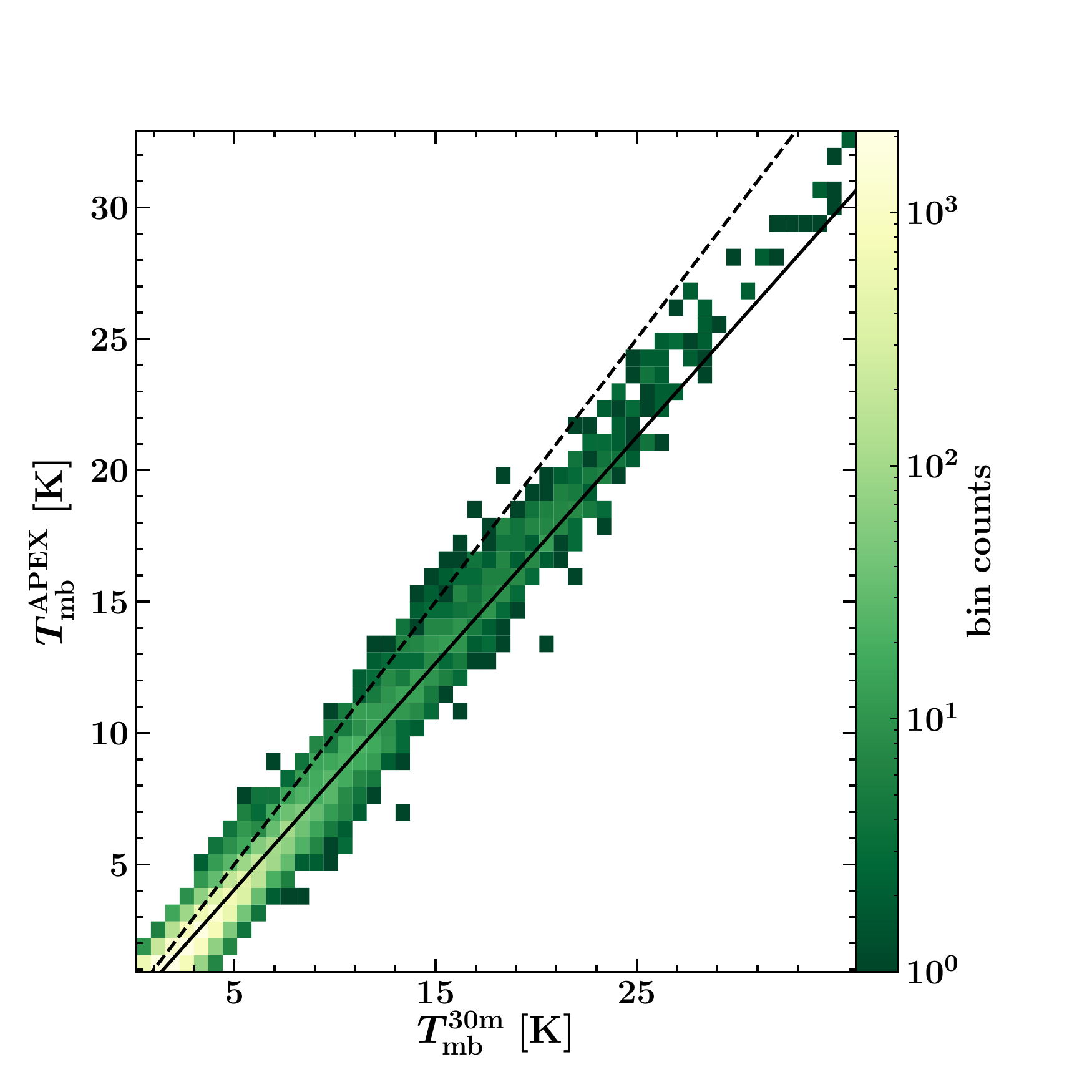}
      \caption{$T^\mathrm{APEX}_\mathrm{mb}$ vs $T_\mathrm{mb}^\mathrm{30m}$ 2D histogram representing the complete comparison between the data cubes obtained with the IRAM-30m and APEX telescopes observations of the transition $^{12}\mathrm{CO}(2-1)$ in the extended G region. The two cubes were resampled to the same spatial and spectral resolution. Every spectral channel in every single pixel is compared and displayed in the histogram. The dashed black line represents the 1:1 relation expected between the two spectral cubes if the data was identical. The solid black line represents represents the empirical relation that is measured by determining the best linear fit corresponding to the data dispersion, using a treshold of 1.5 K in order to best describe the high signal-to-noise bins.}
         \label{fig:hist_TT}
   \end{figure}

\subsubsection{Comparison between the IRAM-30m and APEX telescope data cubes}\label{systematic_errors}
\begin{figure*}[]
\resizebox{\hsize}{!}
{\includegraphics[trim={0cm 0.75cm 0cm 1.5cm},clip]{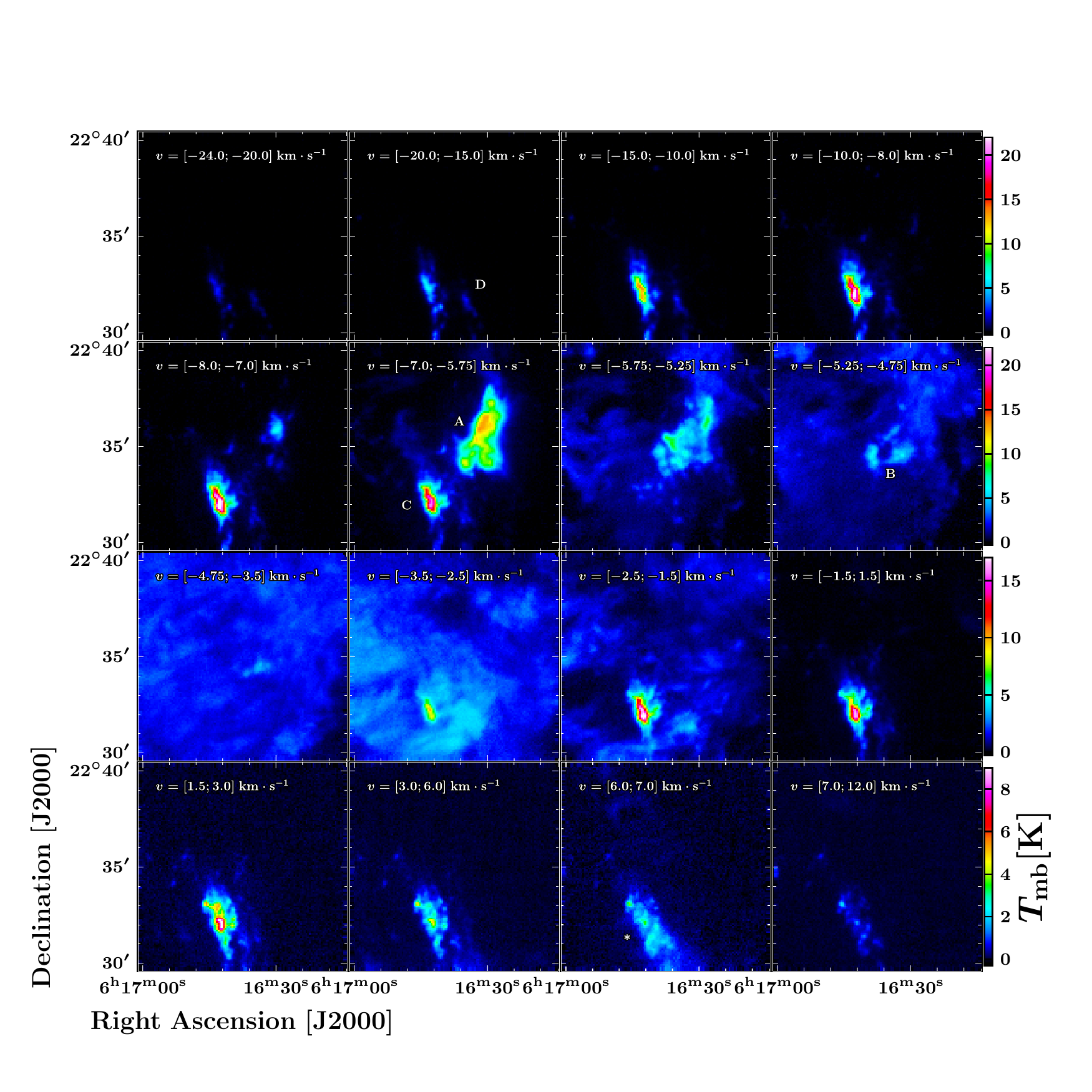}}
\caption{Channel maps of the $^{12}\mathrm{CO}(2-1)$ observations carried out with the IRAM-30m telescope. Each panel represents the emission integrated over an interval of velocity along the line of sight. Velocity intervals are indicated on the top left corner of each panel. Velocity channels represented in this figure are between $\varv=-24~\mathrm{km ~ s^{-1}}$ and $\varv=+12~\mathrm{km ~ s^{-1}}$, covering all the spectral features detected towards the extended G region. Structures described in section \ref{morphology} are indicated with the corresponding letters.}
\label{fig:channelmap}
\end{figure*}

We estimated the systematic errors between the IRAM-30m and APEX based on the data cubes corresponding to the observation of the rotational transition $^{12}\mathrm{CO}(2-1)$ by both telescopes. For visual comparison of the two data cubes, the channel maps are given in Fig. \ref{fig:channelmap} (IRAM-30m) and Fig. \ref{fig:channelmap12co21apex} (APEX). We found no evidence of systematic pointing error between the two data cubes. We also performed a quantitative comparison of the two spectral cubes. First, we resampled the data cubes in order to get the same spatial and spectral resolution. The spatial resolution was set to the nominal resolution of APEX $\theta$ = 28.7$^{\prime \prime}$, and the spectral resolution was set to the nominal resolution of the IRAM-30M $\Delta \varv$ = 0.5 km s$^{-1}$. Then, in each frequency channel and every single pixel of the mosaic we compared the signal detected by the two telescopes where the signal is greater than 3$\sigma$. The results of this complete investigation are represented on a $T^\mathrm{APEX}_\mathrm{mb}$ vs $T^\mathrm{30m}_\mathrm{mb}$ 2D histogram shown in Fig. \ref{fig:hist_TT}. We determined the best linear fit $x \mapsto a \cdot x + b$ corresponding to the data dispersion between the two telescopes using a treshold of 5$\sigma$ in order to best describe the high signal-to-noise data points. The parameters given by the $\chi^2$-minimization are $a$ = 0.88; $b$ = -0.35, indicating a slight overestimate in the measurement of the flux by the IRAM-30m with respect to APEX measurements, at least in the scope of our observations. Our statistical analysis of the two data cubes shows that:
\begin{enumerate}
\item Our measurements are characterized by a systematic error of approximately 12\%. This could be due to inaccurate correction of telescope efficiencies and/or absolute flux calibration. 
\item According to the dispersion around the instrumental linear model, our measurements are also affected by a random error characterized by a standard deviation of 305 mK.
\end{enumerate}

   \begin{figure*}[]
   \centering
   \includegraphics[width=0.5\hsize, trim={1.5cm 2.5cm 3cm 4.25cm},clip]{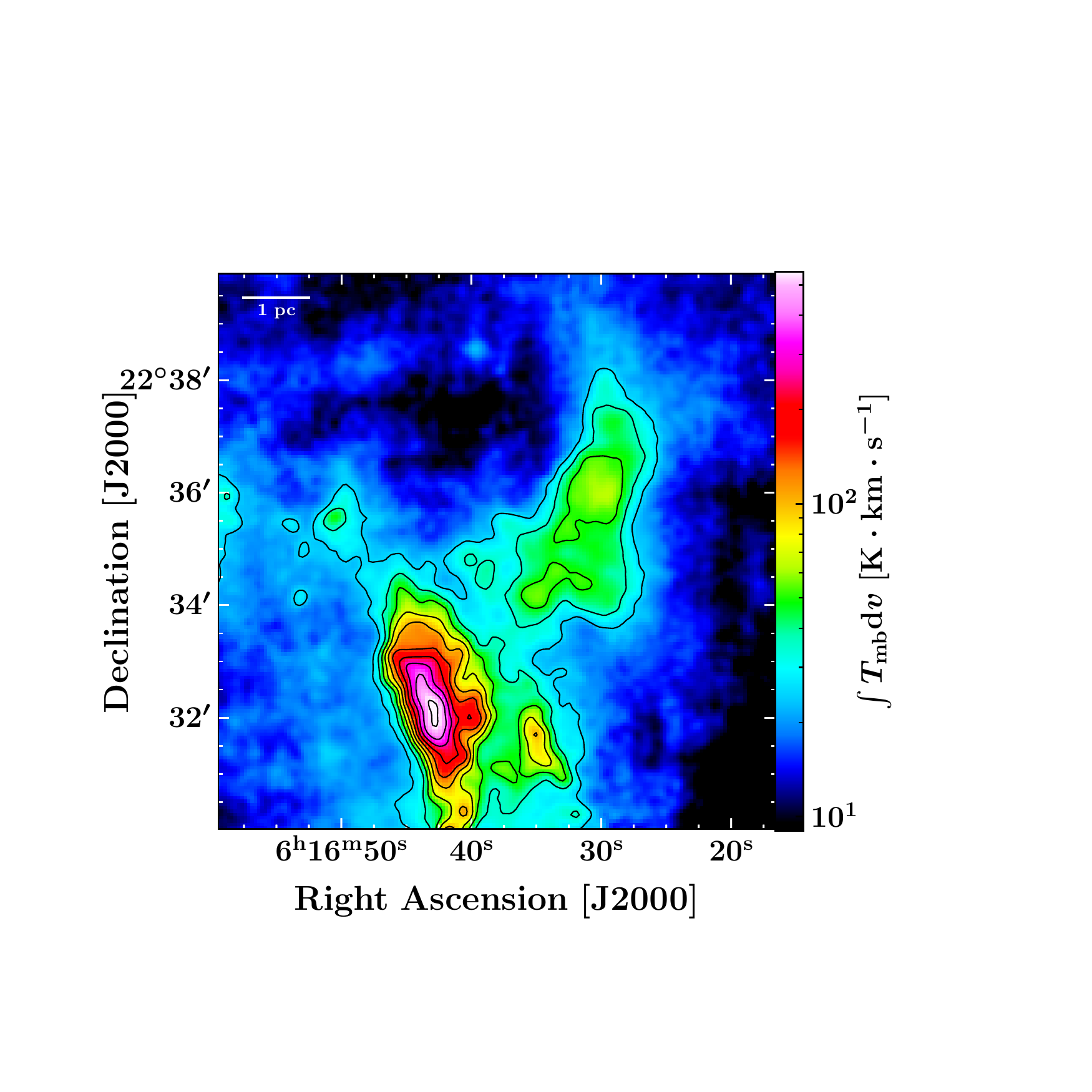}
      \includegraphics[width=0.42\hsize, trim={1.5cm 1.75cm 3.5cm 3.5cm},clip]{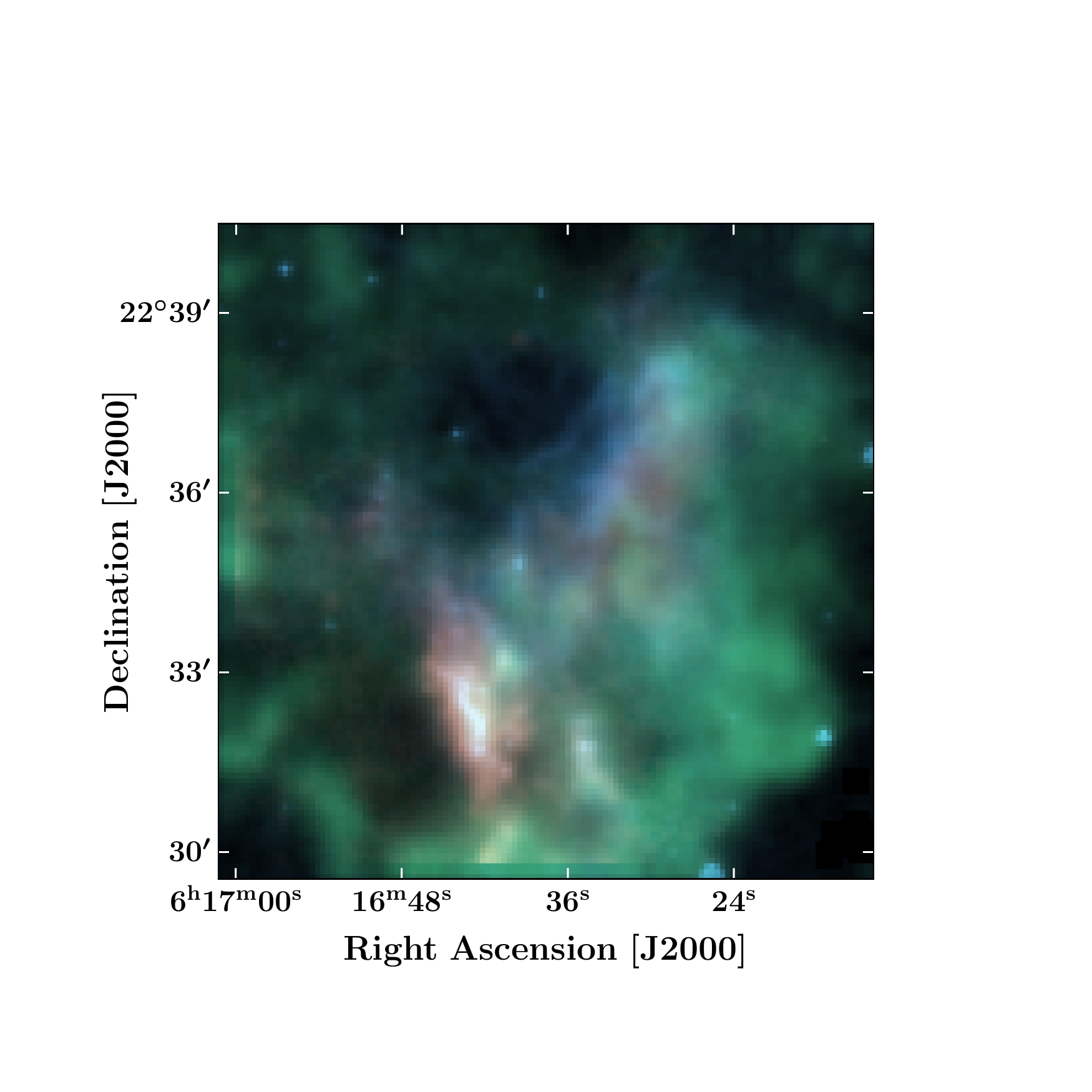}
      \caption{\textit{Left:} 0$^{\mathrm{th}}$ moment of the $\mathrm{^{12}CO}$(2--1) observations carried out with the IRAM-30m over the extended G region. This map corresponds to the signal integrated in the velocity interval [-40; +30] $\mathrm{km~s^{-1}}$. The colorscale used to represent data is logarithmic in order to enhance dynamic range and emphasize the fainter molecular cloudlet. \textit{Right:} composite image of our field of observations in the extended G region, using IRAM-30m data cube, as well as \text{Spitzer}-MIPS data. $^{12}$CO(2--1) ([-40, +30] km s$^{-1}$ as on the left panel) is coded in red, MIPS-24$\upmu$m in blue and $^{13}$CO(1--0) [-4.0, -2.5] km s$^{-1}$, corresponding to the emission spatially and kinematically associated with the ambient cloud described by \citealt{Lee2012}) in green. Colorscale levels are based on the minimum and maximum value of each map.}
         \label{fig:moment0}
   \end{figure*}

\subsection{Morphology of the region}\label{morphology}
Our 10$^\prime \times$10$^\prime$, $\sim$10-30$^{\prime \prime}$ resolution maps of $^{12}$CO and $^{13}$CO in the extended G region provide a detailed picture of the morphology of molecular clumps. Fig. \ref{fig:moment0} (left panel) shows the emission in $^{12}$CO(2--1) integrated between $\varv=-40.0~\mathrm{km ~ s^{-1}}$ and $\varv=+30~\mathrm{km ~ s^{-1}}$ mapped with the IRAM-30m. This wide interval of velocity includes all the components of the signal that are detected within the bandpass of our observations towards the extended G region. This map reveals a structured region with two main molecular structures that are spatially separated. The first structure (bottom-center in Fig. \ref{fig:moment0}) has a peak integrated intensity of 578 $\mathrm{K ~ km ~ s^{-1}}$, a magnitude higher than the peak integrated intensity of the second structure (top-right in Fig. \ref{fig:moment0}) that is around 63 $\mathrm{K ~ km  ~ s^{-1}}$.

Fig. \ref{fig:channelmap} shows the channel maps corresponding to the IRAM-30m observations of the $^{12}$CO J=2--1 transition, which gives the finest spatial resolution of all our observations: $\theta$=11.2$^{\prime \prime}$, or $\sim$0.1pc at the adopted distance of 1.9 kpc. The channel maps corresponding to other transitions of $^{12}$CO and its isotopologue $^{13}$CO are also available in the appendix: 
\begin{itemize}
\item $^{12}$CO(1--0) mapped with the IRAM-30m (Fig. \ref{fig:channelmap12co10})
\item $^{13}$CO(1--0) mapped with the IRAM-30m (Fig. \ref{fig:channelmap13co10}).
\item $^{12}$CO(2--1) mapped with APEX (Fig. \ref{fig:channelmap12co21apex})
\item $^{12}$CO(3--2) mapped with APEX (Fig. \ref{fig:channelmap12co32}) 
\end{itemize}
Several faint and sparse knots are detected over all the field of observations, especially around the systemic velocity of IC443 $\varv_{\mathrm{LSR}}= -4.5~\mathrm{km ~ s^{-1}}$ \citep{Hewitt06}. These structures, noticeable between $\varv=-7.5~\mathrm{km ~ s^{-1}}$ and $\varv=-1.5~\mathrm{km ~ s^{-1}}$ might either be corresponding to a slice of turbulent medium driven by the SN shockwave and/or belonging to the ambient gas associated with the NW-SE molecular cloud in which IC443G is embedded \citep{Lee2012}. Other than that, the description of the region probed by our observations can be divided into six distinct structures:
\begin{enumerate}
\item \textit{Cloudlet.} In the upper part of the field we observe a large ($\sim$5'$\times$2', i.e. $\sim$2.8$\times$1.1 pc) elongated cloudlet detected between $\varv=-7.0~\mathrm{km ~ s^{-1}}$ and $\varv=-5.5~\mathrm{km ~ s^{-1}}$ (indicated by the letter \lq A' on Fig. \ref{fig:channelmap}), which is also detected in $^{12}$CO(3--2). This structure was labelled G1 by \citet{Zhang2010}, as part of the double peaked morphology of the extended G region. The $^{13}$CO J=1--0 counterpart of this structure is much brighter than the other main structures in the field, and it is also detected in the transitions J=2--1 and J=3--2, as well as in C$^{18}$O J=1--0 and J=2--1. This structure was also presented and characterized by \citet{Lee2012} who proposed the label SC 03, among a total of 12 SCs (of size $\sim$1') found in IC443.
\item \textit{Ring-like structure.} A ring-like structure seemingly lying in the center of the field (indicated by the letter \lq B' on Fig. \ref{fig:channelmap}), appearing between $\varv=-5.5~\mathrm{km ~ s^{-1}}$ and $\varv=-4.5~\mathrm{km ~ s^{-1}}$ and also detected in $^{12}$CO(3--2). It has a semi-major axis of 1.5', or 0.8 pc. This structure might be spurious and is likely to be physically connected to the elongated cloudlet as both are spatially contiguous and their emission lines are spectrally close. It is partially detected as well in our observations of $^{13}$CO J=1--0, J=2--1 and J=3--2, and also has a faint, partial counterpart in C$^{18}$O J=1--0 and J=2--1. To understand the nature of this region we searched for counterparts in \textit{Spitzer}-MIPS, WISE, DSS, XMM-Newton, as well as in near-infrared and optical point source catalogues (Sect. \ref{stars}), without success. Due to projection effects, this apparent circular shape could also be explained by an unresolved and clumpy distribution of gas.
\item \textit{Shocked clump.} In the lower part of the field we identify a very bright clump emitting between $\varv=-31.0~\mathrm{km ~ s^{-1}}$ and $\varv=16~\mathrm{km ~ s^{-1}}$. This structure of size $\sim$2'$\times$0.75' ($\sim$1.1$\times$0.4 pc), which is detected in the $^{12}$CO(3--2) transition as well, belongs to the southwestern ridge of the molecular shell of the SNR and has been described as a shocked molecular structure by several studies (\citealt{Vandishoeck93}, \citealt{Cesarsky1999}, \citealt{Snell2005}, \citealt{Shinn2011}, \citealt{Zhang2010}). The core of the shocked clump (indicated by the letter \lq C' on Fig. \ref{fig:channelmap}) is also detected in $^{13}$CO in the transitions J=1--0, J=2--1 and J=3--2, as well as in C$^{18}$O J=1--0 and J=2--1. 
\item \textit{Shocked knot.} An additional shocked knot (indicated by the letter \lq D' on Fig. \ref{fig:channelmap}) is also detected to the west of the previously described structure. This fainter and smaller structure is spatially separated from the main shocked clump.
\item At the same position as the shocked clump and extending southward and westward, we find a faint, elongated clump emitting between $\varv=5.0~\mathrm{km~s^{-1}}$ and $\varv=7.5~\mathrm{km~s^{-1}}$. This structure (indicated by the symbol \lq *' on Fig. \ref{fig:channelmap}) is spatially coinciding with the shocked clump , yet the peak velocity is not the exact same (also see below developments on kinematics of the region in Sect. \ref{sect:kinematics}). It has a faint counterpart in $^{13}$CO(1--0). Observations of the ambient molecular cloud by \citet{Lee2012} indicate this structure as part of a faint NE-SW complex of molecular gas in the velocity range +3 km s$^{-1}$ < $\varv_{\mathrm{LSR}}$ < +10 km s$^{-1}$.
\item Finally, the $^{13}\mathrm{CO}$(1--0) map (Fig. \ref{fig:moment0}, right panel and Fig. \ref{fig:channelmap13co10}) indicates a large clump of gas extending from the bottom-center to the right end of the field, with a bright knot in the bottom-right corner of the field. However, this structure has no bright, well-defined counterpart in any of the $^{12}\mathrm{CO}$ transitions maps. It is spatially and kinematically correlated with the faint and diffuse $^{12}$CO J=1--0 and J=2--1 emission seen in the velocity range -5.5 km s$^{-1}$ < $\varv_{\mathrm{LSR}}$ < -2 km s$^{-1}$. From the comparison with the $^{12}$CO observations of \citet{Lee2012} and $^{13}$CO observations of \citet{Su2014} towards the SNR, we conclude that this structure is part of the western molecular complex observed in the velocity range -10 km s$^{-1}$ < $\varv_{\mathrm{LSR}}$ < 0 km s$^{-1}$.
\end{enumerate}

\begin{figure*}
\begin{center}
    \includegraphics[width=\textwidth, trim={0cm 8.5cm 0.25cm 0cm},clip]{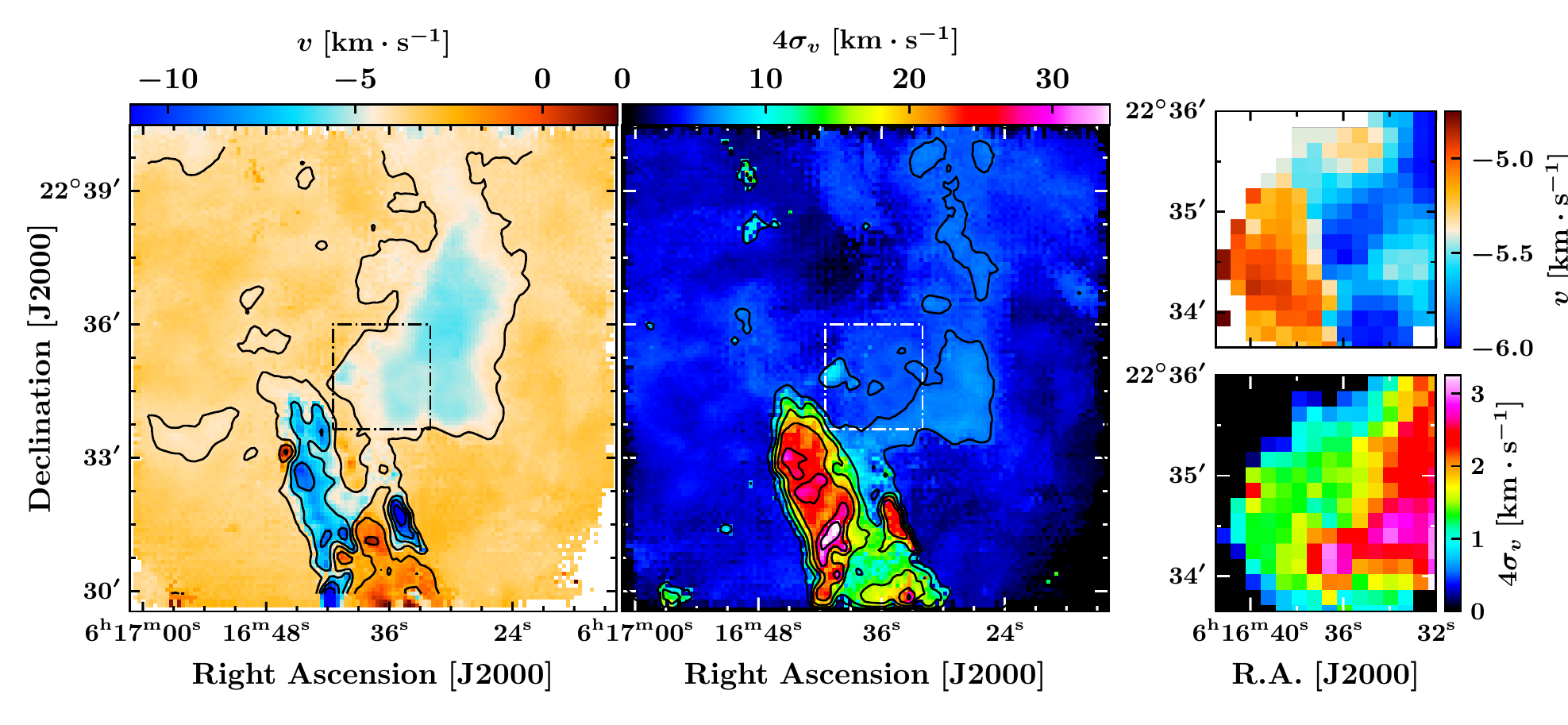}
    \caption{Left: first moment map of the IRAM-30m $^{12}\mathrm{CO}(2-1)$ data cube. Center: second moment map of the same data cube. Right, upper panel: zoom into the dashed box on the first moment map of the APEX $^{12}\mathrm{CO}(3-2)$ data cube to enhance the spectral resolution. Right top and bottom pannels display the first and second moment maps of the APEX $^{12}$CO(3--2) data cube into the dashed box shown in the right and middle pannels, respectively. The colorbar of the left figure is centered on the velocity of IC443 in the local standard of rest, $\varv_{\mathrm{LSR}}=-4.5~\mathrm{km ~ s^{-1}}$).
    } \label{fig:moments}
\end{center}
\end{figure*}

\begin{table*}
\caption{Summary of the spectral characteristics of the lines of $^{12}$CO and $^{13}$CO within the boxes corresponding to each structures, defined in Fig. \ref{fig:boxes}. $T^*_\mathrm{mb}$ is the peak temperature of the line, $\varv_0$ is the centroid and v$_\mathrm{FWHM}$ the linewidth. $T^*_\mathrm{mb}$ and the area are directly measured from reading the spectral data. v$_\mathrm{FWHM}$  and $\varv_0$ are measured by modelling the line either by a single or double gaussian function that best fits the data. When two gaussian functions are needed to model the sum of the signal emitted by the ambient cloud and structure of interest, we discard the contribution of the ambient cloud and only take into account the parameters $\varv_0$ and $\varv_\mathrm{FWHM}$ associated with the broad component. The asymetric high-velocity wings of $^{12}$CO lines corresponding to the shocked clump require two gaussians, thus we measure $\varv_0=(c_1 + c_2)/2$ and $\varv_{\mathrm{FWHM}}=\sqrt{((2.355 ~ \sigma_1)^2 + (2.355 ~ \sigma_2)^2)}$ where $c_1$, $c_2$, $\sigma_1$ and $\sigma_2$ are the centroids and standard deviations corresponding to each gaussian function, respectively.}
\label{table:linemeasures}      
\centering                          
\begin{tabular}{l  c  c  c  c c c c c c c c c}        
\hline           
\hline \\[-1.0em] 
& & & \footnotesize{\hspace{1cm}$^{12}$CO} & & & & & \hspace{1cm}\footnotesize{$^{13}$CO} \\\cline{3-6} \cline{8-11} \\[-0.9em] \hline
\footnotesize{Region} & \footnotesize{J$_\mathrm{u}$-J$_\mathrm{l}$} & \footnotesize{$T^*_\mathrm{mb}$} &  \footnotesize{$\varv_0$} & \footnotesize{$\varv_{\mathrm{FWHM}}$} & \footnotesize{Area} & & \footnotesize{$T^*_\mathrm{mb}$} &  \footnotesize{$\varv_0$} & \footnotesize{$\varv_{\mathrm{FWHM}}$} & \footnotesize{Area} \\   \footnotesize{} 
& \footnotesize{} & \footnotesize{(K)} & \footnotesize{(km s$^{-1}$)} & \footnotesize{(km s$^{-1}$)} & \footnotesize{(K km s$^{-1}$)} & & \footnotesize{(K)} & \footnotesize{(km s$^{-1}$)} & \footnotesize{(km s$^{-1}$)} & \footnotesize{(K km s$^{-1}$)}\\
\hline \hline \\[-1.0em] 
\footnotesize{cloudlet} & \footnotesize{1--0} & \footnotesize{$6.1\pm 0.1$} & \footnotesize{$-5.8\pm0.2$} & \footnotesize{$2.1\pm0.3$} & \footnotesize{$15.5\pm0.3$} & & \footnotesize{$1.7\pm 0.1$} & \footnotesize{$-5.6\pm0.2$} & \footnotesize{$2.0\pm0.2$} & \footnotesize{$4.1\pm0.1$} \\
\footnotesize{$\alpha=6^\mathrm{h}16^\mathrm{m}30^\mathrm{s}$} & \footnotesize{2--1} & \footnotesize{$4.7\pm0.1$} & \footnotesize{$-5.9\pm0.3$} & \footnotesize{$1.9\pm0.3$} & \footnotesize{$9.7\pm0.2$} & & \footnotesize{$1.1\pm0.1$} & \footnotesize{$-5.7\pm0.3$} & \footnotesize{$1.8\pm0.2$} & \footnotesize{$1.7\pm0.3$} \\
\footnotesize{$\delta=22^{\circ}36'00^{\prime \prime}$} & \footnotesize{3--2} & \footnotesize{$3.5\pm0.1$} & \footnotesize{$-6.1\pm0.2$} & \footnotesize{$1.6\pm0.2$} & \footnotesize{$6.9\pm0.2$} & & \footnotesize{$0.8\pm0.1$} & \footnotesize{$-5.8\pm0.3$} & \footnotesize{$2.2\pm0.3$} & \footnotesize{$1.2\pm0.1$} \\
\hline \hline \\[-1.0em] 
\footnotesize{ring} & \footnotesize{1--0} & \footnotesize{$11.7\pm0.2$} & \footnotesize{$-6.0\pm0.2$} & \footnotesize{$1.6\pm0.2$} & \footnotesize{$20.1\pm0.3$} & & \footnotesize{$3.5\pm0.1$} & \footnotesize{$-5.7\pm0.2$} & \footnotesize{$1.7\pm0.2$} & \footnotesize{$5.8\pm0.2$}\\
\footnotesize{$\alpha=6^\mathrm{h}16^\mathrm{m}36^\mathrm{s}$} & \footnotesize{2--1} & \footnotesize{$9.5\pm0.1$} & \footnotesize{$-6.1\pm0.1$} & \footnotesize{$1.4\pm0.2$} & \footnotesize{$14.3\pm0.2$} & & \footnotesize{$2.8\pm0.1$} & \footnotesize{$-5.5\pm0.2$} & \footnotesize{$1.6\pm0.2$} & \footnotesize{$3.6\pm0.1$}\\
\footnotesize{$\delta=22^{\circ}34'40^{\prime \prime}$} & \footnotesize{3--2} & \footnotesize{$9.5\pm0.1$} & \footnotesize{$-5.7\pm0.3$} & \footnotesize{$1.9\pm0.2$} & \footnotesize{$15.0\pm0.3$} & & \footnotesize{$1.8\pm0.1$} & \footnotesize{$-5.5\pm0.3$} & \footnotesize{$1.7\pm0.3$} & \footnotesize{$2.5\pm0.1$} \\
\hline \hline \\[-1.0em] 
\footnotesize{shocked clump} & \footnotesize{1--0} & \footnotesize{$7.1\pm0.1$} & \footnotesize{$-6\pm1$} & \footnotesize{$19\pm1$} & \footnotesize{$49.7\pm0.4$} & & \footnotesize{$2.5\pm0.1$} & \footnotesize{$-3.8\pm0.1$} & \footnotesize{$2.1\pm0.1$} & \footnotesize{$6.5\pm0.1$}  \\
\footnotesize{$\alpha=6^\mathrm{h}16^\mathrm{m}42^\mathrm{s}$} & \footnotesize{2--1} & \footnotesize{$5.2\pm0.1$} & \footnotesize{$-7\pm1$} & \footnotesize{$23\pm1$} & \footnotesize{$71\pm1$} & & \footnotesize{$0.8\pm0.1$} & \footnotesize{$-3.8\pm0.2$} & \footnotesize{$2.4\pm0.2$} & \footnotesize{$3.5\pm0.1$} \\
\footnotesize{$\delta=22^{\circ}32'00^{\prime \prime}$} & \footnotesize{3--2} & \footnotesize{$4.9\pm0.1$} & \footnotesize{$-7\pm1$} & \footnotesize{$24\pm1$} & \footnotesize{$80\pm1$} & & \footnotesize{$0.7\pm0.1$} & \footnotesize{$-4.4\pm0.2$} & \footnotesize{$1.4\pm0.2$} & \footnotesize{$4.5\pm0.2$} \\
\hline \hline \\[-1.0em] 
\footnotesize{shocked knot} & \footnotesize{1--0} & \footnotesize{$8.8\pm0.1$} & \footnotesize{$-8\pm2$} & \footnotesize{$17\pm1$} & \footnotesize{$18.8\pm0.2$} & & \footnotesize{$2.4\pm0.1$} & \footnotesize{$-3.7\pm0.1$} & \footnotesize{$2.4\pm0.1$} & \footnotesize{$7.5\pm0.1$}  \\
\footnotesize{$\alpha=6^\mathrm{h}16^m{m}42^\mathrm{s}$} & \footnotesize{2--1} & \footnotesize{$6.0\pm0.1$} & \footnotesize{$-9\pm2$} & \footnotesize{$21\pm1$} & \footnotesize{$21.1\pm0.3$} & & \footnotesize{$0.7\pm0.1$} & \footnotesize{$-3.5\pm0.2$} & \footnotesize{$2.6\pm0.3$} & \footnotesize{$3.2\pm0.1$} \\
\footnotesize{$\delta=22^{\circ}32'00^{\prime \prime}$} & \footnotesize{3--2} & \footnotesize{$3.4\pm0.1$} & \footnotesize{$-9\pm2$} & \footnotesize{$22\pm1$} & \footnotesize{$24.4\pm0.4$} & & \footnotesize{$0.3\pm0.1$} & \footnotesize{$-3.9\pm0.3$} & \footnotesize{$1.9\pm0.3$} & \footnotesize{$1.3\pm0.5$} \\
\hline
\end{tabular}
\end{table*}

\subsection{Kinematics of the region}\label{sect:kinematics}
Using the nominal spectral resolution of 0.5 km s$^{-1}$ attained with our IRAM-30m $^{12}$CO J=2--1 observations, we identified several velocity components of the molecular gas in IC443G based on the determination of the first and second moment maps (Fig. \ref{fig:moments}, left and center) of the $^{12}$CO(2--1) data cube. We also produced the moments maps of the $^{12}$CO(3--2) data cube towards the ring-like structure (Fig. \ref{fig:moments}, right) to profit from the spectral resolution of 0.1 km s$^{-1}$.

\begin{enumerate}
\item The cloudlet has a mean velocity of about $\varv_{\mathrm{LSR}}=-5.5~\mathrm{km ~ s^{-1}}$ that is remarkably uniform throughout the structure. We measured $\varv_{\mathrm{LSR}}=-5.7\pm0.3~\mathrm{km ~ s^{-1}}$ from the centroid of the $^{13}$CO lines, contrasting with the velocity of IC443 in the local standard of rest by more than 1 km s$^{-1}$. It is likely that this discrepancy is due to a distinct velocity component with respect to the rest of the molecular gas in the extended G region. If this is not the case, this velocity shift could correspond to a maximum displacement of the kinematic distance $\Delta d\approx 300$ pc (following \citealt{Wenger2018}). Yet, the velocity wings of the $^{12}$CO lines are still within the velocity range of the maser source in IC443G. The second moment map reveals a much lower velocity dispersion within the cloudet than for the shocked clump. It varies between $\sim$5 $\mathrm{km ~ s^{-1}}$ and $\sim$7 $\mathrm{km ~ s^{-1}}$, which is slightly higher than the velocity dispersion across the background field, around $\sim$ 4 km s$^{-1}$.
\item The apparent ring-like structure is further analysed in the two right panels of the Fig. \ref{fig:moments} where the first and second moment maps are determined for the $^{12}\mathrm{CO}$(3--2) data cube obtained with APEX. The superior spectral resolution of the APEX data cube offers a better precision in the determination of the moment maps, at the cost of a lower spatial resolution. In the first moment map the mean velocity gradient within the ring is suggesting that the structure is rotating or expanding isotropically, as the mean velocity field varies between $\varv_{\mathrm{LSR}}=-4.7~\mathrm{km ~ s^{-1}}$ and $\varv_{\mathrm{LSR}}=-5.8~\mathrm{km ~ s^{-1}}$ from the western to the eastern arc of the ring. This apparent velocity gradient could be also due to systematic velocity variations between two or more distinct sub-clumps that are not well resolved by our J=3--2 observations ($\theta=19.2^{\prime \prime}$, $\sim$0.2 pc). The velocity dispersion measured within the ring-like structure varies between 1 $\mathrm{km ~ s^{-1}}$ and 3 $\mathrm{km ~ s^{-1}}$, with a positive gradient from the eastern part to the western part of the structure where it spatially connects to the cloudlet.
\item The shocked clump has a mean velocity varying between $\varv_{\mathrm{LSR}}=-6~\mathrm{km ~ s^{-1}}$ and $\varv_{\mathrm{LSR}}=-8.5~\mathrm{km ~ s^{-1}}$ throughout its structure. These mean velocities are subject to caution as the self-absorption and asymmetric wings characterizing the line emissions of $^{12}\mathrm{CO}$  might bias the value of the centroid. In fact, careful measurement of the centroid of $^{13}$CO lines using a single gaussian function favors a velocity centroid of -4.4$\pm$0.2 km s$^{-1}$ for the shocked clump, which is consistent with the velocity v$_\mathrm{LSR}= -4.5$ km s$^{-1}$ of the maser in IC443G \citep{Hewitt06} . The second moment map displays important velocity dispersions, spanning from $\sim$15 $\mathrm{km ~ s^{-1}}$ and up to $\sim$36 $\mathrm{km ~ s^{-1}}$ within the shocked gas, increasing towards the center of the clump.
\item The shocked knot has a mean velocity $v=-9~\mathrm{km~s^{-1}}$ that is slightly shifted with respect to the shocked clump. The second moment map shows an uniform velocity dispersion of $\sim25~\mathrm{km~s^{-1}}$, similar to the dispersion measured within the main shocked molecular structure.
\end{enumerate}
The rest of the field of observation has a quasi-uniform mean velocity of $\sim$-4.1 $\mathrm{km ~ s^{-1}}$ to $\sim$-2.8 $\mathrm{km ~ s^{-1}}$, which is slightly different than the mean velocity of IC443G in the local standard of rest but consistent with the ambient NW-SE molecular cloud in which IC443G is embedded \citep{Lee2012}. The velocity dispersion of this ambient gas is spanning from <1 $\mathrm{km ~ s^{-1}}$ to $\sim$10 $\mathrm{km ~ s^{-1}}$ in a few areas where the velocity dispersion is locally enhanced, with an average of $\sim$4 $\mathrm{km ~ s^{-1}}$.
Excluding the contribution of the shocked structures and localized high-velocity dispersion knots, velocity dispersions measured from the $^{12}$CO J=2--1 line in the extended G field span a range of r.m.s velocity $\sigma_v=0.4$ to 1.3 km s$^{-1}$ in the ambient gas, and $\sigma_v=1.2$ to 1.8 km s$^{-1}$ towards the cloudlet. At a temperature of 10 K, the thermal contribution is $\sigma_v=0.32$ km s$^{-1}$ and it is likely that small-scale motions within the complex of molecular gas contribute to the measured dispersion, hence the ambient cloud is mostly quiescent, with turbulent motions smaller than 1 km s$^{-1}$. The velocity dispersion measured towards the cloudlet with the $^{13}$CO lines is $\sigma_v = 0.8\pm0.1$ km s$^{-1}$, which is consistent with typical molecular condensations \citep{Larson1981}. Thus, we dot not find any kinematic signature of interaction of the cloudlet nor the ambient cloud with the SNR shocks in the extended G region, except for the few localized high-velocity dispersion knots.

\section{Column density and mass measurements}\label{mass}

\subsection{Spatial separation of spectrally uniform structures}\label{sect:spatialsep}

   \begin{figure}[]
   \centering
   \includegraphics[width=\hsize, trim={0cm 1cm 1.5cm 2cm},clip]{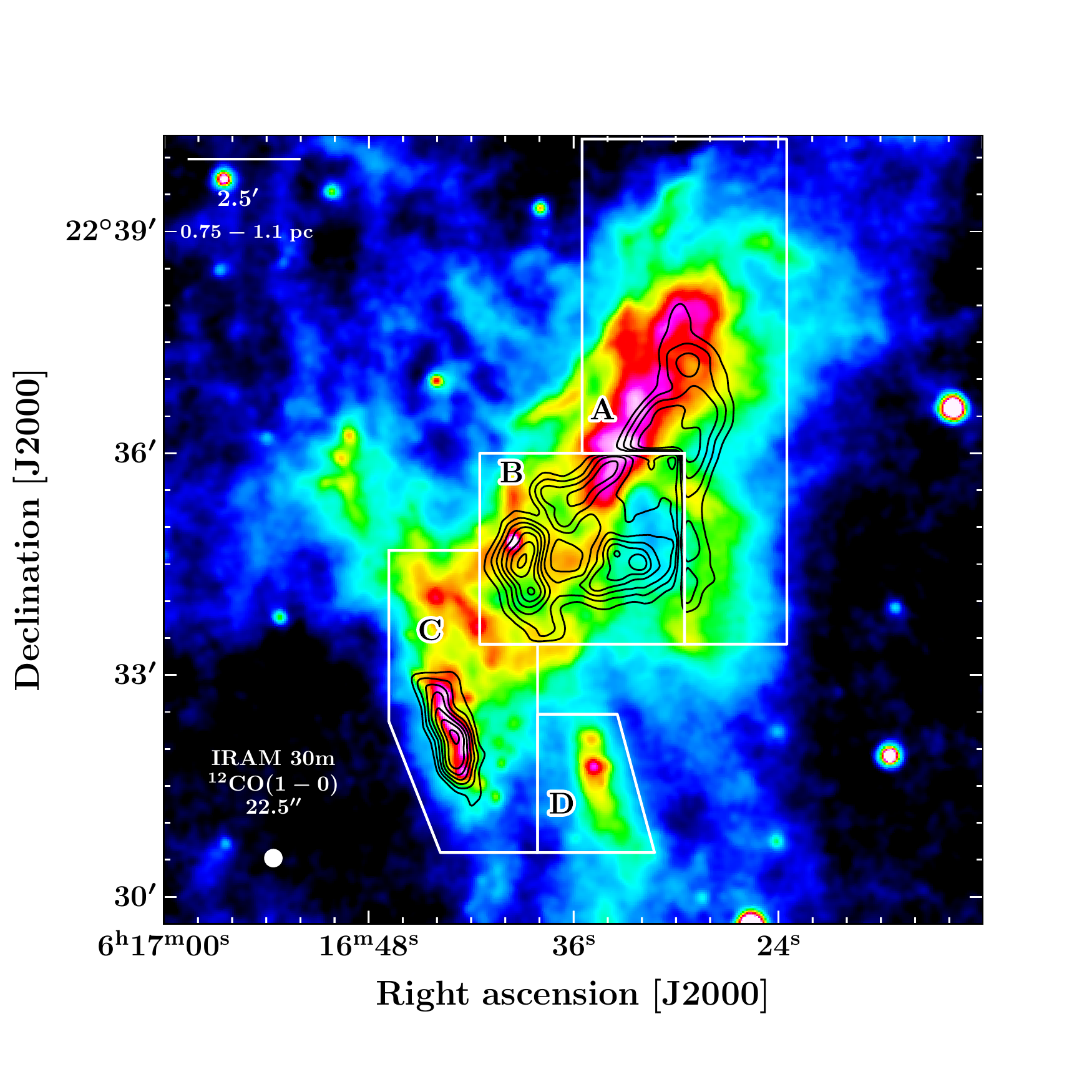}
      \caption{\textit{Spitzer}/MIPS map at 24.0 $\upmu$m. In black contours, the emission of $^{12}$CO(2--1) observed with the IRAM-30m is shown over different intervals of velocities: A. [-7; -5] $\mathrm{km ~ s^{-1}}$ (cloudlet); B. [-5.5; -4.5] $\mathrm{km ~ s^{-1}}$ (ring-like structure); C. and D. [-40; +30] $\mathrm{km ~ s^{-1}}$ (shocked clump and shocked knot). White boxes represent the area where the signal corresponding to each structure is integrated. The beam diameter of the IRAM-30m observations of $^{12}$CO(2--1) is shown in the bottom-left corner.}
         \label{fig:boxes}
   \end{figure}
   
      \begin{figure*}[]
   \centering
   \includegraphics[width=0.7\hsize, trim={0.5cm 2cm 1cm 2.5cm},clip]{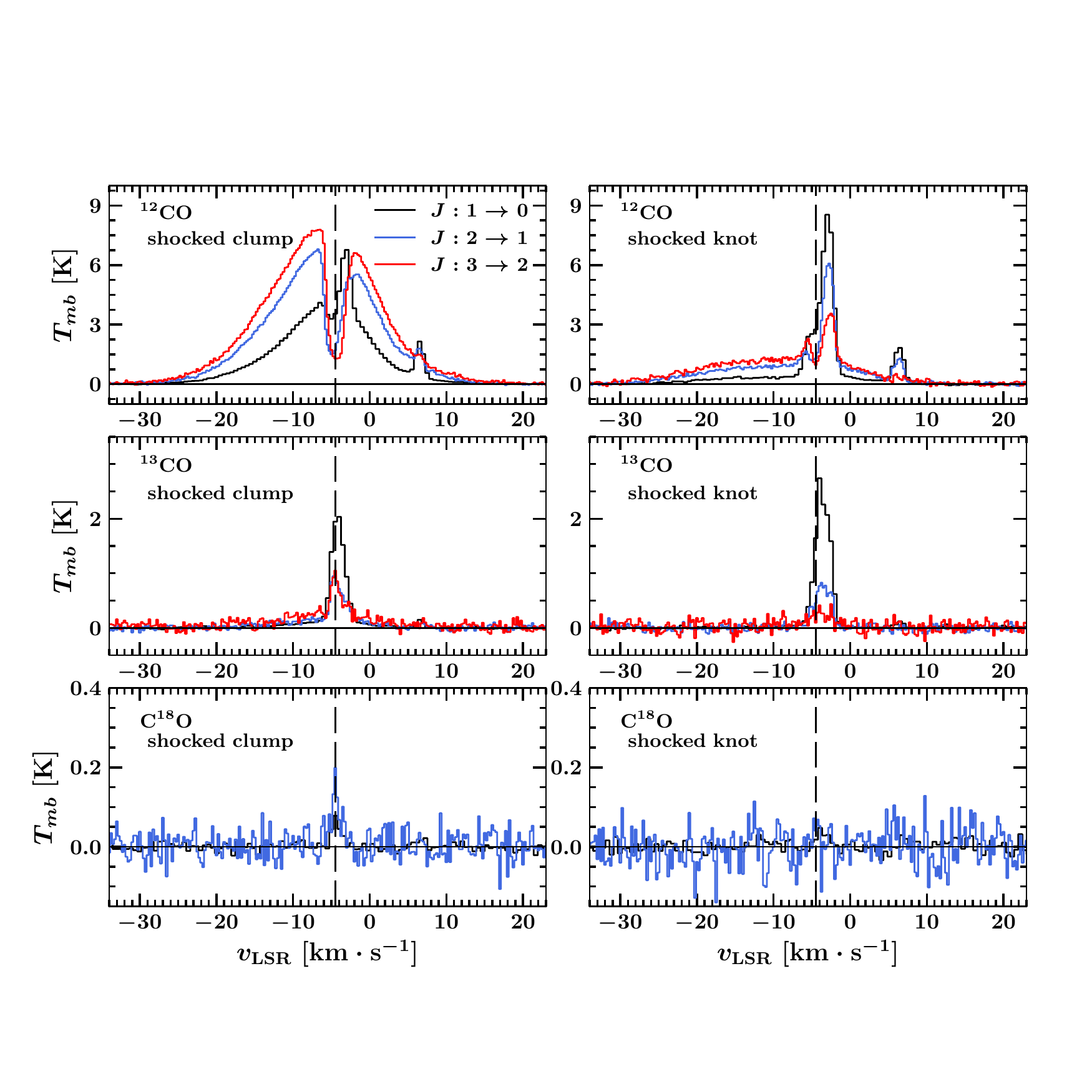}
      \caption{Spectra averaged over the region of the shocked clump (left) and shocked knot (right) defined in Fig. \ref{fig:boxes} for the following lines: $^{12}\mathrm{CO}$(1--0), $^{13}\mathrm{CO}$(1--0) and $\mathrm{C^{18}O}$(1--0) (in black, IRAM-30m); $^{12}\mathrm{CO}$(2--1), $^{13}\mathrm{CO}$(2--1) and $\mathrm{C^{18}O}$(2--1) (in blue, APEX); $^{12}\mathrm{CO}$(3--2) and $^{13}\mathrm{CO}$(3-2) (in gray, APEX). Spectral cubes were resampled to allow direct comparison between the different spectra. Spatial resolutions of all transitions were modified to the nominal resolution of $\mathrm{C^{18}O}$(2--1), $\theta= 30.2^{\prime \prime}$. Spectral resolutions were set to $0.5~\mathrm{km ~ s^{-1}}$ for IRAM-30m data, and $0.25~\mathrm{km ~ s^{-1}}$ for APEX data. On both panels, the $\varv_{\mathrm{LSR}}$ of IC443 is indicated with a vertical dashed line (at $-4.5~\mathrm{km ~ s^{-1}}$).}
         \label{fig:shockspectra}
   \end{figure*}

   \begin{figure*}[]
   \centering
   \includegraphics[width=0.7\hsize, trim={0.5cm 2cm 1cm 2.5cm},clip]{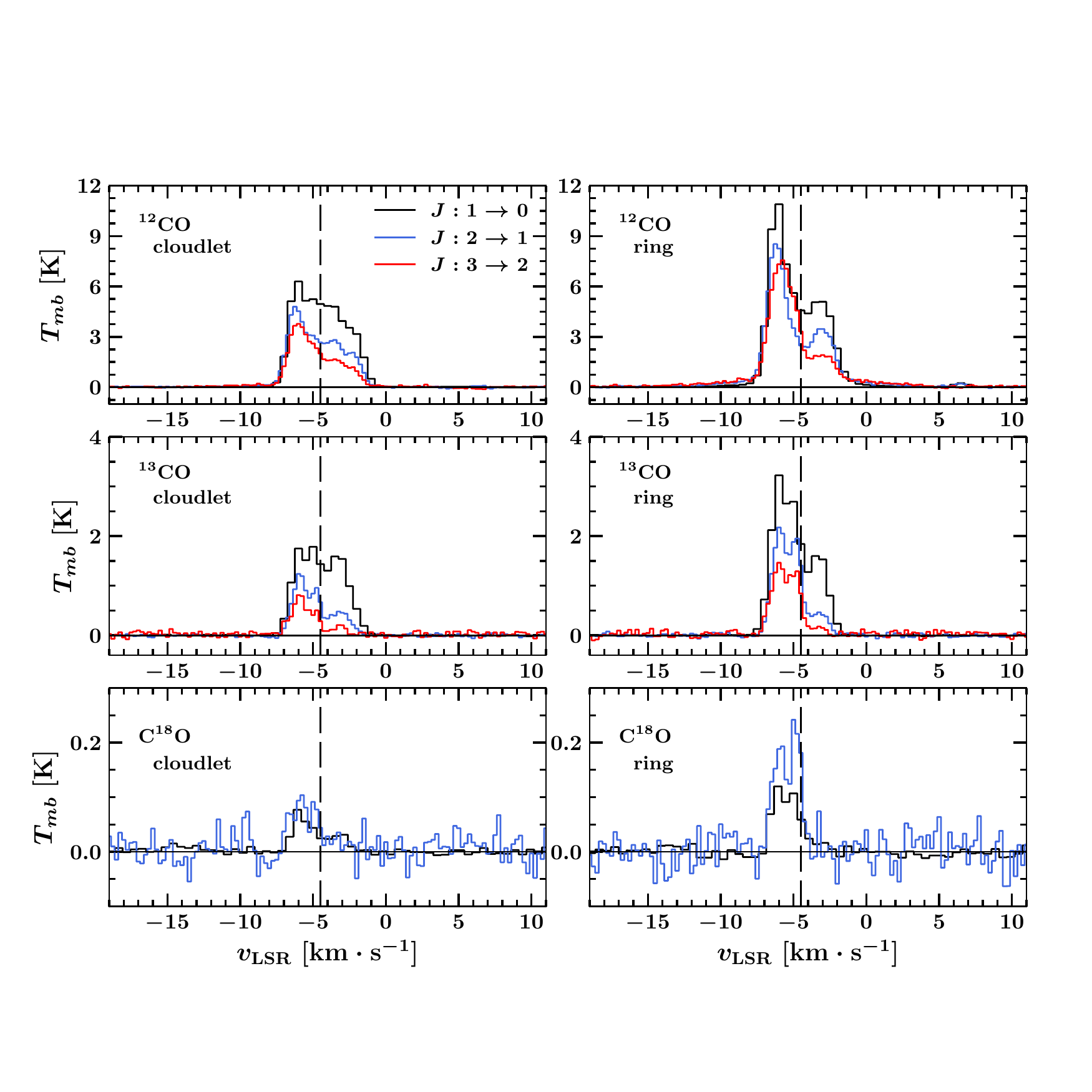}
      \caption{Spectra averaged over the regions of the cloudlet (left panels) and the ring (right panels) defined in Fig. \ref{fig:boxes} for the following lines: $^{12}\mathrm{CO}$(1--0), $^{13}\mathrm{CO}$(1--0) and $\mathrm{C^{18}O}$(1--0) (in black, IRAM-30m); $^{12}\mathrm{CO}$(2--1), $^{13}\mathrm{CO}$(2--1) and $\mathrm{C^{18}O}$(2--1) (in blue, APEX); $^{12}\mathrm{CO}$(3--2) and $^{13}\mathrm{CO}$(3-2) (in gray, APEX). Spectral cubes were resampled to allow direct comparison between the different spectra. Spatial resolutions of all transitions were modified to the nominal resolution of $\mathrm{C^{18}O}$(2--1), $\theta= 30.2^{\prime \prime}$. Spectral resolutions were set to $0.5~\mathrm{km ~ s^{-1}}$ for IRAM-30m data, and $0.25~\mathrm{km ~ s^{-1}}$ for APEX data. On both panels, the $\varv_{\mathrm{LSR}}$ of IC443 is indicated with a vertical dashed line (at $-4.5~\mathrm{km ~ s^{-1}}$).}
         \label{fig:denseringspectra}
   \end{figure*}
   
We aim to measure the mass associated with each molecular structures described in the previous section. We defined spatial boundaries enclosing these structures and independently studied the spectral data corresponding to each sub-region of the field. The spatial boxes defined for the cloudlet (A), ring-like structure (B), shocked clump (C) and shocked knot (D) are shown in Fig. \ref{fig:boxes}, and the average spectra obtained in these boxes are presented in the Fig. \ref{fig:shockspectra} and Fig. \ref{fig:denseringspectra} for every line from CO and its isotopologues that are available in our IRAM-30m and APEX data cubes. The choice of the boundaries is based on our morphological classification, but we carefully checked that the brightest spectral features are coherent across the different boxes that we defined (coordinates of these boxes are given in Tab. \ref{table:boxes}). We performed that selection manually, as the size of our sample is not large enough to apply statistical methods (e.g. clustering, see \citealt{Bron2018}). Based on the analysis of the emission of $^{12}$CO, $^{13}$CO and C$^{18}$O lines, our description of these spectral features is the following:
\begin{enumerate}
\item \textit{Cloudlet.} Towards box A (Fig. \ref{fig:denseringspectra}, left-panel), the line profile of $^{12}\mathrm{CO}$ and $^{13}\mathrm{CO}$ lines are similarly double-peaked, and best modeled by the sum of two gaussian functions centered on the systemic velocities $\varv_\mathrm{LSR}=-5.7\pm0.3$ km s$^{-1}$ (associated with the cloudlet) and $\varv_\mathrm{LSR}=-3.3\pm0.1$ km s$^{-1}$ (associated with the ambient cloud).\\[-1.0em]
\item \textit{Ring-like structure.} Towards box B (Fig. \ref{fig:denseringspectra}, right-panel), the $^{12}\mathrm{CO}$ and $^{13}\mathrm{CO}$ lines are double-peaked as well. The use of two gaussian functions to model the line profile yields the systemic velocities $\varv_\mathrm{LSR}=-5.6\pm0.2$ km s$^{-1}$ (associated with the ring-like structure) and $\varv_\mathrm{LSR}=-3.3\pm0.1$ km s$^{-1}$ (associated with the ambient cloud). The gaussian decomposition is very similar to that of the cloudlet, suggesting that the apparent ring-like structure might be incidental despite its remarkable features in the first moment map (Fig. \ref{fig:moments}).
\item \textit{Shocked clump.} Considering the geometry of the SNR and the locally perpendicular direction of propagation of the SNR shockwave \citep{Vandishoeck93}, the high-velocity emission arises from at least two shock waves, if not a collection of transverse shocks propagating along the molecular shell. In other words, the projection along the line of sight of several distinct shocked knots with distinct systematic velocities could contribute to the broadening of the $^{12}$CO lines. We measure $\varv_\mathrm{s} \simeq 27~\mathrm{km ~ s^{-1}}$ and $\varv_\mathrm{s} \simeq 21~\mathrm{km ~ s^{-1}}$ respectively for the blueshifted and redshifted transverse shocks. Except for the J=1--0 spectrum where the emission of the ambient gas contributes to the average spectra, all spectra of $^{12}\mathrm{CO}$ lines exhibit a significant absorption feature around the $\varv_{\mathrm{LSR}}$ of IC443G, suggesting that there is strong self-absorption of the emission lines. Evidence of line absorption is found in the velocity range -6 km s$^{-1}$ < v$_\mathrm{LSR}$ < -2 km s$^{-1}$, which is where we detect the spatially extended features associated with the NW-SE complex of molecular gas described by \citet{Lee2012}. Hence, it is possible that the foreground cold molecular cloud is at the origin of the absorption of the $^{12}$CO J=1--0 and J=2--1 lines. A faint and thin emission line is detected around $\varv=6.5~\mathrm{km ~ s^{-1}}$ both in the $^{12}\mathrm{CO}$ and $^{13}\mathrm{CO}$ spectra. This signal is associated with the NE-SW complex of molecular gas described in section \ref{morphology}.
\item \textit{Shocked knot.} The shock signature of this line is distinct from the shocked clump. As hinted by the moments map (Fig. \ref{fig:moments}), its fainter high-velocity wings are displaced towards negative velocities. A self-absorption feature is also observed in this structure. Between $v=-5.5~\mathrm{km~s^{-1}}$ and $v=-2~\mathrm{km~s^{-1}}$ a bright and thin feature traces the ambient gas shown on the channel maps on Fig. \ref{fig:channelmap} (second row from bottom; first, second and third panels from left).\\[-1.0em]
\end{enumerate}
The linewidth measured from the $^{13}\mathrm{CO}$ line profiles when we consider only the spectral component that are physically associated with the cloudlet and ring-like structure (discarding the contribution of the ambient cloud) are respectively $2.0\pm0.3~\mathrm{km~s^{-1}}$ and $1.6\pm0.3~\mathrm{km~s^{-1}}$, measured by carefully defining much more constrained spatial boundaries around the structures. From the average spectra presented in this section, there is no spectral evidence for the propagation of molecular shocks and/or outflows within these two structures, except for the faint wings displayed by the $^{12}\mathrm{CO}$ lines in the box associated with the ring-like structure. These extended wings arise from the contamination by the high-velocity emission of the shocked sub-structures that are contained in box B (Fig. \ref{fig:moments}).
We measured the peak temperature, velocity centroid, FWHM and area of the $^{12}$CO and $^{13}$CO lines J=1--0, J=2--1 and J=3--2 in each average spectrum and report our results in Tab. \ref{table:linemeasures}.

\subsection{LTE method}
\subsubsection{2D histograms of CO data}\label{sect:LTEmodels}

\begin{figure*}[]
\begin{center}
\resizebox{0.7\hsize}{!}
{\includegraphics[trim={0.5cm 3.5cm 1.2cm 4.5cm},clip]{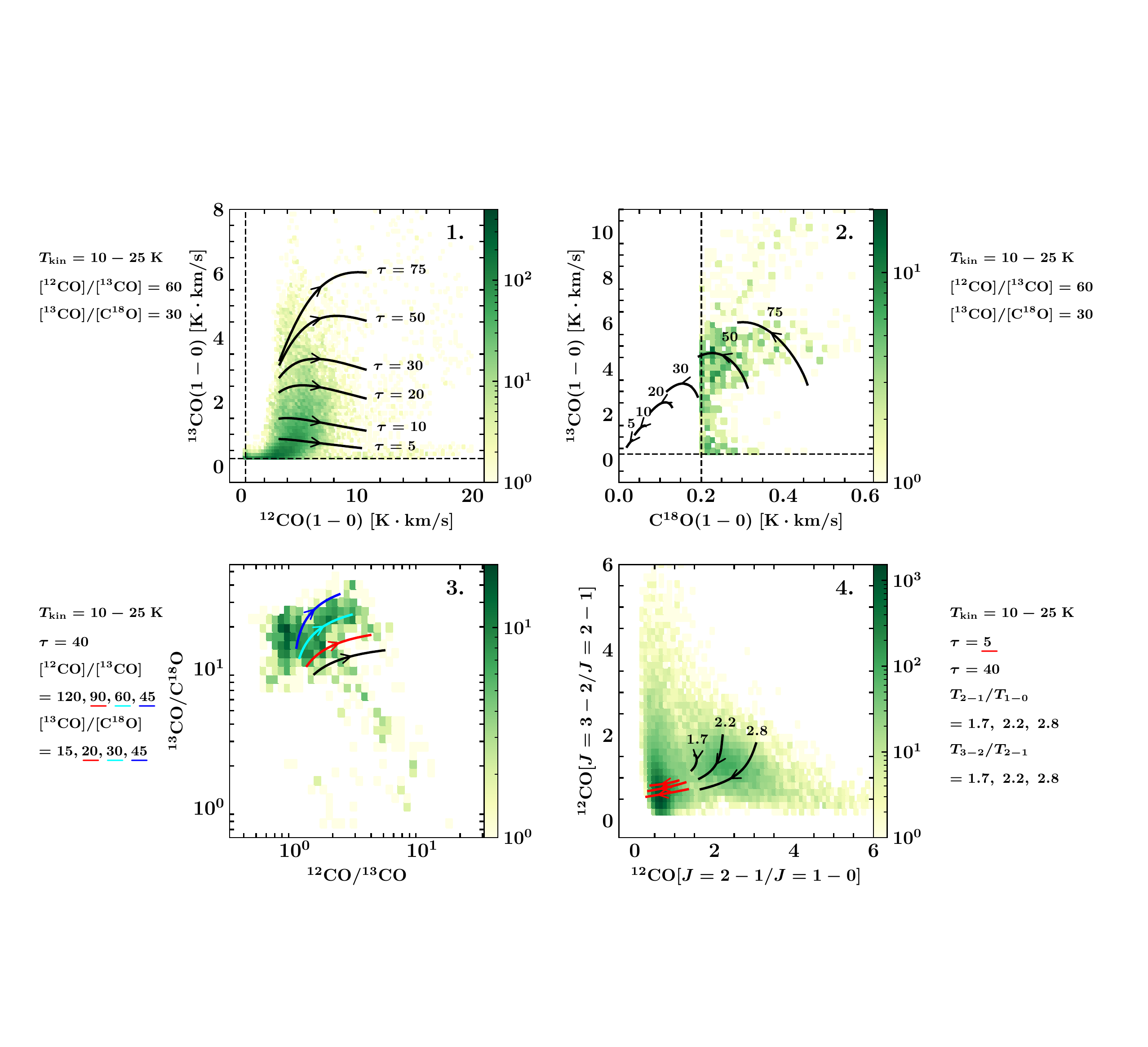}}

\caption{LTE models and data histograms of the emission of the rotational transitions of $^{12}\mathrm{CO}$ and its isotopologues mapped in the extended G field with the IRAM-30m and APEX: 1. $^{13}\mathrm{CO}$(1--0) vs $^{12}\mathrm{CO}$(1--0); 2. $^{13}\mathrm{CO}$(1--0) vs $\mathrm{C^{18}O}$(1--0); 3. [$^{13}\mathrm{CO}$(1--0)] / [$\mathrm{C^{18}O}$(1--0)] vs [$^{12}\mathrm{CO}$(1--0)] / [$^{13}\mathrm{CO}$(1--0)]; 4. $^{12}\mathrm{CO}$(3--2) / $^{12}\mathrm{CO}$(2--1) vs $^{12}\mathrm{CO}$(2--1) / $^{12}\mathrm{CO}$(1--0). The colormap corresponds to the amount of data points (line of sight; velocity channels) that fall into a given bin of the histogram. The curves represent families of models of the LTE intensity as a function of the kinetic temperature of $^{13}\mathrm{CO}$. The arrows indicate the direction in which the kinetic temperature grows along a given curve. Control parameters of the LTE models are given in Tab. \ref{LTEparams} for each histogram. Dashed black lines represent the 3$\sigma$ detection level for each axis (histograms 1 and 2).}
\label{fig:LTEmodels}
\end{center}

\begin{center}
\captionof{table}{Control parameters of the LTE models represented in Fig. \ref{fig:LTEmodels}: $\tau_\mathrm{^{12}CO}$ is the list of optical depths of the lines used to generate the family of curves, $T_\mathrm{^{13}CO}$ and $T_\mathrm{^{12}CO}$ are respectively the ranges of kinetic temperature used for $^{13}$CO and $^{12}$CO. [$^{12}$CO/$^{13}$CO] and [$^{13}$CO/C$^{18}$O] are the adopted isotopic ratios. The ratios of excitation temperatures for different isotopologues or rotational transitions are indicated by the columns $T_a$/$T_b$, and the ratios of opacities for distinct rotational transitions are given in the columns $\tau_a$/$\tau_b$. The colors used for control parameters in this table correspond to the colors of the curves in Fig. \ref{fig:LTEmodels}.} \label{tab:LTEparams}
\label{LTEparams}      
\begin{tabular}{c c c c c c c}        
\hline\hline         \\[-1.0em]        
\footnotesize{Histogram} & \footnotesize{$\tau_{^{12}\mathrm{CO}}(20\mathrm{K})$} & \footnotesize{T$_\mathrm{kin}$ (K)} & \footnotesize{[$^{12}\mathrm{CO}/^{13}\mathrm{CO}$]} & \footnotesize{[$^{13}\mathrm{CO}/\mathrm{C^{18}O}$]} & \footnotesize{T$_{^{12}\mathrm{CO}}$ / T$_{^{13}\mathrm{CO}}$} & \footnotesize{T$_{^{13}\mathrm{CO}}$ / T$_{\mathrm{C^{18}O}}$} \\ 
\hline    \\[-1.0em]                     
    \footnotesize{1. $^{13}$CO(1--0) vs $^{12}$CO(1--0)} & \footnotesize{[5, 10, 20, 30, 50, 75]} & \footnotesize{[10-25]} & \footnotesize{60} & \footnotesize{30} & \footnotesize{1} & \footnotesize{1}\\
    \footnotesize{2. $^{13}$CO(1--0) vs $^{18}$CO(1--0)} & \footnotesize{[5, 10, 20, 30, 50, 75]} & \footnotesize{[10-25]} & \footnotesize{60} & \footnotesize{30} & \footnotesize{1} & \footnotesize{1}\\
    \footnotesize{3. [$^{13}$CO/$^{18}$CO] vs [$^{12}$CO/$^{13}$CO]} & \footnotesize{40} & \footnotesize{[10-25]} & \footnotesize{[120, \color{red}90\color{black}, \color{cyan}60\color{black}, \color{blue} 45\color{black}]} & \footnotesize{[15, \color{red}20\color{black}, \color{cyan}30\color{black}, \color{blue} 45\color{black}]} & \footnotesize{1} & \footnotesize{1}\\
\hline\hline \\[-1.0em]
\footnotesize{Histogram} & \footnotesize{$\tau_{^{12}\mathrm{CO}}(20\mathrm{K})$} & \footnotesize{T$_\mathrm{kin}$ (K)} & \footnotesize{T$_{2-1}$ / T$_{1-0}$} & \footnotesize{T$_{3-2}$ / T$_{2-1}$} & \footnotesize{$\tau_{2-1}$ / $\tau_{1-0}$} & \footnotesize{$\tau_{3-2}$ / $\tau_{2-1}$} \\
\hline \\[-1.0em] 
    \footnotesize{4. $^{12}$CO [3--2/2--1] vs [2--1/1--0]} & \footnotesize{[\color{red}5\color{black}, 40]} & \footnotesize{[10-25]} & [\footnotesize{1.7, 2.2, 2.8]} & \footnotesize{[1.7, 2.2, 2.8]} & \footnotesize{0.2} & \footnotesize{1}\\
    
\hline                                   
\end{tabular}
\end{center}
\end{figure*}

In the next section we aim to build population diagrams in which we correct the effect of optical depth on the column density of upper levels. To measure the optical depth of CO lines, we relied on several strong assumptions, in particular the adopted isotopic ratios and the identity of excitation temperature for $^{12}\mathrm{CO}$ and $^{13}\mathrm{CO}$ (see section \ref{LTE} for a description of our method, and \citealt{Roueff20} for a complete discussion of the corresponding assumptions). To assess the validity of this approach and estimate the key parameters (isotopic ratios, excitation temperature ratios), we built 2D histograms from the $^{12}\mathrm{CO}$ J=1--0, J=2--1, J=3--2 data cubes, as well as $^{13}\mathrm{CO}$ and $\mathrm{C^{18}O}$ J=1--0 data cubes to compare the line intensity from different isotopologues and rotational transitions of CO with modified LTE models \citep{Bron2018}. These assumptions are also discussed in section \ref{LTEdiscussion}. We examined four different 2D data histograms:
\begin{enumerate}
\item J=1--0 line intensity, $^{13}\mathrm{CO}$ vs $^{12}\mathrm{CO}$.\\[-1.0em]
\item J=1--0 line intensity, $^{13}\mathrm{CO}$ vs $\mathrm{C^{18}O}$.\\[-1.0em]
\item J=1--0 line intensity ratio, [$^{13}\mathrm{CO}$/$\mathrm{C^{18}O}$] vs [$^{12}\mathrm{CO}$/$^{13}\mathrm{CO}$].\\[-1.0em]
\item $^{12}$CO line intensity ratio, [3--2]/[2--1] vs [2--1]/[1--0]. 
\end{enumerate}
In order to build the first three data histograms, we convolved all IRAM-30m data cubes to the nominal spatial resolution of $\mathrm{C^{18}O}$(1--0), i.e. 23.6$^{\prime \prime}$ and to the nominal spectral resolution of the FTS backend, i.e. 0.5 km s$^{-1}$. To build the fourth histogram, we resampled IRAM-30m $^{12}$CO(1--0), APEX $^{12}$CO(2--1) and APEX $^{12}$CO(3--2) data cubes to the nominal spatial resolution of $^{12}$CO(1--0), i.e. 22.5$^{\prime \prime}$ and to the nominal spectral resolution given by the FTS backend, i.e. 0.5 km s$^{-1}$. We used a treshold of $3 \sigma$ to select data points where the signal is significantly above the noise level. The resulting 2D data histograms are shown in Fig. \ref{fig:LTEmodels}, where we emphasize the high signal-to-noise areas of the third histogram with black contours at 3$\sigma$ and 4$\sigma$.

\subsection*{Results}
\begin{enumerate}
\item The first histogram is characterized by a high signal-to-noise ratio and represents a large statistical sample ($n=20531$). The $^{13}\mathrm{CO}$(1--0) vs $^{12}\mathrm{CO}$(1--0) relation presents at least two distinct branches. The lower quasi-horizontal branch, tracing bright $^{12}\mathrm{CO}$(1--0) emission associated with faint $^{13}\mathrm{CO}$(1--0) line emission ($T_{13}<1$ K km s$^{-1}$). This branch is spatially correlated with the shock structure and spectrally correlated with the high-velocity wings of the $^{12}$CO line that have no bright $^{13}$CO counterpart because of insufficient integration time. It is spatially correlated mainly with the quiescent molecular gas that is found within the cloudlet and ring-like structure, as well as the ambient cloud.
\item The $^{13}\mathrm{CO}$(1--0) vs $\mathrm{C^{18}O}$(1--0) histogram has a much smaller amount of bins determined with a good signal-to-noise ratio ($n=833$) due to the faint emission of $\mathrm{C^{18}O}$(1--0) that is hardly detected at a $3\sigma$ confidence level within our data cube. Still, we find evidence of at least one stastistically significant branch. Due to the small size of the sample it is not possible to identify any spatial or spectral correlation with certainty.
\item The third histogram has a poor statistical sample for the same reason as the second one ($n=833$). The [$^{13}\mathrm{CO}$(1--0)] / [$\mathrm{C^{18}O}$(1--0)] ratio vs [$^{12}\mathrm{CO}$(1--0)] / [$^{13}\mathrm{CO}$(1--0)] relationship is localized in an area with little dispersion. Hence, for high signal-to-noise measurements the isotopic ratios are almost uniform in the field of observations. Nonetheless, the lower signal-to-noise data bins display a statistically well-defined comet-shaped branch extending from this area. This branch might correspond to distinct physical conditions and/or isotopic ratios for a fraction of the field of observations. The high signal-to-noise ratio area of this branch is spatially correlated with the cloudlet and ring-like structure, whereas the 'tail' of the branch is spatially and spectrally associated with the shocked clump.
\item The fourth histogram is built on a large statistical sample with high signal-to-noise data ($n=9793$), as the rotational lines J=1--0, J=2--1 and J=3--2 are well detected in our data cubes. The $^{12}$CO [3-2]/[2-1] vs [2-1]/[1-0] relationship is clearly bimodal. Indeed, a quasi-vertical branch centered on [2-1]/[1-0] $\simeq$0.5 can be distinguished from a crescent-shaped branch centered on [2-1]/[1-0] $\simeq$2 and [3-2]/[2-1] $\simeq$1.3. The crescent-shaped branch is highly correlated to the high-velocity wings of the $^{12}$CO lines tracing the emission of shocked gas, where it is expected to measure high, enhanced J=2--1 / J=1--0 ratio (e.g. \citealt{Seta1998}).
\end{enumerate}

We compared the observational histograms with synthetic families of curves generated using modified LTE radiative transfer models of the observed rotational transition for $^{12}\mathrm{CO}$, $^{13}\mathrm{CO}$, and $\mathrm{C^{18}O}$.
Assuming that CO is at thermal equilibrium, the intensity of the line integrated over an element of spectral resolution ($\Delta \varv=0.5$ km s $^{-1}$) is given by:
\begin{ceqn}
\begin{equation}
W_{\alpha} = T_{\alpha} \Delta \varv 
\end{equation}
\end{ceqn}
where $\alpha=12,13,18$ (histograms 1-3) or $\alpha=$1--0, 2--1, 3--2 (histogram 4) respectively for the $^{12}\mathrm{CO}$, $^{13}\mathrm{CO}$, $\mathrm{C^{18}O}$ isotopologues and J=1--0, J=2--1, J=3--2 rotational lines. $T_{\alpha}$ is the integrated intensity of the line, defined as:
\begin{ceqn}
\begin{equation}
T_{\alpha} = T_{\alpha}^0 \left( 1 - \mathrm{e}^{-\tau_{\alpha}} \right) \cdot \left( \dfrac{1}{\mathrm{e}^{(T_{\alpha}^0/T_{\alpha}^{\mathrm{exc}})} - 1} - \dfrac{1}{\mathrm{e}^{(T_{\alpha}^0/T_{\mathrm{cmb}})} - 1} \right)
\end{equation}
\end{ceqn}
where $T_{\alpha}^0=h\nu /k_{\mathrm{B}}$ is the temperature of the transition or energy of the upper level $E_{\mathrm{up}}$, given in Tab. \ref{table:1} for all studied lines. $\tau_{\alpha}$ is the optical depth of the line, $T_{\mathrm{cmb}}$ is the cosmic microwave background temperature, and $T_{\alpha}^{\mathrm{exc}}$ is the excitation temperature of the line. The optical depth $\tau_{\alpha}$ and its dependence on the excitation temperature is described by:
\begin{ceqn}
\begin{equation}
\tau_{\alpha} = \tau_{\alpha}^{T_0} \left( \dfrac{T_0}{T_{\alpha}^{\mathrm{exc}}} \right) \left( \dfrac{1-\mathrm{e}^{(T_{\alpha}^0/T_{\alpha}^{\mathrm{exc}})}}{1-\mathrm{e}^{(T_{\alpha}^0/T_0)}} \right)
\end{equation}
\end{ceqn}
where $\tau_{\alpha}^{T_0}$ is the optical depth at a kinetic temperature $T_0$. We use the optical depth at $T_0=20$K as a reference. We made the assumption that the excitation temperature of the different isotopologues ($T_{\mathrm{exc}}^{12}$, $T_{\mathrm{exc}}^{13}$, $T_{\mathrm{exc}}^{18}$) and rotational lines ($T_{\mathrm{exc}}^{1-0}$, $T_{\mathrm{exc}}^{2-1}$, $T_{\mathrm{exc}}^{3-2}$) are distinct and controled their ratios as four supplementary independent parameters of the modified LTE models \citep{Roueff20}. The opacities of $^{12}\mathrm{CO}$ and $\mathrm{C^{18}O}$ isotopologues were determined from the optical depth of $^{13}\mathrm{CO}$ using the corresponding isotopic ratio: 

\begin{ceqn}
\begin{equation}
\dfrac{\tau_{12}}{\tau_{13}}=\dfrac{[^{12}\mathrm{CO}]}{[^{13}\mathrm{CO}]},~\dfrac{\tau_{13}}{\tau_{18}}=\dfrac{[^{13}\mathrm{CO}]}{[\mathrm{C^{18}O}]}
\end{equation}

\end{ceqn}

Lastly, we assumed a hierarchy in the optical depth of the different rotational lines such that $\tau_{1-0}\geq \tau_{2-1} \geq \tau_{3-2}$ and control their ratio as two supplementary parameters of the modified LTE models \citep{Roueff20}. Hence, there are a total of 9 control parameters that we can set: the optical depth $\tau_{13}$, the isotopic ratio $[^{12}\mathrm{CO}] / [^{13}\mathrm{CO}]$, the isotopic ratio $[^{13}\mathrm{CO}] / [\mathrm{C^{18}O}]$, the temperature ratios [$T_{^{12}\mathrm{CO}}$/$T_{^{13}\mathrm{CO}}$], [$T_{^{13}\mathrm{CO}}$/$T_{\mathrm{C^{18}O}}$], [$T_{3-2}$/$T_{2-1}$], [$T_{2-1}$/$T_{1-0}$], and optical depth ratios [$\tau_{3-2}$/$\tau_{2-1}$], [$\tau_{2-1}$/$\tau_{1-0}$]. We used the kinetic temperature of $^{13}$CO as a variable of the parametric equation $I_{ij}(T_{\mathrm{kin}}^{13})$ to synthesize LTE models that we can plot on top of each 2D data histogram. Each curve in Fig. \ref{fig:LTEmodels} corresponds to a given set of control parameters, with the kinetic temperature growing linearly along a curve. We produced a family of LTE models with linearly increasing optical depth and a kinetic temperature varying between 10~K and 25~K, corresponding to the typical temperatures of cold molecular clouds. We adopted the isotopic ratio $[^{12}\mathrm{CO}] / [^{13}\mathrm{CO}]$ and $[^{13}\mathrm{CO}] / [\mathrm{C^{18}O}]$ that offers the best visual match with our data, and similarly fine-tuned the other control parameters to reproduce the branches observed in each 2D data histogram. 

All other parameters remaining constants, each control parameter has an effect on the pattern and location of a curve in the histogram space:
\begin{itemize}
\item An increase in the isotopic ratio $[^{12}\mathrm{CO}] / [^{13}\mathrm{CO}]$ translates the curves downward in the histogram 1, does not modify the histogram 2 and translates the curves rightward in the histogram 3.
\item An increase in the isotopic ratio $[^{13}\mathrm{CO}] / [\mathrm{C^{18}O}]$ does not modify the histogram 1, translates the curves upward in the histogram 2 and translates the curves upward in the histogram 3. 
\item The modification of the temperature ratios alters the shape and orientation of the curves in histograms 3 and 4, and determines the boundaries of the $^{12}$CO and C$^{18}$O curves. 
\end{itemize}
The LTE models that visually best match our data are represented in Fig. \ref{fig:LTEmodels} over the data histograms, and their corresponding set of control parameters are summarized in Tab. \ref{LTEparams}. A kinetic temperature in the range 10-25 K is sufficient to account for most of the data where the sample is statistically significant, and we did not need to assume that [$T_{^{12}\mathrm{CO}}$/$T_{^{13}\mathrm{CO}}$] and [$T_{^{13}\mathrm{CO}}$/$T_{\mathrm{C^{18}O}}$] are different from 1 to find a satisfactory fit. However, to obtain a model that accounts for the high [3--2]/[2--1] and [2--1]/[1--0] data bins (histogram 4) it was necessary to set their temperature ratios as greater than 1, such that we infer an average temperature of up to $\sim$55~K traced by the J=2--1 lines, and up to $\sim$120~K for the J=3--2 lines in the high-velocity wings. This area of the histogram 4 also required lower $\tau_{2-1}/\tau_{1-0}$ and $\tau_{3-2}/\tau_{1-0}$ ratios, that we set to 0.2 for our models, suggesting that optical depth decreases significantly in the high-velocity wings of the lines. Considering histogram 3, our best models indicate a value of the isotopic ratio $[^{12}\mathrm{CO}] / [^{13}\mathrm{CO}] = 45-90$ and $[^{13}\mathrm{CO}] / [\mathrm{C^{18}O}]=20-45$ to account for the emission of the different isotopologues observed in the extended G field. A large range of isotopic ratios can account for the observed data spread in histogram 3, hence this result does not provide a precise determination of the $[^{12}\mathrm{CO}] / [^{13}\mathrm{CO}]$ and $[^{13}\mathrm{CO}] / [\mathrm{C^{18}O}]$ abundances. The higher boundary of our $^{12}$C/$^{13}$C ratio estimate is anomalous with respect to the local interstellar medium average of 62$\pm$4 \citep{Langer93}, however \citet{Wilson1994} provide an estimate of the elemental abundances based on their distance to the Galactic center (GC):
\begin{ceqn}
\begin{align}
\begin{split}
^{12}\mathrm{C} / ^{13}\mathrm{C} = (7.5 \pm 1.9) D_\mathrm{GC} + (7.6 \pm 12.9)
\\
^{16}\mathrm{O} / ^{18}\mathrm{O} = (58.8 \pm 11.8) D_\mathrm{GC} + (37.1 \pm 82.6)
\end{split}
\end{align}
\end{ceqn}
If we adopt the distance of 1.9 kpc for IC443 \citep{Ambrociocruz17} the SNR is located at a distance of $\sim$10 kpc from the GC. Hence our estimate of the $^{12}\mathrm{CO} / ^{13}\mathrm{CO}$ ratio is in agreement with the expected abundance of 80$\pm$30, but our measured $^{13}\mathrm{CO} / \mathrm{C^{18}O}$ is much higher than the predicted ratio of 8$\pm$2. The measured enhancement of the $^{13}\mathrm{CO} / \mathrm{C^{18}O}$ ratio might be the product of the fractionation of carbon monoxide by photodissociation, as the shielding of the various isotopologues of CO provides a mechanism for the rarefaction of C$^{18}$O with respect to the expected abundance (e.g. \citealt{Glassgold1985}). We checked the spatial distribution of data points corresponding to particular areas of the histograms presented in Fig. \ref{fig:LTEmodels}. 
\begin{itemize}
\item \textit{Histogram 3}. Interestingly, the data points corresponding to high ($>20$) $^{13}\mathrm{CO} / \mathrm{C^{18}O}$ line ratio are spatially correlated with the cloudlet only. If no protostars are positively identified towards the cloudlet (section \ref{stars}), radiative decay from X-ray irradiation seen in the SNR cavity with XMM-Newton might provide a source to account for the selective photodissociation of CO within the cloudlet (\citealt{Troja2006}, \citealt{Troja2008}).\\[-1.0em]   
\item \textit{Histogram 4}. Data points corresponding to high ($>2$) $[J=2-1]/[J=1-0]$ and $[J=3-2]/[J=2-1]$ ratios are spatially correlated with the shocked clump, and spectrally correlated with the high-velocity wings of the molecular line. The most striking spatial correlation is seen towards the eastern surface layer of the shocked clump, where the molecular gas seems to be systematically characterized by a large ($>2$) $[J=3-2]/[J=2-1]$ and small ($<1$) $[J=2-1]/[J=1-0]$ ratio apparently tracing the shock front. Although, this combination of line ratios is not predicted by RADEX models and could be due to an inaccurate registration between the data sets.
\end{itemize}

\begin{table}[]
\caption{Velocity ranges $[v_{\mathrm{min}}, v_{\mathrm{max}}]$ and spatial boxes (as defined in Fig. \ref{fig:boxes}) used to measure specific intensity corresponding to each structure. Three velocity ranges are given for the shocked clump, as we distinguish three components within the molecular lines: \textit{i.} the left (blueshifted) high-velocity wing, \textit{ii.} the core, \textit{iii.} the right (redshifted) high-velocity wing.}             
\label{table:velocity_intervals}      
\centering                          
\begin{tabular}{c c c}        
\hline\hline   \\[-1.0em]               
Region & $[v_{\mathrm{min}}, v_{\mathrm{max}}]$ \footnotesize{km s$^{-1}$} & box (fig. \ref{fig:boxes})\\ 
\hline   \\[-1.0em]                     
    \footnotesize{cloudlet} & \footnotesize{[-8.5; -4.5]} & \footnotesize{A} \\
    \footnotesize{ring-like structure} & \footnotesize{[-7.0; -4.5]} & \footnotesize{B}\\
    \footnotesize{shocked clump \textit{i.}} & \footnotesize{[-35; -6.5]} & \footnotesize{C}\\
    \footnotesize{shocked clump \textit{ii.}} & \footnotesize{[-6.5; -1.5]} & \footnotesize{C}\\
    \footnotesize{shocked clump \textit{iii.}} & \footnotesize{[-1.5; +25]} & \footnotesize{C}\\
    \footnotesize{shocked knot} & \footnotesize{[-35; +25]} & \footnotesize{D}\\
    \footnotesize{ambient cloud} & \footnotesize{[-6.5; -1.5]} & \footnotesize{$\mathrm{\overline{A}\cap \overline{B}\cap \overline{C}\cap \overline{D}}$}\\
    
\hline                                   
\end{tabular}
\end{table}

\subsubsection{Population diagrams}\label{LTE}
We determined the physical conditions (column density $N_\mathrm{CO}$, kinetic temperature $T_\mathrm{kin}$) of the molecular gas in the extended G region assuming that the emission lines are thermalized. We used pixel-per-pixel population diagrams \citep{Goldsmith1999} corrected for optical depth for the $^{12}$CO(3-2), $^{12}$CO(2-1) and $^{12}$CO(1-0) transitions to measure the upper level column density $N_\mathrm{up}$ (hereafter level population) for $J_\mathrm{up}=3$, $J_\mathrm{up}=2$, $J_\mathrm{up}=1$. Systematically, the $^{13}$CO data is used to correct the effect of optical depth in the population diagrams, for each element of resolution (i.e. for each of line of sight) and for all velocity channels. To perform this task we choosed to use the APEX data cube over the IRAM-30m data cube to measure the specific intensity corresponding to the $^{12}$CO(2-1) transition. The IRAM-30m has the advantage of having a lower noise (see Tab. \ref{table:2}). However, the $^{13}$CO(2-1) transition was only observed with the APEX telescope due to receiver setup constraints. Hence to avoid intercalibration issues, we used the APEX data for $^{12}$CO(2--1) and $^{13}$CO(2--1). First, for consistency we convolved all maps to the same spatial resolution, using the nominal resolution of $^{13}$CO(2-1), such that we have a beam diameter of $30.1^{\prime \prime}$ for the six maps considered. The spectral resolution was also modified and set to 2 $\mathrm{km ~ s^{-1}}$ for each transition in order to increase the signal-to-noise ratio. Then, the following measurements were performed pixel-per-pixel using a Python\footnote{Python Core Team (2015). Python: A dynamic, open source programming 
  language. Python Software Foundation. \url{https://www.python.org/}.} algorithm:
\begin{enumerate}
\item Sigma-clipping was applied to the spectra of all transitions of the two isotopologues $^{12}$CO and $^{13}$CO, using a threshold of $3\sigma$. 
\item In every single velocity channel remaining after sigma-clipping, the $N_{J=3}$, $N_{J=2}$ and $N_{J=1}$ level populations were measured both for $^{12}$CO and $^{13}$CO to determine the optical depth of the three transitions using the adopted value of isotopic ratio. We measured the optical depths $\tau_{3-2}$, $\tau_{2-1}$ and $\tau_{1-0}$ assuming that $^{13}\mathrm{CO}$ lines are optically thin and that the excitation temperatures of the two isotopologues are approximately equal (this second assumption is supported by the results presented in section \ref{fig:LTEmodels}). Under this assumption we have:
\begin{ceqn}
\begin{equation}
\int I_{\nu , ^{12}\mathrm{CO}} \mathrm{d} \nu = \dfrac{h \nu_{ij,^{12}\mathrm{CO}}}{4 \pi} N_{\mathrm{up},^{12}\mathrm{CO}} A_{ij,^{12}\mathrm{CO}} \cdot \beta_{ij}
\label{eq:intensity12CO}
\end{equation}
\end{ceqn}
\begin{ceqn}
\begin{equation}
\int I_{\nu , ^{13}\mathrm{CO}} \mathrm{d} \nu = \dfrac{h \nu_{ij,^{13}\mathrm{CO}}}{4 \pi} N_{\mathrm{up},^{13}\mathrm{CO}} A_{ij,^{13}\mathrm{CO}}
\label{eq:intensity13CO}
\end{equation}
\end{ceqn}
where $\beta_{ij} = [1 - \mathrm{e}^{-\tau_{ij}}] / \tau_{ij}$ is the escape probability of a photon, $\nu_{ij}$ the frequency of the transition, $N_\mathrm{up}$ the level population, $A_{ij}$ the probability of the transition and $I_{\nu}$ the specific intensity. 
Based on the comparison of our data with modified LTE models (Sect. \ref{sect:LTEmodels}) we adopted different values for the expected isotopic ratio towards each region (indicated in Tab. \ref{table:LTEresults}).
\item The estimates of $^{12}$CO level populations were corrected for optical depth using the correction factor $C_{\tau_{ij}}=\tau_{ij} / (1-e^{-\tau_{ij}})$, such that the corrected column density is given by $N^\mathrm{*}_\mathrm{up}=C_{\tau_{ij}} N_\mathrm{up}$.
\item We put the corrected level populations $N^\mathrm{*}_\mathrm{up}$ and their corresponding energies $E_\mathrm{up}=h \nu_{ij}/k_\mathrm{B}$ into the population diagram $N^\mathrm{*}_\mathrm{up}=f(E_{\mathrm{up}})$, and used a $\chi^2$-minimization algorithm\footnote{The Python module scipy.optimize.curve\_fit uses the Levenberg-Marquardt least square curve fitting algorithm.} to determine the best linear fit $y=a x + b$. 
The excitation temperature $T_{\mathrm{ex}}$ is deduced from the slope, and the total column density $N_{\mathrm{CO}}$ is determined by computing the partition function $Q(T_{\mathrm{ex}})$\footnote{We computed the partition function taking into account the first 40 upper rotational levels of $^{12}$CO taken from the Cologne Database for Molecular Spectroscopy and Jet Propulsion Laboratory database. The thermal energy of the 40$^\mathrm{th}$ level is E$_{40}=4513$K.} and measuring the offset.
\item We finally derived the mass of the gas from the measured total column density  $N_{\mathrm{CO}}$, using the expected $[\mathrm{H_2}] / [^{12}\mathrm{CO}]$ ratio towards dense molecular regions and assuming that the distance of IC443 is 1.9 kpc \citep{Ambrociocruz17}. We adopted the value of $10^4$ for the $\mathrm{H_2}$-to-$^{12}\mathrm{CO}$ abundance ratio (\citealt{Gordon&Burton1976}, \citealt{Frerking1982}), thus we have $M_\mathrm{H_2}=(2 m_\mathrm{H}) \mathcal{A} N_\mathrm{CO} \times 10^4$ where $\mathcal{A}$ is the area of the box and $m_\mathrm{H}$ the mass of hydrogen. 
\end{enumerate}

\textit{Uncertainties.} The errorbars on total CO column density, gas mass and kinetic temperature are determined by: \textit{(i.)} instrumental errors (dominated by the systematic uncertainty on the flux, Sect. \ref{systematic_errors}); \textit{(ii.)} the quality of the linear fit applied to the population diagram; \textit{(iii.)} systematic errors on the adopted distance of IC443 and the $^{12}$CO/$^{13}$CO isotopic ratio. \\

\begin{table*}[]
\caption{Results of the population diagrams analysis for each structure. $\overline{N_{\mathrm{CO}}}$ is the average column density and $M_\odot$ the mass in solar unit. The adopted isotopic ratio $^{12}$CO/$^{13}$CO for each region is indicated in the fourth column. In the mass column, we indicate in parenthesis the 'cold mass' estimated by considering only the upper levels J=1,2. $\overline{T_{\mathrm{exc}}}$  is the average excitation temperature. 
The quality of the linear fits is indicated by the average value of $\chi^2$. $\overline{T_{\mathrm{exc}}^\mathrm{w}}$ is the average excitation temperature of the warm component, obtained by considering the levels J=2,3 separately from the level J=1. $C_{\tau}^{\star}=M^{\star}/M$ gives an indication of the enhancement of the mass estimate when we correct the optical depth of the transition J=1--0 independently of the detection of $^{13}$CO for the transitions J=2--1 and J=3--2. For comparison, we indicate the estimate of the mass obtained by using the CO-to-H$_2$ conversion factor $X_\mathrm{CO}=2\times 10^{20}$ cm$^{-2}$(K km s$^{-1}$)$^{-1}$ applied to the raw J=1--0 data (without optical depth correction).}             
\label{table:LTEresults}      
\centering                          
\begin{tabular}{c c c c c c c c c c c c}        
\hline\hline \\[-0.9em]                
region & $\overline{N_{\mathrm{CO}}}[\mathrm{10^{17} cm^{-2}}]$ & Mass ($M_\odot$) & \footnotesize{$\dfrac{^{12}\mathrm{CO}}{^{13}\mathrm{CO}}$} & $C_\mathrm{\tau}^{\star}$ & $\overline{T_{\mathrm{exc}}}[\mathrm{K}]$ & $\overline{\mathrm{log_{10}(\chi^2)}}$ & $\overline{T_{\mathrm{exc}}^\mathrm{w}}[\mathrm{K}]$ & $M_{X_\mathrm{CO}}^{J=1-0}$ ($M_\odot$)\\
\\[-0.9em] 
\hline          \\[-1.0em]              
    \footnotesize{cloudlet (A)} & \footnotesize{1.4-2.3 (3.7)} & \footnotesize{130-210 (330)} & \footnotesize{80$\pm$20} & \footnotesize{1.1} & \footnotesize{9$^{+1}_{-3}$} & \footnotesize{0.34}  & \footnotesize{12$\pm$1} & \footnotesize{450$\pm$140} \\ \\[-1.0em]
    \footnotesize{ring-like structure (B)} & \footnotesize{2.3-3.8 (4.4)} & \footnotesize{60-110 (130)} & \footnotesize{60$\pm$20} & \footnotesize{1.1} & \footnotesize{11$^{+1}_{-2}$} & \footnotesize{-0.68} & \footnotesize{13$\pm$1} & \footnotesize{200$\pm$60} \\ \\[-1.0em]
    \footnotesize{shocked clump (C) (i.)} & \footnotesize{0.21-0.32 (0.40)} & \footnotesize{12-17 (22)} & \footnotesize{45$\pm$10} & \footnotesize{1.4} & \footnotesize{28$\pm$2} & \footnotesize{-0.85	} & \footnotesize{32$\pm$2} & \footnotesize{190$\pm$60} \\ \\[-1.0em]
    \footnotesize{shocked clump (C) (ii.)} & \footnotesize{1.6-2.8 (3.4)} & \footnotesize{80-150 (230)} & \footnotesize{45$\pm$10} & \footnotesize{3.5} & \footnotesize{12$^{+1}_{-5}$} & \footnotesize{-0.18} & \footnotesize {19$\pm$2} & \footnotesize{240$\pm$70} \\ \\[-1.0em]
    \footnotesize{shocked clump (C) (iii.)} & \footnotesize{0.12-0.14 (0.23)} & \footnotesize{6-7 (12)} & \footnotesize{45$\pm$10} & \footnotesize{1.4} & \footnotesize{22$\pm$2} & \footnotesize{-0.08} & \footnotesize{28$\pm$2} & \footnotesize{110$\pm$40}\\ \\[-1.0em]
    \footnotesize{shocked knot (D)} & \footnotesize{0.19-0.24} & \footnotesize{2-4 (6)} & \footnotesize{45$\pm$10} & \footnotesize{1.7} &  \footnotesize{21$\pm$2} & \footnotesize{-0.38} & \footnotesize{30$\pm$2} & \footnotesize{30$\pm$10} \\ \\[-1.0em]
    \footnotesize{ambient cloud} & \footnotesize{1.1-1.9 (3.0)} & \footnotesize{400-700 (1800)} & \footnotesize{60$\pm$20} & \footnotesize{13} &  \footnotesize{9$^{+1}_{-5}$} & \footnotesize{0.85} & \footnotesize{10$\pm$1} & \footnotesize{1300$\pm$400} \\ \\[-1.0em]
    \footnotesize{foreground clump} & \footnotesize{0.02-0.03 (0.07)} & \footnotesize{3-5 (14)} & \footnotesize{45$\pm$10} & \footnotesize{-} & \footnotesize{9$\pm$1} & \footnotesize{0.07} & \footnotesize{11$\pm$1} & \footnotesize{38$\pm$11}\\
\hline\hline\\[-1.0em]
    \footnotesize{IC443G (extended)} & \footnotesize{1.6-2.8 (4.0)} & \footnotesize{900-1600 (3100)} & - & \footnotesize{3.6} &  \footnotesize{7$\pm$1 / 9$\pm$1} & \footnotesize{0.89} & \footnotesize{11$\pm$1} & \footnotesize{2600$\pm$800} \\
\hline
\end{tabular}
\end{table*}

   \begin{figure}[]
   \centering
   \includegraphics[width=\hsize, trim={0.5cm 0.2cm 1.5cm 0.2cm},clip]{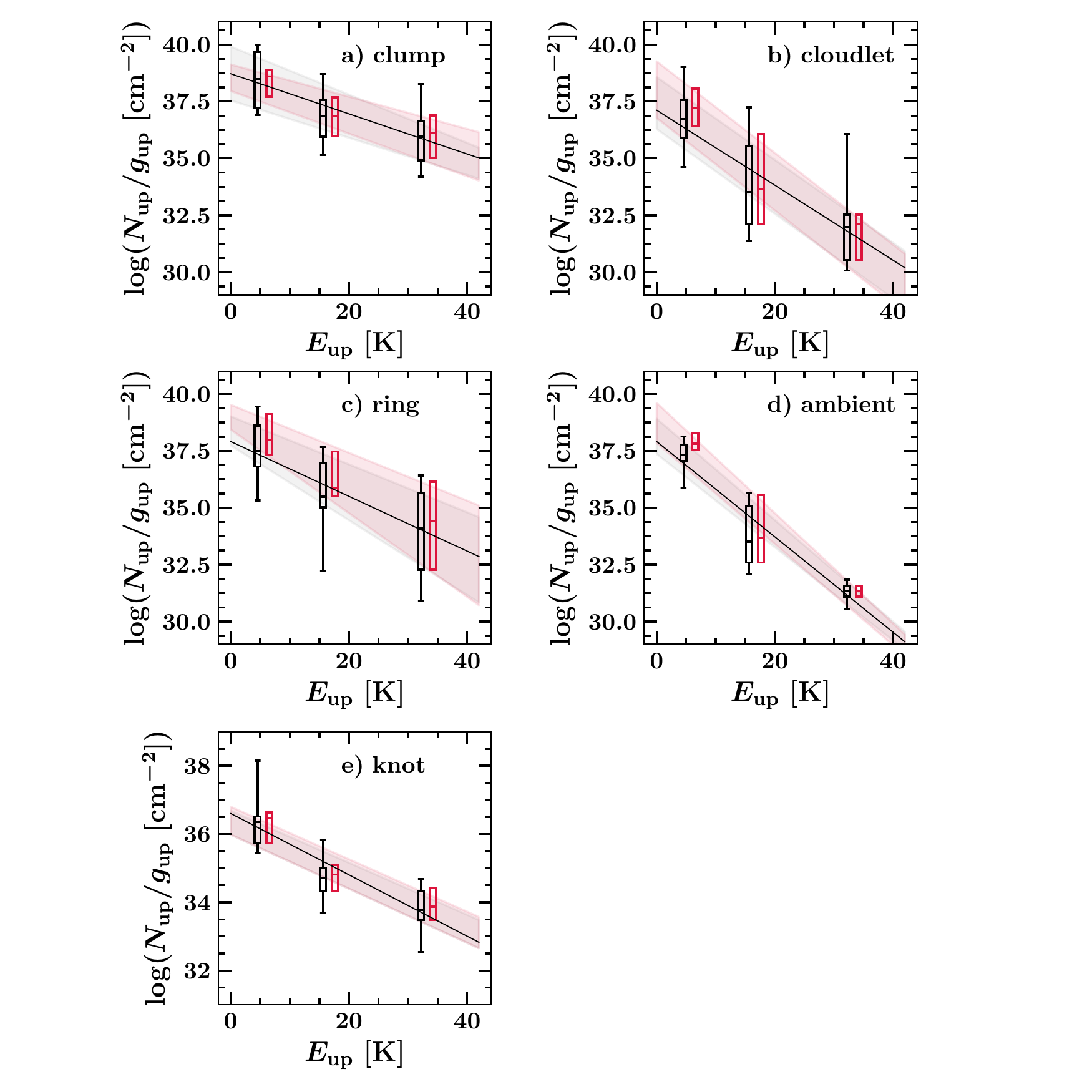}
      \caption{Average population diagrams of $\mathrm{^{12}}$CO corrected for optical depth. The measured average, first quartile, third quartile, 5$^\mathrm{th}$ percentile and 95$^\mathrm{th}$ percentile associated with our sample of level populations corresponding to $J_{\mathrm{up}} = 1$, $J_{\mathrm{up}} = 2$ and $J_{\mathrm{up}} = 3$ are indicated for two adopted values of the $^{12}$CO/$^{13}$CO isotopic ratio ($^{12}$CO/$^{13}$CO = 60 in black, $^{12}$CO/$^{13}$CO = 100 in red). The black solid line represents the linear fit of the black data points. The boundaries of the filled areas are defined by the linear fits corresponding to the first and third quartiles for each value of the isotopic ratio adopted. The 5$^\mathrm{th}$ percentile and 95$^\mathrm{th}$ percentile are indicated by the caps, the first quartile and third quartile are indicated by the ends of the box and the average is indicated by the cap inside of the box. Each diagram accounts for a sample of measurements performed in one of the spatial boxes defined in Fig. \ref{fig:boxes} and integration ranges defined in Tab. \ref{table:velocity_intervals}: a) the shocked clump, b) the cloudlet, c) the ring, d) the ambient cloud and e) the shocked knot.}
         \label{fig:excdiag}
   \end{figure}
   
   \begin{figure}[]
   \centering
   \includegraphics[width=\hsize, trim={0.5cm 0.2cm 1.5cm 0.2cm},clip]{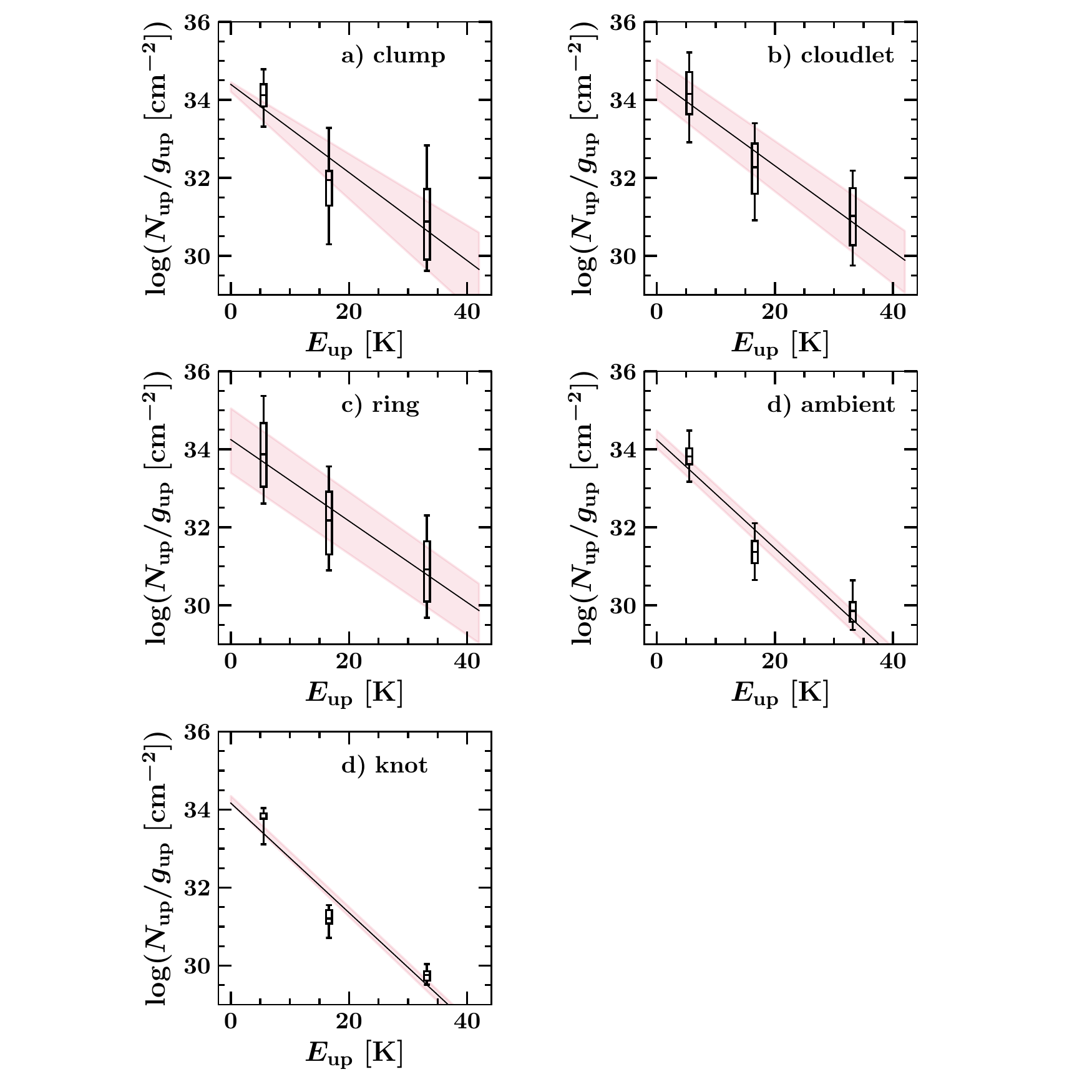}
      \caption{Average population diagrams of $\mathrm{^{13}}$CO. The measured average, first quartile, third quartile, 5$^\mathrm{th}$ percentile and 95$^\mathrm{th}$ percentile associated with our sample of measured level populations are indicated for each level. The black solid line represents the linear fit of the average data points. The boundaries of the filled areas are defined by the linear fits corresponding to the first and third quartiles. Each diagram accounts for a sample of measurements performed in one of the spatial boxes, as in Fig. \ref{fig:excdiag}.}
         \label{fig:excdiag-13co}
   \end{figure}

\textit{Results.} Average population diagrams are presented on Fig. \ref{fig:excdiag} and Fig. \ref{fig:excdiag-13co}. The measurements obtained are presented in Tab. \ref{table:LTEresults}, in which the minimum and maximum boundaries are given for the total CO column density, gas mass and kinetic temperature. We measured the column densities and masses for the regions A (cloudlet), B (ring-like structure), C (shocked clump) and D (shocked knot) as indicated in Fig. \ref{fig:boxes}, and also for the ambient cloud by averaging the signal over the entire field (regions A, B, C and D excluded). 
We also measured the mass of the foreground clump that is spatially coinciding with the shocked clump using a gaussian model for the average CO lines in the velocity range [2.5; 12] km s$^{-1}$. The result of this measurement is presented at the end of Tab. \ref{table:LTEresults}.

\subsubsection{Discussion}\label{LTEdiscussion}
As a first-order verification we compared our measurement of the mass based on the population diagram with a rough estimate of the mass using the CO(J=1--0)-to-H$_2$ conversion factor $X_\mathrm{CO}$ (\citealt{Dame2001}, \citealt{Bolatto2013}). We adopted the following conversion factor, with a $\pm$30\% uncertainty:
\begin{ceqn}
\begin{equation}
X_\mathrm{CO}=2\times10^{20}~\mathrm{cm}^{-2}(\mathrm{K~km~s}^{-1})^{-1}
\end{equation}
\end{ceqn}
The H$_2$ column density was determined by the product of $X_\mathrm{CO}$ with the area of the raw $^{12}$CO J=1--0 line (without optical depth correction), and the mass was inferred from N$_\mathrm{H_2}$ in the same manner as described previously. We report the results of this measurement in the last column of Tab. \ref{table:LTEresults} where they are referred to as $M_{X_\mathrm{CO}}^{J=1-0}$. Within errorbars, almost all our measurements are consistent with this rough estimate of the mass, except for the line wings towards the shocked clump. With respect to the population diagram method, this method systematically overestimates the mass. The enhanced J=2--1/J=1--0 emission of the high-velocity gas accounts for a lower mass in the population diagram, as it traces a warmer gas for which the $X_\mathrm{CO}$ conversion factor yields an overestimate.

We built a single population diagram for each spatial box, using a single value of $N_\mathrm{up}$ that is the spatial and spectral average of our sample of measurements. The resulting population diagrams are shown in Fig. \ref{fig:excdiag}, as well as the statistical informations on the spread of the sample around these average values of $N_\mathrm{up}$. 
As an additional information on the thermalization of carbon monoxide, in Fig. \ref{fig:excdiag-13co} we show the average population diagrams obtained using the $^{13}$CO lines with the same method, without correction of the optical depth.
\begin{enumerate}
\item \textit{Thermalization.} At a temperature of 10~K, the critical density $n_{ij}=A_{ij}/C_{ij}$ of the $^{12}$CO J=1--0, J=2--1 and J=3--2 lines are respectively $n_{1-0}=2.2\times10^3$ cm$^{-3}$, $n_{2-1}=2.3\times10^4$ cm$^{-3}$ and $n_{3-2}=3.5\times10^4$ cm$^{-3}$\footnote{The Einstein coefficients $A_{ij}$ are given in Tab. \ref{table:1}, and the collisional excitation rates $C_{ij}$ are taken from the Cologne Database for Molecular Spectroscopy and Jet Propulsion Laboratory database.}. Towards the shocked clump, it is likely that these critical densities are attained \citep{Cesarsky1999}, hence the lines should be thermalized. In the population diagrams displayed in Fig. \ref{fig:excdiag}, the average data points are generally in satisfactory agreement with the assumption that the emission lines are thermalized, as there is no significant divergence from the Boltzmann distribution for any of the structure studied. The highest value of $\chi^2$ is obtained towards the ambient cloud that also presents an abnormally low kinetic temperature, down to $\sim$5~K. This might indicate that the J=2--1 and J=3--2 are subthermal, which is expected if some parts of the NW-SE molecular cloud we are probing have a density lower than $10^4$ cm$^{-3}$. There is a systematic discrepancy between the linear fit and the measured column density of the upper level J=2, which is lower than expected in each structure. This anomaly can be solved if we adopt two distinct kinetic temperature to model the relative distribution of the level populations. In the high-velocity wings of the lines, it is expected and it was hinted by our LTE analysis (Sect. \ref{sect:LTEmodels}) that the J=2--1 and J=3--2 lines trace a warmer gas than the J=1--0 line that is primarily tracing the quiescent and cold phase. In this case, the distribution in these population diagrams should be modeled by a linear fit of kinetic temperature $T_1$ for the upper levels J$_\mathrm{up}$=1,2 (cold component) and a linear fit of temperature $T_2$ for levels J$_\mathrm{up}$=2,3 (warm component), hence assuming that along a line of sight we observe two layer of gas thermalized at distinct temperatures, with $T_1<T_2$. In Tab. \ref{table:LTEresults} we present the temperature obtained for the warm component, as well as the mass of the cold component. The statistical spread of our sample of measured level populations around the average linear distribution provides a strong motive for the use of pixel-per-pixel, channel-per-channel population diagrams, as it is evidence of a large range of physical conditions that require distinct LTE models. \\[-1.0em]
\item \textit{Filling factor.} The filling factor used to infer column densities from the main beam temperature measured during our observations is set to 1. From the morphology of the gas mapped in $\mathrm{^{12}}$CO(2--1) with the IRAM-30m with a nominal resolution of 11.2$^{\prime \prime}$ (Fig. \ref{fig:moment0}, Fig. \ref{fig:channelmap}) we consider that we are probing extended, clumpy structures with dimensions that are greater than the beam diameter characterizing our observations, between and 19.2$^{\prime \prime}$ and 30.1$^{\prime \prime}$ for the data cubes used to measure the gas mass. \\[-1.0em]
\item \textit{Optical depth.} The measurement of the optical depth from the comparison of the fluxes between $\mathrm{^{12}}$CO and $\mathrm{^{13}}$CO is displayed in the appendix , as well as the measurement based on the $^{12}$CO/C$^{18}$O isotopic ratio (Fig. \ref{fig:opticaldepth13co}). The optical depth is non negligible both in the ring-like structure and cloudlet. It is important in the self-absorbed component of the emission lines corresponding to the shocked clump where the cold, ambient cloud intervenes, and much lower in the high-velocity wings. We checked the assumption that $^{13}$CO is optically thin using the C$^{18}$O J=1--0 and J=2--1 lines to estimate its optical depth based on the expected $^{13}$CO/C$^{18}$O isotopic ratio. In our pixel-per-pixel, channel-per-channel population diagrams the correction of the optical depth is partially biased, as the detection of $^{13}$CO is limited by the noise level. At a certainty level of $5\sigma$, we cannot measure a flux lower than 125 mK, 420 mK and 400 mK respectively for the J=1--0, J=2--1 and J=3--2 transitions. As a consequence, towards a significant number of line of sights and velocity channels the optical depth is measured for the J=1--0 transition but not for J=2--1 and J=3--2. In the population diagram this results in the biased displacement of the level J=1 with respect to the levels J=2 and J=3, hence the slope of the linear fit is artifically enhanced and the excitation temperature is lowered (Fig. \ref{fig:excdiag}, Tab. \ref{table:LTEresults}).\\[-1.0em]
\item \textit{Self-absorption of CO lines towards the shocked clump.} We performed an independant measure of the temperature and column density of the absorbing envelope of the shocked clump from the absorption features of CO lines. Towards the position ($6^{\mathrm{h}}16^{\mathrm{m}}42.5^{\mathrm{s}}$, $+22^\circ32'12^{\prime \prime}$), we renormalized the signal in each frequency channel with respect to the profile of the emission lines and measured the optical depth for all transitions in a beam of radius 22.5$^{\prime \prime}$. From the equivalent widths $W$ of the transitions J=1--2 and J=0--1, we measured the temperature of the gas solving the following equation:
\begin{ceqn}
\begin{equation}
F \left( \dfrac{T_0}{T_\mathrm{ex}} \right) = \dfrac{W_{1 \rightarrow 2}}{4 W_{0 \rightarrow 1}}
\end{equation}
\end{ceqn}
Where $T_0 = h \nu / k$ and $F(u)=\mathrm{e}^{-3u} \mathrm{cosh}(u)$ (see appendix \ref{thermometry}). We obtained an excitation temperature of $18\pm 3$K. We then built a population diagram of the lower population $N_l$ levels from the measures of the optical depth, using the following relation:
\begin{ceqn}
\begin{equation}
\tau = 1.1 \cdot 10^{-15} f_{ul} \left( \dfrac{N_l}{\mathrm{cm^{-2}}} \right) \left( \dfrac{\lambda}{\AA} \right) \left( \dfrac{\sigma_v}{\mathrm{km \cdot s^{-1}}} \right)
\end{equation}
\end{ceqn}
Where $f_{ul}$ is the oscillator force, $\lambda$ the wavelength of the transition and $\sigma_v$ the velocity dispersion of the absorbing medium, which we assume equal to the velocity dispersion of the lines associated with the shocked clump that we measured in Tab. \ref{table:linemeasures}. This population diagram yield a total column density of the absorbing medium $N_{\mathrm{CO}}=3.4\times 10^{16}$ $\mathrm{cm^{-2}}$, corresponding to a total mass of 20 M$_\odot$ for the absorbing envelope of the shocked clump.

\end{enumerate}
\subsection{LVG method}

\begin{figure*}[]
\begin{center}
\resizebox{0.85\hsize}{!}
{\includegraphics[trim={0cm 0cm 0cm 0cm},clip]{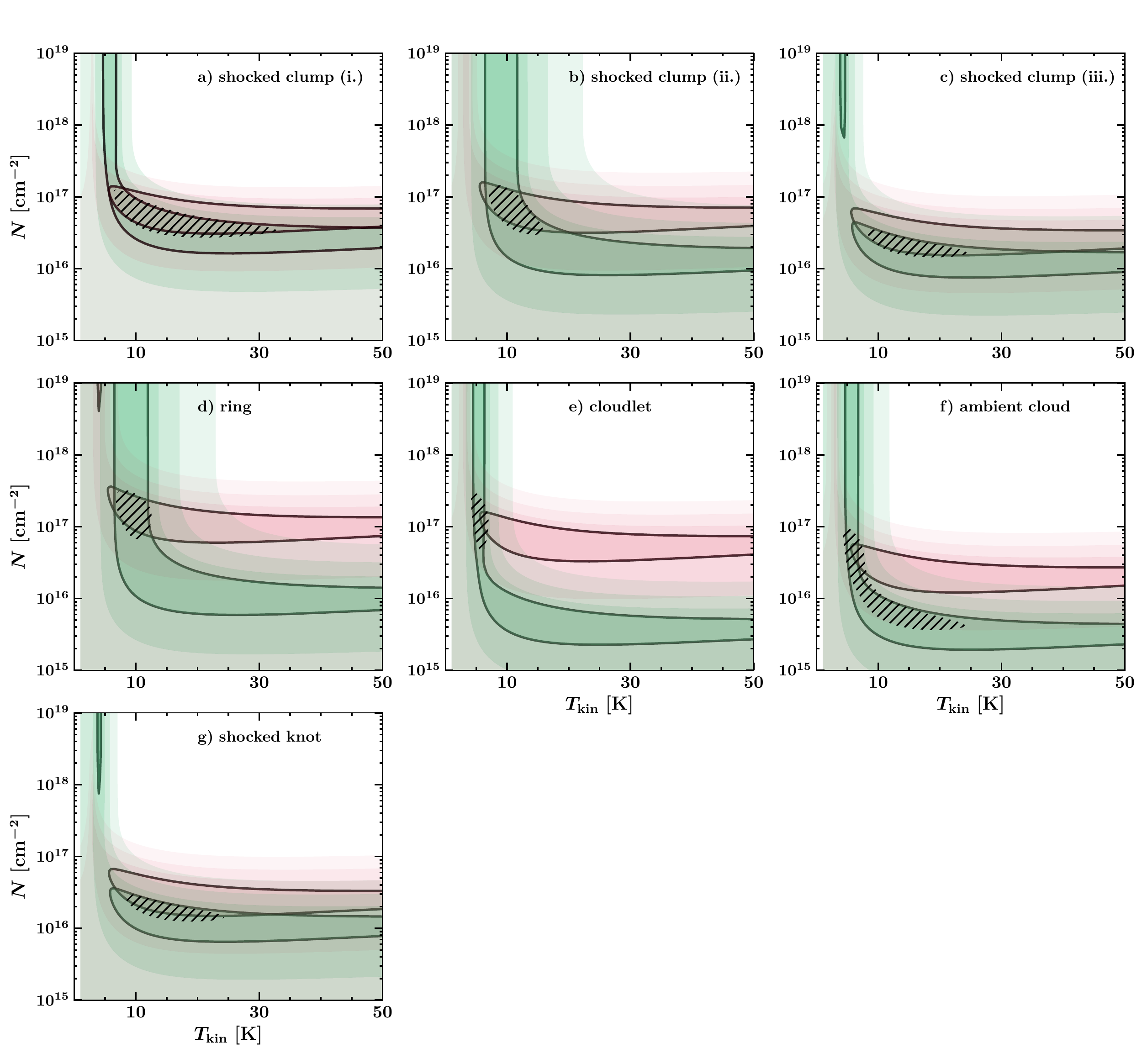}}
\caption{$\chi^2$-minimization diagrams of the comparison between 1$\times$200$\times$200 LVG grids modeled with sets of parameters ($n_\mathrm{H_2}$, $N_\mathrm{CO}$, $T$) and observations of $^{12}\mathrm{CO}$ and $^{13}\mathrm{CO}$ lines towards the boxes corresponding to the shocked clump (a, b, c), ring-like structure (d), cloudlet (e), ambient cloud (f) and shocked knot (g) defined in Tab. \ref{table:boxes}. The shades of green represent the areas of minimum $\chi^2$ corresponding to the minimization of the specific intensity of the $^{12}\mathrm{CO}$(3--2), $^{12}\mathrm{CO}$(2--1) and $^{12}\mathrm{CO}$(1--0) lines (first term introduced in Eq. \ref{eq:chi2}), and the shades of pink represent the areas of minimum $\chi^2$ corresponding to the minimization of $^{13}\mathrm{CO}$ lines (second term introduced in Eq. \ref{eq:chi2}. The filled contour levels are $\mathrm{log_{10}}(\chi^2)=[0.6, 1, 1.5, 2]$, with the first contour level highlighted by a solid black line. Hatched areas represent the intersection between the previously described sets of contours, defined by $\mathrm{log_{10}}(\chi^2_\mathrm{RADEX}) \leq [0.5, 0.45, 0.45, 0.5, 0.75, 0.75, 0.5]$ (Eq. \ref{eq:chi2}) respectively for the boxes a, b, c, d, e, f and g.}
\label{fig:LVGgrid}
\end{center}
\end{figure*}

\begin{table*}
\caption{Summary of the results and parameters used in the LVG analysis. We show ranges of values corresponding to the uncertainty interval. $N_{\mathrm{CO}}$ is the column density of $^{12}$CO, $T_{\mathrm{kin}}$ the kinetic temperature of the gas, $n$ the local density of $\mathrm{H_{2}}$, $\mathrm{log_{10}}(\chi^2)$ gives the minimum value of $\chi^2$ within the minimization diagrams presented in Fig. \ref{fig:LVGgrid}, $M$ is the mass measured using the [$\mathrm{H_2}$]/[$\mathrm{^{12}CO}$] ratio. $N_{\mathrm{grid}}$ is the range of column densities parsed across the grid of models, $T_{\mathrm{grid}}$ the range of kinetic temperatures, $T_{\mathrm{b}}$ and $\Delta \varv$ are respectively the background temperature and the width of the emission line used by RADEX for all models. We indicate the ratio $M_\mathrm{LVG}/M_\mathrm{LTE}$ of the estimate of the mass with the LVG approach with respect to the results obtained with population diagrams (Sect. \ref{LTE}, Tab. \ref{table:LTEresults}).}
\label{table:5}      
\centering                          
\begin{tabular}{l  c  c  c  c c c c c}        
\hline           
\hline
\footnotesize{Regions} &  \footnotesize{cloudlet (A)} &  \footnotesize{ring-like (B)} & & \footnotesize{shocked clump (C)} & & \footnotesize{shocked knot (D)} & \footnotesize{ambient}\\ \hline
& & & \footnotesize{blue-shifted} & \footnotesize{core} & \footnotesize{red-shifted} &\\
\hline \\[-1.0em] 
\footnotesize{$N_{\mathrm{CO}}~[10^{17}\mathrm{cm^{-2}}]$} & \footnotesize{1.4-2.4} & \footnotesize{1.7-2.3} & \footnotesize{0.68-0.99} & \footnotesize{0.9-1.48} & \footnotesize{0.24-0.45} & \footnotesize{0.17-0.22} & \footnotesize{0.44-0.89} \\
\footnotesize{$T_{\mathrm{kin}}~[\mathrm{K}]$} & \footnotesize{6$\pm$1} & \footnotesize{10$\pm$1} & \footnotesize{16$\pm$7} & \footnotesize{12$\pm$2} & \footnotesize{17$\pm$9} & \footnotesize{13$\pm$6} & \footnotesize{13$\pm$5} \\
\footnotesize{$n_\mathrm{H_2}~[\mathrm{cm^{-3}}]$} & \footnotesize{$\sim 7\times10^5$} & 
\footnotesize{$>10^4$} & \footnotesize{$>10^4$} & \footnotesize{$>10^4$} & \footnotesize{$>10^4$} & \footnotesize{$ > 10^4$} & \footnotesize{$\sim 3\times10^4$} \\
\footnotesize{$\mathrm{log_{10}}(\chi^2)$} & \footnotesize{-0.04} & \footnotesize{-0.16} & \footnotesize{0.23} & \footnotesize{0.04} & \footnotesize{0.48} & \footnotesize{0.12} & \footnotesize{0.47} \\
 \\[-1.0em] 
\footnotesize{$M~[\mathrm{M_{\odot}}]$} & \footnotesize{120-220} & \footnotesize{40-70} & \footnotesize{32-52} & \footnotesize{42-80} & \footnotesize{19-26} & \footnotesize{2-3} & \footnotesize{210-350} \\
\footnotesize{$M_\mathrm{LVG}/M_\mathrm{LTE}$} & \footnotesize{1} & \footnotesize{0.64} & \footnotesize{2.90} & \footnotesize{0.53} & \footnotesize{3.46} & \footnotesize{0.83} & \footnotesize{0.51} \\
\hline \\[-1.0em] 
\footnotesize{$N_{\mathrm{^{12}CO}}~[\mathrm{cm^{-2}]}$} & \footnotesize{[$10^{15}-10^{19}$]} & \footnotesize{[$10^{15}-10^{19}$]} & \footnotesize{[$10^{15}-10^{19}$]} & \footnotesize{[$10^{15}-10^{19}$]} & \footnotesize{[$10^{15}-10^{19}$]} & \footnotesize{[$10^{15}-10^{19}$]} & \footnotesize{[$10^{15}-10^{19}$]} \\
\footnotesize{$[\mathrm{^{12}CO/^{13}CO}]$} & \footnotesize{$80$} & \footnotesize{$60$} & \footnotesize{$45$} & \footnotesize{$45$} & \footnotesize{$45$} & \footnotesize{$45$} & \footnotesize{$60$} \\
\footnotesize{$T_{\mathrm{grid}}~[\mathrm{K}]$} & \footnotesize{[1-100]} & \footnotesize{[1-100]} & \footnotesize{[1-100]} & \footnotesize{[1-100]} & \footnotesize{[1-100]} & \footnotesize{[1-100]} & \footnotesize{[1-100]}   \\
\footnotesize{$T_{\mathrm{b}}~[\mathrm{K}]$} & \footnotesize{2.7} & \footnotesize{2.7} & \footnotesize{2.7} & \footnotesize{2.7} & \footnotesize{2.7} & \footnotesize{2.7} & \footnotesize{2.7} \\
\footnotesize{$\Delta \varv~[\mathrm{km~s^{-1}}]$} & \footnotesize{4.0} & \footnotesize{3.5} & \footnotesize{25} & \footnotesize{5.0} & \footnotesize{25} & \footnotesize{35} & \footnotesize{3.0}  \\
\hline
\end{tabular}
\end{table*}

The results obtained in the previous section are based on several strong assumptions, in particular that the excitation temperature of $^{12}\mathrm{CO}$ and $^{13}\mathrm{CO}$ are equal. To adress this issue, we investigated the application of a second method that does not require this assumption to determine the mass of the gas towards our field of observations using CO lines. We used the LVG model for an expanding spherical shell (e.g. \citealt{Sobolev1960}, \citealt{Surdej77}) to model the lines observed. This model require a local, large velocity gradient $\nabla v$ along the line of sight. We consider that this assumption is verified throughout the field of observation as the linewidth of $^{12}$CO lines suggest typical turbulent broadening (Sect. \ref{sect:kinematics}). We used the non-LTE radiative transfer RADEX code \citep{VanderTak2007}, a computer program solving the radiative transfer equation based on the escape probability formulation in an isothermal and homogeneous medium characterized by a large velocity gradient. Three geometries are available, we selected the expanding spherical shell for which the escape probability is related to the optical depth $\tau$ by the following formula \citep{Mihalas1978}:
\begin{ceqn}
\begin{equation}
\beta_\mathrm{LVG}=\frac{1-e^{-\tau}}{\tau}.
\end{equation}
\end{ceqn}
In the framework of the LVG assumption, the value of the optical depth at line center determined by RADEX  is proportional to $N$/$\Delta \varv$ where $N$ is the column density and $\Delta \varv$ the full width at half-maximum of the line profile. Hence when using RADEX the choice of the parameter $\Delta \varv$ is critical to model the expected intensities of atomic and molecular lines. The program solves the statistical equilibrium equations taking into account up to seven collisional partners, allowing to analyze spectral line observations if the molecular collisional data is known. RADEX models require a total of 7 input parameters: the column density $N$, local density $n_{H_2}$, kinetic temperature $T_{kin}$, line width $\Delta \varv$ and background temperature $T_b$. In return, RADEX estimates the excitation temperature $T_\mathrm{ex}$, optical depth $\tau_{ij}$, peak temperature $T_\mathrm{R}$, upper and lower level populations $N_\mathrm{up}$, $N_\mathrm{low}$, and flux $\int T \mathrm{d}v$ of an arbitrary number of transitions defined by the user. 

We applied the LVG method to average spectra measured over the extent of each box defined in Fig. \ref{fig:boxes} and velocity ranges defined in Tab. \ref{table:velocity_intervals}. Using a $\chi^2$-minimization, we compared RADEX outputs with our measurements in order to constrain the physical conditions. We compared the line intensities of the three first rotational transitions of $^{12}\mathrm{CO}$ and $^{13}\mathrm{CO}$ with a grid of RADEX models. In the same manner as in the population diagram approach, for consistency we convolved all maps to the same spatial resolution, using the nominal resolution of $^{13}$CO(2-1), such that we have a beam diameter of $30.1^{\prime \prime}$ for the six maps considered. The spectral resolution was also set to 2 $\mathrm{km ~ s^{-1}}$ for each transition in order to increase the signal-to-noise ratio. Then we produced several three-dimensional grids of RADEX models with the varying parameters ($N$, $n_\mathrm{H_2}$, $T$) and fixed parameters ($T_{b}$, $\Delta \varv$). These parameters are specified in Tab.~\ref{table:5} for the different grids used. We use large ranges of column densities and kinetic temperature in order to probe a consistent fraction of the space of parameters, corresponding to a variety of physical conditions as wide as possible.
We set the value of the parameter $\Delta \varv$ based on the equivalent linewidth that we measured on average spectra using bigaussian functions to model the line profiles. The following steps were followed for each average spectrum using a Python algorithm:
\begin{enumerate}
\item We applied sigma-clipping to the spectra of all transitions of $^{12}$CO and $^{13}$CO, using a threshold of $3\sigma$.
\item We measured the specific intensity of each transition on the velocity range corresponding to each spectral feature (e.g. the wings of the shocked clump are treated separately from the core of the line). Then, in each element of the model grid we computed the following quantity, corresponding to a combined reduced $\chi^2$ statistical estimator: 
\begin{ceqn}
\begin{equation}
\label{eq:chi2}
\chi^2_{\mathrm{RADEX}}= \dfrac{1}{2} \sum\limits_{J_{\mathrm{up}}=1}^{3} \left[ \dfrac{(s_{12}^{\mathrm{RADEX}} - s_{12}^{\mathrm{obs}})^2}{3 \sigma(s_{12}^{\mathrm{obs}})^{2}} + \dfrac{(s_{13}^{\mathrm{RADEX}} - s_{13}^{\mathrm{obs}})^2}{3 \sigma(s_{13}^{\mathrm{obs}})^{2}} \right]
\end{equation}
\end{ceqn} 
In this equation, $J_{\mathrm{up}}$ indicates the upper levels of each transition, $s^{\mathrm{RADEX}}_{12}$ and $s^{\mathrm{RADEX}}_{13}$ are respectively the specific intensity of $^{12}$CO and $^{13}$CO returned by a given RADEX model, $s^{\mathrm{obs}}$ the observational intensity and $\sigma$ the uncertainty associated with the measurements of the intensity $s$.
\item The physical conditions for each average spectrum were deduced from the minimization of the quantity $\chi^2$, i.e. we localized the minimum element of the resulting $\chi^2_{\mathrm{RADEX}}$ grid and infer the quantities ($N$, $n_\mathrm{H_2}$, $T$) from the corresponding RADEX input.
\end{enumerate}

The errorbars on column density and kinetic temperature were estimated from the uncertainties on the flux and more importantly from the uncertainty on the isotopic ratio $^{12}$CO/$^{13}$CO.

\textit{Results.} The results of our LVG analysis are presented in Tab. \ref{table:5}, in which the minimum and maximum boundaries are given for the CO column density and gas mass for the shocked clump, shocked knot, ring-like structure, cloudlet and ambient cloud. As previously, we measured the molecular mass assuming that the H$_2$-to-$^{12}$CO abundance ratio is equal to 10$^4$. The $\chi^2$-minimization was successfully attempted for the major part of our analysis. We present the $\chi^2$ diagrams for each region in Fig. \ref{fig:LVGgrid}, where the first and second term of Eq. \ref{eq:chi2} are independently represented by two sets of filled contours. For most structures, we were not able to determine a precise measurement of the local density from the $\chi^2$-minimization, as the variation of $\chi^2$ with respect to the choice of the input $n_\mathrm{H_2}$ does not strongly favor any LVG model. In all cases we observe that $\chi^2$ increases significantly for densities $n_\mathrm{H_2}<10^4$ cm$^{-3}$, hence our analysis suggests that \textit{(i.)} the $^{12}$CO lines are thermalized; \textit{(ii.)} the local density is greater than $10^4$ cm$^{-3}$ across the field of observations. 
\subsection{Discussion}
The gas masses measured with the LVG approach are systematically lower than the masses obtained using population diagrams corrected for optical depth (Tab. \ref{table:LTEresults}), except for the high-velocity wings where the LVG estimate is higher by a factor $\sim 3$ and for the cloudlet where both methods yield the same result. Deviation from a single excitation temperature is not sufficient to account for this discrepancy, since the LVG models that fit our data do not predict a disagreement higher than $\sim$1 K for the excitation temperatures of $^{12}$CO and $^{13}$CO. This discrepancy is mainly due to the fact that the LVG method is applied to average spectra, whereas the LTE approach is applied pixel-per-pixel and channel-per-channel. The channel-per-channel measures of optical depth are particularly different from the average measures towards the shocked clump where the optical depth varies strongly between the center of the line and the high-velocity wings. The highest estimates of the mass are obtained using the CO-to-H$_2$ conversion factor, based on the emission of $^{12}$CO J=1--0 only. In comparison to previous measures of the molecular mass in the extended G region:
\begin{itemize}
\item \citet{Dickman1992} measured a mass of 41.6 M$_\odot$ for the clump G assuming that $^{12}$CO J=1--0 emission is optically thin. 
\item Using a CO-to-H$_2$ conversion factor, \citet{Lee2012} measured a mass of 57.7$\pm$0.9 M$_\odot$ for the cloudlet.
\item \citet{Xu11} measured a mass of 2.06$\times$10$^3$ M$_\odot$ for the 'cloud G' from the $^{12}$CO J=3--2 line. Their measure included a larger field, and they lacked sufficient data to correctly estimate the optical depth.
\end{itemize}
Our new measurements of the mass are crucial for the interpretation of the interaction of CRs with the ISM. The extended G region is likely to be the main target of interaction with CRs, hence the source of TeV $\gamma$-ray emission in IC443. Our results put constraints on the amount of molecular mass that is available to interact via bremsstrahlung and pion decay mechanisms. \citet{Torres2010} showed that the characteristics of the $\gamma$-ray spectra in IC443 suggest the interaction with two distinct molecular structures: \textit{i.} a lower mass cloud ($\sim$350 M$_\odot$) at distance of 4 pc from the SNR, and a higher mass cloud ($\sim$4000 M$_\odot$) at a distance of 10 pc. Their results were obtained from the analysis of the $\gamma$-ray spectra of the SNR in a larger field, yet they are consistent with out findings if we consider that the mass of the ambient gas that we measured could contribute to the distant component. As it has been suggested by \citet{Lee2012}, either the shocked clump or cloudlet and ring-like structure could correspond to the closer component, while the distant component might correspond to the ambient cloud. Within errorbars, the mass we measured for these structures could fit with this scenario.

\section{Protostar candidates}\label{stars}

In order to characterize the local star formation over the extent of the molecular structures characterized in the previous section, we studied the distribution of optical, infrared and near-infrared point sources in the extended G region. We aimed to check if these infrared point sources can be identified as protostars and if so, to constrain their evolutionary stage based on their infrared fluxes. Finally, we aimed to study their spatial distribution and their association with the molecular clumps found in the extended G region.

\subsection{Origin of the data}

\begin{table}[]
\caption{Photometric parameters for each band of the telescopes WISE, 2MASS and Gaia. $\lambda$ is the band center, $\Delta \lambda$ the band width and FWHM is the full width half maximum of the point spread function (corresponding to the seeing in the case of 2MASS.)}             
\label{table:pointsourcecatalogue}      
\centering                          
\begin{tabular}{c c c c}        
\hline\hline     \\[-1.0em]         
band & $\lambda$ ($\upmu$m) & $\Delta \lambda$ ($\upmu$m) & \footnotesize{FWHM}($^{\prime \prime}$) \\
\hline                        
    \footnotesize{WISE $W_1$} & \footnotesize{3.35} & \footnotesize{0.66} & \footnotesize{6.1}  \\
    \footnotesize{WISE $W_2$} & \footnotesize{4.60} & \footnotesize{1.04} & \footnotesize{6.4}  \\
    \footnotesize{WISE $W_3$} & \footnotesize{11.56} & \footnotesize{5.51} & \footnotesize{6.5}  \\
    \footnotesize{WISE $W_4$} & \footnotesize{22.09} & \footnotesize{4.10} & \footnotesize{12.0}  \\
\hline \\[-1.0em]
    \footnotesize{2MASS J} & \footnotesize{1.235} & \footnotesize{0.162} &  \footnotesize{2.5}\\
    \footnotesize{2MASS H} & \footnotesize{1.662}  & \footnotesize{0.251} & \footnotesize{2.5} \\
    \footnotesize{2MASS K} & \footnotesize{2.159}  & \footnotesize{0.262} & \footnotesize{2.5} \\
\hline
\hline \\[-1.0em]
band & $\lambda$ (nm) & $\Delta \lambda$ (nm) & \footnotesize{FWHM} ($^{\prime \prime}$) \\
 \hline \\[-1.0em]
    \footnotesize{Gaia G} & \footnotesize{673}  & 440 & 0.4 \\
    \footnotesize{Gaia $\mathrm{G_{BP}}$} & \footnotesize{532}  & \footnotesize{253} & \footnotesize{0.4} \\
    \footnotesize{Gaia $\mathrm{G_{RP}}$} & \footnotesize{797}  & \footnotesize{296} & \footnotesize{0.4} \\
    \footnotesize{Gaia $\mathrm{G_{RVS}}$} & \footnotesize{860}  & \footnotesize{28} & \footnotesize{0.4} \\
\hline                                   
\end{tabular}
\end{table}

Our field was fully observed both by 2MASS and WISE. 2MASS operates between $1.235~\mathrm{\mu m}$ and $2.159~\mathrm{\mu m}$, and WISE between $3.4~\mathrm{\mu m}$ and $22~\mathrm{\mu m}$. The exact photometric parameters for each band of 2MASS and WISE are given in Tab. \ref{table:pointsourcecatalogue}.

To recover point sources from these two catalogues, we used the NASA/IPAC infrared science archive to obtain all entries in a $10~\mathrm{arcmin}$ sized square box around the field center $\alpha_{[\rm{J}2000]}$=$6^{\mathrm{h}}16^{\mathrm{m}}37.5^{\mathrm{s}}$, $\delta_{[\rm{J}2000]}$=$+22^\circ35'00^{\prime \prime}$, corresponding to the same field that was mapped with IRAM-30m and APEX. A total of 487 point sources in the 2MASS All-Sky Point Source Catalog (PSC) and 515 point sources in the AllWISE Source Catalog were found using this query within the extended G region. Both catalogues provide position coordinates, photometric measurements for each band and their uncertainties, signal-to-noise ratio, as well as several flags specifying contamination by extended emission, quality of the PSF profile-fit and other possible sources of bias.

\subsection{Selection of relevant IR point sources}

In order to reject false positives, we applied several selection criteria to our primary catalogues of point sources detected by 2MASS and WISE:
\begin{enumerate}
\item We required a complete detection in the WISE bands W$_1$, W$_2$ and 2MASS bands J, H and K (\textit{i.e.} the measurement is not an upper limit). \\[-1.0em]
\item We selected only the sources characterized by a signal-to-noise ratio greater than 2 for the photometric bands W$_1$, W$_2$, W$_3$, J, H and K.
\item We reject the WISE point sources that were flagged for confusion and/or contamination of the photometric bands by image artifacts.
\item We rejected the 2MASS point sources that were flagged for low quality photometric measurements.
\end{enumerate}

After this selection, 214/515 point sources remain for the AllWISE Source Catalog and 328/487 for the 2MASS PSC. A total of 99 point sources were detected both by 2MASS and WISE. AllWISE point sources marked with a value of \textit{ext\_flg} flag different than '0' are indicated as extended sources. Either their morphology is not consistent with the point spread function of any band or they are spatially associated with a known extended source of the 2MASS Extended Source Catalog (XSC). 2MASS point sources with \textit{gal\_contam} = 0 are also sources that fall within the elliptical profile of a known extended source. A search in the 2MASS XSC catalog shows that 17 extended sources are found in our 10$^\prime \times$10$^\prime$ field. The shocked clump is particularly crowded, suggesting that the extended emission of bright knot of shocked material are detected by the survey in this area. The photometric measurements are likely to be contaminated by this extended emission, hence the identification of these sources as protostar candidates is uncertain. Nonetheless we do not completely rule out these entries in the catalog since this extended emission might be produced by outflows.
   
   \begin{figure*}[]
   \centering
   \resizebox{0.45\hsize}{!}
   {\includegraphics[width=\hsize, trim={0cm 0cm 0cm 1.5cm},clip]{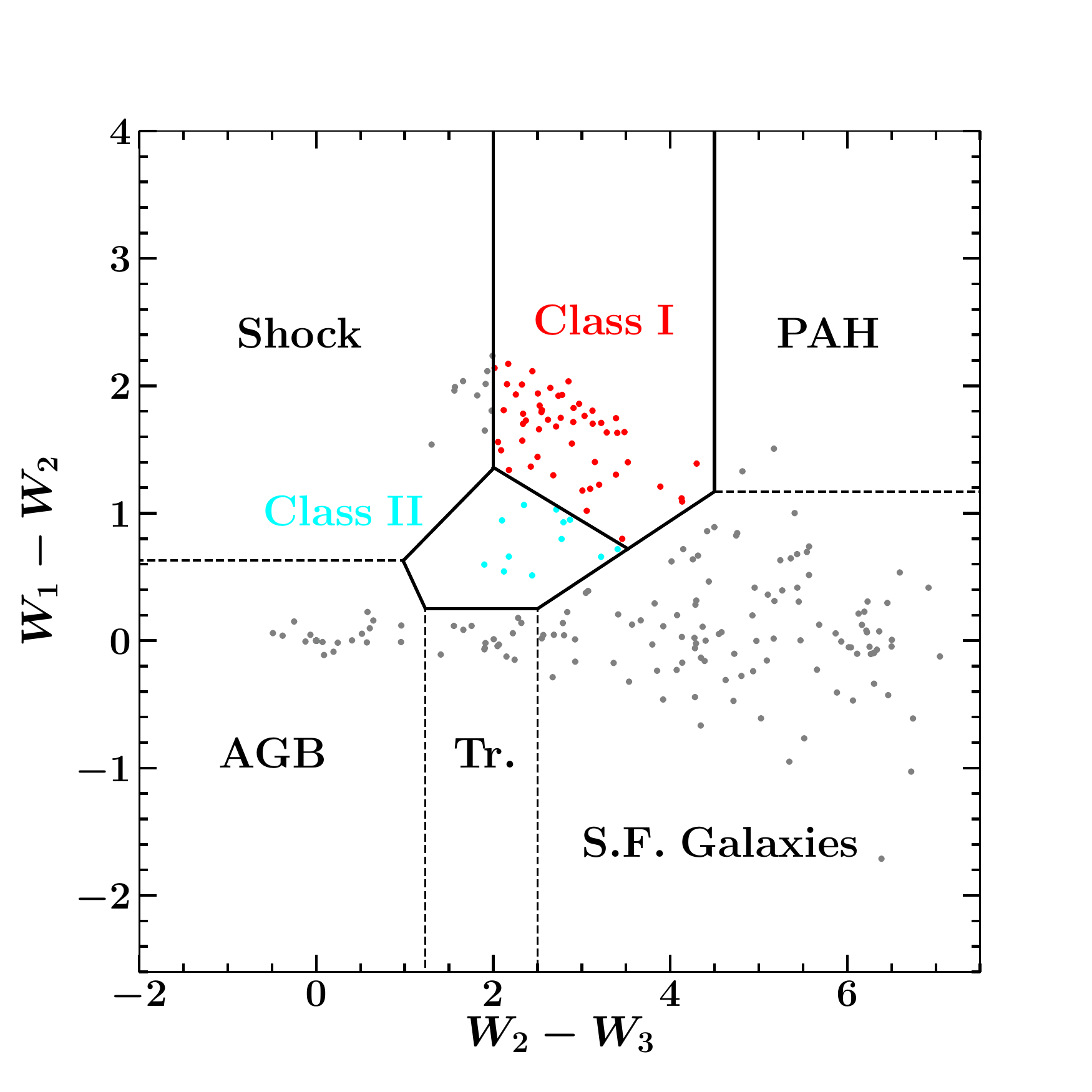}}
   \resizebox{0.45\hsize}{!}
   {\includegraphics[width=\hsize, trim={0cm 0cm 0cm 1.5cm},clip]{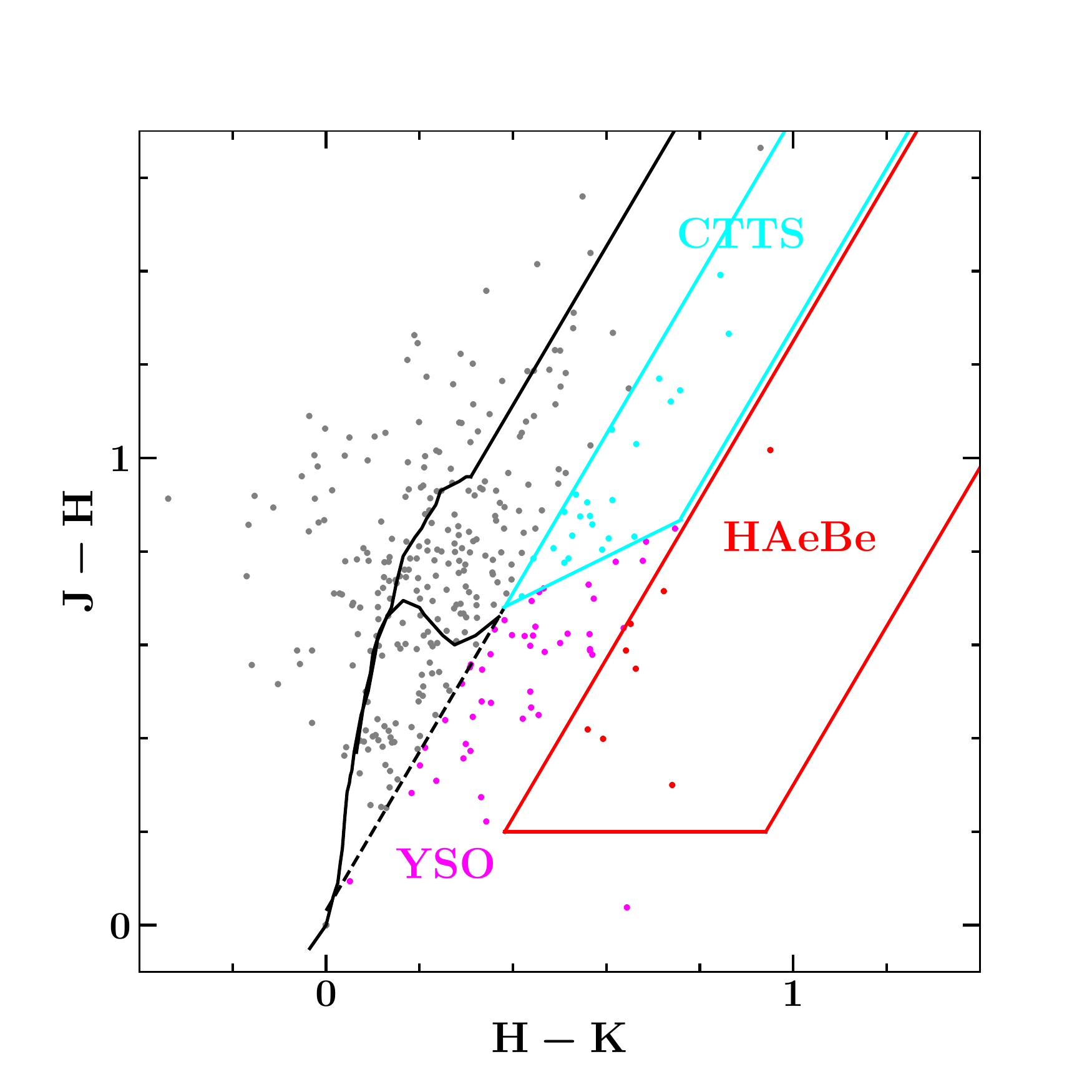}}
      \caption{\textit{Left}: WISE census of protostar candidates in the extended G region. Cyan dots are Class II candidates, red dots are Class I candidates and gray dots are point sources detected by WISE and rejected by our color-color filtering based on photometric bands W$_1$, W$_2$ and W$_3$. \textit{Right}: 2MASS census of protostar candidates in the extended G region. Cyan dots are CTTS candidates, red dots are HaeBe candidates, magenta dots are YSO candidates and gray dots are point sources detected by 2MASS and rejected by our JHK color-color filtering.}
         \label{fig:colordiagrams}
   \end{figure*}

\subsection{Color-color filtering}

In order to select only the point sources that could possibly be young stellar objects, we applied a color-color criteria to our near-infrared point source catalogues. To identify the nature of a protostar candidate detected by 2MASS, we compared the relative flux in the $J$, $K$ and $H$ photometric bands. We used the following empirical color criteria (\citealt{Xu11}), based on the idea that protostars have an infrared excess in the 1.235-2.159 $\upmu$m range that determines their position in the JHK color-color space and that is directly related to their evolutionary stage  (\citealt{Lada&Adams92}):
\begin{ceqn}
\begin{equation}
\begin{aligned}
(J-H) - 1,7 \cdot (H -K) + 0,450 = 0 \\
(J-H) - 0,493 \cdot (H -K) - 0,439 = 0 \\
(J-H) - 1,7 \cdot (H -K) + 1,400 = 0 \\
(J-H) = 0,2
\end{aligned}
\end{equation}
\end{ceqn}
Within the color-color diagram, this system of equations defines the color-color domains which mark out the different types of sources (see Fig. \ref{fig:colordiagrams}). This method allows to filter the sample of point sources and to produce a subset of different types of candidate young stellar objects (YSO):
\begin{itemize}
\item Classical T Tauri stars (CTTS).
\item Herbig Ae/Be stars (HAeBe).
\item Other YSOs.
\end{itemize}
Similarly, we used the following color-color criteria \citep{Koenig&Leisawitz14} to characterize the point sources from the relative flux in the three first bands $W_1$, $W_2$ and $W_3$. \citet{Fischer2016} proposed a slightly modified version of these filtering criteria that excludes an area of the color-color diagram that would be considered as Class II in the original diagram. That area is instead interpreted as shock emission, together with all the point sources that are beyond the left branches of the Class II and Class I areas. In this version of the diagram, there are two well-defined domains of color-color space: \\~\\
1.~ A first region is defined by the following system of 4 equations that constrain the infrared excess observed in Class I young stellar objects:
\begin{ceqn}
\begin{equation}
\begin{aligned}
\mathrm{W_2} - \mathrm{W_3} > 2,0 \\
\mathrm{W_1} - \mathrm{W_2} > -0,42 \cdot ( \mathrm{W_2} - \mathrm{W_3} ) + 2,2 \\
\mathrm{W_1} - \mathrm{W_2} > 0,46 \cdot ( \mathrm{W_2} - \mathrm{W_3} ) - 0,9 \\
\mathrm{W_2} - \mathrm{W_3} < 4,5
\end{aligned}
\end{equation}
\end{ceqn}
2.~ A second, adjacent region is defined by the following system of 5 equations that constrain the infrared excess observed in Class II young stellar objects:
\begin{ceqn}
\begin{equation}
\begin{aligned}
\mathrm{W_1} - \mathrm{W_2} > 0,25 \\
\mathrm{W_1} - \mathrm{W_2} < 0,71 \cdot ( \mathrm{W_2} - \mathrm{W_3} ) - 0,07 \\
\mathrm{W_1} - \mathrm{W_2} > -1,5 \cdot ( \mathrm{W_2} - \mathrm{W_3} ) + 2,1 \\
\mathrm{W_1} - \mathrm{W_2} > 0,46 \cdot ( \mathrm{W_2} - \mathrm{W_3} ) - 0,9 \\
\mathrm{W_2} - \mathrm{W_3} < 4,5
\end{aligned}
\end{equation}
\end{ceqn}

In the color-color diagram represented in Fig. \ref{fig:colordiagrams}, this system of 9 equations allows to distinguish two samples from the rest of the catalogue and enables to sort the YSO candidates into two distinct expected evolutionary stage based on their infrared excess:
\begin{itemize}
\item Class I protostar.
\item Class II protostar.
\end{itemize}

Additionnaly to the shocked emission area of the color-color diagram, \citet{Fischer2016} added the labels 'Polycyclic Aromatic Hydrocarbon (PAH) emission', 'AGB stars', 'Tr. (Transition) disks' and 'Star-forming galaxies' to differents parts of the diagram based on the expected emission of these objects in the WISE photometric bands. We defined arbitrary branches to separate the areas corresponding to each label and identified point sources falling into one of these area.

After the color-color filtering of the catalogues, the total amount of remaining point sources is 79/328 for 2MASS and 65/214 for WISE. 9 point sources are detected both by 2MASS and WISE. 1 point source is detected by Gaia and detected as a Class II protostar by WISE. The uncertainty due to extended emission and their identification in the color-color space is the following:  \\
\textit{2MASS}. 12.7\% of the protostar candidates found in the 2MASS PSC are contaminated by extended emission. 23 CTTS candidates, 8 HAeBe candidates and 48 other YSO candidates are identified based on their JHK photometric measurements. \\
\textit{WISE}. 76.9\% of the protostar candidates found in the AllWISE catalog are contaminated by extended emission towards the crowded shocked clump. 53 class I protostar candidates and 12 class II protostar candidates are identified in our field.

\subsection{Spectral index $\alpha$ of protostar candidates in the range 3.4-12 $\upmu$m}

We measured the spectral index $\alpha=d~\mathrm{log}(\lambda F_{\lambda})/d~\mathrm{log}\lambda$ of our protostar candidates detected by 2MASS and WISE to derive an identification based on their infrared spectral energy distribution (SED) \citep{Adams1987}. We used the classification system of \citet{Greene1994} to attribute an evolutionary stages to each point source and rule out the sources characterized by a flat SED:
\begin{itemize}
\item Class I: $\alpha \geq 0.3$
\item Flat-SED: $-0.3 \leq \alpha \leq 0.3$
\item Class II: $-1.6 \leq \alpha \leq -0.3$
\item Class III: $\alpha < -1.6$
\end{itemize}
The photometric fluxes of the bands W$_1$, W$_2$ and W$_3$ were used to compute the slope of the SED in the range 3.4-12 $\upmu$m for each protostar candidate found precedently and given in Tab. \ref{table:WISEphot} where we compare the classification based on color-color diagram with the results obtained with the measurement of $\alpha$. We show the SEDs of 17 protostar candidates in Fig. \ref{fig:SEDs}. With respect to the value of the infrared spectral index $\alpha$, 33 Flat-SED point sources are found within our sample of 65 AllWISE protostar candidates (ID 9-15 and ID 40-65). Every single point sources classified as Class I by our measurement of $\alpha$ were also identified as Class I using the color-color diagram (ID 1-4 and ID 16-34). On the contrary, there is only partial agreement between the two methods for the identification of Class II protostar candidates: excluding Flat-SED sources, 75\% of the identifications were confirmed by both approaches (ID 5-8 and ID 35-39). No Class III were identified in our sample.

   \begin{figure*}[]
   \centering
   \resizebox{0.45\hsize}{!}
   {\includegraphics[width=\hsize, trim={0cm 1.5cm 0cm 1.5cm},clip]{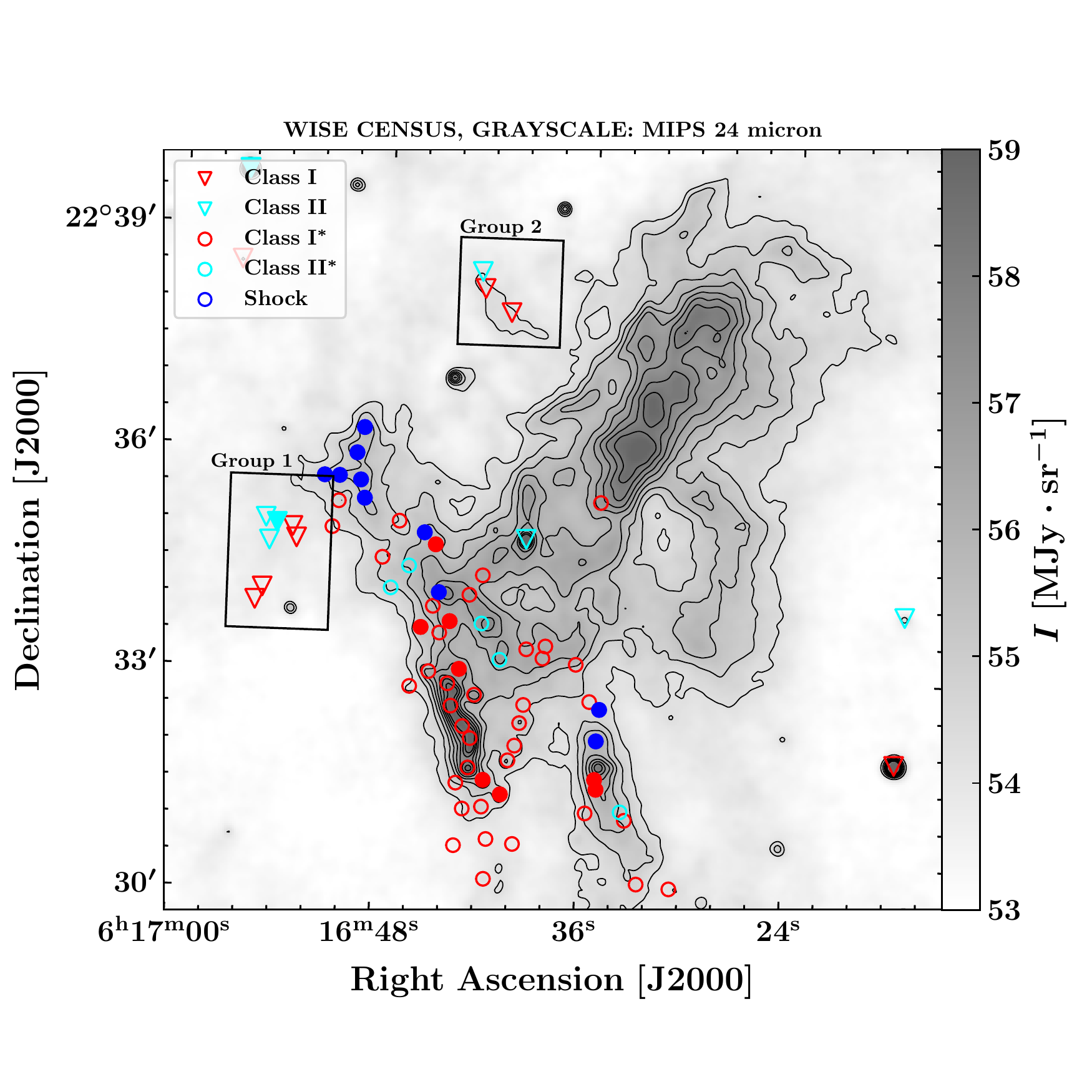}}
   \resizebox{0.45\hsize}{!}
   {\includegraphics[width=\hsize, trim={0cm 1.5cm 0cm 1.9cm},clip]{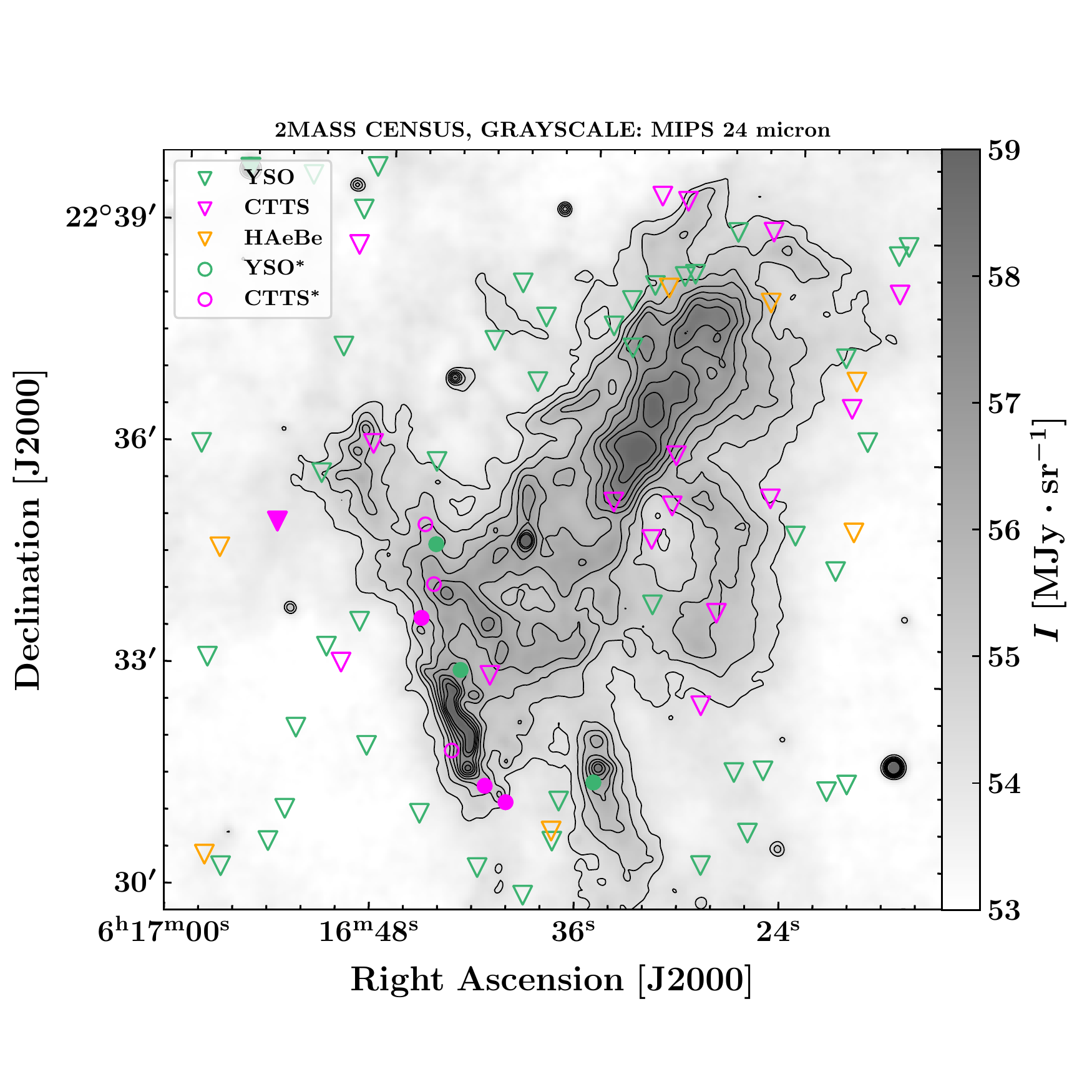}}
   \resizebox{0.45\hsize}{!}
   {\includegraphics[width=\hsize, trim={0.3cm 1.2cm 0cm 1.5cm},clip]{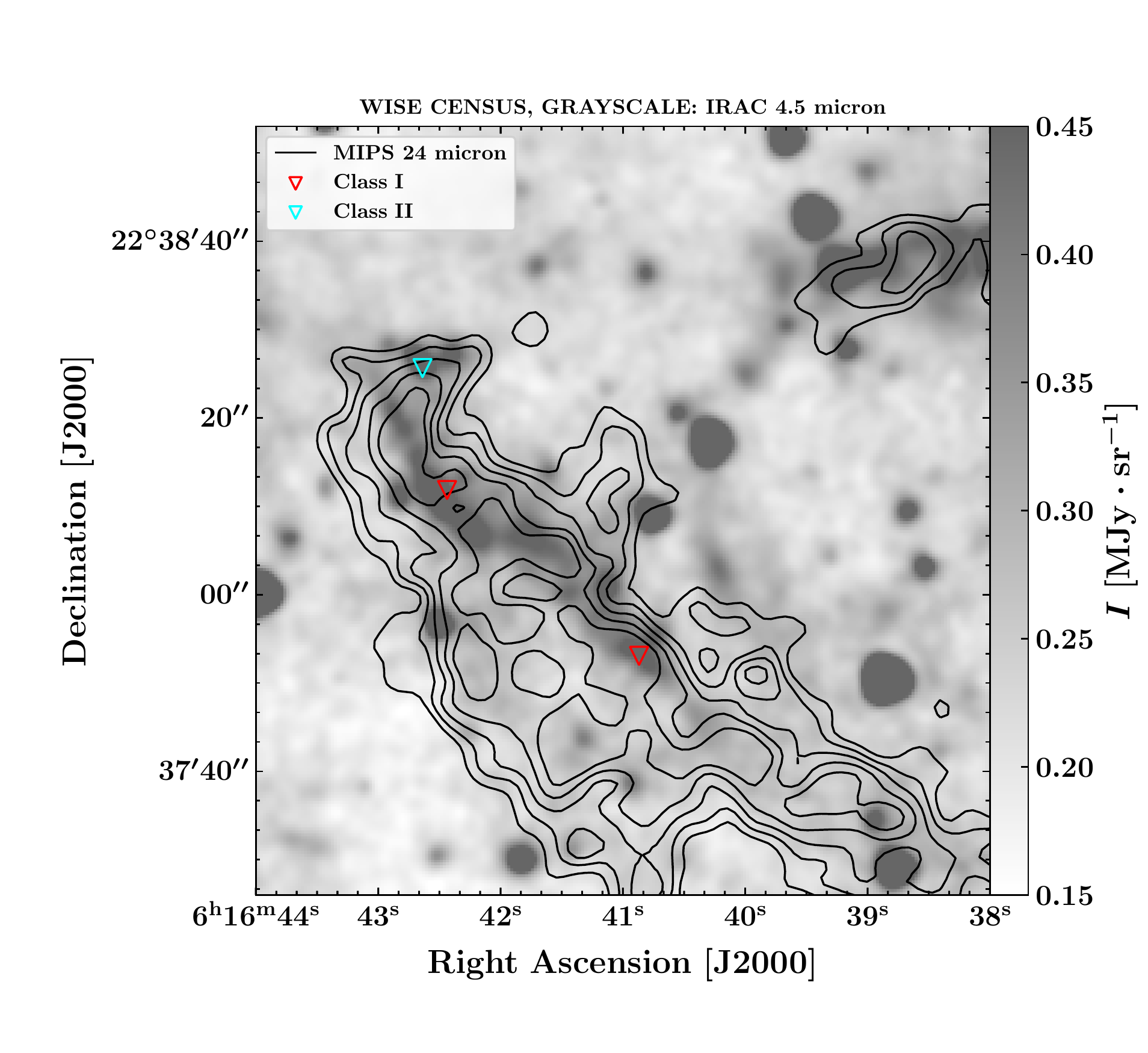}}
   \resizebox{0.45\hsize}{!}
   {\includegraphics[width=\hsize, trim={0cm 1.5cm 0cm 1.9cm},clip]{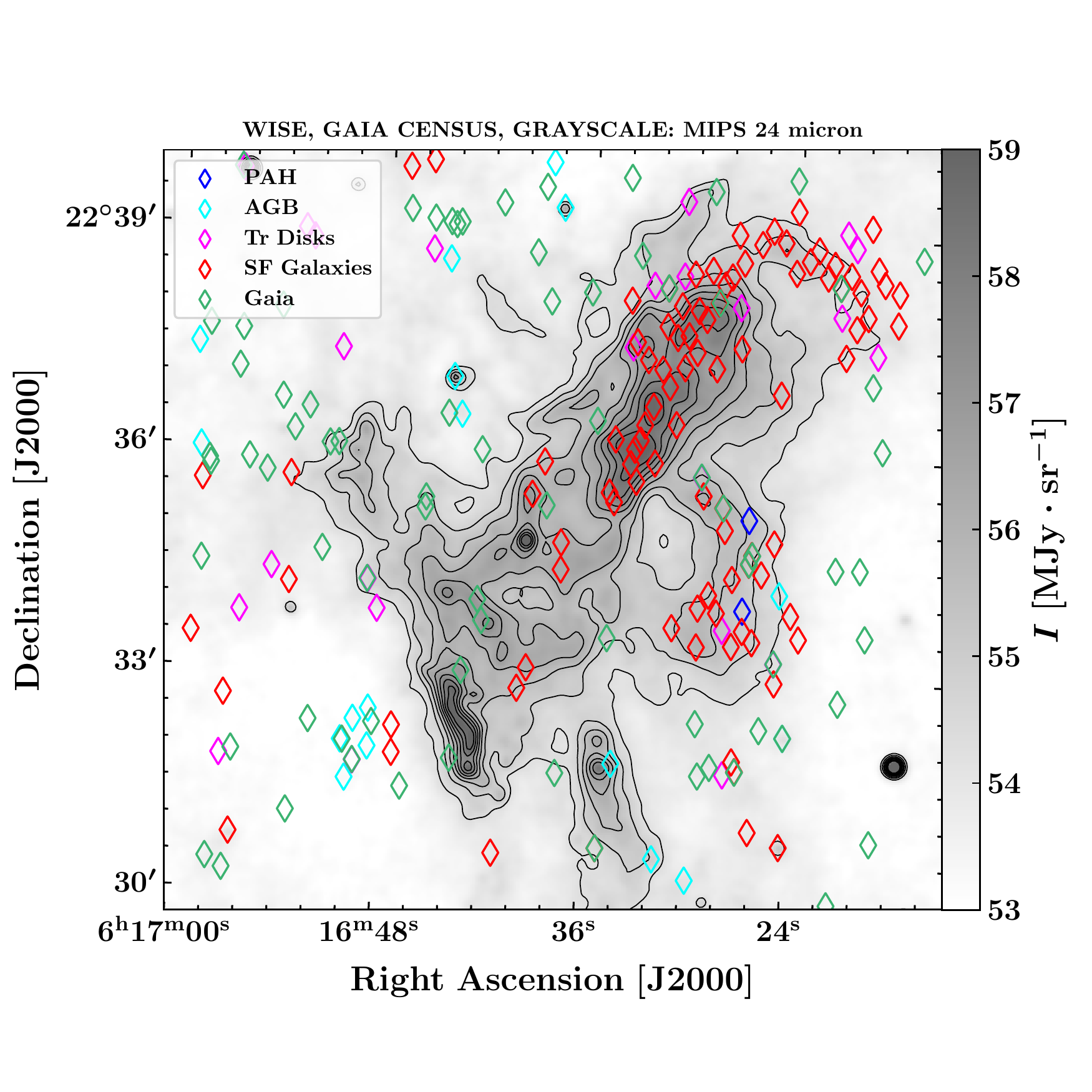}}
   \resizebox{0.45\hsize}{!}
   {\includegraphics[width=\hsize, trim={3cm 1.2cm 0cm 1cm},clip]{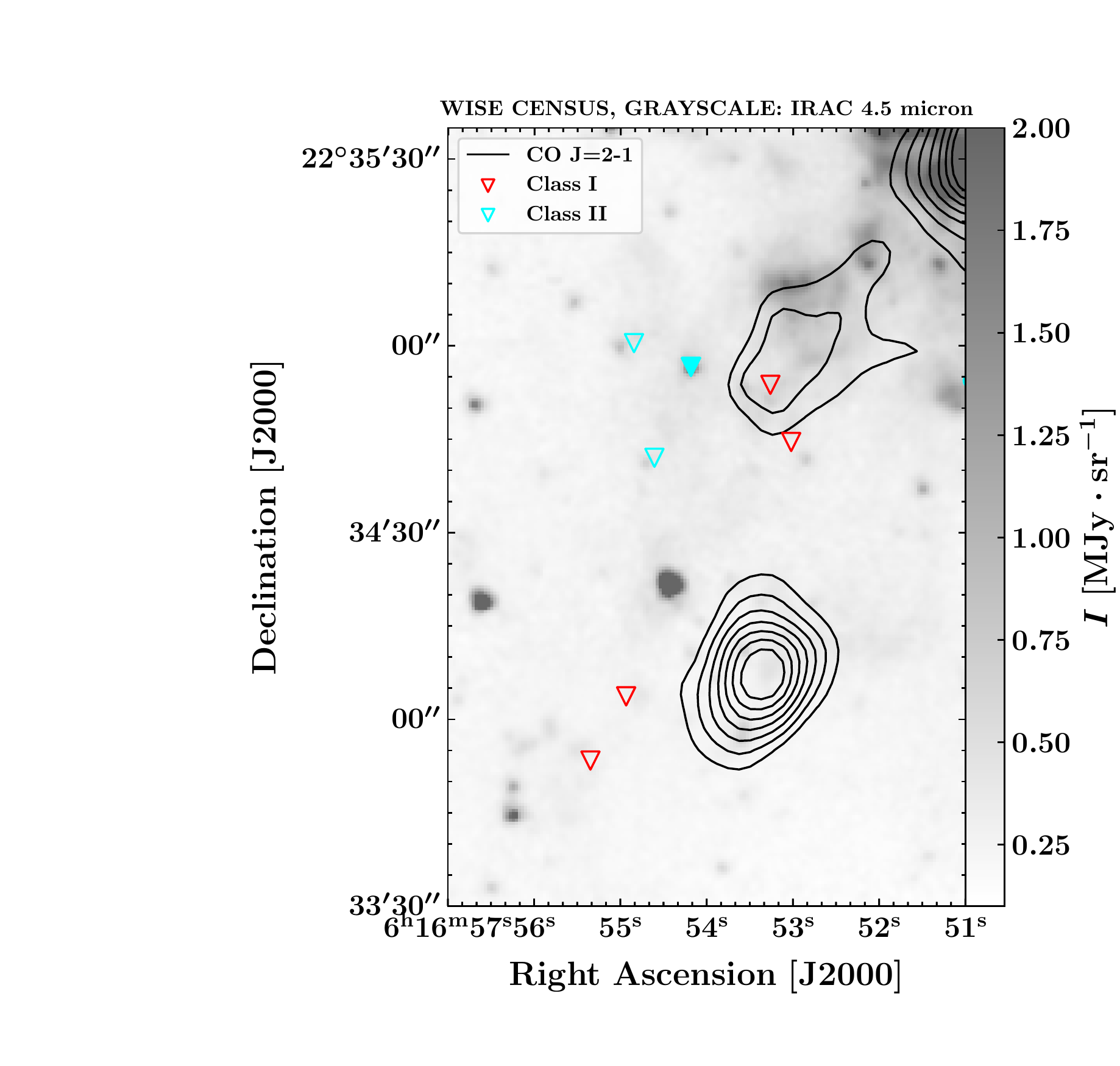}}
      \caption{\textit{Upper-left panel:} WISE census of Class I and Class II protostar candidates, with shock emission indicated. Based on the selection criteria by \citet{Fischer2016}. \textit{Upper-right panel:} 2MASS census of YSO, CTTS and HAeBe candidates. Based on the selection criteria by \citet{Xu11}. \textit{Center-left panel:} Close-up on the box corresponding to group 2. \textit{Center-right panel:} Transition disks, AGB star, star-forming galaxies emission and PAH emission candidates, based on the selection criteria by \citet{Fischer2016}; and Gaia point sources. \textit{Lower panel:} Close-up on the box corresponding to group 1. On all panels, triangle-shaped markers correspond to well-detected point sources, and circle-shaped markers (labels marked with a \lq*')  correspond to point sources that were flagged for contamination by extended emission. Filled markers represent point sources that were detected both by 2MASS and WISE.}
         \label{fig:protostars-scatter}
   \end{figure*}

\subsection{Gaia point sources}

We used Gaia data (see Tab. \ref{table:pointsourcecatalogue} for exact photometric parameters) to complete our point source census with optical sources and search for multiple detection by Gaia, WISE and 2MASS.  We used the ESA Gaia science archive to obtain all entries in a $10~\mathrm{arcmin}$ sized square box around the field center. A total of 468 point sources were found within our 10$^\prime \times$10$^\prime$ field of observations. In order to estimate the amount of optical point sources that might be evolved stars physically associated with the SNR, we also used a distance criterion to filter the result. Assuming that the distance of IC443 is $\sim$1.9 kpc, we rejected the point sources with relative distance greater than 2500 pc or lower than 1500 pc, based on the lower and upper bound on the confidence interval for the estimated distance determined by parallax measurements \citet{BailerJones18}. There are 16 point sources in the interval 1500-2500 pc. We repeated the same process with a different distance criterion in order to assess the variability of the result with respect to the amplitude of distance interval applied as a filter. We found a total of 40 optical point sources which relative distance is greater than 1250 pc and lower than 2750 pc. We checked if our samples of optical point sources is correlated with candidate protostars found in our WISE and 2MASS census:
\begin{enumerate}
\item Among the 16 Gaia point sources found with the first distance check ($1500\leq d \leq 2500$), 3 point sources were also detected by 2MASS, and 1 was spatially correlated with a WISE point source.
\item For the 40 Gaia point sources detected with the second distance check ($1250\leq d \leq 2750$), 5 point sources were also detected by 2MASS, and 5 with a WISE point source.
\end{enumerate}
Hence a fraction of the YSO detections by WISE and 2MASS are confirmed by Gaia, and the distance estimates allow to strenghten the assumption that these sources are physically associated with the extended G region

\subsection{Discussion}

A higher concentration of protostar candidates was found within the shocked clump by our WISE census, suggesting that this might be a star-forming region. However, the third photometric band of WISE W$_3$ ($\lambda$=11.56 $\upmu$m, $\Delta \lambda$=5.51 $\upmu$m) is partially sensitive to three pure rotational transitions of $\mathrm{H_2}$, namely the transitions S(2) ($\lambda$=12.28 $\upmu$m), S(3) ($\lambda$=9.665 $\upmu$m) and S(4) ($\lambda$=8.026 $\upmu$m). Moreover, several H$_2$ rovibrational transitions are detected by WISE bands. Hence our detections might be contaminated by the emission of warm $\mathrm{H_2}$ clumps that are excited by the propagation of the shock. \citet{Gutermuth2009} reported that when trying to build YSO samples, unresolved knots of shock emission are regularly detected in all IRAC bands (3.6 - 8 $\upmu$m). As a consequence, warm and shocked condensations of $\mathrm{H_2}$ with sizes smaller than the spatial resolution might be detected as infrared point sources by the survey. \citet{Fischer2016} provides a quantitative distinction between shock emission and the signature of protostar candidates in the WISE color-color diagram, yet none of the point sources located along the bright shocked clump are identified as shock emission, casting doubts about the reliability of this criterion to effectively detect shocks. Also, several point sources associated with faint, extended \textit{Spitzer}-MIPS 24 $\upmu$m northeast of the brightest shocked clump are identified as shock emission. The clumpy and filamentary morphology of the structure in which these point sources are embedded, seen in \textit{Spitzer}-IRAC 4.5 $\upmu$m, is similar to that of the main shocked clump. This spatial disparity in the identification of point sources might indicate that this region is dominated by shock emission, whereas the point sources detected in the shocked clump are dominated by protostar emission despite the contamination by extended shock structures. The 2MASS census is also biased by the emission of warm H$_2$, as the band K ($\lambda$=2.159 $\upmu$m, $\Delta \lambda$=0.262 $\upmu$m) is sensitive to the rovibrational line $\varv=1-0$ S(1) at 2.12 $\upmu$m. The clumpy and bright extended emission displayed in the band K map of the southern ridge makes it clear that the flux of point sources might be contaminated by extended emission, in particular along the bright shocked clump. \citet{Rho2001} confirmed that the band K extended emission from this region of the remnant is dominated by H$_2$ emission, and to a lesser extent the bands J and H. On the one hand, if the flux of the band K is overestimated due to H$_2$ emission, then the correction would lead to a displacement along the H-K axis, which means that more protostar candidates would be detected towards the shocked clump where the emission of warm H$_2$ is significant. This might explain why the higher density of protostar candidates found in the WISE census is not reproduced by our 2MASS census. On the other hand, it is possible that a number of point sources detected by 2MASS are unresolved knots of shocked H$_2$, in a similar fashion as what is expected for WISE.

A greater amount of point sources is found on the edges of the 24$\upmu$m bright knot that is spatially associated with the eastern edge of the quiescent molecular cloudlet. Excluding the shocked clump, there is an anti-correlation between both 2MASS and WISE sources and the MIPS-24$\upmu$m flux map, in particular in the vicinity of the cloudlet. We suppose that the absence of optical point sources within this region is due to the extinction caused by this massive dark cloud. No IR point sources are found at the center of the gap within the ring-like structure. 

Interestingly, the few point sources that were not flagged for contamination by extended sources are primarily organized in two clusters in the eastern and north-eastern part of the field (group 1 and group 2, Fig. \ref{fig:protostars-scatter}). All these candidates were positively identified by both methods as Class I and Class II protostars. The Class I protostar candidates of group 2 are well associated with \textit{Spitzer}-IRAC filamentary structures at 4.5 $\upmu$m, at the edge of a faint structure detected by \textit{Spitzer}-MIPS at 24 $\upmu$m. This bright filament can be either part of the SNR shock or unresolved outflows. Most IR sources with protostar signatures are spatially correlated with the molecular shell in IC443G, suggesting enhanced star formation along the rims. Given the protostellar collapse phase timescale of $\sim$10$^5$ yr \citep{Lefloch&Lazareff1994} we dot not establish any causal relation between the SNR shocks and the formation of Class I and Class II protostars. Based on our color-color census for the 65 protostar candidates detected by WISE, the Class II / Class I ratio is equal to 0.23. Following \citet{Dunham2015}, we assumed a Class II duration of 2 Myr and used the standard method to infer the age of a stellar population from this ratio (\citealt{Wilking1989}, \citealt{Evans2009}). We obtain an age of $\sim$500 kyr for the YSO population found in the extended G region, which is not consistent with the scenario in which the formation of these YSOs was triggered by interaction of the SNR with its environment. It is likely that the collapse of molecular clouds in the vicinity of the high-mass progenitor was triggered by the compression driven by stellar winds \citep{Xu11}. To find evidence of enhanced star formation in the SNR shocks would require to detect deeply embedded prestellar cores, which is not attainable with the data at our disposal. In the future, high-resolution ($\sim$1$^{\prime \prime}$) sub-mm and mm observations will be required to disentangle the kinematics of large-scale shock structure and outflows, and to eventually uncover prestellar cores in the extended G region (e.g. as \citet{Motte18} did for the W43-MM1 star forming region). Still, our sample of protostar candidates provides a unique opportunity to study star formation in an environment that is subject to intense $\gamma$-ray flux and shock dynamical feedback. The protons accelerated by young protostars might also provide a source of fresh mid-energy CRs (up to 10-13 GeV and 26-37 GeV respectively for jet acceleration and protostellar surface shock acceleration in lower-mass stars, \citealt{Padovani16}). With jet velocities up to 1000 km s$^{-1}$, high-mass protostellar shocks could contribute to the TeV $\gamma$-ray peak in the extended G region.

Point sources identified as star-forming galaxies from their location in the color-color diagram display a significant correlation with the spiral-shaped, bright and extended infrared features mapped by \textit{Spitzer}-MIPS at 24 $\upmu$m. Most of these sources might as well be identified as PAH emission, as our arbitrary criterion of selection puts a significant uncertainty on the cut between the two categories of sources, in particular along  the W$_1$ - W$_2$ axis \citep{Koenig2012}. In fact, it is more likely that these point sources are associated with PAH, stellar or protostellar emission. With our current criterion, only 2 sources identified as PAH emission are found and located in the same area as point sources that were labelled 'star-forming galaxies'. If most of these detections are actually arising from PAH emission, this result would support the assumption that the bright infrared extended emission arise from a dark dust lane that is spatially associated with the molecular cloudlet detected by our CO observations.

In comparison to previous work on the star formation in IC443:
\begin{itemize}
\item Based on 2MASS point source catalogues, \citet{Xu11} found a total of 1666 YSO candidates, 154 CTTS and 419 HAeBe stars in a search circle around IC443 within a 25$^\prime$ radius. Their candidates were mostly concentrated around the CO molecular shell, in particular towards \textit{i.} the clump C; \textit{ii.} the extended G region where their YSO candidates are spatially correlated with the shocked clump. They proposed that the formation of these YSOs have been triggered by the stellar wind of the IC443 progenitor.
\item Based on 2MASS and WISE point source catalogues, \citet{Su2014} found a total of 98 YSO candidates. They proposed a sample of 62 YSO candidates concentrated along the boundary of the radio shell. They also proposed that the formation of these protostars is likely to have been triggered by the stellar winds of the SNR progenitor. In contrast with our results and the findings of \citet{Xu11}, their distribution does not show a strong correlation with the molecular structures in the extended G region.
\end{itemize}

\section{Summary}\label{conclusion}
\begin{enumerate}
\item In this work we report on new $10^\prime \times 10 ^\prime$ fully sampled maps of the extended IC443G region for the first three rotational transitions of both $^{12}$CO and $^{13}$CO, as well as the two first rotational transitions of C$^{18}$O obtained with the IRAM-30m and APEX telescope. These extensive maps allow to probe the position of the TeV $\gamma$-ray peak in IC443G and its surroundings with unprecedented spectral and angular resolution at millimeter and sub-millimeter wavelengths. 
\item We proposed a description of the morphology and kinematics of the extended G region based on the definition of four main molecular structures: \textit{(i.)} a shocked clump; \textit{(ii.)} a quiescent, "ring-like" structure that might be spurious; \textit{(iii.)} a quiescent cloudlet that is spatially connected to the ring-like structure and \textit{(iv.)} a shocked knot.
\item The comparison of our data histograms with modified LTE radiative transfer models revealed enhanced $^{12}$CO J=2--1/1--0  and J=3--2/2--1 ratios towards the shocked clump, in particular within the high-velocity wings of the emission lines that are tracing warm CO (40 - 120 K). We provide a rough estimate of the isotopic ratio based on the best match between modified LTE models and the measured intensity of the $^{12}$CO, $^{13}$CO and C$^{18}$O J=1--0 line.
\item We measured the mass of the molecular gas in the extended G region using pixel-per-pixel, channel-per-channel population diagrams corrected for optical depth. We obtained a total molecular mass estimate of $\sim$230 $M_{\odot}$, $\sim$90 $M_{\odot}$, $\sim$210 $M_{\odot}$ and $\sim$4 $M_{\odot}$ respectively for the cloudlet, ring-like structure, shocked clump and shocked knot. We also measured a mass of $\sim$1100 $M_{\odot}$ for the ambient gas in the [-6.5, -1.5] km s$^{-1}$ $\varv_{\mathrm{LSR}}$ range. The estimate of the mass depends on the adopted $^{12}$CO/$^{13}$CO isotopic ratio, but it is established that a total molecular mass 0.9-3.1$\times 10^3$ $M_{\odot}$ is available to interact with CRs via pion-decay in the extended G region.
\item  We proposed a second estimate of the mass using the LVG assumption with a grid of RADEX models. The $\chi^2$-minimization of the grid of models with respect to data in each structure yields results that are in partial agreement with the previous measurements, with an estimate that is systematically lower than the LTE estimate, except for the mass measurements of the high-velocity wings towards the shocked clump, which are higher by a factor $\sim$3 .
\item We studied the spectral energy distribution of infrared and optical point sources using the 2MASS, WISE and Gaia catalogues of point sources. Using color-color diagrams we determined a sample of protostar candidates in the extended G region. 144 protostar candidates are found in the region, and 16 stars are found in the Gaia census but it is worth to note that a large fraction of these point sources are contaminated by extended emission (12.7\% for 2MASS, and 76.9\% for WISE). Based on the spatial distribution of these candidates, we propose three sites of possible star formation in IC443G (groups 1 and 2 in Fig. \ref{fig:protostars-scatter}, and the shocked clump). These protostars might constitute a source of fresh CRs.
\end{enumerate}

\begin{acknowledgements}
This work was supported by the Programme National "Physique et
Chimie du Milieu Interstellaire" (PCMI) of CNRS/INSU with INC/INP co-funded
by CEA and CNES.

This publication is based on observations made with ESO Telescopes at the Atacama Pathfinder EXperiment under programme M-0102.F-9508A-2018.

This work has made use of data from the European Space Agency (ESA) mission {\it Gaia} (\url{https://www.cosmos.esa.int/gaia}), processed by the {\it Gaia} Data Processing and Analysis Consortium (DPAC, \url{https://www.cosmos.esa.int/web/gaia/dpac/consortium}). Funding for the DPAC has been provided by national institutions, in particular the institutions participating in the {\it Gaia} Multilateral Agreement. 

This research has made use of the NASA/ IPAC Infrared Science Archive, which is operated by the Jet Propulsion Laboratory, California Institute of Technology, under contract with the National Aeronautics and Space Administration. 

This publication makes use of data products from the Wide-field Infrared Survey Explorer, which is a joint project of the University of California, Los Angeles, and the Jet Propulsion Laboratory/California Institute of Technology, funded by the National Aeronautics and Space Administration. 

This publication makes use of data products from the Two Micron All Sky Survey, which is a joint project of the University of Massachusetts and the Infrared Processing and Analysis Center/California Institute of Technology, funded by the National Aeronautics and Space Administration and the National Science Foundation. 
     
This research made use of Astropy,\footnote{http://www.astropy.org} a community-developed core Python package for Astronomy \citep{AstropyI, AstropyII}. 
    
This research has made use of NASA's Astrophysics Data System. 

M.P. acknowledges funding from the INAF PRIN-SKA 2017 program 1.05.01.88.04.

      \nocite{Matplotlib07}
      \nocite{Numpy11}
      \nocite{Scipy01}
\end{acknowledgements}

\bibliographystyle{aa}
\nocite{*}
\bibliography{biblio}

\begin{thebibliography}{242}
\expandafter\ifx\csname natexlab\endcsname\relax\def\natexlab#1{#1}\fi

\bibitem[{{Abdo} {et~al.}(2010){Abdo}, {Ackermann}, {Ajello}, {Baldini},
  {Ballet}, {Barbiellini}, {Bastieri}, {Baughman}, {Bechtol}, {Bellazzini},
  {Berenji}, {Blandford}, {Bloom}, {Bonamente}, {Borgland}, {Bregeon}, {Brez},
  {Brigida}, {Bruel}, {Burnett}, {Buson}, {Caliandro}, {Cameron}, {Caraveo},
  {Casand jian}, {Cecchi}, {{\c{C}}elik}, {Chekhtman}, {Cheung}, {Chiang},
  {Cillis}, {Ciprini}, {Claus}, {Cohen-Tanugi}, {Cominsky}, {Conrad}, {Cutini},
  {Dermer}, {de Angelis}, {de Palma}, {Silva}, {Drell}, {Drlica-Wagner},
  {Dubois}, {Dumora}, {Farnier}, {Favuzzi}, {Fegan}, {Focke}, {Fortin},
  {Frailis}, {Fukazawa}, {Funk}, {Fusco}, {Gargano}, {Gasparrini}, {Gehrels},
  {Germani}, {Giavitto}, {Giebels}, {Giglietto}, {Giordano}, {Glanzman},
  {Godfrey}, {Grenier}, {Grondin}, {Grove}, {Guillemot}, {Guiriec}, {Hanabata},
  {Harding}, {Hayashida}, {Hughes}, {Jackson}, {J{\'o}hannesson}, {Johnson},
  {Johnson}, {Johnson}, {Kamae}, {Katagiri}, {Kataoka}, {Kawai}, {Kerr},
  {Kn{\"o}dlseder}, {Kocian}, {Kuss}, {Land e}, {Latronico}, {Lee},
  {Lemoine-Goumard}, {Longo}, {Loparco}, {Lott}, {Lovellette}, {Lubrano},
  {Madejski}, {Makeev}, {Mazziotta}, {Meurer}, {Michelson}, {Mitthumsiri},
  {Moiseev}, {Monte}, {Monzani}, {Morselli}, {Moskalenko}, {Murgia},
  {Nakamori}, {Nolan}, {Norris}, {Nuss}, {Ohsugi}, {Orlando}, {Ormes}, {Ozaki},
  {Paneque}, {Panetta}, {Parent}, {Pelassa}, {Pepe}, {Pesce-Rollins}, {Piron},
  {Porter}, {Rain{\`o}}, {Rando}, {Razzano}, {Reimer}, {Reimer}, {Reposeur},
  {Rochester}, {Rodriguez}, {Romani}, {Roth}, {Ryde}, {Sadrozinski}, {Sanchez},
  {Sander}, {Saz Parkinson}, {Scargle}, {Sgr{\`o}}, {Siskind}, {Smith},
  {Smith}, {Spandre}, {Spinelli}, {Strickman}, {Strong}, {Suson}, {Tajima},
  {Takahashi}, {Takahashi}, {Tanaka}, {Thayer}, {Thayer}, {Thompson},
  {Tibaldo}, {Torres}, {Tosti}, {Tramacere}, {Uchiyama}, {Usher}, {Van Etten},
  {Vasileiou}, {Venter}, {Vilchez}, {Vitale}, {Waite}, {Wang}, {Winer}, {Wood},
  {Ylinen}, \& {Ziegler}}]{Abdo2010}
{Abdo}, A.~A., {Ackermann}, M., {Ajello}, M., {et~al.} 2010, \apj, 712, 459

\bibitem[{{Acciari} {et~al.}(2009){Acciari}, {Aliu}, {Arlen}, {Aune},
  {Bautista}, {Beilicke}, {Benbow}, {Bradbury}, {Buckley}, {Bugaev}, {Butt},
  {Byrum}, {Cannon}, {Celik}, {Cesarini}, {Chow}, {Ciupik}, {Cogan}, {Colin},
  {Cui}, {Daniel}, {Dickherber}, {Duke}, {Dwarkadas}, {Ergin}, {Fegan},
  {Finley}, {Finnegan}, {Fortin}, {Fortson}, {Furniss}, {Gall}, {Gibbs},
  {Gillanders}, {Godambe}, {Grube}, {Guenette}, {Gyuk}, {Hanna}, {Hays},
  {Holder}, {Horan}, {Hui}, {Humensky}, {Imran}, {Kaaret}, {Karlsson},
  {Kertzman}, {Kieda}, {Kildea}, {Konopelko}, {Krawczynski}, {Krennrich},
  {Lang}, {LeBohec}, {Maier}, {McCann}, {McCutcheon}, {Millis}, {Moriarty},
  {Ong}, {Otte}, {Pandel}, {Perkins}, {Pohl}, {Quinn}, {Ragan}, {Reyes},
  {Reynolds}, {Roache}, {Rose}, {Schroedter}, {Sembroski}, {Smith}, {Steele},
  {Swordy}, {Theiling}, {Toner}, {Valcarcel}, {Varlotta}, {Vassiliev},
  {Vincent}, {Wagner}, {Wakely}, {Ward}, {Weekes}, {Weinstein}, {Weisgarber},
  {Williams}, {Wissel}, {Wood}, \& {Zitzer}}]{Acciari2009}
{Acciari}, V.~A., {Aliu}, E., {Arlen}, T., {et~al.} 2009, \apjl, 698, L133

\bibitem[{{Ackermann} {et~al.}(2013){Ackermann}, {Ajello}, {Allafort},
  {Baldini}, {Ballet}, {Barbiellini}, {Baring}, {Bastieri}, {Bechtol},
  {Bellazzini}, {Bland ford}, {Bloom}, {Bonamente}, {Borgland }, {Bottacini},
  {Brandt}, {Bregeon}, {Brigida}, {Bruel}, {Buehler}, {Busetto}, {Buson},
  {Caliandro}, {Cameron}, {Caraveo}, {Casandjian}, {Cecchi}, {{\c{C}}elik},
  {Charles}, {Chaty}, {Chaves}, {Chekhtman}, {Cheung}, {Chiang}, {Chiaro},
  {Cillis}, {Ciprini}, {Claus}, {Cohen-Tanugi}, {Cominsky}, {Conrad}, {Corbel},
  {Cutini}, {D'Ammando}, {de Angelis}, {de Palma}, {Dermer}, {do Couto e
  Silva}, {Drell}, {Drlica-Wagner}, {Falletti}, {Favuzzi}, {Ferrara},
  {Franckowiak}, {Fukazawa}, {Funk}, {Fusco}, {Gargano}, {Germani},
  {Giglietto}, {Giommi}, {Giordano}, {Giroletti}, {Glanzman}, {Godfrey},
  {Grenier}, {Grondin}, {Grove}, {Guiriec}, {Hadasch}, {Hanabata}, {Harding},
  {Hayashida}, {Hayashi}, {Hays}, {Hewitt}, {Hill}, {Hughes}, {Jackson},
  {Jogler}, {J{\'o}hannesson}, {Johnson}, {Kamae}, {Kataoka}, {Katsuta},
  {Kn{\"o}dlseder}, {Kuss}, {Lande}, {Larsson}, {Latronico}, {Lemoine-Goumard},
  {Longo}, {Loparco}, {Lovellette}, {Lubrano}, {Madejski}, {Massaro}, {Mayer},
  {Mazziotta}, {McEnery}, {Mehault}, {Michelson}, {Mignani}, {Mitthumsiri},
  {Mizuno}, {Moiseev}, {Monzani}, {Morselli}, {Moskalenko}, {Murgia},
  {Nakamori}, {Nemmen}, {Nuss}, {Ohno}, {Ohsugi}, {Omodei}, {Orienti},
  {Orlando}, {Ormes}, {Paneque}, {Perkins}, {Pesce-Rollins}, {Piron}, {Pivato},
  {Rain{\`o}}, {Rando}, {Razzano}, {Razzaque}, {Reimer}, {Reimer}, {Ritz},
  {Romoli}, {S{\'a}nchez-Conde}, {Schulz}, {Sgr{\`o}}, {Simeon}, {Siskind},
  {Smith}, {Spand re}, {Spinelli}, {Stecker}, {Strong}, {Suson}, {Tajima},
  {Takahashi}, {Takahashi}, {Tanaka}, {Thayer}, {Thayer}, {Thompson},
  {Thorsett}, {Tibaldo}, {Tibolla}, {Tinivella}, {Troja}, {Uchiyama}, {Usher},
  {Vandenbroucke}, {Vasileiou}, {Vianello}, {Vitale}, {Waite}, {Werner},
  {Winer}, {Wood}, {Wood}, {Yamazaki}, {Yang}, \& {Zimmer}}]{Ackermann2013}
{Ackermann}, M., {Ajello}, M., {Allafort}, A., {et~al.} 2013, Science, 339, 807

\bibitem[{{Adams} {et~al.}(1987){Adams}, {Lada}, \& {Shu}}]{Adams1987}
{Adams}, F.~C., {Lada}, C.~J., \& {Shu}, F.~H. 1987, \apj, 312, 788

\bibitem[{{Aharonian} {et~al.}(2006){Aharonian}, {Akhperjanian}, {Bazer-Bachi},
  {Beilicke}, {Benbow}, {Berge}, {Bernl{\"o}hr}, {Boisson}, {Bolz}, {Borrel},
  {Braun}, {Brown}, {B{\"u}hler}, {B{\"u}sching}, {Carrigan}, {Chadwick},
  {Chounet}, {Cornils}, {Costamante}, {Degrange}, {Dickinson},
  {Djannati-Ata{\"\i}}, {O'C. Drury}, {Dubus}, {Egberts}, {Emmanoulopoulos},
  {Espigat}, {Feinstein}, {Ferrero}, {Fiasson}, {Fontaine}, {Funk}, {Funk},
  {F{\"u}{\ss}ling}, {Gallant}, {Giebels}, {Glicenstein}, {Goret},
  {Hadjichristidis}, {Hauser}, {Hauser}, {Heinzelmann}, {Henri}, {Hermann},
  {Hinton}, {Hoffmann}, {Hofmann}, {Holleran}, {Horns}, {Jacholkowska}, {de
  Jager}, {Kendziorra}, {Kh{\'e}lifi}, {Komin}, {Konopelko}, {Kosack},
  {Latham}, {Le Gallou}, {Lemi{\`e}re}, {Lemoine-Goumard}, {Lohse}, {Martin},
  {Martineau-Huynh}, {Marcowith}, {Masterson}, {Maurin}, {McComb}, {Moulin},
  {de Naurois}, {Nedbal}, {Nolan}, {Noutsos}, {Orford}, {Osborne}, {Ouchrif},
  {Panter}, {Pelletier}, {Pita}, {P{\"u}hlhofer}, {Punch}, {Raubenheimer},
  {Raue}, {Rayner}, {Reimer}, {Reimer}, {Ripken}, {Rob}, {Rolland }, {Rowell},
  {Sahakian}, {Santangelo}, {Saug{\'e}}, {Schlenker}, {Schlickeiser},
  {Schr{\"o}der}, {Schwanke}, {Schwarzburg}, {Shalchi}, {Sol}, {Spangler},
  {Spanier}, {Steenkamp}, {Stegmann}, {Superina}, {Tavernet}, {Terrier},
  {Th{\'e}oret}, {Tluczykont}, {van Eldik}, {Vasileiadis}, {Venter}, {Vincent},
  {V{\"o}lk}, {Wagner}, \& {Ward}}]{Aharonian2006}
{Aharonian}, F., {Akhperjanian}, A.~G., {Bazer-Bachi}, A.~R., {et~al.} 2006,
  \aap, 460, 365

\bibitem[{{Akabane}(1966)}]{Akabane1966}
{Akabane}, K. 1966, \pasj, 18, 96

\bibitem[{{Alarie} \& {Drissen}(2019)}]{Alarie&Drissen2019}
{Alarie}, A. \& {Drissen}, L. 2019, \mnras, 489, 3042

\bibitem[{{Albert} {et~al.}(2007){Albert}, {Aliu}, {Anderhub}, {Antoranz},
  {Armada}, {Baixeras}, {Barrio}, {Bartko}, {Bastieri}, {Becker}, {Bednarek},
  {Berger}, {Bigongiari}, {Biland}, {Bock}, {Bordas}, {Bosch-Ramon}, {Bretz},
  {Britvitch}, {Camara}, {Carmona}, {Chilingarian}, {Coarasa}, {Commichau},
  {Contreras}, {Cortina}, {Costado}, {Curtef}, {Danielyan}, {Dazzi}, {De
  Angelis}, {Delgado}, {de los Reyes}, {De Lotto}, {Domingo-Santamar{\'\i}a},
  {Dorner}, {Doro}, {Errando}, {Fagiolini}, {Ferenc}, {Fern{\'a}ndez}, {Firpo},
  {Flix}, {Fonseca}, {Font}, {Fuchs}, {Galante}, {Garc{\'\i}a-L{\'o}pez},
  {Garczarczyk}, {Gaug}, {Giller}, {Goebel}, {Hakobyan}, {Hayashida},
  {Hengstebeck}, {Herrero}, {H{\"o}hne}, {Hose}, {Hsu}, {Jacon}, {Jogler},
  {Kosyra}, {Kranich}, {Kritzer}, {Laille}, {Lindfors}, {Lombardi}, {Longo},
  {L{\'o}pez}, {L{\'o}pez}, {Lorenz}, {Majumdar}, {Maneva}, {Mannheim},
  {Mansutti}, {Mariotti}, {Mart{\'\i}nez}, {Mazin}, {Merck}, {Meucci}, {Meyer},
  {Mirand a}, {Mirzoyan}, {Mizobuchi}, {Moralejo}, {Nieto}, {Nilsson},
  {Ninkovic}, {O{\~n}a-Wilhelmi}, {Otte}, {Oya}, {Paneque}, {Panniello},
  {Paoletti}, {Paredes}, {Pasanen}, {Pascoli}, {Pauss}, {Pegna}, {Persic},
  {Peruzzo}, {Piccioli}, {Prandini}, {Puchades}, {Raymers}, {Rhode},
  {Rib{\'o}}, {Rico}, {Rissi}, {Robert}, {R{\"u}gamer}, {Saggion}, {Saito},
  {S{\'a}nchez}, {Sartori}, {Scalzotto}, {Scapin}, {Schmitt}, {Schweizer},
  {Shayduk}, {Shinozaki}, {Shore}, {Sidro}, {Sillanp{\"a}{\"a}}, {Sobczynska},
  {Stamerra}, {Stark}, {Takalo}, {Temnikov}, {Tescaro}, {Teshima}, {Torres},
  {Turini}, {Vankov}, {Vitale}, {Wagner}, {Wibig}, {Wittek}, {Zandanel},
  {Zanin}, \& {Zapatero}}]{Albert2007}
{Albert}, J., {Aliu}, E., {Anderhub}, H., {et~al.} 2007, \apjl, 664, L87

\bibitem[{{Ambrocio-Cruz} {et~al.}(2017){Ambrocio-Cruz}, {Rosado}, {de la
  Fuente}, {Silva}, \& {Blanco-Pi{\~n}on}}]{Ambrociocruz17}
{Ambrocio-Cruz}, P., {Rosado}, M., {de la Fuente}, E., {Silva}, R., \&
  {Blanco-Pi{\~n}on}, A. 2017, \mnras, 472, 51

\bibitem[{{Asaoka} \& {Aschenbach}(1994)}]{Asakoa1994}
{Asaoka}, I. \& {Aschenbach}, B. 1994, \aap, 284, 573

\bibitem[{{Astropy Collaboration} {et~al.}(2018){Astropy Collaboration},
  {Price-Whelan}, {Sip{\H o}cz}, {G{\"u}nther}, {Lim}, {Crawford}, {Conseil},
  {Shupe}, {Craig}, {Dencheva}, {Ginsburg}, {VanderPlas}, {Bradley},
  {P{\'e}rez-Su{\'a}rez}, {de Val-Borro}, {Aldcroft}, {Cruz}, {Robitaille},
  {Tollerud}, {Ardelean}, {Babej}, {Bach}, {Bachetti}, {Bakanov}, {Bamford},
  {Barentsen}, {Barmby}, {Baumbach}, {Berry}, {Biscani}, {Boquien}, {Bostroem},
  {Bouma}, {Brammer}, {Bray}, {Breytenbach}, {Buddelmeijer}, {Burke},
  {Calderone}, {Cano Rodr{\'{\i}}guez}, {Cara}, {Cardoso}, {Cheedella},
  {Copin}, {Corrales}, {Crichton}, {D'Avella}, {Deil}, {Depagne}, {Dietrich},
  {Donath}, {Droettboom}, {Earl}, {Erben}, {Fabbro}, {Ferreira}, {Finethy},
  {Fox}, {Garrison}, {Gibbons}, {Goldstein}, {Gommers}, {Greco}, {Greenfield},
  {Groener}, {Grollier}, {Hagen}, {Hirst}, {Homeier}, {Horton}, {Hosseinzadeh},
  {Hu}, {Hunkeler}, {Ivezi{\'c}}, {Jain}, {Jenness}, {Kanarek}, {Kendrew},
  {Kern}, {Kerzendorf}, {Khvalko}, {King}, {Kirkby}, {Kulkarni}, {Kumar},
  {Lee}, {Lenz}, {Littlefair}, {Ma}, {Macleod}, {Mastropietro}, {McCully},
  {Montagnac}, {Morris}, {Mueller}, {Mumford}, {Muna}, {Murphy}, {Nelson},
  {Nguyen}, {Ninan}, {N{\"o}the}, {Ogaz}, {Oh}, {Parejko}, {Parley}, {Pascual},
  {Patil}, {Patil}, {Plunkett}, {Prochaska}, {Rastogi}, {Reddy Janga},
  {Sabater}, {Sakurikar}, {Seifert}, {Sherbert}, {Sherwood-Taylor}, {Shih},
  {Sick}, {Silbiger}, {Singanamalla}, {Singer}, {Sladen}, {Sooley},
  {Sornarajah}, {Streicher}, {Teuben}, {Thomas}, {Tremblay}, {Turner},
  {Terr{\'o}n}, {van Kerkwijk}, {de la Vega}, {Watkins}, {Weaver}, {Whitmore},
  {Woillez}, {Zabalza}, \& {Astropy Contributors}}]{AstropyII}
{Astropy Collaboration}, {Price-Whelan}, A.~M., {Sip{\H o}cz}, B.~M., {et~al.}
  2018, \aj, 156, 123

\bibitem[{{Astropy Collaboration} {et~al.}(2013){Astropy Collaboration},
  {Robitaille}, {Tollerud}, {Greenfield}, {Droettboom}, {Bray}, {Aldcroft},
  {Davis}, {Ginsburg}, {Price-Whelan}, {Kerzendorf}, {Conley}, {Crighton},
  {Barbary}, {Muna}, {Ferguson}, {Grollier}, {Parikh}, {Nair}, {Unther},
  {Deil}, {Woillez}, {Conseil}, {Kramer}, {Turner}, {Singer}, {Fox}, {Weaver},
  {Zabalza}, {Edwards}, {Azalee Bostroem}, {Burke}, {Casey}, {Crawford},
  {Dencheva}, {Ely}, {Jenness}, {Labrie}, {Lim}, {Pierfederici}, {Pontzen},
  {Ptak}, {Refsdal}, {Servillat}, \& {Streicher}}]{AstropyI}
{Astropy Collaboration}, {Robitaille}, T.~P., {Tollerud}, E.~J., {et~al.} 2013,
  \aap, 558, A33

\bibitem[{Bailer-Jones {et~al.}(2018)Bailer-Jones, Rybizki, Fouesneau,
  Mantelet, \& Andrae}]{BailerJones18}
Bailer-Jones, C. A.~L., Rybizki, J., Fouesneau, M., Mantelet, G., \& Andrae, R.
  2018, The Astronomical Journal, 156, 58

\bibitem[{{Barnard}(1894)}]{Barnard1894}
{Barnard}, E.~E. 1894, Astronomy and Astro-Physics (formerly The Sidereal
  Messenger), 13, 177

\bibitem[{{Bocchino} \& {Bykov}(2000)}]{Bocchino00}
{Bocchino}, F. \& {Bykov}, A.~M. 2000, \aap, 362, L29

\bibitem[{{Bocchino} \& {Bykov}(2001{\natexlab{a}})}]{Bocchino01}
{Bocchino}, F. \& {Bykov}, A.~M. 2001{\natexlab{a}}, \aap, 376, 248

\bibitem[{{Bocchino} \& {Bykov}(2001{\natexlab{b}})}]{Bocchino&Bykov2001}
{Bocchino}, F. \& {Bykov}, A.~M. 2001{\natexlab{b}}, \aap, 376, 248

\bibitem[{{Bocchino} \& {Bykov}(2003)}]{Bocchino&Bykov2003}
{Bocchino}, F. \& {Bykov}, A.~M. 2003, \aap, 400, 203

\bibitem[{{Bolatto} {et~al.}(2013){Bolatto}, {Wolfire}, \&
  {Leroy}}]{Bolatto2013}
{Bolatto}, A.~D., {Wolfire}, M., \& {Leroy}, A.~K. 2013, \araa, 51, 207

\bibitem[{{Braun} \& {Strom}(1986)}]{Braun&Strom1986}
{Braun}, R. \& {Strom}, R.~G. 1986, \aap, 164, 193

\bibitem[{{Bron} {et~al.}(2018){Bron}, {Daudon}, {Pety}, {Levrier}, {Gerin},
  {Gratier}, {Orkisz}, {Guzman}, {Bardeau}, {Goicoechea}, {Liszt}, {{\"O}berg},
  {Peretto}, {Sievers}, \& {Tremblin}}]{Bron2018}
{Bron}, E., {Daudon}, C., {Pety}, J., {et~al.} 2018, \aap, 610, A12

\bibitem[{{Burton} {et~al.}(1988){Burton}, {Geballe}, {Brand}, \&
  {Webster}}]{Burton1988}
{Burton}, M.~G., {Geballe}, T.~R., {Brand}, P.~W.~J.~L., \& {Webster}, A.~S.
  1988, \mnras, 231, 617

\bibitem[{{Burton} {et~al.}(1990){Burton}, {Hollenbach}, {Haas}, \&
  {Erickson}}]{Burton1990}
{Burton}, M.~G., {Hollenbach}, D.~J., {Haas}, M.~R., \& {Erickson}, E.~F. 1990,
  \apj, 355, 197

\bibitem[{{Bykov} {et~al.}(2005){Bykov}, {Bocchino}, \& {Pavlov}}]{Bykov05}
{Bykov}, A.~M., {Bocchino}, F., \& {Pavlov}, G.~G. 2005, \apjl, 624, L41

\bibitem[{{Bykov} {et~al.}(2018{\natexlab{a}}){Bykov}, {Ellison}, {Marcowith},
  \& {Osipov}}]{Bykov18}
{Bykov}, A.~M., {Ellison}, D.~C., {Marcowith}, A., \& {Osipov}, S.~M.
  2018{\natexlab{a}}, \ssr, 214, 41

\bibitem[{{Bykov} {et~al.}(2018{\natexlab{b}}){Bykov}, {Ellison}, {Marcowith},
  \& {Osipov}}]{Bykov2018}
{Bykov}, A.~M., {Ellison}, D.~C., {Marcowith}, A., \& {Osipov}, S.~M.
  2018{\natexlab{b}}, \ssr, 214, 41

\bibitem[{{Cardillo} {et~al.}(2016){Cardillo}, {Amato}, \&
  {Blasi}}]{Cardillo16}
{Cardillo}, M., {Amato}, E., \& {Blasi}, P. 2016, \aap, 595, A58

\bibitem[{{Carter} {et~al.}(2012){Carter}, {Lazareff}, {Maier}, {Chenu},
  {Fontana}, {Bortolotti}, {Boucher}, {Navarrini}, {Blanchet}, {Greve}, {John},
  {Kramer}, {Morel}, {Navarro}, {Pe{\~n}alver}, {Schuster}, \& {Thum}}]{EMIR12}
{Carter}, M., {Lazareff}, B., {Maier}, D., {et~al.} 2012, \aap, 538, A89

\bibitem[{{Castelletti} {et~al.}(2011){Castelletti}, {Dubner}, {Clarke}, \&
  {Kassim}}]{Castelletti2011}
{Castelletti}, G., {Dubner}, G., {Clarke}, T., \& {Kassim}, N.~E. 2011, \aap,
  534, A21

\bibitem[{{Cazzoli} {et~al.}(2004){Cazzoli}, {Puzzarini}, \&
  {Lapinov}}]{Cazzoli2004}
{Cazzoli}, G., {Puzzarini}, C., \& {Lapinov}, A.~V. 2004, \apj, 611, 615

\bibitem[{{Celli} {et~al.}(2019){Celli}, {Morlino}, {Gabici}, \&
  {Aharonian}}]{Celli19}
{Celli}, S., {Morlino}, G., {Gabici}, S., \& {Aharonian}, F.~A. 2019, \mnras,
  487, 3199

\bibitem[{{Cesarsky} {et~al.}(1999){Cesarsky}, {Cox}, {Pineau des For{\^e}ts},
  {van Dishoeck}, {Boulanger}, \& {Wright}}]{Cesarsky1999}
{Cesarsky}, D., {Cox}, P., {Pineau des For{\^e}ts}, G., {et~al.} 1999, \aap,
  348, 945

\bibitem[{{Cherchneff}(2014)}]{Cherchneff14}
{Cherchneff}, I. 2014, arXiv e-prints, arXiv:1405.1216

\bibitem[{{Chevalier}(1974)}]{Chevalier1974}
{Chevalier}, R.~A. 1974, \apj, 188, 501

\bibitem[{{Chevalier}(1999)}]{Chevalier1999}
{Chevalier}, R.~A. 1999, \apj, 511, 798

\bibitem[{{Claussen} {et~al.}(1997{\natexlab{a}}){Claussen}, {Frail}, {Goss},
  \& {Gaume}}]{Claussen1997}
{Claussen}, M.~J., {Frail}, D.~A., {Goss}, W.~M., \& {Gaume}, R.~A.
  1997{\natexlab{a}}, \apj, 489, 143

\bibitem[{{Claussen} {et~al.}(1997{\natexlab{b}}){Claussen}, {Frail}, {Goss},
  \& {Gaume}}]{Claussen97}
{Claussen}, M.~J., {Frail}, D.~A., {Goss}, W.~M., \& {Gaume}, R.~A.
  1997{\natexlab{b}}, \apj, 489, 143

\bibitem[{{Claussen} {et~al.}(1999){Claussen}, {Goss}, {Frail}, \&
  {Seta}}]{Claussen99}
{Claussen}, M.~J., {Goss}, W.~M., {Frail}, D.~A., \& {Seta}, M. 1999, \aj, 117,
  1387

\bibitem[{{Cornett} {et~al.}(1977){Cornett}, {Chin}, \& {Knapp}}]{Cornett1977}
{Cornett}, R.~H., {Chin}, G., \& {Knapp}, G.~R. 1977, \aap, 54, 889

\bibitem[{{Dame} {et~al.}(2001){Dame}, {Hartmann}, \& {Thaddeus}}]{Dame2001}
{Dame}, T.~M., {Hartmann}, D., \& {Thaddeus}, P. 2001, \apj, 547, 792

\bibitem[{{Davies} {et~al.}(1972){Davies}, {Lyne}, \& {Seiradakis}}]{Davies72}
{Davies}, J.~G., {Lyne}, A.~G., \& {Seiradakis}, J.~H. 1972, \nat, 240, 229

\bibitem[{{Denoyer}(1977)}]{Denoyer1977}
{Denoyer}, L.~K. 1977, \apj, 212, 416

\bibitem[{{Denoyer}(1978)}]{Denoyer78}
{Denoyer}, L.~K. 1978, \mnras, 183, 187

\bibitem[{{Denoyer}(1979{\natexlab{a}})}]{Denoyer79b}
{Denoyer}, L.~K. 1979{\natexlab{a}}, \apjl, 232, L165

\bibitem[{{Denoyer}(1979{\natexlab{b}})}]{Denoyer79a}
{Denoyer}, L.~K. 1979{\natexlab{b}}, \apjl, 228, L41

\bibitem[{{Denoyer} \& {Frerking}(1981)}]{Denoyer1981}
{Denoyer}, L.~K. \& {Frerking}, M.~A. 1981, \apjl, 246, L37

\bibitem[{{Dickel}(1973)}]{Dickel73}
{Dickel}, J.~R. 1973, \aplett, 15, 61

\bibitem[{{Dickel} {et~al.}(1989){Dickel}, {Williamson}, {Mufson}, \&
  {Wood}}]{Dickel1989}
{Dickel}, J.~R., {Williamson}, C.~E., {Mufson}, S.~L., \& {Wood}, C.~A. 1989,
  \aj, 98, 1363

\bibitem[{{Dickman} {et~al.}(1992){Dickman}, {Snell}, {Ziurys}, \&
  {Huang}}]{Dickman1992}
{Dickman}, R.~L., {Snell}, R.~L., {Ziurys}, L.~M., \& {Huang}, Y.-L. 1992,
  \apj, 400, 203

\bibitem[{{Draine}(2011)}]{Draine11}
{Draine}, B.~T. 2011, {Physics of the Interstellar and Intergalactic Medium}

\bibitem[{{Dubner} \& {Giacani}(2015)}]{Dubner15}
{Dubner}, G. \& {Giacani}, E. 2015, \aapr, 23, 3

\bibitem[{{Duin} \& {van der Laan}(1975)}]{Duin1975}
{Duin}, R.~M. \& {van der Laan}, H. 1975, \aap, 40, 111

\bibitem[{{Dunham} {et~al.}(2015){Dunham}, {Allen}, {Evans},
  {Broekhoven-Fiene}, {Cieza}, {Di Francesco}, {Gutermuth}, {Harvey},
  {Hatchell}, {Heiderman}, {Huard}, {Johnstone}, {Kirk}, {Matthews}, {Miller},
  {Peterson}, \& {Young}}]{Dunham2015}
{Dunham}, M.~M., {Allen}, L.~E., {Evans}, Neal~J., I., {et~al.} 2015, \apjs,
  220, 11

\bibitem[{{Egron} {et~al.}(2017){Egron}, {Pellizzoni}, {Iacolina}, {Loru},
  {Marongiu}, {Righini}, {Cardillo}, {Giuliani}, {Mulas}, {Murtas}, {Simeone},
  {Concu}, {Melis}, {Trois}, {Pilia}, {Navarrini}, {Vacca}, {Ricci}, {Serra},
  {Bachetti}, {Buttu}, {Perrodin}, {Buffa}, {Deiana}, {Gaudiomonte}, {Fara},
  {Ladu}, {Loi}, {Marongiu}, {Migoni}, {Pisanu}, {Poppi}, {Saba}, {Urru},
  {Valente}, \& {Vargiu}}]{Egron2017}
{Egron}, E., {Pellizzoni}, A., {Iacolina}, M.~N., {et~al.} 2017, \mnras, 470,
  1329

\bibitem[{{Elitzur}(1976)}]{Elitzur76}
{Elitzur}, M. 1976, \apj, 203, 124

\bibitem[{{Elmegreen}(1998)}]{Elmegreen98}
{Elmegreen}, B.~G. 1998, Astronomical Society of the Pacific Conference Series,
  Vol. 148, {Observations and Theory of Dynamical Triggers for Star Formation},
  ed. C.~E. {Woodward}, J.~M. {Shull}, \& J.~{Thronson}, Harley~A., 150

\bibitem[{{Endres} {et~al.}(2016){Endres}, {Schlemmer}, {Schilke}, {Stutzki},
  \& {M{\"u}ller}}]{Endres2016}
{Endres}, C.~P., {Schlemmer}, S., {Schilke}, P., {Stutzki}, J., \&
  {M{\"u}ller}, H. S.~P. 2016, Journal of Molecular Spectroscopy, 327, 95

\bibitem[{{Esposito} {et~al.}(1996){Esposito}, {Hunter}, {Kanbach}, \&
  {Sreekumar}}]{Esposito1996}
{Esposito}, J.~A., {Hunter}, S.~D., {Kanbach}, G., \& {Sreekumar}, P. 1996,
  \apj, 461, 820

\bibitem[{{Evans} {et~al.}(2009){Evans}, {Dunham}, {J{\o}rgensen}, {Enoch},
  {Mer{\'\i}n}, {van Dishoeck}, {Alcal{\'a}}, {Myers}, {Stapelfeldt}, {Huard},
  {Allen}, {Harvey}, {van Kempen}, {Blake}, {Koerner}, {Mundy}, {Padgett}, \&
  {Sargent}}]{Evans2009}
{Evans}, Neal~J., I., {Dunham}, M.~M., {J{\o}rgensen}, J.~K., {et~al.} 2009,
  \apjs, 181, 321

\bibitem[{{Fazio} \& {Hora}(2004)}]{Fazio2004}
{Fazio}, G. \& {Hora}, J. 2004, {Studying Stellar Ejecta on the Large Scale
  using SIRTF-IRAC}, Spitzer Proposal

\bibitem[{{Fesen}(1984)}]{Fesen84}
{Fesen}, R.~A. 1984, \apj, 281, 658

\bibitem[{{Fesen} \& {Kirshner}(1980)}]{Fesen80}
{Fesen}, R.~A. \& {Kirshner}, R.~P. 1980, \apj, 242, 1023

\bibitem[{Fischer {et~al.}(2016)Fischer, Padgett, Stapelfeldt, \&
  Sewiło}]{Fischer2016}
Fischer, W.~J., Padgett, D.~L., Stapelfeldt, K.~L., \& Sewiło, M. 2016, The
  Astrophysical Journal, 827, 96

\bibitem[{{Fish} {et~al.}(2007){Fish}, {Sjouwerman}, \&
  {Pihlstr{\"o}m}}]{Fish07}
{Fish}, V.~L., {Sjouwerman}, L.~O., \& {Pihlstr{\"o}m}, Y.~M. 2007, \apjl, 670,
  L117

\bibitem[{{Flagey} {et~al.}(2009){Flagey}, {Noriega-Crespo}, {Boulanger},
  {Carey}, {Brooke}, {Falgarone}, {Huard}, {McCabe}, {Miville-Desch{\^e}nes},
  {Padgett}, {Paladini}, \& {Rebull}}]{Flagey2009}
{Flagey}, N., {Noriega-Crespo}, A., {Boulanger}, F., {et~al.} 2009, \apj, 701,
  1450

\bibitem[{{Flower} \& {Pineau des
  For{\^e}ts}(2003)}]{Flower&PineauDesForets2003}
{Flower}, D.~R. \& {Pineau des For{\^e}ts}, G. 2003, \mnras, 343, 390

\bibitem[{{Flower} \& {Pineau des
  For{\^e}ts}(2015)}]{Flower&PineauDesForets2015}
{Flower}, D.~R. \& {Pineau des For{\^e}ts}, G. 2015, \aap, 578, A63

\bibitem[{{Frail} {et~al.}(1996){Frail}, {Goss}, {Reynoso}, {Giacani}, {Green},
  \& {Otrupcek}}]{Frail96}
{Frail}, D.~A., {Goss}, W.~M., {Reynoso}, E.~M., {et~al.} 1996, \aj, 111, 1651

\bibitem[{{Fran{\c{c}}ois} {et~al.}(2004){Fran{\c{c}}ois}, {Matteucci},
  {Cayrel}, {Spite}, {Spite}, \& {Chiappini}}]{Francois04}
{Fran{\c{c}}ois}, P., {Matteucci}, F., {Cayrel}, R., {et~al.} 2004, \aap, 421,
  613

\bibitem[{{Frerking} {et~al.}(1982){Frerking}, {Langer}, \&
  {Wilson}}]{Frerking1982}
{Frerking}, M.~A., {Langer}, W.~D., \& {Wilson}, R.~W. 1982, \apj, 262, 590

\bibitem[{{Gabici} {et~al.}(2009{\natexlab{a}}){Gabici}, {Aharonian}, \&
  {Casanova}}]{Gabici09}
{Gabici}, S., {Aharonian}, F.~A., \& {Casanova}, S. 2009{\natexlab{a}}, \mnras,
  396, 1629

\bibitem[{{Gabici} {et~al.}(2009{\natexlab{b}}){Gabici}, {Aharonian}, \&
  {Casanova}}]{Gabici2009}
{Gabici}, S., {Aharonian}, F.~A., \& {Casanova}, S. 2009{\natexlab{b}}, \mnras,
  396, 1629

\bibitem[{{Gabici} {et~al.}(2019){Gabici}, {Evoli}, {Gaggero}, {Lipari},
  {Mertsch}, {Orlando}, {Strong}, \& {Vittino}}]{Gabici19}
{Gabici}, S., {Evoli}, C., {Gaggero}, D., {et~al.} 2019, arXiv e-prints,
  arXiv:1903.11584

\bibitem[{{Gaensler} {et~al.}(2006){Gaensler}, {Chatterjee}, {Slane}, {van der
  Swaluw}, {Camilo}, \& {Hughes}}]{Gaensler06}
{Gaensler}, B.~M., {Chatterjee}, S., {Slane}, P.~O., {et~al.} 2006, \apj, 648,
  1037

\bibitem[{{Gaia Collaboration} {et~al.}(2018){Gaia Collaboration}, {Brown},
  {Vallenari}, {Prusti}, {de Bruijne}, {Babusiaux}, {Bailer-Jones}, {Biermann},
  {Evans}, {Eyer}, \& et~al.}]{Gaia18}
{Gaia Collaboration}, {Brown}, A.~G.~A., {Vallenari}, A., {et~al.} 2018, \aap,
  616, A1

\bibitem[{{Gaia Collaboration} {et~al.}(2016){Gaia Collaboration}, {Brown},
  {Vallenari}, {Prusti}, {de Bruijne}, {Mignard}, {Drimmel}, {Babusiaux},
  {Bailer-Jones}, {Bastian}, \& et~al.}]{Gaia16}
{Gaia Collaboration}, {Brown}, A.~G.~A., {Vallenari}, A., {et~al.} 2016, \aap,
  595, A2

\bibitem[{{Georgelin}(1975)}]{Georgelin75}
{Georgelin}, Y.~M. 1975, PhD thesis, Université de Provence, Observatoire de
  Marseille

\bibitem[{{Giacconi} {et~al.}(1971){Giacconi}, {Kellogg}, {Gorenstein},
  {Gursky}, \& {Tananbaum}}]{Giacconi1971}
{Giacconi}, R., {Kellogg}, E., {Gorenstein}, P., {Gursky}, H., \& {Tananbaum},
  H. 1971, \apjl, 165, L27

\bibitem[{{Giacconi} {et~al.}(1974){Giacconi}, {Murray}, {Gursky}, {Kellogg},
  {Schreier}, {Matilsky}, {Koch}, \& {Tananbaum}}]{Giacconi1974}
{Giacconi}, R., {Murray}, S., {Gursky}, H., {et~al.} 1974, \apjs, 27, 37

\bibitem[{{Giovanelli} \& {Haynes}(1979)}]{Giovanelli1979}
{Giovanelli}, R. \& {Haynes}, M.~P. 1979, \apj, 230, 404

\bibitem[{{Girichidis} {et~al.}(2016){Girichidis}, {Naab}, {Walch}, {Hanasz},
  {Mac Low}, {Ostriker}, {Gatto}, {Peters}, {W{\"u}nsch}, {Glover}, {Klessen},
  {Clark}, \& {Baczynski}}]{Girichidis16}
{Girichidis}, P., {Naab}, T., {Walch}, S., {et~al.} 2016, \apjl, 816, L19

\bibitem[{{Glassgold} {et~al.}(1985){Glassgold}, {Huggins}, \&
  {Langer}}]{Glassgold1985}
{Glassgold}, A.~E., {Huggins}, P.~J., \& {Langer}, W.~D. 1985, \apj, 290, 615

\bibitem[{{Godard} {et~al.}(2019){Godard}, {Pineau des For{\^e}ts}, {Lesaffre},
  {Lehmann}, {Gusdorf}, \& {Falgarone}}]{Godard2019}
{Godard}, B., {Pineau des For{\^e}ts}, G., {Lesaffre}, P., {et~al.} 2019, \aap,
  622, A100

\bibitem[{{Goldsmith} \& {Langer}(1999)}]{Goldsmith1999}
{Goldsmith}, P.~F. \& {Langer}, W.~D. 1999, \apj, 517, 209

\bibitem[{{Gordon} \& {Burton}(1976)}]{Gordon&Burton1976}
{Gordon}, M.~A. \& {Burton}, W.~B. 1976, \apj, 208, 346

\bibitem[{{Goss}(1968)}]{Goss68}
{Goss}, W.~M. 1968, \apjs, 15, 131

\bibitem[{{Graham} \& {Hunt}(1973)}]{Graham73}
{Graham}, D. \& {Hunt}, G.~C. 1973, Nature Physical Science, 242, 86

\bibitem[{{Graham} {et~al.}(1987){Graham}, {Wright}, \&
  {Longmore}}]{Graham1987}
{Graham}, J.~R., {Wright}, G.~S., \& {Longmore}, A.~J. 1987, \apj, 313, 847

\bibitem[{{Greco} {et~al.}(2018){Greco}, {Miceli}, {Orlando}, {Peres}, {Troja},
  \& {Bocchino}}]{Greco2018}
{Greco}, E., {Miceli}, M., {Orlando}, S., {et~al.} 2018, \aap, 615, A157

\bibitem[{{Green}(1986)}]{Green1986}
{Green}, D.~A. 1986, \mnras, 221, 473

\bibitem[{{Green}(1989)}]{Green89}
{Green}, D.~A. 1989, \mnras, 238, 737

\bibitem[{{Greene} {et~al.}(1994){Greene}, {Wilking}, {Andre}, {Young}, \&
  {Lada}}]{Greene1994}
{Greene}, T.~P., {Wilking}, B.~A., {Andre}, P., {Young}, E.~T., \& {Lada},
  C.~J. 1994, \apj, 434, 614

\bibitem[{{Grenier} {et~al.}(2015){Grenier}, {Black}, \& {Strong}}]{Grenier15}
{Grenier}, I.~A., {Black}, J.~H., \& {Strong}, A.~W. 2015, \araa, 53, 199

\bibitem[{{Gutermuth} {et~al.}(2009){Gutermuth}, {Megeath}, {Myers}, {Allen},
  {Pipher}, \& {Fazio}}]{Gutermuth2009}
{Gutermuth}, R.~A., {Megeath}, S.~T., {Myers}, P.~C., {et~al.} 2009, \apjs,
  184, 18

\bibitem[{{Hanabata} {et~al.}(2014){Hanabata}, {Katagiri}, {Hewitt}, {Ballet},
  {Fukazawa}, {Fukui}, {Hayakawa}, {Lemoine-Goumard}, {Pedaletti}, {Strong},
  {Torres}, \& {Yamazaki}}]{Hanabata14}
{Hanabata}, Y., {Katagiri}, H., {Hewitt}, J.~W., {et~al.} 2014, \apj, 786, 145

\bibitem[{{Harris}(1962)}]{Harris62}
{Harris}, D.~E. 1962, \apj, 135, 661

\bibitem[{{Herbst} \& {Assousa}(1977)}]{Herbst77}
{Herbst}, W. \& {Assousa}, G.~E. 1977, \apj, 217, 473

\bibitem[{{Hester}(1987)}]{Hester1987}
{Hester}, J.~J. 1987, \apj, 314, 187

\bibitem[{{Hewitt} {et~al.}(2008){Hewitt}, {Yusef-Zadeh}, \&
  {Wardle}}]{Hewitt08}
{Hewitt}, J.~W., {Yusef-Zadeh}, F., \& {Wardle}, M. 2008, \apj, 683, 189

\bibitem[{{Hewitt} {et~al.}(2009){Hewitt}, {Yusef-Zadeh}, \&
  {Wardle}}]{Hewitt09}
{Hewitt}, J.~W., {Yusef-Zadeh}, F., \& {Wardle}, M. 2009, \apjl, 706, L270

\bibitem[{{Hewitt} {et~al.}(2006){Hewitt}, {Yusef-Zadeh}, {Wardle}, {Roberts},
  \& {Kassim}}]{Hewitt06}
{Hewitt}, J.~W., {Yusef-Zadeh}, F., {Wardle}, M., {Roberts}, D.~A., \&
  {Kassim}, N.~E. 2006, \apj, 652, 1288

\bibitem[{{Heyminck} {et~al.}(2006){Heyminck}, {Kasemann}, {G{\"u}sten}, {de
  Lange}, \& {Graf}}]{Heyminck06}
{Heyminck}, S., {Kasemann}, C., {G{\"u}sten}, R., {de Lange}, G., \& {Graf},
  U.~U. 2006, \aap, 454, L21

\bibitem[{{Hezareh} {et~al.}(2013){Hezareh}, {Wiesemeyer}, {Houde}, {Gusdorf},
  \& {Siringo}}]{Hezareh13}
{Hezareh}, T., {Wiesemeyer}, H., {Houde}, M., {Gusdorf}, A., \& {Siringo}, G.
  2013, \aap, 558, A45

\bibitem[{{Hildebrand}(1983)}]{Hildebrand1983}
{Hildebrand}, R.~H. 1983, \qjras, 24, 267

\bibitem[{{Hill}(1972)}]{Hill1972}
{Hill}, L.~E. 1972, \mnras, 157, 419

\bibitem[{{Hoffman} {et~al.}(2003){Hoffman}, {Goss}, {Brogan}, {Claussen}, \&
  {Richards}}]{Hoffman03}
{Hoffman}, I.~M., {Goss}, W.~M., {Brogan}, C.~L., {Claussen}, M.~J., \&
  {Richards}, A.~M.~S. 2003, \apj, 583, 272

\bibitem[{{Hogg}(1964)}]{Hogg64}
{Hogg}, D.~E. 1964, \apj, 140, 992

\bibitem[{{Hollenbach} \& {McKee}(1989{\natexlab{a}})}]{Hollenbach&McKee1989}
{Hollenbach}, D. \& {McKee}, C.~F. 1989{\natexlab{a}}, \apj, 342, 306

\bibitem[{{Hollenbach} \& {McKee}(1989{\natexlab{b}})}]{Hollenbach89}
{Hollenbach}, D. \& {McKee}, C.~F. 1989{\natexlab{b}}, \apj, 342, 306

\bibitem[{{Houde} {et~al.}(2013){Houde}, {Hezareh}, {Jones}, \&
  {Rajabi}}]{Houde2013}
{Houde}, M., {Hezareh}, T., {Jones}, S., \& {Rajabi}, F. 2013, \apj, 764, 24

\bibitem[{{Howard} \& {Dickel}(1963)}]{Howard63}
{Howard}, William~E., I. \& {Dickel}, H.~R. 1963, \pasp, 75, 149

\bibitem[{{Huang} {et~al.}(1986){Huang}, {Dickman}, \& {Snell}}]{Huang86}
{Huang}, Y.~L., {Dickman}, R.~L., \& {Snell}, R.~L. 1986, \apjl, 302, L63

\bibitem[{{Humensky} \& {VERITAS Collaboration}(2015)}]{Humensky15}
{Humensky}, B. \& {VERITAS Collaboration}. 2015, in International Cosmic Ray
  Conference, Vol.~34, 34th International Cosmic Ray Conference (ICRC2015), 875

\bibitem[{Hunter(2007)}]{Matplotlib07}
Hunter, J.~D. 2007, Computing In Science \& Engineering, 9, 90

\bibitem[{{Indriolo} {et~al.}(2010){Indriolo}, {Blake}, {Goto}, {Usuda}, {Oka},
  {Geballe}, {Fields}, \& {McCall}}]{Indriolo10}
{Indriolo}, N., {Blake}, G.~A., {Goto}, M., {et~al.} 2010, \apj, 724, 1357

\bibitem[{Jones {et~al.}(2001--)Jones, Oliphant, Peterson, {et~al.}}]{Scipy01}
Jones, E., Oliphant, T., Peterson, P., {et~al.} 2001--, {SciPy}: Open source
  scientific tools for {Python}

\bibitem[{{Kawasaki} {et~al.}(2002){Kawasaki}, {Ozaki}, {Nagase}, {Masai},
  {Ishida}, \& {Petre}}]{Kawasaki2002}
{Kawasaki}, M.~T., {Ozaki}, M., {Nagase}, F., {et~al.} 2002, \apj, 572, 897

\bibitem[{{Keohane} {et~al.}(1997){Keohane}, {Petre}, {Gotthelf}, {Ozaki}, \&
  {Koyama}}]{Keohane97}
{Keohane}, J.~W., {Petre}, R., {Gotthelf}, E.~V., {Ozaki}, M., \& {Koyama}, K.
  1997, \apj, 484, 350

\bibitem[{{Klapper} {et~al.}(2000{\natexlab{a}}){Klapper}, {Lewen}, {Belova},
  \& {Winnewisser}}]{Klapper2000b}
{Klapper}, G., {Lewen}, F., {Belova}, S.~P., \& {Winnewisser}, G.
  2000{\natexlab{a}}, Zeitschrift Naturforschung Teil A, 55, 441

\bibitem[{{Klapper} {et~al.}(2000{\natexlab{b}}){Klapper}, {Lewen},
  {Gendriesch}, {Belov}, \& {Winnewisser}}]{Klapper2000a}
{Klapper}, G., {Lewen}, F., {Gendriesch}, R., {Belov}, S.~P., \& {Winnewisser},
  G. 2000{\natexlab{b}}, Journal of Molecular Spectroscopy, 201, 124

\bibitem[{{Klapper} {et~al.}(2003){Klapper}, {Surin}, {Lewen}, {M{\"u}ller},
  {Pak}, \& {Winnewisser}}]{Klapper2003}
{Klapper}, G., {Surin}, L., {Lewen}, F., {et~al.} 2003, \apj, 582, 262

\bibitem[{{Klein} {et~al.}(2012){Klein}, {Hochg\"urtel}, {Kr\"amer}, {Bell},
  {H\"ubers}, \& {G\"usten}}]{Klein12}
{Klein}, B., {Hochg\"urtel}, S., {Kr\"amer}, I., {et~al.} 2012, \aap, this
  volume

\bibitem[{{Koenig} \& {Leisawitz}(2014)}]{Koenig&Leisawitz14}
{Koenig}, X.~P. \& {Leisawitz}, D.~T. 2014, \apj, 791, 131

\bibitem[{{Koenig} {et~al.}(2012){Koenig}, {Leisawitz}, {Benford}, {Rebull},
  {Padgett}, \& {Assef}}]{Koenig2012}
{Koenig}, X.~P., {Leisawitz}, D.~T., {Benford}, D.~J., {et~al.} 2012, \apj,
  744, 130

\bibitem[{{Koo} {et~al.}(2010){Koo}, {Heiles}, {Stanimirovi{\'c}}, \&
  {Troland}}]{Koo10}
{Koo}, B.-C., {Heiles}, C., {Stanimirovi{\'c}}, S., \& {Troland}, T. 2010, \aj,
  140, 262

\bibitem[{{Koo} {et~al.}(2008{\natexlab{a}}){Koo}, {McKee}, {Lee}, {Lee},
  {Lee}, {Moon}, {Hong}, {Kaneda}, \& {Onaka}}]{Koo08}
{Koo}, B.-C., {McKee}, C.~F., {Lee}, J.-J., {et~al.} 2008{\natexlab{a}}, \apjl,
  673, L147

\bibitem[{{Koo} {et~al.}(2008{\natexlab{b}}){Koo}, {McKee}, {Lee}, {Lee},
  {Lee}, {Moon}, {Hong}, {Kaneda}, \& {Onaka}}]{Koo2008}
{Koo}, B.-C., {McKee}, C.~F., {Lee}, J.-J., {et~al.} 2008{\natexlab{b}}, \apjl,
  673, L147

\bibitem[{{Lada} \& {Adams}(1992)}]{Lada&Adams92}
{Lada}, C.~J. \& {Adams}, F.~C. 1992, \apj, 393, 278

\bibitem[{{Langer} \& {Penzias}(1993)}]{Langer93}
{Langer}, W.~D. \& {Penzias}, A.~A. 1993, \apj, 408, 539

\bibitem[{{Larson}(1981)}]{Larson1981}
{Larson}, R.~B. 1981, \mnras, 194, 809

\bibitem[{{Lee} {et~al.}(2012){Lee}, {Koo}, {Snell}, {Yun}, {Heyer}, \&
  {Burton}}]{Lee2012}
{Lee}, J.-J., {Koo}, B.-C., {Snell}, R.~L., {et~al.} 2012, \apj, 749, 34

\bibitem[{{Lee} {et~al.}(2008){Lee}, {Koo}, {Yun}, {Stanimirovi{\'c}},
  {Heiles}, \& {Heyer}}]{Lee2008}
{Lee}, J.-J., {Koo}, B.-C., {Yun}, M.~S., {et~al.} 2008, \aj, 135, 796

\bibitem[{{Lefloch} \& {Lazareff}(1994)}]{Lefloch&Lazareff1994}
{Lefloch}, B. \& {Lazareff}, B. 1994, \aap, 289, 559

\bibitem[{{Lesaffre} {et~al.}(2013){Lesaffre}, {Pineau des For{\^e}ts},
  {Godard}, {Guillard}, {Boulanger}, \& {Falgarone}}]{Lesaffre2013}
{Lesaffre}, P., {Pineau des For{\^e}ts}, G., {Godard}, B., {et~al.} 2013, \aap,
  550, A106

\bibitem[{{Levine} {et~al.}(1979){Levine}, {Petre}, {Rappaport}, {Smith},
  {Evans}, \& {Rolf}}]{Levine79}
{Levine}, A., {Petre}, R., {Rappaport}, S., {et~al.} 1979, \apjl, 228, L99

\bibitem[{{Li} \& {Draine}(2001)}]{Li&Draine2001}
{Li}, A. \& {Draine}, B.~T. 2001, \apj, 554, 778

\bibitem[{{Litovchenko} {et~al.}(2011){Litovchenko}, {Alakoz}, {Val'Tts}, \&
  {Larionov}}]{Litovchenko11}
{Litovchenko}, I.~D., {Alakoz}, A.~V., {Val'Tts}, I.~E., \& {Larionov}, G.~M.
  2011, Astronomy Reports, 55, 978

\bibitem[{{Locke} {et~al.}(1964){Locke}, {Galt}, \& {Costain}}]{Locke1964}
{Locke}, J.~L., {Galt}, J.~A., \& {Costain}, C.~H. 1964, \apj, 139, 1071

\bibitem[{{Lockett} {et~al.}(1999){Lockett}, {Gauthier}, \&
  {Elitzur}}]{Lockett99}
{Lockett}, P., {Gauthier}, E., \& {Elitzur}, M. 1999, \apj, 511, 235

\bibitem[{{Loru} {et~al.}(2019){Loru}, {Pellizzoni}, {Egron}, {Righini},
  {Iacolina}, {Mulas}, {Cardillo}, {Marongiu}, {Ricci}, {Bachetti}, {Pilia},
  {Trois}, {Ingallinera}, {Petruk}, {Murtas}, {Serra}, {Buffa}, {Concu},
  {Gaudiomonte}, {Melis}, {Navarrini}, {Perrodin}, \& {Valente}}]{Loru2019}
{Loru}, S., {Pellizzoni}, A., {Egron}, E., {et~al.} 2019, \mnras, 482, 3857

\bibitem[{{Louvet} {et~al.}(2016){Louvet}, {Motte}, {Gusdorf}, {Nguy{\^e}n
  Luong}, {Lesaffre}, {Duarte-Cabral}, {Maury}, {Schneider}, {Hill}, {Schilke},
  \& {Gueth}}]{Louvet16}
{Louvet}, F., {Motte}, F., {Gusdorf}, A., {et~al.} 2016, \aap, 595, A122

\bibitem[{{Lozinskaia}(1975)}]{Lozinskaya75}
{Lozinskaia}, T.~A. 1975, Soviet Astronomy Letters, 1, 35

\bibitem[{{Lozinskaya}(1969)}]{Lozinskaya69}
{Lozinskaya}, T.~A. 1969, \sovast, 13, 192

\bibitem[{{Mac Low} \& {Klessen}(2004)}]{McLow04}
{Mac Low}, M.-M. \& {Klessen}, R.~S. 2004, Reviews of Modern Physics, 76, 125

\bibitem[{{Madau} \& {Dickinson}(2014)}]{Madau14}
{Madau}, P. \& {Dickinson}, M. 2014, \araa, 52, 415

\bibitem[{{Malina} {et~al.}(1976){Malina}, {Lampton}, \& {Bowyer}}]{Malina76}
{Malina}, R., {Lampton}, M., \& {Bowyer}, S. 1976, \apj, 207, 894

\bibitem[{{McDonnell} {et~al.}(2008){McDonnell}, {Wardle}, \&
  {Vaughan}}]{Mcdonnell08}
{McDonnell}, K.~E., {Wardle}, M., \& {Vaughan}, A.~E. 2008, \mnras, 390, 49

\bibitem[{{Meaburn} {et~al.}(1990){Meaburn}, {Whitehead}, {Raymond}, {Clayton},
  \& {Marston}}]{Meaburn1990}
{Meaburn}, J., {Whitehead}, M.~J., {Raymond}, J.~C., {Clayton}, C.~A., \&
  {Marston}, A.~P. 1990, \aap, 227, 191

\bibitem[{{Micelotta} {et~al.}(2018){Micelotta}, {Matsuura}, \&
  {Sarangi}}]{Micellota18}
{Micelotta}, E.~R., {Matsuura}, M., \& {Sarangi}, A. 2018, \ssr, 214, 53

\bibitem[{{Mihalas}(1978)}]{Mihalas1978}
{Mihalas}, D. 1978, {Stellar atmospheres}

\bibitem[{{Milne}(1971)}]{Milne1971}
{Milne}, D.~K. 1971, Australian Journal of Physics, 24, 429

\bibitem[{{Minkowski}(1946)}]{Minkowski46}
{Minkowski}, R. 1946, \pasp, 58, 305

\bibitem[{{Minkowski}(1959)}]{Minkowski59}
{Minkowski}, R. 1959, in IAU Symposium, Vol.~9, URSI Symp. 1: Paris Symposium
  on Radio Astronomy, ed. R.~N. {Bracewell}, 315

\bibitem[{{Montmerle}(1979{\natexlab{a}})}]{Montmerle79}
{Montmerle}, T. 1979{\natexlab{a}}, \apj, 231, 95

\bibitem[{{Montmerle}(1979{\natexlab{b}})}]{Montmerle1979}
{Montmerle}, T. 1979{\natexlab{b}}, \apj, 231, 95

\bibitem[{{Moorhouse} {et~al.}(1991){Moorhouse}, {Brand}, {Geballe}, \&
  {Burton}}]{Moorhouse1991}
{Moorhouse}, A., {Brand}, P.~W.~J.~L., {Geballe}, T.~R., \& {Burton}, M.~G.
  1991, \mnras, 253, 662

\bibitem[{{Motte} {et~al.}(2018){Motte}, {Nony}, {Louvet}, {Marsh}, {Bontemps},
  {Whitworth}, {Men'shchikov}, {Nguyen Luong}, {Csengeri}, {Maury}, {Gusdorf},
  {Chapillon}, {K{\"o}nyves}, {Schilke}, {Duarte-Cabral}, {Didelon}, \&
  {Gaudel}}]{Motte18}
{Motte}, F., {Nony}, T., {Louvet}, F., {et~al.} 2018, Nature Astronomy, 2, 478

\bibitem[{{Muders} {et~al.}(2006){Muders}, {Hafok}, {Wyrowski}, {Polehampton},
  {Belloche}, {K{\"o}nig}, {Schaaf}, {Schuller}, {Hatchell}, \& {van der
  Tak}}]{Muders06}
{Muders}, D., {Hafok}, H., {Wyrowski}, F., {et~al.} 2006, \aap, 454, L25

\bibitem[{{Mufson} {et~al.}(1986){Mufson}, {McCollough}, {Dickel}, {Petre},
  {White}, \& {Chevalier}}]{Mufson1986}
{Mufson}, S.~L., {McCollough}, M.~L., {Dickel}, J.~R., {et~al.} 1986, \aj, 92,
  1349

\bibitem[{{M{\"u}ller} {et~al.}(2005){M{\"u}ller}, {Schl{\"o}der}, {Stutzki},
  \& {Winnewisser}}]{Muller2005}
{M{\"u}ller}, H. S.~P., {Schl{\"o}der}, F., {Stutzki}, J., \& {Winnewisser}, G.
  2005, Journal of Molecular Structure, 742, 215

\bibitem[{{M{\"u}ller} {et~al.}(2001){M{\"u}ller}, {Thorwirth}, {Roth}, \&
  {Winnewisser}}]{Muller2001}
{M{\"u}ller}, H.~S.~P., {Thorwirth}, S., {Roth}, D.~A., \& {Winnewisser}, G.
  2001, \aap, 370, L49

\bibitem[{{Neufeld} {et~al.}(2004){Neufeld}, {Bergin}, {Hollenbach}, {Kaufman},
  {Melnick}, \& {Snell}}]{Neufeld2004}
{Neufeld}, D., {Bergin}, E., {Hollenbach}, D., {et~al.} 2004, {IRS Spectroscopy
  of Shocked Molecular Gas in Supernova Remnants: Probing the Interaction of a
  Supernova with a Molecular Cloud}, Spitzer Proposal

\bibitem[{{Neufeld} {et~al.}(2007){Neufeld}, {Hollenbach}, {Kaufman}, {Snell},
  {Melnick}, {Bergin}, \& {Sonnentrucker}}]{Neufeld2007}
{Neufeld}, D.~A., {Hollenbach}, D.~J., {Kaufman}, M.~J., {et~al.} 2007, \apj,
  664, 890

\bibitem[{{Neugebauer} {et~al.}(1984){Neugebauer}, {Habing}, {van Duinen},
  {Aumann}, {Baud}, {Beichman}, {Beintema}, {Boggess}, {Clegg}, {de Jong},
  {Emerson}, {Gautier}, {Gillett}, {Harris}, {Hauser}, {Houck}, {Jennings},
  {Low}, {Marsden}, {Miley}, {Olnon}, {Pottasch}, {Raimond}, {Rowan-Robinson},
  {Soifer}, {Walker}, {Wesselius}, \& {Young}}]{Neugebauer1984}
{Neugebauer}, G., {Habing}, H.~J., {van Duinen}, R., {et~al.} 1984, \apjl, 278,
  L1

\bibitem[{{Noriega-Crespo} {et~al.}(2009){Noriega-Crespo}, {Hines}, {Gordon},
  {Marleau}, {Rieke}, {Rho}, \& {Latter}}]{Noriega-crespo2009}
{Noriega-Crespo}, A., {Hines}, D.~C., {Gordon}, K., {et~al.} 2009, in The
  Evolving ISM in the Milky Way and Nearby Galaxies, 46

\bibitem[{{Odenwald} \& {Shivanandan}(1985)}]{Odenwald85}
{Odenwald}, S.~F. \& {Shivanandan}, K. 1985, \apj, 292, 460

\bibitem[{{Olbert} {et~al.}(2001{\natexlab{a}}){Olbert}, {Clearfield},
  {Williams}, {Keohane}, \& {Frail}}]{Olbert01}
{Olbert}, C.~M., {Clearfield}, C.~R., {Williams}, N.~E., {Keohane}, J.~W., \&
  {Frail}, D.~A. 2001{\natexlab{a}}, \apjl, 554, L205

\bibitem[{{Olbert} {et~al.}(2001{\natexlab{b}}){Olbert}, {Clearfield},
  {Williams}, {Keohane}, \& {Frail}}]{Olbert2001}
{Olbert}, C.~M., {Clearfield}, C.~R., {Williams}, N.~E., {Keohane}, J.~W., \&
  {Frail}, D.~A. 2001{\natexlab{b}}, in American Institute of Physics
  Conference Series, Vol. 565, Young Supernova Remnants, ed. S.~S. {Holt} \&
  U.~{Hwang}, 341--344

\bibitem[{{Oliva} {et~al.}(1999){Oliva}, {Lutz}, {Drapatz}, \&
  {Moorwood}}]{Oliva1999}
{Oliva}, E., {Lutz}, D., {Drapatz}, S., \& {Moorwood}, A.~F.~M. 1999, \aap,
  341, L75

\bibitem[{{Ossenkopf}(2007)}]{Ossenkopf2007}
{Ossenkopf}, V. 2007, {KPGT\_vossenko\_1: The warm and dense ISM}, Herschel
  Space Observatory Proposal

\bibitem[{{Padovani} {et~al.}(2015){Padovani}, {Hennebelle}, {Marcowith}, \&
  {Ferri{\`e}re}}]{Padovani15}
{Padovani}, M., {Hennebelle}, P., {Marcowith}, A., \& {Ferri{\`e}re}, K. 2015,
  \aap, 582, L13

\bibitem[{{Padovani} {et~al.}(2016){Padovani}, {Marcowith}, {Hennebelle}, \&
  {Ferri{\`e}re}}]{Padovani16}
{Padovani}, M., {Marcowith}, A., {Hennebelle}, P., \& {Ferri{\`e}re}, K. 2016,
  \aap, 590, A8

\bibitem[{{Padovani} {et~al.}(2019){Padovani}, {Marcowith},
  {S{\'a}nchez-Monge}, {Meng}, \& {Schilke}}]{Padovani2019}
{Padovani}, M., {Marcowith}, A., {S{\'a}nchez-Monge}, {\'A}., {Meng}, F., \&
  {Schilke}, P. 2019, \aap, 630, A72

\bibitem[{{Parker}(1963)}]{Parker63}
{Parker}, R. A.~R. 1963, PhD thesis, California Institute of Technology

\bibitem[{{Parker}(1964)}]{Parker64}
{Parker}, R. A.~R. 1964, \apj, 139, 493

\bibitem[{{Parkes} {et~al.}(1977){Parkes}, {Charles}, {Culhane}, \&
  {Ives}}]{Parkes1977}
{Parkes}, G.~E., {Charles}, P.~A., {Culhane}, J.~L., \& {Ives}, J.~C. 1977,
  \mnras, 179, 55

\bibitem[{{Petre} {et~al.}(1988){Petre}, {Szymkowiak}, {Seward}, \&
  {Willingale}}]{Petre1988}
{Petre}, R., {Szymkowiak}, A.~E., {Seward}, F.~D., \& {Willingale}, R. 1988,
  \apj, 335, 215

\bibitem[{{Pickett} {et~al.}(1998){Pickett}, {Poynter}, {Cohen}, {Delitsky},
  {Pearson}, \& {M{\"u}ller}}]{Pickett1998}
{Pickett}, H.~M., {Poynter}, R.~L., {Cohen}, E.~A., {et~al.} 1998, \jqsrt, 60,
  883

\bibitem[{{Pihlstr{\"o}m} {et~al.}(2008){Pihlstr{\"o}m}, {Fish}, {Sjouwerman},
  {Zschaechner}, {Lockett}, \& {Elitzur}}]{Pihlstrom08}
{Pihlstr{\"o}m}, Y.~M., {Fish}, V.~L., {Sjouwerman}, L.~O., {et~al.} 2008,
  \apj, 676, 371

\bibitem[{{Pihlstr{\"o}m} {et~al.}(2014){Pihlstr{\"o}m}, {Sjouwerman}, {Frail},
  {Claussen}, {Mesler}, \& {McEwen}}]{Pihlstrom14}
{Pihlstr{\"o}m}, Y.~M., {Sjouwerman}, L.~O., {Frail}, D.~A., {et~al.} 2014,
  \aj, 147, 73

\bibitem[{{Pinheiro Gon{\c{c}}alves} {et~al.}(2011){Pinheiro Gon{\c{c}}alves},
  {Noriega-Crespo}, {Paladini}, {Martin}, \& {Carey}}]{Pinheiro11}
{Pinheiro Gon{\c{c}}alves}, D., {Noriega-Crespo}, A., {Paladini}, R., {Martin},
  P.~G., \& {Carey}, S.~J. 2011, \aj, 142, 47

\bibitem[{{Reach} {et~al.}(2019){Reach}, {Tram}, {Richter}, {Gusdorf}, \&
  {DeWitt}}]{Reach2019}
{Reach}, W.~T., {Tram}, L.~N., {Richter}, M., {Gusdorf}, A., \& {DeWitt}, C.
  2019, \apj, 884, 81

\bibitem[{{Rho} {et~al.}(2001){Rho}, {Jarrett}, {Cutri}, \& {Reach}}]{Rho2001}
{Rho}, J., {Jarrett}, T.~H., {Cutri}, R.~M., \& {Reach}, W.~T. 2001, \apj, 547,
  885

\bibitem[{{Rho} \& {Petre}(1998)}]{Rho1998}
{Rho}, J. \& {Petre}, R. 1998, \apjl, 503, L167

\bibitem[{{Richter} {et~al.}(1995{\natexlab{a}}){Richter}, {Graham}, \&
  {Wright}}]{Richter1995a}
{Richter}, M.~J., {Graham}, J.~R., \& {Wright}, G.~S. 1995{\natexlab{a}}, \apj,
  454, 277

\bibitem[{{Richter} {et~al.}(1995{\natexlab{b}}){Richter}, {Graham}, {Wright},
  {Kelly}, \& {Lacy}}]{Richter1995c}
{Richter}, M.~J., {Graham}, J.~R., {Wright}, G.~S., {Kelly}, D.~M., \& {Lacy},
  J.~H. 1995{\natexlab{b}}, \apjl, 449, L83

\bibitem[{{Richter} {et~al.}(1995{\natexlab{c}}){Richter}, {Graham}, {Wright},
  {Kelly}, \& {Lacy}}]{Richter1995b}
{Richter}, M.~J., {Graham}, J.~R., {Wright}, G.~S., {Kelly}, D.~M., \& {Lacy},
  J.~H. 1995{\natexlab{c}}, \apss, 233, 67

\bibitem[{{Roueff} {et~al.}(2020){Roueff}, {Gerin}, {Gratier}, {Levrier},
  {Pety}, {Gaudel}, {Goicoechea}, {Orkisz}, {de Souza Magalhaes}, {Vono},
  {Bardeau}, {Bron}, {Chanussot}, {Chainais}, {Guzman}, {Hughes},
  {Kainulainen}, {Languignon}, {Le Bourlot}, {Le Petit}, {Liszt}, {Marchal},
  {Miville-Deschenes}, {Peretto}, {Roueff}, \& {Sievers}}]{Roueff20}
{Roueff}, A., {Gerin}, M., {Gratier}, P., {et~al.} 2020, arXiv e-prints,
  arXiv:2005.08317

\bibitem[{{Sarangi} {et~al.}(2018){Sarangi}, {Matsuura}, \&
  {Micelotta}}]{Sarangi18}
{Sarangi}, A., {Matsuura}, M., \& {Micelotta}, E.~R. 2018, \ssr, 214, 63

\bibitem[{{Savage} \& {Mathis}(1979)}]{Savage1979}
{Savage}, B.~D. \& {Mathis}, J.~S. 1979, \araa, 17, 73

\bibitem[{{Scoville} {et~al.}(1977){Scoville}, {Irvine}, {Wannier}, \&
  {Predmore}}]{Scoville1977}
{Scoville}, N.~Z., {Irvine}, W.~M., {Wannier}, P.~G., \& {Predmore}, C.~R.
  1977, \apj, 216, 320

\bibitem[{{Seta} {et~al.}(1998){Seta}, {Hasegawa}, {Dame}, {Sakamoto}, {Oka},
  {Handa}, {Hayashi}, {Morino}, {Sorai}, \& {Usuda}}]{Seta1998}
{Seta}, M., {Hasegawa}, T., {Dame}, T.~M., {et~al.} 1998, \apj, 505, 286

\bibitem[{{Sharpless}(1965)}]{Sharpless65}
{Sharpless}, S. 1965, {Distribution of Associations, Emission Regions, Galactic
  Clusters and Supergiants}, 131

\bibitem[{{Shinn} {et~al.}(2011){Shinn}, {Koo}, {Seon}, \& {Lee}}]{Shinn2011}
{Shinn}, J.-H., {Koo}, B.-C., {Seon}, K.-I., \& {Lee}, H.-G. 2011, \apj, 732,
  124

\bibitem[{{Skrutskie} {et~al.}(2006){Skrutskie}, {Cutri}, {Stiening},
  {Weinberg}, {Schneider}, {Carpenter}, {Beichman}, {Capps}, {Chester},
  {Elias}, {Huchra}, {Liebert}, {Lonsdale}, {Monet}, {Price}, {Seitzer},
  {Jarrett}, {Kirkpatrick}, {Gizis}, {Howard}, {Evans}, {Fowler}, {Fullmer},
  {Hurt}, {Light}, {Kopan}, {Marsh}, {McCallon}, {Tam}, {Van Dyk}, \&
  {Wheelock}}]{Skrutskie06}
{Skrutskie}, M.~F., {Cutri}, R.~M., {Stiening}, R., {et~al.} 2006, \aj, 131,
  1163

\bibitem[{{Snell} {et~al.}(2005){Snell}, {Hollenbach}, {Howe}, {Neufeld},
  {Kaufman}, {Melnick}, {Bergin}, \& {Wang}}]{Snell2005}
{Snell}, R.~L., {Hollenbach}, D., {Howe}, J.~E., {et~al.} 2005, \apj, 620, 758

\bibitem[{{Sobolev}(1960)}]{Sobolev1960}
{Sobolev}, V.~V. 1960, {Moving envelopes of stars}

\bibitem[{{Straal} \& {van Leeuwen}(2019)}]{Straal19}
{Straal}, S.~M. \& {van Leeuwen}, J. 2019, \aap, 623, A90

\bibitem[{{Su} {et~al.}(2014){Su}, {Fang}, {Yang}, {Zhou}, \& {Chen}}]{Su2014}
{Su}, Y., {Fang}, M., {Yang}, J., {Zhou}, P., \& {Chen}, Y. 2014, \apj, 788,
  122

\bibitem[{{Surdej}(1977)}]{Surdej77}
{Surdej}, J. 1977, \aap, 60, 303

\bibitem[{{Surdej}(1978)}]{Surdej1978}
{Surdej}, J. 1978, \aap, 66, 45

\bibitem[{{Swartz} {et~al.}(2015){Swartz}, {Pavlov}, {Clarke}, {Castelletti},
  {Zavlin}, {Bucciantini}, {Karovska}, {van der Horst}, {Yukita}, \&
  {Weisskopf}}]{Swartz15}
{Swartz}, D.~A., {Pavlov}, G.~G., {Clarke}, T., {et~al.} 2015, \apj, 808, 84

\bibitem[{{Tang}(2019)}]{Tang19}
{Tang}, X. 2019, \mnras, 482, 3843

\bibitem[{{Tatischeff} \& {Gabici}(2018)}]{Tatischeff18}
{Tatischeff}, V. \& {Gabici}, S. 2018, Annual Review of Nuclear and Particle
  Science, 68, 377

\bibitem[{{Tauber} {et~al.}(1994){Tauber}, {Snell}, {Dickman}, \&
  {Ziurys}}]{Tauber1994}
{Tauber}, J.~A., {Snell}, R.~L., {Dickman}, R.~L., \& {Ziurys}, L.~M. 1994,
  \apj, 421, 570

\bibitem[{{Tavani} {et~al.}(2010){Tavani}, {Giuliani}, {Chen}, {Argan},
  {Barbiellini}, {Bulgarelli}, {Caraveo}, {Cattaneo}, {Cocco}, {Contessi},
  {D'Ammand o}, {Costa}, {De Paris}, {Del Monte}, {Di Cocco}, {Donnarumma},
  {Evangelista}, {Ferrari}, {Feroci}, {Fuschino}, {Galli}, {Gianotti},
  {Labanti}, {Lapshov}, {Lazzarotto}, {Lipari}, {Longo}, {Marisaldi},
  {Mastropietro}, {Mereghetti}, {Morelli}, {Moretti}, {Morselli}, {Pacciani},
  {Pellizzoni}, {Perotti}, {Piano}, {Picozza}, {Pilia}, {Pucella}, {Prest},
  {Rapisarda}, {Rappoldi}, {Scalise}, {Rubini}, {Sabatini}, {Striani},
  {Soffitta}, {Trifoglio}, {Trois}, {Vallazza}, {Vercellone}, {Vittorini},
  {Zambra}, {Zanello}, {Pittori}, {Verrecchia}, {Santolamazza}, {Giommi},
  {Colafrancesco}, {Antonelli}, \& {Salotti}}]{Tavani2010}
{Tavani}, M., {Giuliani}, A., {Chen}, A.~W., {et~al.} 2010, \apjl, 710, L151

\bibitem[{{Torres} {et~al.}(2010){Torres}, {Marrero}, \& {de Cea Del
  Pozo}}]{Torres2010}
{Torres}, D.~F., {Marrero}, A. Y.~R., \& {de Cea Del Pozo}, E. 2010, \mnras,
  408, 1257

\bibitem[{{Torres} {et~al.}(2003){Torres}, {Romero}, {Dame}, {Combi}, \&
  {Butt}}]{Torres2003}
{Torres}, D.~F., {Romero}, G.~E., {Dame}, T.~M., {Combi}, J.~A., \& {Butt},
  Y.~M. 2003, \physrep, 382, 303

\bibitem[{{Treffers}(1979)}]{Treffers1979}
{Treffers}, R.~R. 1979, \apjl, 233, L17

\bibitem[{{Trimble}(1991)}]{Trimble91}
{Trimble}, V. 1991, \aapr, 3, 1

\bibitem[{{Troja} {et~al.}(2008){Troja}, {Bocchino}, {Miceli}, \&
  {Reale}}]{Troja2008}
{Troja}, E., {Bocchino}, F., {Miceli}, M., \& {Reale}, F. 2008, \aap, 485, 777

\bibitem[{{Troja} {et~al.}(2006){Troja}, {Bocchino}, \& {Reale}}]{Troja2006}
{Troja}, E., {Bocchino}, F., \& {Reale}, F. 2006, \apj, 649, 258

\bibitem[{{Turner}(1982)}]{Turner82}
{Turner}, B.~E. 1982, Astrophysics and Space Science Library, Vol.~93, {VLA
  observations of OH masers and associated ultracompact continuum sources}, ed.
  R.~S. {Roger} \& P.~E. {Dewdney}, 425--431

\bibitem[{{Turner} {et~al.}(1992){Turner}, {Chan}, {Green}, \&
  {Lubowich}}]{Turner1992}
{Turner}, B.~E., {Chan}, K.-W., {Green}, S., \& {Lubowich}, D.~A. 1992, \apj,
  399, 114

\bibitem[{{van der Laan}(1962{\natexlab{a}})}]{vanderLaan1962}
{van der Laan}, H. 1962{\natexlab{a}}, \mnras, 124, 125

\bibitem[{{van der Laan}(1962{\natexlab{b}})}]{vanderLaan1962b}
{van der Laan}, H. 1962{\natexlab{b}}, \mnras, 124, 179

\bibitem[{{van der Tak} {et~al.}(2007){van der Tak}, {Black}, {Sch{\"o}ier},
  {Jansen}, \& {van Dishoeck}}]{VanderTak2007}
{van der Tak}, F.~F.~S., {Black}, J.~H., {Sch{\"o}ier}, F.~L., {Jansen}, D.~J.,
  \& {van Dishoeck}, E.~F. 2007, \aap, 468, 627

\bibitem[{{van der Walt} {et~al.}(2011){van der Walt}, {Colbert}, \&
  {Varoquaux}}]{Numpy11}
{van der Walt}, S., {Colbert}, S.~C., \& {Varoquaux}, G. 2011, Computing in
  Science and Engineering, 13, 22

\bibitem[{{van Dishoeck} {et~al.}(1993){van Dishoeck}, {Jansen}, \&
  {Phillips}}]{Vandishoeck93}
{van Dishoeck}, E.~F., {Jansen}, D.~J., \& {Phillips}, T.~G. 1993, \aap, 279,
  541

\bibitem[{{Vaupré} {et~al.}(2014){Vaupré}, {Hily-Blant}, {Ceccarelli},
  {Dubus}, {Gabici}, \& {Montmerle}}]{Vaupre2014}
{Vaupré}, S., {Hily-Blant}, P., {Ceccarelli}, C., {et~al.} 2014, \aap, 568,
  A50

\bibitem[{{Wang} \& {Scoville}(1992)}]{Wang&Scoville1992}
{Wang}, Z. \& {Scoville}, N.~Z. 1992, \apj, 386, 158

\bibitem[{{Wang} {et~al.}(1992){Wang}, {Asaoka}, {Hayakawa}, \&
  {Koyama}}]{Wang1992}
{Wang}, Z.~R., {Asaoka}, I., {Hayakawa}, S., \& {Koyama}, K. 1992, \pasj, 44,
  303

\bibitem[{{Wardle}(1999)}]{Wardle99}
{Wardle}, M. 1999, \apjl, 525, L101

\bibitem[{{Wardle} \& {Yusef-Zadeh}(2002)}]{Wardle02}
{Wardle}, M. \& {Yusef-Zadeh}, F. 2002, Science, 296, 2350

\bibitem[{{Welsh} \& {Sallmen}(2003)}]{Welsh03}
{Welsh}, B.~Y. \& {Sallmen}, S. 2003, \aap, 408, 545

\bibitem[{{Wenger} {et~al.}(2018){Wenger}, {Balser}, {Anderson}, \&
  {Bania}}]{Wenger2018}
{Wenger}, T.~V., {Balser}, D.~S., {Anderson}, L.~D., \& {Bania}, T.~M. 2018,
  \apj, 856, 52

\bibitem[{{White} {et~al.}(1987){White}, {Rainey}, {Hayashi}, \&
  {Kaifu}}]{White1987}
{White}, G.~J., {Rainey}, R., {Hayashi}, S.~S., \& {Kaifu}, N. 1987, \aap, 173,
  337

\bibitem[{{Wilking} {et~al.}(1989){Wilking}, {Lada}, \& {Young}}]{Wilking1989}
{Wilking}, B.~A., {Lada}, C.~J., \& {Young}, E.~T. 1989, \apj, 340, 823

\bibitem[{{Wilson} \& {Matteucci}(1992)}]{Wilson92}
{Wilson}, T.~L. \& {Matteucci}, F. 1992, \aapr, 4, 1

\bibitem[{{Wilson} \& {Rood}(1994)}]{Wilson1994}
{Wilson}, T.~L. \& {Rood}, R. 1994, \araa, 32, 191

\bibitem[{{Winkler} \& {Clark}(1974)}]{Winkler74}
{Winkler}, P.~Frank, J. \& {Clark}, G.~W. 1974, \apjl, 191, L67

\bibitem[{{Woltjer}(1972)}]{Woltjer72}
{Woltjer}, L. 1972, \araa, 10, 129

\bibitem[{{Wood} {et~al.}(1991){Wood}, {Mufson}, \& {Dickel}}]{Wood91}
{Wood}, C.~A., {Mufson}, S.~L., \& {Dickel}, J.~R. 1991, \aj, 102, 224

\bibitem[{{Woodall} \& {Gray}(2007)}]{Woodall07}
{Woodall}, J.~M. \& {Gray}, M.~D. 2007, \mnras, 378, L20

\bibitem[{{Wright} {et~al.}(2010){Wright}, {Eisenhardt}, {Mainzer}, {Ressler},
  {Cutri}, {Jarrett}, {Kirkpatrick}, {Padgett}, {McMillan}, {Skrutskie},
  {Stanford}, {Cohen}, {Walker}, {Mather}, {Leisawitz}, {Gautier}, {McLean},
  {Benford}, {Lonsdale}, {Blain}, {Mendez}, {Irace}, {Duval}, {Liu}, {Royer},
  {Heinrichsen}, {Howard}, {Shannon}, {Kendall}, {Walsh}, {Larsen}, {Cardon},
  {Schick}, {Schwalm}, {Abid}, {Fabinsky}, {Naes}, \& {Tsai}}]{Wright10}
{Wright}, E.~L., {Eisenhardt}, P.~R.~M., {Mainzer}, A.~K., {et~al.} 2010, \aj,
  140, 1868

\bibitem[{{Xu} {et~al.}(2011){Xu}, {Wang}, \& {Miller}}]{Xu11}
{Xu}, J.-L., {Wang}, J.-J., \& {Miller}, M. 2011, \apj, 727, 81

\bibitem[{{Yamaguchi} {et~al.}(2009){Yamaguchi}, {Ozawa}, {Koyama}, {Masai},
  {Hiraga}, {Ozaki}, \& {Yonetoku}}]{Yamaguchi2009}
{Yamaguchi}, H., {Ozawa}, M., {Koyama}, K., {et~al.} 2009, \apjl, 705, L6

\bibitem[{{York} {et~al.}(2000){York}, {Adelman}, {Anderson}, {Anderson},
  {Annis}, {Bahcall}, {Bakken}, {Barkhouser}, {Bastian}, {Berman}, {Boroski},
  {Bracker}, {Briegel}, {Briggs}, {Brinkmann}, {Brunner}, {Burles}, {Carey},
  {Carr}, {Castander}, {Chen}, {Colestock}, {Connolly}, {Crocker}, {Csabai},
  {Czarapata}, {Davis}, {Doi}, {Dombeck}, {Eisenstein}, {Ellman}, {Elms},
  {Evans}, {Fan}, {Federwitz}, {Fiscelli}, {Friedman}, {Frieman}, {Fukugita},
  {Gillespie}, {Gunn}, {Gurbani}, {de Haas}, {Haldeman}, {Harris}, {Hayes},
  {Heckman}, {Hennessy}, {Hindsley}, {Holm}, {Holmgren}, {Huang}, {Hull},
  {Husby}, {Ichikawa}, {Ichikawa}, {Ivezi{\'c}}, {Kent}, {Kim}, {Kinney},
  {Klaene}, {Kleinman}, {Kleinman}, {Knapp}, {Korienek}, {Kron}, {Kunszt},
  {Lamb}, {Lee}, {Leger}, {Limmongkol}, {Lindenmeyer}, {Long}, {Loomis},
  {Loveday}, {Lucinio}, {Lupton}, {MacKinnon}, {Mannery}, {Mantsch}, {Margon},
  {McGehee}, {McKay}, {Meiksin}, {Merelli}, {Monet}, {Munn}, {Narayanan},
  {Nash}, {Neilsen}, {Neswold}, {Newberg}, {Nichol}, {Nicinski}, {Nonino},
  {Okada}, {Okamura}, {Ostriker}, {Owen}, {Pauls}, {Peoples}, {Peterson},
  {Petravick}, {Pier}, {Pope}, {Pordes}, {Prosapio}, {Rechenmacher}, {Quinn},
  {Richards}, {Richmond}, {Rivetta}, {Rockosi}, {Ruthmansdorfer}, {Sand ford},
  {Schlegel}, {Schneider}, {Sekiguchi}, {Sergey}, {Shimasaku}, {Siegmund},
  {Smee}, {Smith}, {Snedden}, {Stone}, {Stoughton}, {Strauss}, {Stubbs},
  {SubbaRao}, {Szalay}, {Szapudi}, {Szokoly}, {Thakar}, {Tremonti}, {Tucker},
  {Uomoto}, {Vanden Berk}, {Vogeley}, {Waddell}, {Wang}, {Watanabe},
  {Weinberg}, {Yanny}, {Yasuda}, \& {SDSS Collaboration}}]{York2000}
{York}, D.~G., {Adelman}, J., {Anderson}, John~E., J., {et~al.} 2000, \aj, 120,
  1579

\bibitem[{{Yuan} \& {Neufeld}(2011)}]{Yuan&Neufeld2011}
{Yuan}, Y. \& {Neufeld}, D.~A. 2011, \apj, 726, 76

\bibitem[{{Yusef-Zadeh} {et~al.}(2003){Yusef-Zadeh}, {Wardle}, {Rho}, \&
  {Sakano}}]{Yusefzadeh03}
{Yusef-Zadeh}, F., {Wardle}, M., {Rho}, J., \& {Sakano}, M. 2003, \apj, 585,
  319

\bibitem[{{Zhang} {et~al.}(2010){Zhang}, {Gao}, \& {Wang}}]{Zhang2010}
{Zhang}, Z., {Gao}, Y., \& {Wang}, J. 2010, Science China Physics, Mechanics,
  and Astronomy, 53, 1357

\bibitem[{{Ziurys} {et~al.}(1988){Ziurys}, {Snell}, \& {Dickman}}]{Ziurys1988}
{Ziurys}, L.~M., {Snell}, R.~L., \& {Dickman}, R.~L. 1988, {Recent Molecular
  Studies of Supernova Remnant IC443 - Some New Results for Shock Chemistry},
  ed. R.~L. {Dickman}, R.~L. {Snell}, \& J.~S. {Young}, Vol. 315, 184

\end{thebibliography}

\clearpage

\begin{appendix}

\section{Coordinates of the boxes}

\begin{table}[h!]
\caption{Coordinates of the vertices defining the boxes corresponding to dense cloudlet, ring-like structure and shocked clump (Fig. \ref{fig:boxes}, Tab. \ref{table:boxes}).}             
\label{table:boxes}      
\centering                          
\begin{tabular}{c c c}        
\hline\hline  \\[-1.0em]                
region & r.a. [J2000] & dec. [J2000]\\
\hline   \\[-1.0em]                     
    \footnotesize{} & \footnotesize{$6^{\mathrm{h}}16^{\mathrm{m}}37.0^{\mathrm{s}}$} & \footnotesize{$+22^\circ40'24.0^{\prime \prime}$} \\
    \footnotesize{} & \footnotesize{$6^{\mathrm{h}}16^{\mathrm{m}}23.1^{\mathrm{s}}$} & \footnotesize{$+22^\circ40'24.0^{\prime \prime}$} \\
    \footnotesize{dense cloudlet} & \footnotesize{$6^{\mathrm{h}}16^{\mathrm{m}}23.1^{\mathrm{s}}$} & \footnotesize{$+22^\circ33'39.8^{\prime \prime}$} \\
    \footnotesize{(A)} & \footnotesize{$6^{\mathrm{h}}16^{\mathrm{m}}30.3^{\mathrm{s}}$} & \footnotesize{$+22^\circ33'39.8^{\prime \prime}$} \\
    \footnotesize{} & \footnotesize{$6^{\mathrm{h}}16^{\mathrm{m}}30.3^{\mathrm{s}}$} & \footnotesize{$+22^\circ36'09.5''$} \\
    \footnotesize{} & \footnotesize{$6^{\mathrm{h}}16^{\mathrm{m}}37.0^{\mathrm{s}}$} & \footnotesize{$+22^\circ36'09.5''$} \\
\hline \\[-1.0em] 
    \footnotesize{} & \footnotesize{$6^{\mathrm{h}}16^{\mathrm{m}}41.2^{\mathrm{s}}$} & \footnotesize{$+22^\circ36'09.5''$} \\
    \footnotesize{ring-like structure} & \footnotesize{$6^{\mathrm{h}}16^{\mathrm{m}}30.3^{\mathrm{s}}$} & \footnotesize{$+22^\circ36'09.5''$} \\
    \footnotesize{(B)} & \footnotesize{$6^{\mathrm{h}}16^{\mathrm{m}}30.3^{\mathrm{s}}$} & \footnotesize{$+22^\circ33'39.8''$} \\
    \footnotesize{} & \footnotesize{$6^{\mathrm{h}}16^{\mathrm{m}}41.2^{\mathrm{s}}$} & \footnotesize{$+22^\circ33'39.8''$} \\
\hline \\[-1.0em] 
    \footnotesize{} & \footnotesize{$6^{\mathrm{h}}16^{\mathrm{m}}47.1^{\mathrm{s}}$} & \footnotesize{$+22^\circ34'30.7''$}   \\
    \footnotesize{} & \footnotesize{$6^{\mathrm{h}}16^{\mathrm{m}}40.2^{\mathrm{s}}$} & \footnotesize{$+22^\circ34'30.7''$} \\
    \footnotesize{} & \footnotesize{$6^{\mathrm{h}}16^{\mathrm{m}}40.2^{\mathrm{s}}$} & \footnotesize{$+22^\circ32'32.4''$} \\
    \footnotesize{} & \footnotesize{$6^{\mathrm{h}}16^{\mathrm{m}}33.8^{\mathrm{s}}$} & \footnotesize{$+22^\circ32'32.4''$} \\
    \footnotesize{} & \footnotesize{$6^{\mathrm{h}}16^{\mathrm{m}}33.8^{\mathrm{s}}$} & \footnotesize{$+22^\circ31'30.7''$} \\
    \footnotesize{shocked clump} & \footnotesize{$6^{\mathrm{h}}16^{\mathrm{m}}31.9^{\mathrm{s}}$} & \footnotesize{$+22^\circ31'30.7''$} \\
    \footnotesize{(C)} & \footnotesize{$6^{\mathrm{h}}16^{\mathrm{m}}31.9^{\mathrm{s}}$} & \footnotesize{$+22^\circ29'36.6''$} \\
    \footnotesize{} & \footnotesize{$6^{\mathrm{h}}16^{\mathrm{m}}44.7^{\mathrm{s}}$} & \footnotesize{$+22^\circ29'36.6''$} \\
    \footnotesize{} & \footnotesize{$6^{\mathrm{h}}16^{\mathrm{m}}44.7^{\mathrm{s}}$} & \footnotesize{$+22^\circ31'06.4''$} \\
    \footnotesize{} & \footnotesize{$6^{\mathrm{h}}16^{\mathrm{m}}46.2^{\mathrm{s}}$} & \footnotesize{$+22^\circ31'06.4''$} \\
    \footnotesize{} & \footnotesize{$6^{\mathrm{h}}16^{\mathrm{m}}46.2^{\mathrm{s}}$} & \footnotesize{$+22^\circ32'21.3''$} \\
    \footnotesize{} & \footnotesize{$6^{\mathrm{h}}16^{\mathrm{m}}47.1^{\mathrm{s}}$} & \footnotesize{$+22^\circ32'21.3''$} \\
\hline                                   
\end{tabular}
\end{table}

\section{Thermometry of absorption lines}\label{thermometry}

Considering a transition between an upper level $u$ and a lower level $l$ , the extinction coefficient $\kappa_{\nu}$ is given by:
\begin{equation}
\kappa_{\nu} = \dfrac{c^2}{8 \pi \nu_{ul}^2} A_{ul} n_l \dfrac{g_u}{g_l} \left[ 1 - \mathrm{exp} \left( \dfrac{h \nu_{ul}}{k T_{ex}} \right) \right] \phi(\nu)
\end{equation}
Where $T_{ex}$ is the excitation temperature, $\phi (\nu)$ the profile of the line , $\nu_{ul}$ the frequency of the transition, $A_{ul}$ the Einstein coefficient of spontaneous emission, $n_l$ the density of the lower level and $g_u$ and $g_l$ are the statistical weights.
If $u \leftrightarrow J + 1$ and $l \leftrightarrow J$, then the Einstein coefficient is defined by:
\begin{equation}
A_{J + 1 \rightarrow J} = \dfrac{64 \pi ^4 \mu^2}{3h c^3} \dfrac{J + 1}{2J + 3} \nu_{J}^3
\end{equation}
Where $\mu$ is the dipole moment of the considered molecule.
Thus, by using this definition together with the definition of the statistical weights we get:
\begin{equation}
\kappa_{\nu}^{J \rightarrow J +1} = \dfrac{8 \pi^3 \mu^2}{3hc} \dfrac{J+1}{2J +1} \nu_{J} n_{J} \left[ 1 - \mathrm{exp} \left( - \dfrac{h \nu_J}{kT_{ex}} \right) \right]
\end{equation}
If we introduce the partition function we get a relation linking $n_J$ to the total density $n$:
\begin{equation}
n_J = n \dfrac{g_J}{Q(T_{ex})} \mathrm{exp} \left(- \dfrac{E_J}{k T_{ex}} \right)
\end{equation}
Where $E_J$ and $E_{J+1}$ are the level energies.
So that:
\begin{equation}
\kappa_{\nu}^{J \rightarrow J +1} = \dfrac{8 \pi^3 \mu^2}{3hc} \nu_J n \dfrac{J+1}{Q(T_{ex})} \mathrm{exp} \left( - \dfrac{E_J}{k T_{ex}} \right) \left[ 1 - \mathrm{exp} \left( \dfrac{E_J - E_{J+1}}{k T_{ex}} \right) \right] \phi (\nu)
\end{equation}
Thus the total column density $N$ can be obtained by computing the following integral over the line of sight:
\begin{eqnarray*}
N &=& \int \mathrm{d}s \int n \phi (\nu) \mathrm{d} \nu  \\
&=& \int \mathrm{d} s \int \dfrac{\kappa_{\nu}^{J \rightarrow J +1} \dfrac{3hc}{8 \pi ^3 \mu^2} \dfrac{1}{\nu_J}\dfrac{Q(T_{ex})}{J+1}\mathrm{exp}\left( \dfrac{E_J}{kT_{ex}}\right)}{\left[ 1 - \mathrm{exp} \left( \dfrac{E_J - E_{J+1}}{k T_{ex}	}\right)\right]} \mathrm{d} \nu
\end{eqnarray*}
Using the definition of the equivalent width $W = \int \tau \mathrm{d} v$ and the variable $\mathrm{d} \nu = \nu_J \mathrm{d} v / c$, we finally get the relation between column density and equivalent width for a given excitation temperature:
\begin{equation}
N = \Lambda \dfrac{Q(T_{ex})}{1 - \mathrm{exp} \left( \dfrac{-h \nu_{ij}}{k T_{ex}} \right)} W_{ij}
\end{equation}
Where $\Lambda = 3h/8 \pi^3 \mu^2$, $h$ is the planck constant and $\nu_{ij}$ the frequency of the transition.
Applied to the transitions J=1--0 and J=2--1, this relation can be written for two different values of the equivalent width $W$ corresponding to each line. As the total column density is constant in each equation, the ratio of these two relations yields:
\begin{equation}
1 = \dfrac{1}{2} \mathrm{exp} \left( \dfrac{h \nu_{ij}}{k T_{ex}} \right) \dfrac{1 - \mathrm{exp}\left( \dfrac{- h \nu_{ij}}{k T_{ex}}\right)}{1 - \mathrm{exp}\left( \dfrac{- 2 h \nu_{ij}}{k T_{ex}}\right)} \dfrac{W_{1 \rightarrow 2}}{W _{0 \rightarrow 1}}
\end{equation}
If we define the temperature of the transition $T_0=h \nu_{ij}/k$ and use the variable $u = T_0 /T_{ex}$ then we have:
\begin{equation}
2 = \mathrm{e}^{2u} \dfrac{1 - \mathrm{e}^{-2u}}{1 - \mathrm{e}^{-4u}} \dfrac{W_{1 \rightarrow 2}}{W _{0 \rightarrow 1}} = \mathrm{e}^{3u} \dfrac{\mathrm{e}^u - \mathrm{e}^{-u}}{\mathrm{e}^{2u} - \mathrm{e}^{-2u}} \dfrac{W_{1 \rightarrow 2}}{W _{0 \rightarrow 1}} 
\end{equation}
Or:
\begin{equation}
2 = \dfrac{\mathrm{e}^{3u}}{2 \mathrm{cosh}(u)} \dfrac{W_{1 \rightarrow 2}}{W _{0 \rightarrow 1}} 
\end{equation}
and, finally we get the following relation which permits to determine the temperature from a measure of the ratio of the two equivalent widths:
\begin{equation}
F \left( \dfrac{T_0}{T_{ex}} \right) = \dfrac{W_{1 \rightarrow 2}}{ 4 W_{0 \rightarrow 1}}
\end{equation}
Where $F(u) = \mathrm{e}^{-3u} \mathrm{cosh}(u)$.

\newpage
\onecolumn

\section{Spatial distribution of YSO candidates}

\begin{figure*}[h]
\begin{center}
\resizebox{0.46\hsize}{!}
{\includegraphics[trim={0cm 0.75cm 0cm 1.cm},clip]{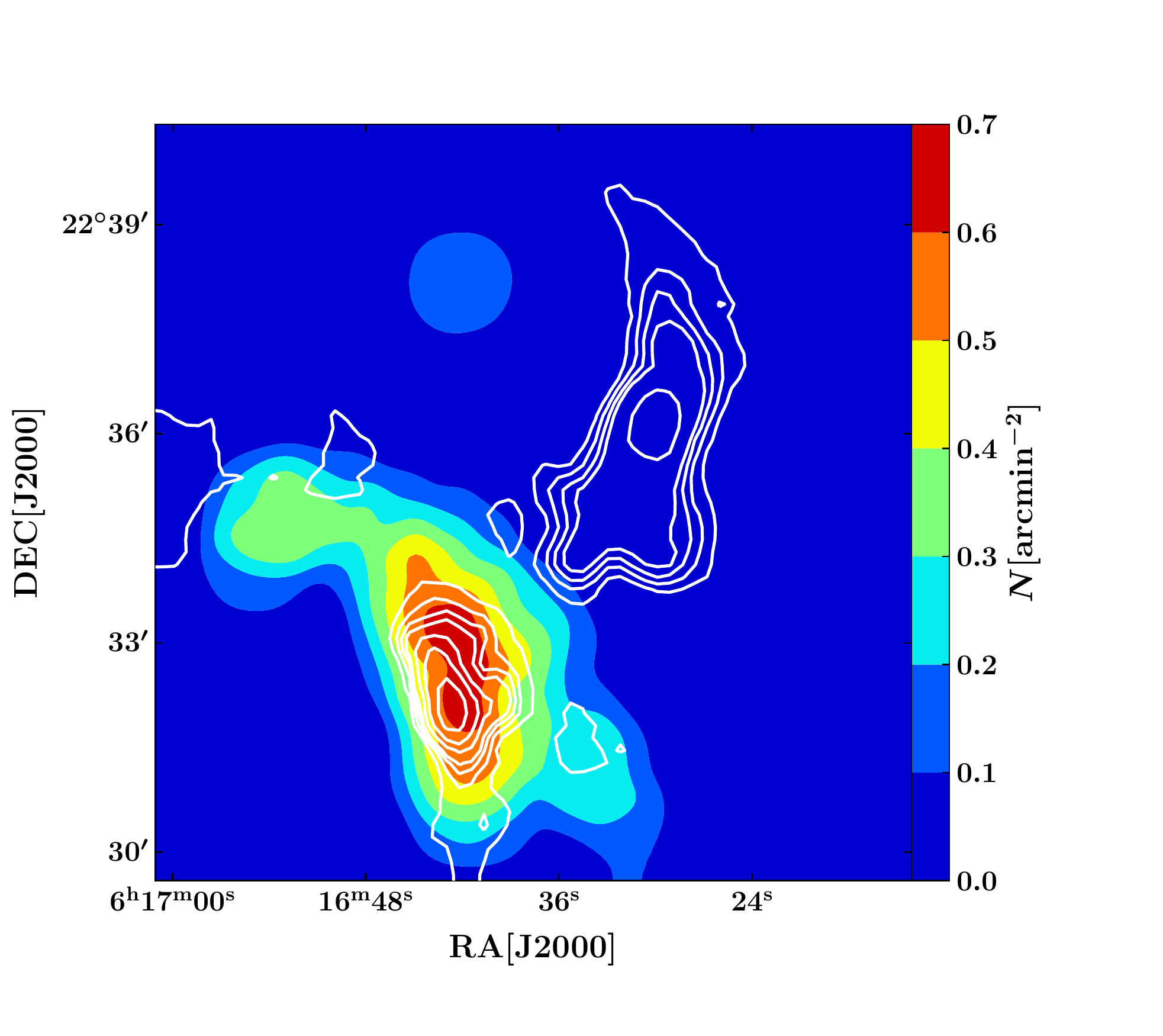}}
\resizebox{0.48\hsize}{!}
{\includegraphics[trim={0cm 0.75cm 0cm 1.8cm},clip]{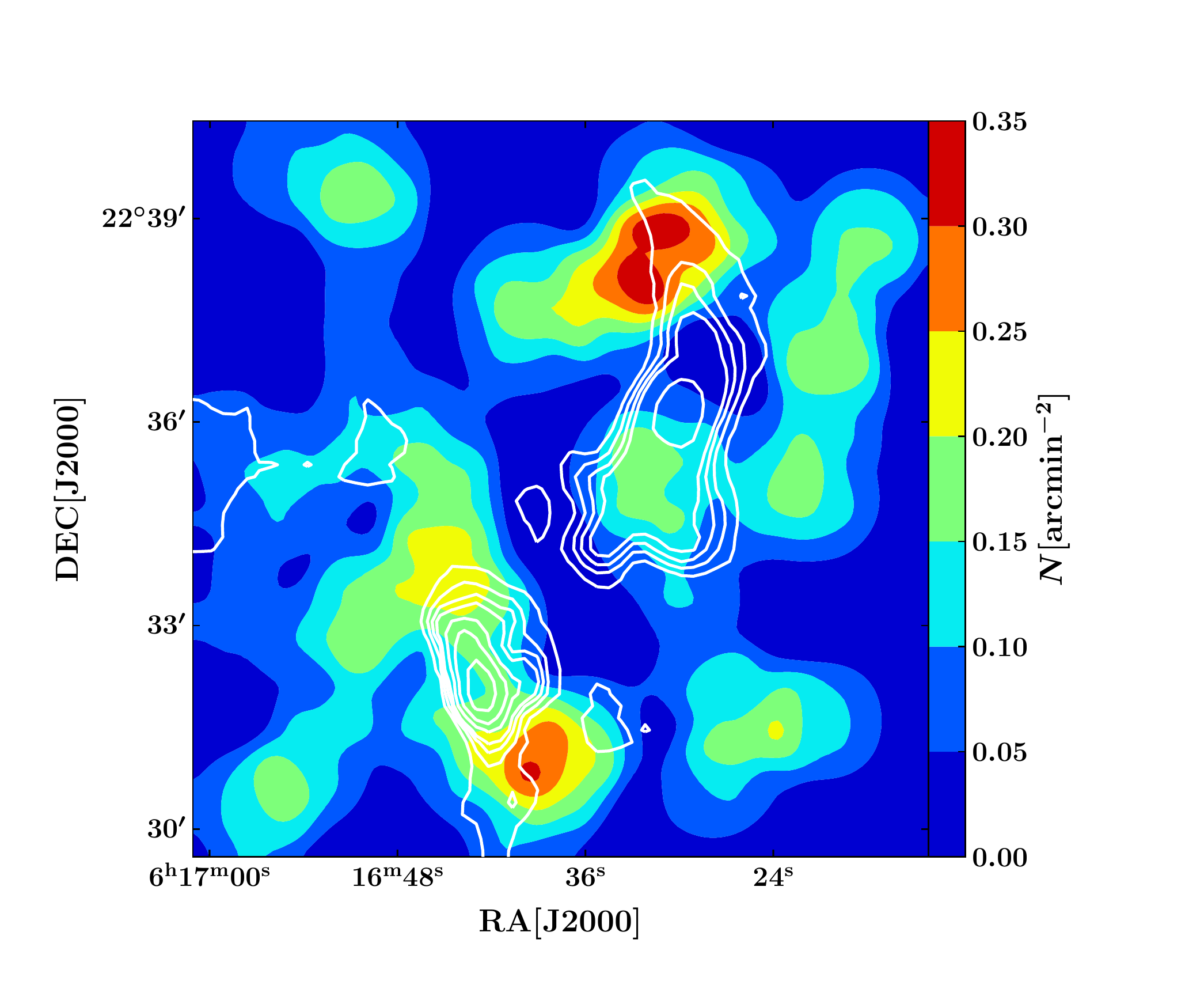}}
\caption{Spatial distribution of protostar candidates in our 10$^\prime\times$10$^\prime$ field of observations. \textit{Left}: Class I and Class II candidates detected by WISE. \textit{Right}: YSO candidates detected by 2MASS. White contours represent the CO(2-1) emission in the velocity range [-7, +3] km/s.}
\label{fig:YSOdist}
\end{center}
\end{figure*}

\section{Complementary channel maps}

\begin{figure*}[h]
\begin{center}
\resizebox{0.61\hsize}{!}
{\includegraphics[trim={0cm 0.75cm 0cm 1.5cm},clip]{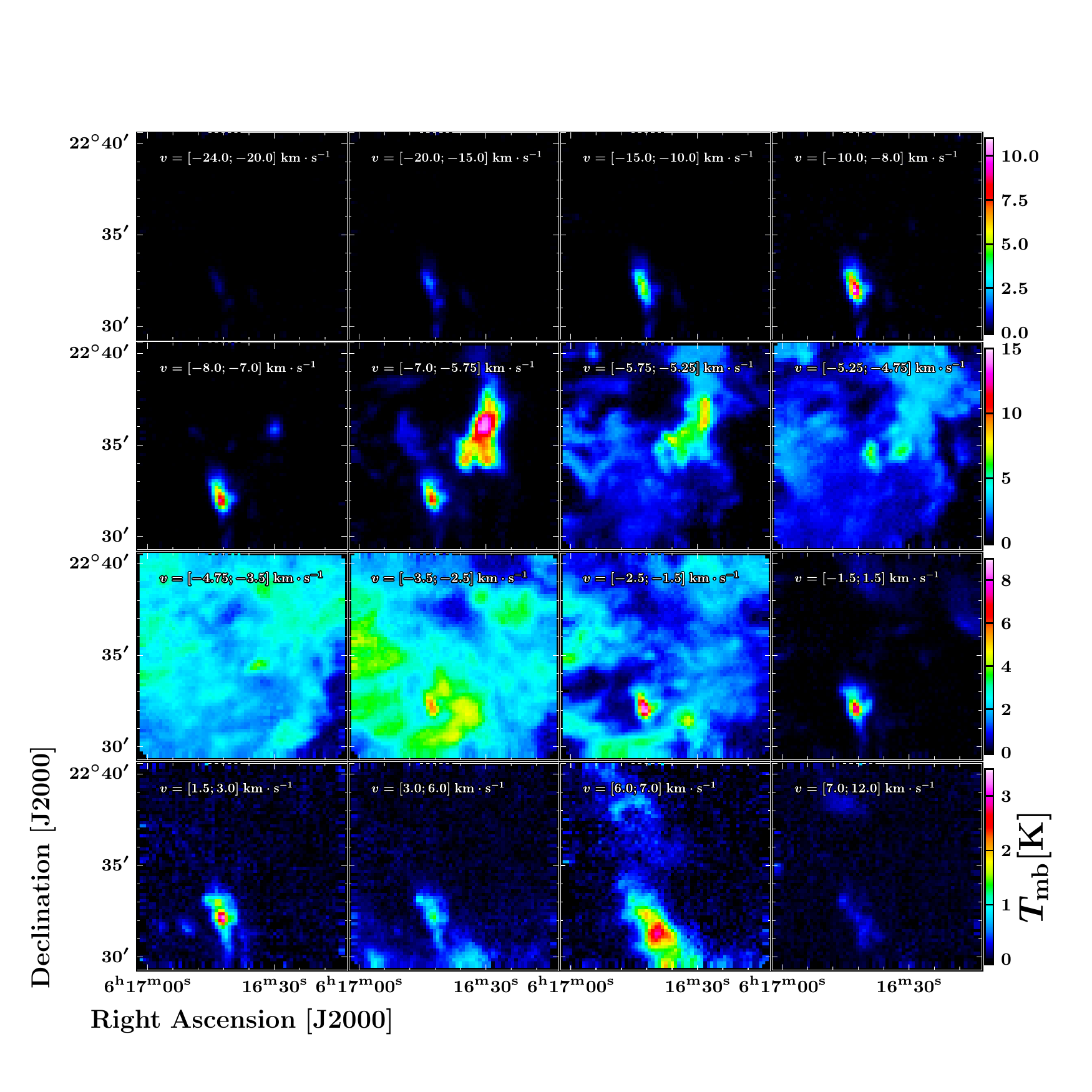}}
\caption{Channel map of the $^{12}\mathrm{CO}(1-0)$ observations carried out with the IRAM-30m telescope. The $10^\prime \times 10^\prime$ field is entirely displayed. Each panel represent the emission integrated over an interval of $0.5~\mathrm{km ~ s^{-1}}$ around a given velocity along the line of sight. Velocities are indicated on the top left corner of each panel. Velocity channels represented on this figure are between $\varv=-20~\mathrm{km ~ s^{-1}}$ and $\varv=8~\mathrm{km ~ s^{-1}}$, covering all the spectral features detected towards IC443G.}
\label{fig:channelmap12co10}
\end{center}
\end{figure*}

\begin{figure*}[h]
\begin{center}
\resizebox{0.61\hsize}{!}
{\includegraphics[trim={0cm 0.75cm 0cm 1.5cm},clip]{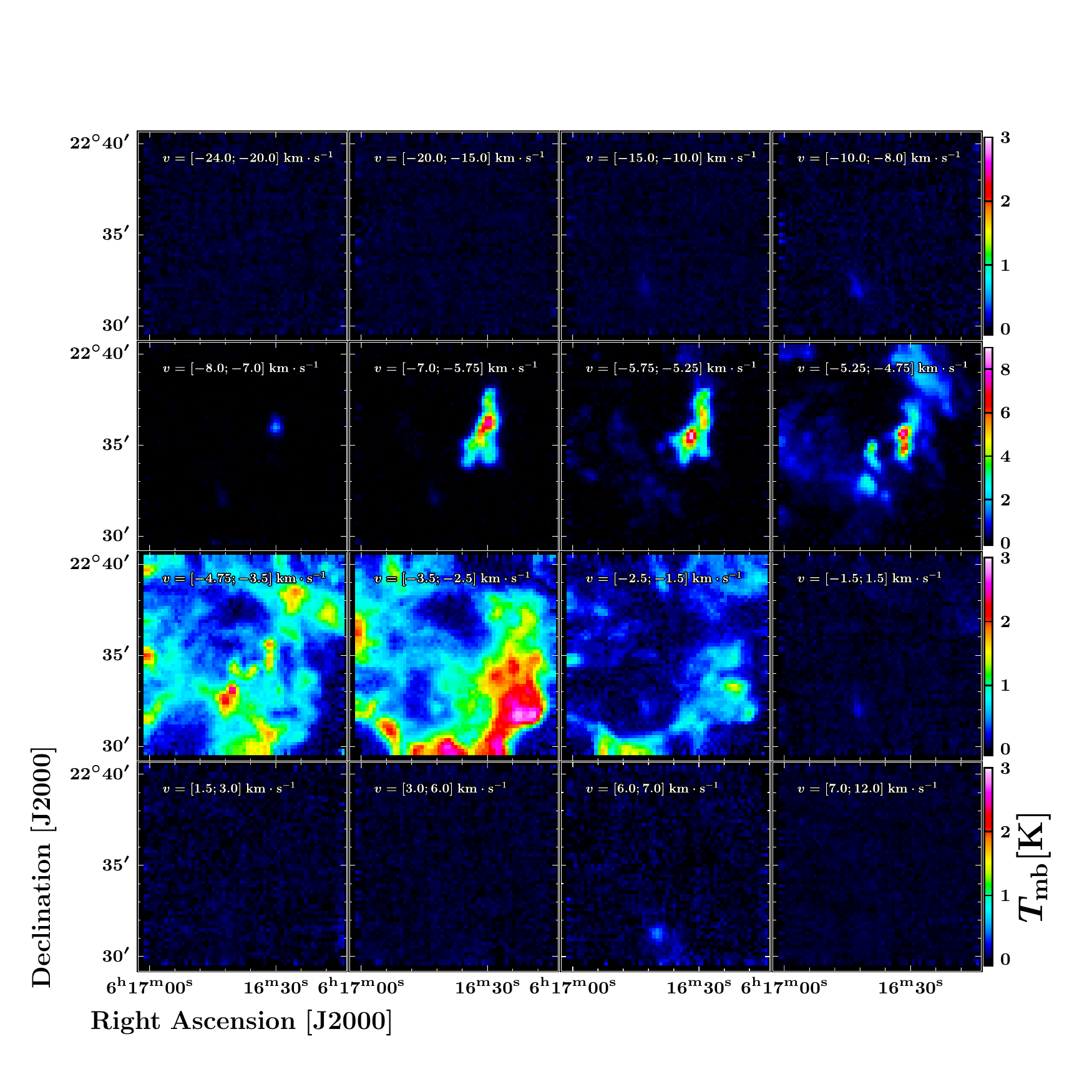}}
\caption{Channel map of the $^{13}\mathrm{CO}(1-0)$ observations carried out with the IRAM-30m telescope. The $10^\prime \times 10^\prime$ field is entirely displayed. Each panel represent the emission integrated over an interval of velocity along the line of sight. Velocities are indicated on the top left corner of each panel. Velocity channels represented in this figure are between $\varv=-24~\mathrm{km ~ s^{-1}}$ and $\varv=+12~\mathrm{km ~ s^{-1}}$, covering all the spectral features detected towards the extended G region.}
\label{fig:channelmap13co10}
\end{center}
\end{figure*}

\begin{figure*}[h]
\begin{center}
\resizebox{0.61\hsize}{!}
{\includegraphics[trim={0cm 0.75cm 0cm 1.5cm},clip]{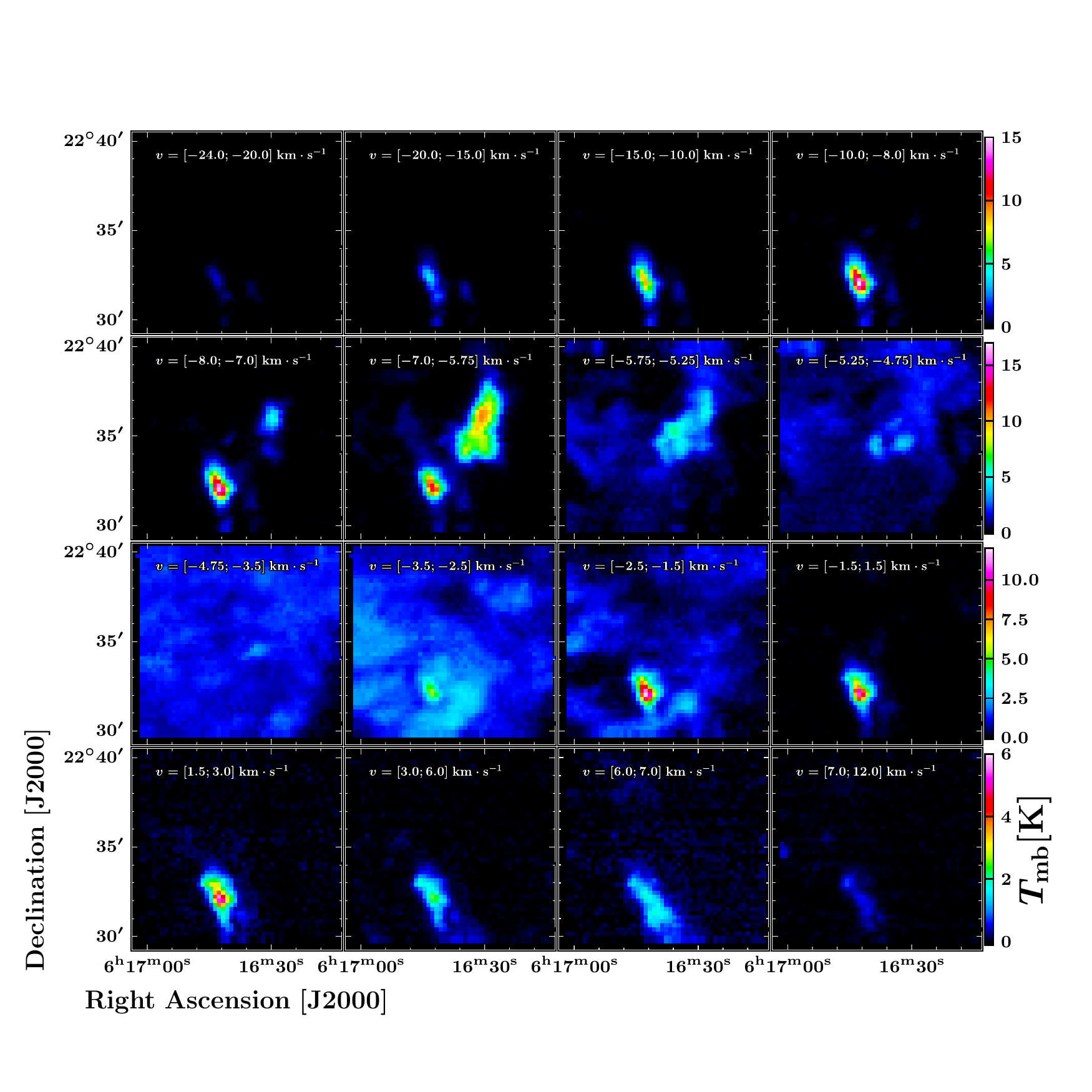}}
\caption{Channel map of the $^{12}\mathrm{CO}(2-1)$ observations carried out with the  APEX. The $10^\prime \times 10^\prime$ field is entirely displayed. Each panel represent the emission integrated over an interval of velocity along the line of sight. Velocities are indicated on the top left corner of each panel. Velocity channels represented in this figure are between $\varv=-24~\mathrm{km ~ s^{-1}}$ and $\varv=+12~\mathrm{km ~ s^{-1}}$, covering all the spectral features detected towards the extended G region.}
\label{fig:channelmap12co21apex}
\end{center}
\end{figure*}

\begin{figure*}[h]
\begin{center}
\resizebox{0.61\hsize}{!}
{\includegraphics[trim={0cm 0.75cm 0cm 1.5cm},clip]{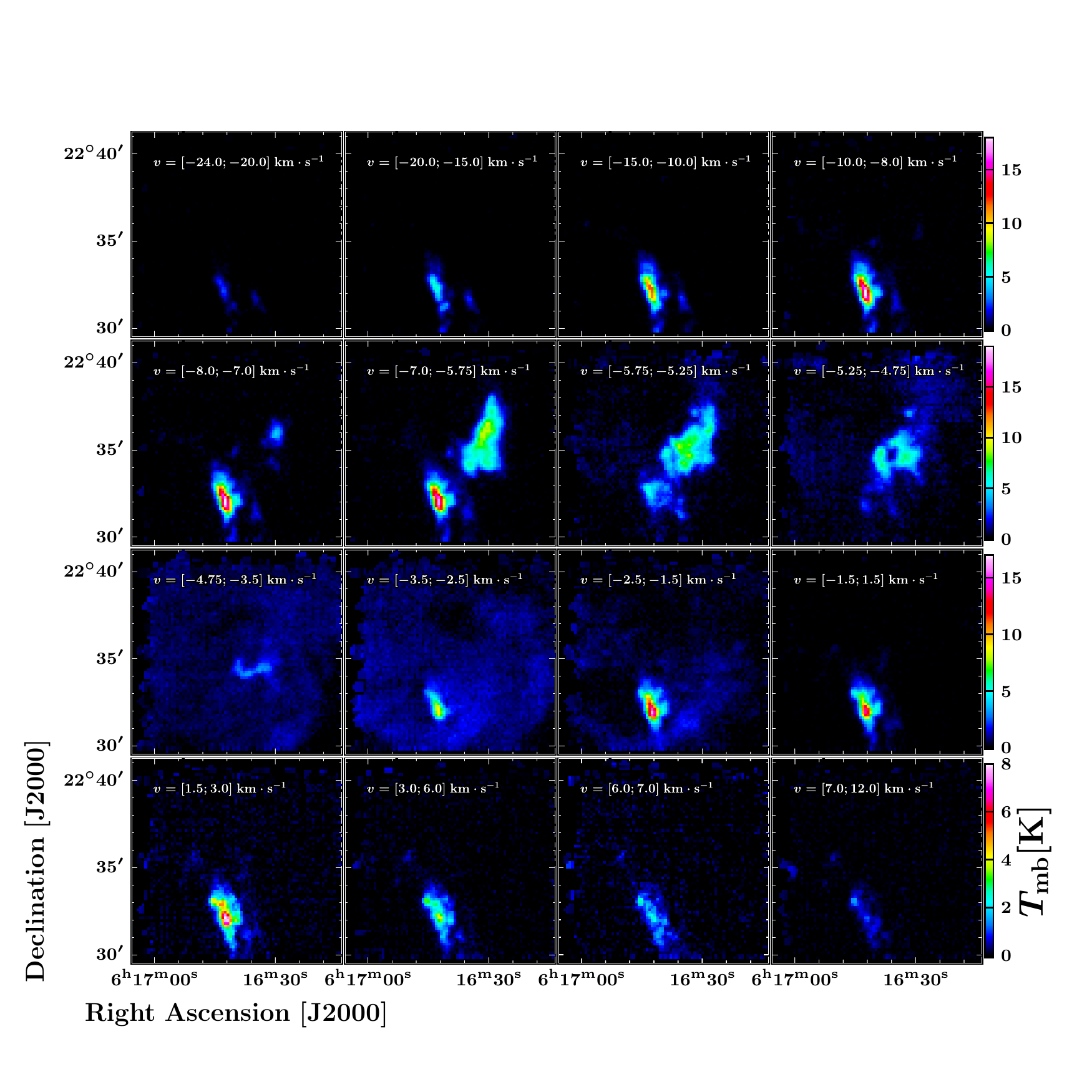}}
\caption{Channel map of the $^{12}\mathrm{CO}(3-2)$ observations carried out with APEX. The $10^\prime \times 10^\prime$ field is entirely displayed. Each panel represent the emission integrated over an interval of velocity along the line of sight. Velocities are indicated on the top left corner of each panel. Velocity channels represented in this figure are between $\varv=-24~\mathrm{km ~ s^{-1}}$ and $\varv=+12~\mathrm{km ~ s^{-1}}$, covering all the spectral features detected towards the extended G region.}
\label{fig:channelmap12co32}
\end{center}
\end{figure*}

\section{Spectral energy distributions of protostar candidates}

\begin{figure*}[h]
\begin{center}
\resizebox{0.8\hsize}{!}
{\includegraphics[trim={0cm 0.75cm 0cm 1.5cm},clip]{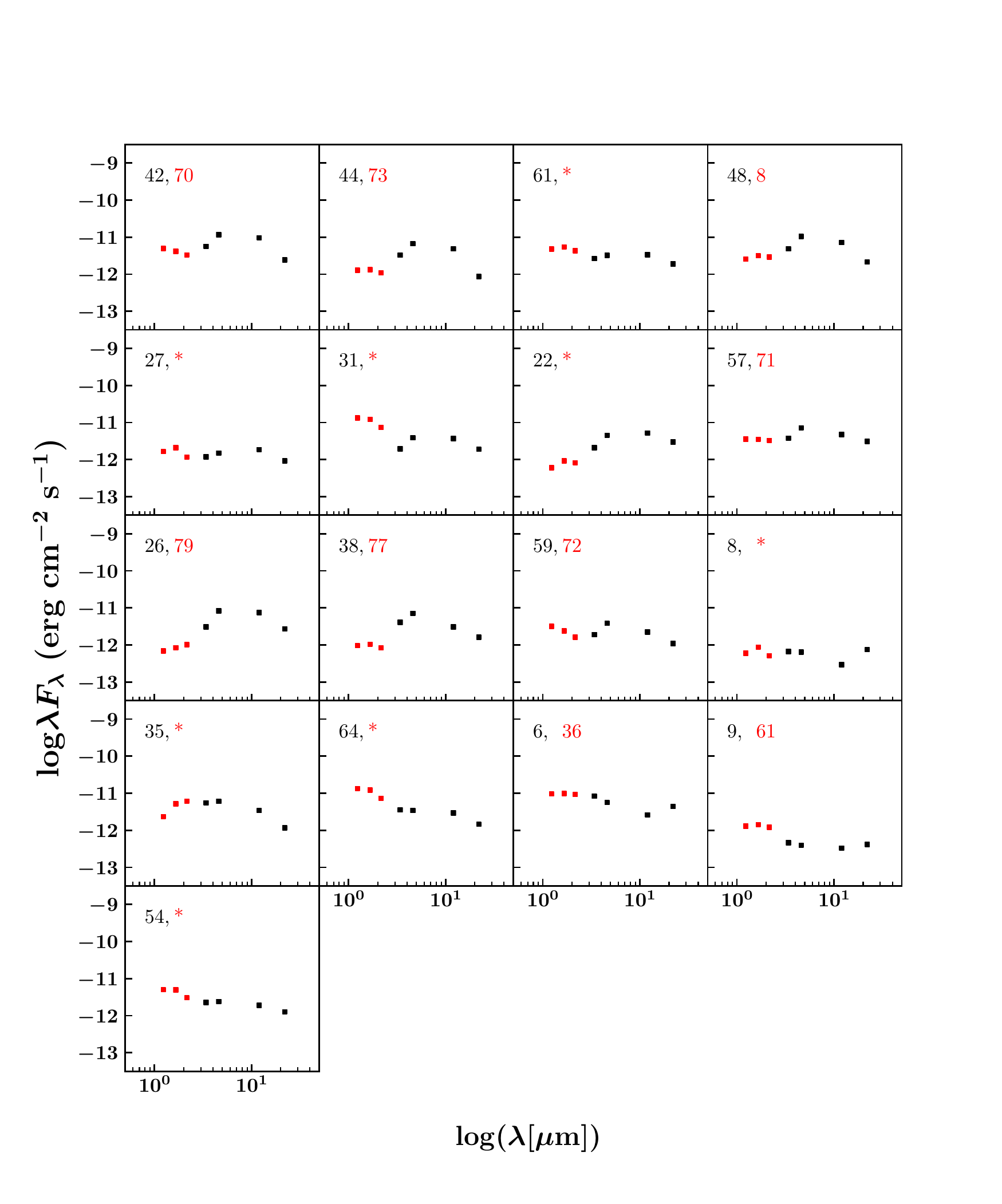}}
\caption{Spectral energy distributions of 17 protostar candidates. Their corresponding ID in Tab. \ref{table:2MASSphot} and Tab. \ref{table:WISEphot} ared indicated in the upper-left corner of each boxes when available (marked with a '*' if not corresponding to any point sources identified by a label). Red markers correspond to 2MASS photometric measures and black markers correspond to WISE measures.}
\label{fig:SEDs}
\end{center}
\end{figure*}

\section{Observational measure of the optical depth}
 
   \begin{figure*}[h]
   \centering
   \includegraphics[width=\hsize, trim={0cm 2cm 1cm 2.5cm},clip]{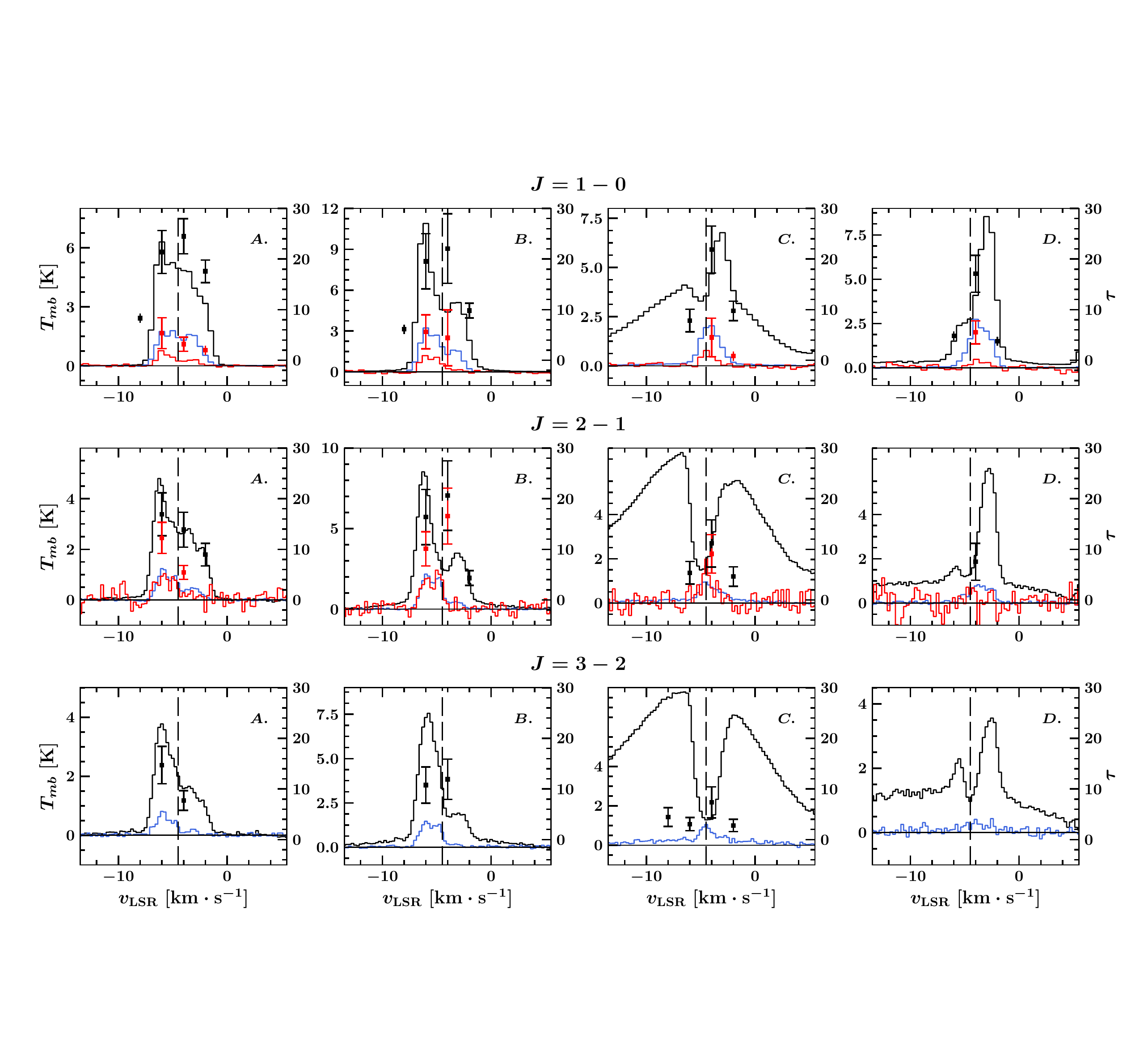}
      \caption{Spectra obtained over the region of the four boxes defined in Fig. \ref{fig:boxes}, for the following lines: $^{12}\mathrm{CO}$ (in black), $^{13}\mathrm{CO}$ (in blue) and $\mathrm{C^{18}O}$ (in red). First row presents the (1--0) transition, second row presents the (2--1) transition and third row presents the (3--2) transition for: A, the cloudlet; B, the ring-like structure; C, the shocked clump and D, the shocked knot. Spectral cubes were resampled to allow direct comparison between the different spectra. Spatial resolutions of all transitions were modified to the nominal resolution of $\mathrm{C^{18}O}$(2-1), $\theta= 30.2^{\prime \prime}$. Spectral resolution were set to $0.5~\mathrm{km ~ s^{-1}}$ for IRAM-30m data, and $0.25~\mathrm{km ~ s^{-1}}$ for APEX data. On both panels, the $\varv_{\mathrm{LSR}}$ of IC443 is indicated with a vertical dashed line (at $-4.5~\mathrm{km ~ s^{-1}}$). The black data points with error bars represent the measurements of the optical depth $\tau$ corresponding to each channel of these spectra resampled with a spectral resolution of 2 km s$^{-1}$, using the [$^{12}$CO]/[$^{13}$CO] ratio as a tracer, and red data points represent the measurements of the optical depth obtained using the [$^{12}$CO]/[C$^{18}$O] ratio. The adopted values for the isotopic ratios are based on the results of section \ref{sect:LTEmodels} and indicated in Tab. \ref{table:LTEresults}.}
         \label{fig:opticaldepth13co}
   \end{figure*}
   
\newpage
   

\section{WISE and 2MASS based classification of infrared point sources in the extended G region}

\onecolumn      
\begin{longtable}{c c c c c c c c c}         
\caption{Infrared photometric magnitudes of YSO candidates detected by WISE in the extended G region.}      \\ 
\label{table:WISEphot}    \\   
\hline\hline                 
\footnotesize{ID} & \footnotesize{Catalog name} & \footnotesize{W1(3.4 $\upmu$m)} & \footnotesize{W2(4.6 $\upmu$m)} & \footnotesize{W3(12 $\upmu$m)} & \footnotesize{W4(22 $\upmu$m)}  & \footnotesize{Fig. \ref{fig:colordiagrams} label} & \footnotesize{$\alpha$} & \footnotesize{FLAG} \\
 &  & \footnotesize{(mag)} & \footnotesize{(mag)} & \footnotesize{(mag)} & \footnotesize{(mag)} \\
\hline                        
\footnotesize{1} & \footnotesize{J061654.93+223403.6} & \footnotesize{17.132 $\pm$ 0.205} & \footnotesize{16.04 $\pm$ 0.25} & \footnotesize{11.906 $\pm$ 0.458} & \footnotesize{8.09} & \footnotesize{Class I} & \footnotesize{0.99} & \footnotesize{0} \\
\footnotesize{2} & \footnotesize{J061655.34+223353.3} & \footnotesize{16.925 $\pm$ 0.206} & \footnotesize{15.716 $\pm$ 0.225} & \footnotesize{11.828 $\pm$ 0.52} & \footnotesize{8.654 $\pm$ 0.508} & \footnotesize{Class I} & \footnotesize{0.86} & \footnotesize{0} \\
\footnotesize{3} & \footnotesize{J061653.26+223453.7} & \footnotesize{15.348 $\pm$ 0.053} & \footnotesize{13.717 $\pm$ 0.042} & \footnotesize{10.316 $\pm$ 0.085} & \footnotesize{8.234} & \footnotesize{Class I} & \footnotesize{0.71} & \footnotesize{0} \\
\footnotesize{4} & \footnotesize{J061656.75+223829.3} & \footnotesize{15.752 $\pm$ 0.068} & \footnotesize{14.049 $\pm$ 0.05} & \footnotesize{10.926 $\pm$ 0.142} & \footnotesize{8.257} & \footnotesize{Class I} & \footnotesize{0.53} & \footnotesize{0} \\
\hline \\[-1.0em]
\footnotesize{5} & \footnotesize{J061617.53+223155.6} & \footnotesize{12.36 $\pm$ 0.025} & \footnotesize{11.021 $\pm$ 0.021} & \footnotesize{8.844 $\pm$ 0.03} & \footnotesize{5.952 $\pm$ 0.049} & \footnotesize{Class I} & \footnotesize{-0.41} & \footnotesize{0} \\
\footnotesize{6} & \footnotesize{J061656.48+223943.2} & \footnotesize{11.276 $\pm$ 0.029} & \footnotesize{10.733 $\pm$ 0.027} & \footnotesize{8.612 $\pm$ 0.036} & \footnotesize{6.027 $\pm$ 0.090} & \footnotesize{Class II} & \footnotesize{-0.9} & \footnotesize{0} \\
\footnotesize{7} & \footnotesize{J061642.63+223825.6} & \footnotesize{15.152 $\pm$ 0.048} & \footnotesize{14.492 $\pm$ 0.066} & \footnotesize{12.316 $\pm$ 0.489} & \footnotesize{8.392 $\pm$ 0.431} & \footnotesize{Class II} & \footnotesize{-0.79} & \footnotesize{0} \\
\footnotesize{8} & \footnotesize{J061617.19+223356.0} & \footnotesize{14.033 $\pm$ 0.031} & \footnotesize{13.089 $\pm$ 0.032} & \footnotesize{10.99 $\pm$ 0.197} & \footnotesize{7.944 $\pm$ 0.219} & \footnotesize{Class II} & \footnotesize{-0.69} & \footnotesize{0} \\
\hline \\[-1.0em]
\footnotesize{9} & \footnotesize{J061654.18+223456.6} & \footnotesize{14.42 $\pm$ 0.034} & \footnotesize{13.622 $\pm$ 0.038} & \footnotesize{10.85 $\pm$ 0.132} & \footnotesize{8.582 $\pm$ 0.398} & \footnotesize{Class II} & \footnotesize{-0.24} & \footnotesize{0} \\
\footnotesize{10} & \footnotesize{J061653.02+223444.5} & \footnotesize{15.494 $\pm$ 0.056} & \footnotesize{14.695 $\pm$ 0.077} & \footnotesize{11.238 $\pm$ 0.244} & \footnotesize{8.687} & \footnotesize{Class I} & \footnotesize{0.29} & \footnotesize{0} \\
\footnotesize{11} & \footnotesize{J061639.55+223449.3} & \footnotesize{13.393 $\pm$ 0.042} & \footnotesize{12.444 $\pm$ 0.029} & \footnotesize{9.577 $\pm$ 0.371} & \footnotesize{6.398 $\pm$ 0.141} & \footnotesize{Class II} & \footnotesize{-0.09} & \footnotesize{0} \\
\footnotesize{12} & \footnotesize{J061642.43+223811.8} & \footnotesize{15.687 $\pm$ 0.077} & \footnotesize{14.321 $\pm$ 0.062} & \footnotesize{11.897 $\pm$ 0.407} & \footnotesize{8.415 $\pm$ 0.377} & \footnotesize{Class I} & \footnotesize{-0.2} & \footnotesize{0} \\
\footnotesize{13} & \footnotesize{J061654.60+223442.0} & \footnotesize{15.628 $\pm$ 0.056} & \footnotesize{14.91 $\pm$ 0.094} & \footnotesize{11.505 $\pm$ 0.258} & \footnotesize{8.369 $\pm$ 0.310} & \footnotesize{Class II} & \footnotesize{0.21} & \footnotesize{0} \\
\footnotesize{14} & \footnotesize{J061654.84+223500.4} & \footnotesize{15.476 $\pm$ 0.054} & \footnotesize{14.818 $\pm$ 0.08} & \footnotesize{11.6 $\pm$ 0.299} & \footnotesize{8.268} & \footnotesize{Class II} & \footnotesize{0.03} & \footnotesize{0} \\
\footnotesize{15} & \footnotesize{J061640.86+223753.0} & \footnotesize{15.845 $\pm$ 0.069} & \footnotesize{14.826 $\pm$ 0.082} & \footnotesize{11.77 $\pm$ 0.382} & \footnotesize{8.51 $\pm$ 0.493} & \footnotesize{Class I} & \footnotesize{0.1} & \footnotesize{0} \\
\hline \\[-1.0em]
\footnotesize{16} & \footnotesize{J061638.20+223322.9} & \footnotesize{14.555 $\pm$ 0.049} & \footnotesize{12.846 $\pm$ 0.035} & \footnotesize{9.626 $\pm$ 0.113} & \footnotesize{7.84 $\pm$ 0.366} & \footnotesize{Class I} & \footnotesize{0.61} & \footnotesize{1} \\
\footnotesize{17} & \footnotesize{J061638.34+223313.0} & \footnotesize{14.453 $\pm$ 0.059} & \footnotesize{13.053 $\pm$ 0.035} & \footnotesize{9.533 $\pm$ 0.249} & \footnotesize{7.877 $\pm$ 0.323} & \footnotesize{Class I} & \footnotesize{0.68} & \footnotesize{1} \\
\footnotesize{18} & \footnotesize{J061643.64+223232.1} & \footnotesize{11.158 $\pm$ 0.028} & \footnotesize{9.394 $\pm$ 0.021} & \footnotesize{6.363 $\pm$ 0.015} & \footnotesize{6.11 $\pm$ 0.145} & \footnotesize{Class I} & \footnotesize{0.49} & \footnotesize{3} \\
\footnotesize{19} & \footnotesize{J061646.10+223246.9} & \footnotesize{13.34 $\pm$ 0.028} & \footnotesize{11.706 $\pm$ 0.023} & \footnotesize{8.424 $\pm$ 0.029} & \footnotesize{8.667} & \footnotesize{Class I} & \footnotesize{0.62} & \footnotesize{3} \\
\footnotesize{20} & \footnotesize{J061641.30+223044.8} & \footnotesize{15.073 $\pm$ 0.053} & \footnotesize{13.683 $\pm$ 0.044} & \footnotesize{9.385 $\pm$ 0.087} & \footnotesize{8.617} & \footnotesize{Class I} & \footnotesize{1.28} & \footnotesize{3} \\
\footnotesize{21} & \footnotesize{J061642.50+223142.4} & \footnotesize{11.168 $\pm$ 0.025} & \footnotesize{9.246 $\pm$ 0.021} & \footnotesize{6.51 $\pm$ 0.015} & \footnotesize{6.217 $\pm$ 0.090} & \footnotesize{Class I} & \footnotesize{0.35} & \footnotesize{3} \\
\footnotesize{22} & \footnotesize{J061634.95+223128.1} & \footnotesize{12.786 $\pm$ 0.036} & \footnotesize{10.981 $\pm$ 0.024} & \footnotesize{7.86 $\pm$ 0.027} & \footnotesize{6.443 $\pm$ 0.071} & \footnotesize{Class I} & \footnotesize{0.59} & \footnotesize{3} \\
\footnotesize{23} & \footnotesize{J061639.57+223219.9} & \footnotesize{13.203 $\pm$ 0.045} & \footnotesize{11.801 $\pm$ 0.037} & \footnotesize{8.653 $\pm$ 0.13} & \footnotesize{8.272} & \footnotesize{Class I} & \footnotesize{0.39} & \footnotesize{5} \\
\footnotesize{24} & \footnotesize{J061632.40+223012.2} & \footnotesize{14.098 $\pm$ 0.036} & \footnotesize{12.352 $\pm$ 0.028} & \footnotesize{8.965 $\pm$ 0.037} & \footnotesize{7.88 $\pm$ 0.305} & \footnotesize{Class I} & \footnotesize{0.76} & \footnotesize{1} \\
\footnotesize{25} & \footnotesize{J061642.44+223206.3} & \footnotesize{11.228 $\pm$ 0.028} & \footnotesize{9.299 $\pm$ 0.023} & \footnotesize{6.521 $\pm$ 0.018} & \footnotesize{6.171 $\pm$ 0.103} & \footnotesize{Class I} & \footnotesize{0.39} & \footnotesize{5} \\
\footnotesize{26} & \footnotesize{J061643.86+223340.7} & \footnotesize{12.358 $\pm$ 0.048} & \footnotesize{10.323 $\pm$ 0.027} & \footnotesize{7.473 $\pm$ 0.053} & \footnotesize{6.553} & \footnotesize{Class I} & \footnotesize{0.5} & \footnotesize{3} \\
\footnotesize{27} & \footnotesize{J061642.75+223108.9} & \footnotesize{13.4 $\pm$ 0.026} & \footnotesize{12.176 $\pm$ 0.026} & \footnotesize{8.98 $\pm$ 0.035} & \footnotesize{7.724 $\pm$ 0.211} & \footnotesize{Class I} & \footnotesize{0.32} & \footnotesize{3} \\
\footnotesize{28} & \footnotesize{J061641.63+223111.0} & \footnotesize{12.829 $\pm$ 0.027} & \footnotesize{11.003 $\pm$ 0.023} & \footnotesize{8.096 $\pm$ 0.029} & \footnotesize{7.607 $\pm$ 0.289} & \footnotesize{Class I} & \footnotesize{0.43} & \footnotesize{5} \\
\footnotesize{29} & \footnotesize{J061635.24+223521.1} & \footnotesize{14.723 $\pm$ 0.057} & \footnotesize{13.606 $\pm$ 0.042} & \footnotesize{9.478 $\pm$ 0.202} & \footnotesize{7.199 $\pm$ 0.156} & \footnotesize{Class I} & \footnotesize{1.0} & \footnotesize{1} \\
\footnotesize{30} & \footnotesize{J061639.81+223201.5} & \footnotesize{12.88 $\pm$ 0.031} & \footnotesize{10.896 $\pm$ 0.023} & \footnotesize{8.251 $\pm$ 0.083} & \footnotesize{8.371} & \footnotesize{Class I} & \footnotesize{0.31} & \footnotesize{3} \\
\footnotesize{31} & \footnotesize{J061633.23+223103.7} & \footnotesize{12.862 $\pm$ 0.027} & \footnotesize{11.146 $\pm$ 0.023} & \footnotesize{8.242 $\pm$ 0.043} & \footnotesize{6.932 $\pm$ 0.098} & \footnotesize{Class I} & \footnotesize{0.37} & \footnotesize{3} \\
\footnotesize{32} & \footnotesize{J061635.54+223108.5} & \footnotesize{13.757 $\pm$ 0.03} & \footnotesize{12.12 $\pm$ 0.026} & \footnotesize{8.638 $\pm$ 0.039} & \footnotesize{7.456 $\pm$ 0.162} & \footnotesize{Class I} & \footnotesize{0.78} & \footnotesize{3} \\
\footnotesize{33} & \footnotesize{J061642.91+223215.9} & \footnotesize{11.159 $\pm$ 0.024} & \footnotesize{9.3 $\pm$ 0.02} & \footnotesize{6.33 $\pm$ 0.014} & \footnotesize{5.976 $\pm$ 0.084} & \footnotesize{Class I} & \footnotesize{0.5} & \footnotesize{3} \\
\footnotesize{34} & \footnotesize{J061639.38+223234.8} & \footnotesize{14.358 $\pm$ 0.046} & \footnotesize{13.055 $\pm$ 0.038} & \footnotesize{9.669 $\pm$ 0.23} & \footnotesize{7.606} & \footnotesize{Class I} & \footnotesize{0.52} & \footnotesize{3} \\
\hline \\[-1.0em]
\footnotesize{35} & \footnotesize{J061640.86+223310.8} & \footnotesize{11.727 $\pm$ 0.023} & \footnotesize{10.662 $\pm$ 0.02} & \footnotesize{8.315 $\pm$ 0.036} & \footnotesize{7.473 $\pm$ 0.244} & \footnotesize{Class II} & \footnotesize{-0.43} & \footnotesize{5} \\
\footnotesize{36} & \footnotesize{J061647.39+223406.4} & \footnotesize{13.909 $\pm$ 0.03} & \footnotesize{13.397 $\pm$ 0.033} & \footnotesize{10.958 $\pm$ 0.212} & \footnotesize{7.292} & \footnotesize{Class II} & \footnotesize{-0.66} & \footnotesize{3} \\
\footnotesize{37} & \footnotesize{J061642.00+223339.7} & \footnotesize{10.74 $\pm$ 0.024} & \footnotesize{10.143 $\pm$ 0.02} & \footnotesize{8.244 $\pm$ 0.085} & \footnotesize{6.939 $\pm$ 0.193} & \footnotesize{Class II} & \footnotesize{-1.04} & \footnotesize{3} \\
\footnotesize{38} & \footnotesize{J061645.55+223335.1} & \footnotesize{12.046 $\pm$ 0.025} & \footnotesize{10.487 $\pm$ 0.021} & \footnotesize{8.433 $\pm$ 0.032} & \footnotesize{7.121} & \footnotesize{Class I} & \footnotesize{-0.39} & \footnotesize{5} \\
\footnotesize{39} & \footnotesize{J061635.51+223239.1} & \footnotesize{14.932 $\pm$ 0.044} & \footnotesize{13.438 $\pm$ 0.035} & \footnotesize{11.349 $\pm$ 0.366} & \footnotesize{8.485} & \footnotesize{Class I} & \footnotesize{-0.4} & \footnotesize{3} \\
\hline \\[-1.0em]
\footnotesize{40} & \footnotesize{J061644.88+223352.6} & \footnotesize{12.305 $\pm$ 0.039} & \footnotesize{10.132 $\pm$ 0.02} & \footnotesize{7.963 $\pm$ 0.031} & \footnotesize{7.454 $\pm$ 0.442} & \footnotesize{Class I} & \footnotesize{0.04} & \footnotesize{3} \\
\footnotesize{41} & \footnotesize{J061642.28+223241.6} & \footnotesize{11.956 $\pm$ 0.034} & \footnotesize{10.024 $\pm$ 0.023} & \footnotesize{7.769 $\pm$ 0.105} & \footnotesize{6.777 $\pm$ 0.128} & \footnotesize{Class I} & \footnotesize{-0.02} & \footnotesize{3} \\
\footnotesize{42} & \footnotesize{J061643.22+223302.2} & \footnotesize{11.709 $\pm$ 0.027} & \footnotesize{9.96 $\pm$ 0.021} & \footnotesize{7.198 $\pm$ 0.018} & \footnotesize{6.666 $\pm$ 0.254} & \footnotesize{Class I} & \footnotesize{0.27} & \footnotesize{3} \\
\footnotesize{43} & \footnotesize{J061647.94+223431.0} & \footnotesize{13.987 $\pm$ 0.032} & \footnotesize{12.795 $\pm$ 0.028} & \footnotesize{9.7 $\pm$ 0.078} & \footnotesize{7.89} & \footnotesize{Class I} & \footnotesize{0.23} & \footnotesize{3} \\
\footnotesize{44} & \footnotesize{J061640.55+223121.5} & \footnotesize{12.291 $\pm$ 0.025} & \footnotesize{10.557 $\pm$ 0.023} & \footnotesize{7.939 $\pm$ 0.026} & \footnotesize{7.795 $\pm$ 0.259} & \footnotesize{Class I} & \footnotesize{0.15} & \footnotesize{5} \\
\footnotesize{45} & \footnotesize{J061630.48+223009.3} & \footnotesize{14.817 $\pm$ 0.041} & \footnotesize{13.375 $\pm$ 0.042} & \footnotesize{10.875 $\pm$ 0.17} & \footnotesize{8.639} & \footnotesize{Class I} & \footnotesize{-0.1} & \footnotesize{1} \\
\footnotesize{46} & \footnotesize{J061643.19+223038.8} & \footnotesize{15.019 $\pm$ 0.043} & \footnotesize{13.721 $\pm$ 0.04} & \footnotesize{11.043 $\pm$ 0.184} & \footnotesize{8.752} & \footnotesize{Class I} & \footnotesize{-0.04} & \footnotesize{1} \\
\footnotesize{47} & \footnotesize{J061650.62+223515.5} & \footnotesize{13.783 $\pm$ 0.032} & \footnotesize{11.77 $\pm$ 0.024} & \footnotesize{9.615 $\pm$ 0.062} & \footnotesize{8.559 $\pm$ 0.490} & \footnotesize{Class I} & \footnotesize{-0.06} & \footnotesize{1} \\
\footnotesize{48} & \footnotesize{J061641.59+223132.8} & \footnotesize{11.874 $\pm$ 0.029} & \footnotesize{10.064 $\pm$ 0.023} & \footnotesize{7.515 $\pm$ 0.022} & \footnotesize{6.815 $\pm$ 0.146} & \footnotesize{Class I} & \footnotesize{0.14} & \footnotesize{3} \\
\footnotesize{49} & \footnotesize{J061640.18+223149.4} & \footnotesize{12.626 $\pm$ 0.031} & \footnotesize{10.845 $\pm$ 0.029} & \footnotesize{8.508 $\pm$ 0.088} & \footnotesize{7.898} & \footnotesize{Class I} & \footnotesize{-0.04} & \footnotesize{5} \\
\footnotesize{50} & \footnotesize{J061639.32+223320.1} & \footnotesize{14.693 $\pm$ 0.061} & \footnotesize{12.991 $\pm$ 0.035} & \footnotesize{10.655 $\pm$ 0.284} & \footnotesize{7.322 $\pm$ 0.192} & \footnotesize{Class I} & \footnotesize{-0.09} & \footnotesize{1} \\
\footnotesize{51} & \footnotesize{J061636.38+223308.9} & \footnotesize{13.925 $\pm$ 0.038} & \footnotesize{12.355 $\pm$ 0.027} & \footnotesize{10.027 $\pm$ 0.396} & \footnotesize{7.67} & \footnotesize{Class I} & \footnotesize{-0.17} & \footnotesize{1} \\
\footnotesize{52} & \footnotesize{J061642.01+223418.9} & \footnotesize{14.356 $\pm$ 0.057} & \footnotesize{12.241 $\pm$ 0.026} & \footnotesize{9.799 $\pm$ 0.462} & \footnotesize{7.823 $\pm$ 0.482} & \footnotesize{Class I} & \footnotesize{0.23} & \footnotesize{3} \\
\footnotesize{53} & \footnotesize{J061639.74+223041.6} & \footnotesize{14.382 $\pm$ 0.036} & \footnotesize{12.701 $\pm$ 0.034} & \footnotesize{9.991 $\pm$ 0.347} & \footnotesize{8.641} & \footnotesize{Class I} & \footnotesize{0.2} & \footnotesize{1} \\
\footnotesize{54} & \footnotesize{J061646.36+223424.7} & \footnotesize{12.69 $\pm$ 0.027} & \footnotesize{11.661 $\pm$ 0.024} & \footnotesize{8.949 $\pm$ 0.045} & \footnotesize{7.379 $\pm$ 0.251} & \footnotesize{Class II} & \footnotesize{-0.16} & \footnotesize{3} \\
\footnotesize{55} & \footnotesize{J061645.01+223259.5} & \footnotesize{11.93 $\pm$ 0.024} & \footnotesize{10.383 $\pm$ 0.02} & \footnotesize{7.495 $\pm$ 0.019} & \footnotesize{7.311 $\pm$ 0.138} & \footnotesize{Class I} & \footnotesize{0.26} & \footnotesize{3} \\
\footnotesize{56} & \footnotesize{J061643.87+223250.1} & \footnotesize{11.07 $\pm$ 0.024} & \footnotesize{9.276 $\pm$ 0.02} & \footnotesize{6.732 $\pm$ 0.016} & \footnotesize{6.358 $\pm$ 0.194} & \footnotesize{Class I} & \footnotesize{0.13} & \footnotesize{5} \\
\footnotesize{57} & \footnotesize{J061635.06+223135.9} & \footnotesize{12.134 $\pm$ 0.026} & \footnotesize{10.475 $\pm$ 0.023} & \footnotesize{7.958 $\pm$ 0.025} & \footnotesize{6.41 $\pm$ 0.073} & \footnotesize{Class I} & \footnotesize{0.03} & \footnotesize{3} \\
\footnotesize{58} & \footnotesize{J061647.01+223500.8} & \footnotesize{14.047 $\pm$ 0.035} & \footnotesize{11.907 $\pm$ 0.024} & \footnotesize{9.892 $\pm$ 0.103} & \footnotesize{7.989} & \footnotesize{Class I} & \footnotesize{-0.1} & \footnotesize{3} \\
\footnotesize{59} & \footnotesize{J061644.84+223442.6} & \footnotesize{12.885 $\pm$ 0.029} & \footnotesize{11.157 $\pm$ 0.026} & \footnotesize{8.788 $\pm$ 0.035} & \footnotesize{7.539 $\pm$ 0.240} & \footnotesize{Class I} & \footnotesize{-0.05} & \footnotesize{3} \\
\footnotesize{60} & \footnotesize{J061641.36+223012.4} & \footnotesize{13.6 $\pm$ 0.037} & \footnotesize{11.791 $\pm$ 0.027} & \footnotesize{9.673 $\pm$ 0.122} & \footnotesize{7.951} & \footnotesize{Class I} & \footnotesize{-0.2} & \footnotesize{3} \\
\footnotesize{61} & \footnotesize{J061643.18+223129.7} & \footnotesize{12.525 $\pm$ 0.022} & \footnotesize{11.347 $\pm$ 0.023} & \footnotesize{8.34 $\pm$ 0.024} & \footnotesize{6.933 $\pm$ 0.111} & \footnotesize{Class I} & \footnotesize{0.15} & \footnotesize{3} \\
\footnotesize{62} & \footnotesize{J061650.94+223454.3} & \footnotesize{14.969 $\pm$ 0.049} & \footnotesize{12.959 $\pm$ 0.032} & \footnotesize{10.635 $\pm$ 0.144} & \footnotesize{8.476 } & \footnotesize{Class I} & \footnotesize{0.07} & \footnotesize{1} \\
\footnotesize{63} & \footnotesize{J061644.45+223331.1} & \footnotesize{12.396 $\pm$ 0.035} & \footnotesize{10.456 $\pm$ 0.022} & \footnotesize{7.952 $\pm$ 0.027} & \footnotesize{7.341} & \footnotesize{Class I} & \footnotesize{0.18} & \footnotesize{3} \\
\footnotesize{64} & \footnotesize{J061633.49+223110.4} & \footnotesize{12.208 $\pm$ 0.024} & \footnotesize{11.279 $\pm$ 0.027} & \footnotesize{8.486 $\pm$ 0.039} & \footnotesize{7.219 $\pm$ 0.144} & \footnotesize{Class II} & \footnotesize{-0.16} & \footnotesize{3} \\
\footnotesize{65} & \footnotesize{J061642.76+223402.6} & \footnotesize{12.266 $\pm$ 0.028} & \footnotesize{10.421 $\pm$ 0.021} & \footnotesize{7.897 $\pm$ 0.071} & \footnotesize{7.022 $\pm$ 0.162} & \footnotesize{Class I} & \footnotesize{0.14} & \footnotesize{3} \\
\hline                                   
\end{longtable}

\begin{longtable}{c c c c c c c c c}         
\caption{Infrared photometric magnitudes of YSO candidates detected by 2MASS in the extended G region.}      \\ 
\label{table:2MASSphot}    \\   
\hline\hline                 
\footnotesize{ID} & \footnotesize{r.a.} & \footnotesize{dec.} & \footnotesize{J(1.25 $\upmu$m)} & \footnotesize{H(1.65 $\upmu$m)} & \footnotesize{K(2.17 $\upmu$m)}  & \footnotesize{Fig. \ref{fig:colordiagrams} label} & \footnotesize{FLAG} \\
 & \footnotesize{(J2000)} & \footnotesize{(J2000)} & \footnotesize{(mag)} & \footnotesize{(mag)} & \footnotesize{(mag)}\\
\hline
1 & \footnotesize{$6^\mathrm{h}16^\mathrm{m}21.16^\mathrm{s}$} & \footnotesize{$22^\mathrm{d}37^\mathrm{m}25.37^\mathrm{s}$} & \footnotesize{16.073 $\pm$ 0.078} & \footnotesize{15.449 $\pm$ 0.108} & \footnotesize{14.932 $\pm$ 0.093} & \footnotesize{YSO} & \footnotesize{0} \\
2 & \footnotesize{$6^\mathrm{h}16^\mathrm{m}21.35^\mathrm{s}$} & \footnotesize{$22^\mathrm{d}34^\mathrm{m}32.27^\mathrm{s}$} & \footnotesize{15.3 $\pm$ 0.048} & \footnotesize{14.92 $\pm$ 0.087} & \footnotesize{14.708 $\pm$ 0.075} & \footnotesize{YSO} & \footnotesize{0} \\
3 & \footnotesize{$6^\mathrm{h}16^\mathrm{m}32.01^\mathrm{s}$} & \footnotesize{$22^\mathrm{d}33^\mathrm{m}59.85^\mathrm{s}$} & \footnotesize{16.83 $\pm$ 0.140} & \footnotesize{15.981 $\pm$ 0.182} & \footnotesize{15.234 $\pm$ 0.120} & \footnotesize{YSO} & \footnotesize{0} \\
4 & \footnotesize{$6^\mathrm{h}16^\mathrm{m}23.77^\mathrm{s}$} & \footnotesize{$22^\mathrm{d}34^\mathrm{m}59.89^\mathrm{s}$} & \footnotesize{15.138 $\pm$ 0.055} & \footnotesize{14.519 $\pm$ 0.064} & \footnotesize{14.094 $\pm$ 0.046} & \footnotesize{YSO} & \footnotesize{0} \\
5 & \footnotesize{$6^\mathrm{h}16^\mathrm{m}19.73^\mathrm{s}$} & \footnotesize{$22^\mathrm{d}36^\mathrm{m}17.92^\mathrm{s}$} & \footnotesize{15.439 $\pm$ 0.074} & \footnotesize{14.997 $\pm$ 0.094} & \footnotesize{14.576 $\pm$ 0.070} & \footnotesize{YSO} & \footnotesize{0} \\
6 & \footnotesize{$6^\mathrm{h}16^\mathrm{m}21.42^\mathrm{s}$} & \footnotesize{$22^\mathrm{d}31^\mathrm{m}33.23^\mathrm{s}$} & \footnotesize{15.4 $\pm$ 0.052} & \footnotesize{14.815 $\pm$ 0.069} & \footnotesize{14.347 $\pm$ 0.067} & \footnotesize{YSO} & \footnotesize{0} \\
7 & \footnotesize{$6^\mathrm{h}16^\mathrm{m}20.25^\mathrm{s}$} & \footnotesize{$22^\mathrm{d}31^\mathrm{m}39.30^\mathrm{s}$} & \footnotesize{16.75 $\pm$ 0.150} & \footnotesize{16.051 $\pm$ 0.192} & \footnotesize{15.478 $\pm$ 0.136} & \footnotesize{YSO} & \footnotesize{0} \\
8 & \footnotesize{$6^\mathrm{h}16^\mathrm{m}54.05^\mathrm{s}$} & \footnotesize{$22^\mathrm{d}30^\mathrm{m}37.07^\mathrm{s}$} & \footnotesize{16.742 $\pm$ 0.165} & \footnotesize{16.013 $\pm$ 0.188} & \footnotesize{15.451 $\pm$ 0.145} & \footnotesize{YSO} & \footnotesize{0} \\
9 & \footnotesize{$6^\mathrm{h}16^\mathrm{m}39.00^\mathrm{s}$} & \footnotesize{$22^\mathrm{d}30^\mathrm{m}00.26^\mathrm{s}$} & \footnotesize{16.22 $\pm$ 0.109} & \footnotesize{15.587 $\pm$ 0.144} & \footnotesize{15.226 $\pm$ 0.123} & \footnotesize{YSO} & \footnotesize{0} \\
10 & \footnotesize{$6^\mathrm{h}16^\mathrm{m}41.73^\mathrm{s}$} & \footnotesize{$22^\mathrm{d}30^\mathrm{m}21.36^\mathrm{s}$} & \footnotesize{15.163 $\pm$ 0.049} & \footnotesize{14.663 $\pm$ 0.067} & \footnotesize{14.226 $\pm$ 0.054} & \footnotesize{YSO} & \footnotesize{0} \\
11 & \footnotesize{$6^\mathrm{h}16^\mathrm{m}45.23^\mathrm{s}$} & \footnotesize{$22^\mathrm{d}31^\mathrm{m}03.71^\mathrm{s}$} & \footnotesize{16.753 $\pm$ 0.170} & \footnotesize{16.174 $\pm$ 0.222} & \footnotesize{15.604 $\pm$ 0.168} & \footnotesize{YSO} & \footnotesize{0} \\
12 & \footnotesize{$6^\mathrm{h}16^\mathrm{m}37.41^\mathrm{s}$} & \footnotesize{$22^\mathrm{d}30^\mathrm{m}45.04^\mathrm{s}$} & \footnotesize{16.45 $\pm$ 0.120} & \footnotesize{16.011 $\pm$ 0.188} & \footnotesize{15.756 $\pm$ 0.189} & \footnotesize{YSO} & \footnotesize{0} \\
13 & \footnotesize{$6^\mathrm{h}16^\mathrm{m}28.65^\mathrm{s}$} & \footnotesize{$22^\mathrm{d}30^\mathrm{m}29.60^\mathrm{s}$} & \footnotesize{15.273 $\pm$ 0.061} & \footnotesize{14.65 $\pm$ 0.090} & \footnotesize{14.086 $\pm$ 0.059} & \footnotesize{YSO} & \footnotesize{0} \\
14 & \footnotesize{$6^\mathrm{h}16^\mathrm{m}25.18^\mathrm{s}$} & \footnotesize{$22^\mathrm{d}31^\mathrm{m}48.32^\mathrm{s}$} & \footnotesize{13.487 $\pm$ 0.027} & \footnotesize{13.011 $\pm$ 0.038} & \footnotesize{12.658 $\pm$ 0.025} & \footnotesize{YSO} & \footnotesize{0} \\
15 & \footnotesize{$6^\mathrm{h}16^\mathrm{m}25.96^\mathrm{s}$} & \footnotesize{$22^\mathrm{d}30^\mathrm{m}57.38^\mathrm{s}$} & \footnotesize{15.913 $\pm$ 0.072} & \footnotesize{15.315 $\pm$ 0.105} & \footnotesize{14.878 $\pm$ 0.107} & \footnotesize{YSO} & \footnotesize{0} \\
16 & \footnotesize{$6^\mathrm{h}16^\mathrm{m}26.89^\mathrm{s}$} & \footnotesize{$22^\mathrm{d}31^\mathrm{m}46.11^\mathrm{s}$} & \footnotesize{15.278 $\pm$ 0.044} & \footnotesize{14.761 $\pm$ 0.075} & \footnotesize{14.47 $\pm$ 0.068} & \footnotesize{YSO} & \footnotesize{0} \\
17 & \footnotesize{$6^\mathrm{h}16^\mathrm{m}37.10^\mathrm{s}$} & \footnotesize{$22^\mathrm{d}31^\mathrm{m}17.74^\mathrm{s}$} & \footnotesize{16.608 $\pm$ 0.148} & \footnotesize{15.972 $\pm$ 0.194} & \footnotesize{15.335 $\pm$ 0.134} & \footnotesize{YSO} & \footnotesize{0} \\
18 & \footnotesize{$6^\mathrm{h}16^\mathrm{m}18.27^\mathrm{s}$} & \footnotesize{$22^\mathrm{d}38^\mathrm{m}49.88^\mathrm{s}$} & \footnotesize{14.252 $\pm$ 0.034} & \footnotesize{13.7 $\pm$ 0.034} & \footnotesize{13.392 $\pm$ 0.030} & \footnotesize{YSO} & \footnotesize{0} \\
19 & \footnotesize{$6^\mathrm{h}16^\mathrm{m}17.72^\mathrm{s}$} & \footnotesize{$22^\mathrm{d}38^\mathrm{m}57.65^\mathrm{s}$} & \footnotesize{16.62 $\pm$ 0.132} & \footnotesize{16.029 $\pm$ 0.190} & \footnotesize{15.464 $\pm$ 0.145} & \footnotesize{YSO} & \footnotesize{0} \\
20 & \footnotesize{$6^\mathrm{h}16^\mathrm{m}27.76^\mathrm{s}$} & \footnotesize{$22^\mathrm{d}39^\mathrm{m}04.81^\mathrm{s}$} & \footnotesize{16.348 $\pm$ 0.106} & \footnotesize{15.627 $\pm$ 0.162} & \footnotesize{15.161 $\pm$ 0.121} & \footnotesize{YSO} & \footnotesize{0} \\
21 & \footnotesize{$6^\mathrm{h}16^\mathrm{m}39.20^\mathrm{s}$} & \footnotesize{$22^\mathrm{d}36^\mathrm{m}57.77^\mathrm{s}$} & \footnotesize{17.06 $\pm$ 0.183} & \footnotesize{16.282 $\pm$ 0.235} & \footnotesize{15.662 $\pm$ 0.170} & \footnotesize{YSO} & \footnotesize{0} \\
22 & \footnotesize{$6^\mathrm{h}16^\mathrm{m}38.84^\mathrm{s}$} & \footnotesize{$22^\mathrm{d}37^\mathrm{m}50.51^\mathrm{s}$} & \footnotesize{13.85 $\pm$ 0.027} & \footnotesize{13.508 $\pm$ 0.039} & \footnotesize{13.307 $\pm$ 0.031} & \footnotesize{YSO} & \footnotesize{0} \\
23 & \footnotesize{$6^\mathrm{h}16^\mathrm{m}34.85^\mathrm{s}$} & \footnotesize{$22^\mathrm{d}37^\mathrm{m}45.31^\mathrm{s}$} & \footnotesize{13.714 $\pm$ 0.031} & \footnotesize{13.268 $\pm$ 0.038} & \footnotesize{12.954 $\pm$ 0.027} & \footnotesize{YSO} & \footnotesize{0} \\
24 & \footnotesize{$6^\mathrm{h}16^\mathrm{m}32.52^\mathrm{s}$} & \footnotesize{$22^\mathrm{d}38^\mathrm{m}19.35^\mathrm{s}$} & \footnotesize{14.745 $\pm$ 0.036} & \footnotesize{14.165 $\pm$ 0.055} & \footnotesize{13.813 $\pm$ 0.039} & \footnotesize{YSO} & \footnotesize{0} \\
25 & \footnotesize{$6^\mathrm{h}16^\mathrm{m}30.79^\mathrm{s}$} & \footnotesize{$22^\mathrm{d}38^\mathrm{m}27.47^\mathrm{s}$} & \footnotesize{14.896 $\pm$ 0.036} & \footnotesize{14.202 $\pm$ 0.034} & \footnotesize{13.762 $\pm$ 0.039} & \footnotesize{YSO} & \footnotesize{0} \\
26 & \footnotesize{$6^\mathrm{h}16^\mathrm{m}30.19^\mathrm{s}$} & \footnotesize{$22^\mathrm{d}38^\mathrm{m}29.64^\mathrm{s}$} & \footnotesize{15.996 $\pm$ 0.079} & \footnotesize{15.408 $\pm$ 0.114} & \footnotesize{14.843 $\pm$ 0.087} & \footnotesize{YSO} & \footnotesize{0} \\
27 & \footnotesize{$6^\mathrm{h}16^\mathrm{m}33.70^\mathrm{s}$} & \footnotesize{$22^\mathrm{d}37^\mathrm{m}27.99^\mathrm{s}$} & \footnotesize{14.368 $\pm$ 0.032} & \footnotesize{13.821 $\pm$ 0.047} & \footnotesize{13.487 $\pm$ 0.031} & \footnotesize{YSO} & \footnotesize{0} \\
28 & \footnotesize{$6^\mathrm{h}16^\mathrm{m}33.83^\mathrm{s}$} & \footnotesize{$22^\mathrm{d}38^\mathrm{m}06.52^\mathrm{s}$} & \footnotesize{15.865 $\pm$ 0.072} & \footnotesize{15.212 $\pm$ 0.087} & \footnotesize{14.83 $\pm$ 0.078} & \footnotesize{YSO} & \footnotesize{0} \\
29 & \footnotesize{$6^\mathrm{h}16^\mathrm{m}56.77^\mathrm{s}$} & \footnotesize{$22^\mathrm{d}30^\mathrm{m}15.22^\mathrm{s}$} & \footnotesize{15.112 $\pm$ 0.042} & \footnotesize{14.755 $\pm$ 0.070} & \footnotesize{14.461 $\pm$ 0.060} & \footnotesize{YSO} & \footnotesize{0} \\
30 & \footnotesize{$6^\mathrm{h}16^\mathrm{m}53.13^\mathrm{s}$} & \footnotesize{$22^\mathrm{d}31^\mathrm{m}03.76^\mathrm{s}$} & \footnotesize{15.879 $\pm$ 0.082} & \footnotesize{15.605 $\pm$ 0.149} & \footnotesize{15.273 $\pm$ 0.131} & \footnotesize{YSO} & \footnotesize{0} \\
31 & \footnotesize{$6^\mathrm{h}16^\mathrm{m}57.99^\mathrm{s}$} & \footnotesize{$22^\mathrm{d}33^\mathrm{m}04.98^\mathrm{s}$} & \footnotesize{14.739 $\pm$ 0.037} & \footnotesize{14.366 $\pm$ 0.057} & \footnotesize{14.057 $\pm$ 0.048} & \footnotesize{YSO} & \footnotesize{0} \\
32 & \footnotesize{$6^\mathrm{h}16^\mathrm{m}48.47^\mathrm{s}$} & \footnotesize{$22^\mathrm{d}31^\mathrm{m}57.35^\mathrm{s}$} & \footnotesize{11.835 $\pm$ 0.023} & \footnotesize{11.741 $\pm$ 0.030} & \footnotesize{11.69 $\pm$ 0.023} & \footnotesize{YSO} & \footnotesize{0} \\
33 & \footnotesize{$6^\mathrm{h}16^\mathrm{m}49.15^\mathrm{s}$} & \footnotesize{$22^\mathrm{d}33^\mathrm{m}37.91^\mathrm{s}$} & \footnotesize{16.26 $\pm$ 0.100} & \footnotesize{15.794 $\pm$ 0.152} & \footnotesize{15.355 $\pm$ 0.129} & \footnotesize{YSO} & \footnotesize{0} \\
34 & \footnotesize{$6^\mathrm{h}16^\mathrm{m}51.04^\mathrm{s}$} & \footnotesize{$22^\mathrm{d}33^\mathrm{m}16.69^\mathrm{s}$} & \footnotesize{16.806 $\pm$ 0.178} & \footnotesize{16.093 $\pm$ 0.223} & \footnotesize{15.637 $\pm$ 0.169} & \footnotesize{YSO} & \footnotesize{0} \\
35 & \footnotesize{$6^\mathrm{h}16^\mathrm{m}52.65^\mathrm{s}$} & \footnotesize{$22^\mathrm{d}32^\mathrm{m}10.02^\mathrm{s}$} & \footnotesize{15.536 $\pm$ 0.056} & \footnotesize{14.916 $\pm$ 0.077} & \footnotesize{14.473 $\pm$ 0.063} & \footnotesize{YSO} & \footnotesize{0} \\
36 & \footnotesize{$6^\mathrm{h}16^\mathrm{m}56.50^\mathrm{s}$} & \footnotesize{$22^\mathrm{d}39^\mathrm{m}43.29^\mathrm{s}$} & \footnotesize{14.008 $\pm$ 0.038} & \footnotesize{13.187 $\pm$ 0.041} & \footnotesize{12.502 $\pm$ 0.028} & \footnotesize{YSO} & \footnotesize{0} \\
37 & \footnotesize{$6^\mathrm{h}16^\mathrm{m}49.04^\mathrm{s}$} & \footnotesize{$22^\mathrm{d}39^\mathrm{m}47.80^\mathrm{s}$} & \footnotesize{14.418 $\pm$ 0.033} & \footnotesize{14.196 $\pm$ 0.048} & \footnotesize{13.853 $\pm$ 0.044} & \footnotesize{YSO} & \footnotesize{0} \\
38 & \footnotesize{$6^\mathrm{h}16^\mathrm{m}49.76^\mathrm{s}$} & \footnotesize{$22^\mathrm{d}39^\mathrm{m}12.79^\mathrm{s}$} & \footnotesize{15.869 $\pm$ 0.078} & \footnotesize{15.481 $\pm$ 0.123} & \footnotesize{15.182 $\pm$ 0.117} & \footnotesize{YSO} & \footnotesize{0} \\
39 & \footnotesize{$6^\mathrm{h}16^\mathrm{m}58.80^\mathrm{s}$} & \footnotesize{$22^\mathrm{d}35^\mathrm{m}58.44^\mathrm{s}$} & \footnotesize{11.262 $\pm$ 0.022} & \footnotesize{10.979 $\pm$ 0.030} & \footnotesize{10.796 $\pm$ 0.021} & \footnotesize{YSO} & \footnotesize{0} \\
40 & \footnotesize{$6^\mathrm{h}16^\mathrm{m}52.79^\mathrm{s}$} & \footnotesize{$22^\mathrm{d}39^\mathrm{m}39.43^\mathrm{s}$} & \footnotesize{16.178 $\pm$ 0.103} & \footnotesize{16.14 $\pm$ 0.228} & \footnotesize{15.496 $\pm$ 0.164} & \footnotesize{YSO} & \footnotesize{0} \\
41 & \footnotesize{$6^\mathrm{h}16^\mathrm{m}51.69^\mathrm{s}$} & \footnotesize{$22^\mathrm{d}35^\mathrm{m}37.42^\mathrm{s}$} & \footnotesize{15.277 $\pm$ 0.052} & \footnotesize{14.827 $\pm$ 0.074} & \footnotesize{14.372 $\pm$ 0.081} & \footnotesize{YSO} & \footnotesize{0} \\
42 & \footnotesize{$6^\mathrm{h}16^\mathrm{m}40.28^\mathrm{s}$} & \footnotesize{$22^\mathrm{d}38^\mathrm{m}17.54^\mathrm{s}$} & \footnotesize{15.355 $\pm$ 0.056} & \footnotesize{14.797 $\pm$ 0.082} & \footnotesize{14.487 $\pm$ 0.068} & \footnotesize{YSO} & \footnotesize{0} \\
43 & \footnotesize{$6^\mathrm{h}16^\mathrm{m}41.82^\mathrm{s}$} & \footnotesize{$22^\mathrm{d}37^\mathrm{m}30.09^\mathrm{s}$} & \footnotesize{16.673 $\pm$ 0.158} & \footnotesize{16.052 $\pm$ 0.215} & \footnotesize{15.654 $\pm$ 0.167} & \footnotesize{YSO} & \footnotesize{0} \\
44 & \footnotesize{$6^\mathrm{h}16^\mathrm{m}44.95^\mathrm{s}$} & \footnotesize{$22^\mathrm{d}35^\mathrm{m}49.96^\mathrm{s}$} & \footnotesize{15.919 $\pm$ 0.081} & \footnotesize{15.28 $\pm$ 0.100} & \footnotesize{14.832 $\pm$ 0.086} & \footnotesize{YSO} & \footnotesize{0} \\
45 & \footnotesize{$6^\mathrm{h}16^\mathrm{m}50.66^\mathrm{s}$} & \footnotesize{$22^\mathrm{d}37^\mathrm{m}20.93^\mathrm{s}$} & \footnotesize{14.637 $\pm$ 0.040} & \footnotesize{14.328 $\pm$ 0.057} & \footnotesize{14.092 $\pm$ 0.055} & \footnotesize{YSO} & \footnotesize{0} \\
\hline \\[-1.0em]
46 & \footnotesize{$6^\mathrm{h}16^\mathrm{m}18.14^\mathrm{s}$} & \footnotesize{$22^\mathrm{d}38^\mathrm{m}18.74^\mathrm{s}$} & \footnotesize{16.236 $\pm$ 0.102} & \footnotesize{15.326 $\pm$ 0.118} & \footnotesize{14.713 $\pm$ 0.069} & \footnotesize{CTTS} & \footnotesize{0} \\
47 & \footnotesize{$6^\mathrm{h}16^\mathrm{m}20.71^\mathrm{s}$} & \footnotesize{$22^\mathrm{d}36^\mathrm{m}44.49^\mathrm{s}$} & \footnotesize{16.83 $\pm$ 0.163} & \footnotesize{16.054 $\pm$ 0.190} & \footnotesize{15.544 $\pm$ 0.151} & \footnotesize{CTTS} & \footnotesize{0} \\
48 & \footnotesize{$6^\mathrm{h}16^\mathrm{m}31.07^\mathrm{s}$} & \footnotesize{$22^\mathrm{d}35^\mathrm{m}20.79^\mathrm{s}$} & \footnotesize{17.069 $\pm$ 0.203} & \footnotesize{15.948 $\pm$ 0.175} & \footnotesize{15.21 $\pm$ 0.128} & \footnotesize{CTTS} & \footnotesize{0} \\
49 & \footnotesize{$6^\mathrm{h}16^\mathrm{m}28.25^\mathrm{s}$} & \footnotesize{$22^\mathrm{d}33^\mathrm{m}54.94^\mathrm{s}$} & \footnotesize{15.992 $\pm$ 0.094} & \footnotesize{15.117 $\pm$ 0.090} & \footnotesize{14.573 $\pm$ 0.072} & \footnotesize{CTTS} & \footnotesize{0} \\
50 & \footnotesize{$6^\mathrm{h}16^\mathrm{m}32.20^\mathrm{s}$} & \footnotesize{$22^\mathrm{d}34^\mathrm{m}52.98^\mathrm{s}$} & \footnotesize{17.311 $\pm$ 0.228} & \footnotesize{15.919 $\pm$ 0.169} & \footnotesize{15.075 $\pm$ 0.112} & \footnotesize{CTTS} & \footnotesize{0} \\
51 & \footnotesize{$6^\mathrm{h}16^\mathrm{m}25.32^\mathrm{s}$} & \footnotesize{$22^\mathrm{d}35^\mathrm{m}29.40^\mathrm{s}$} & \footnotesize{16.402 $\pm$ 0.098} & \footnotesize{15.57 $\pm$ 0.136} & \footnotesize{14.91 $\pm$ 0.093} & \footnotesize{CTTS} & \footnotesize{0} \\
52 & \footnotesize{$6^\mathrm{h}16^\mathrm{m}41.39^\mathrm{s}$} & \footnotesize{$22^\mathrm{d}32^\mathrm{m}57.89^\mathrm{s}$} & \footnotesize{16.827 $\pm$ 0.180} & \footnotesize{15.657 $\pm$ 0.153} & \footnotesize{14.944 $\pm$ 0.111} & \footnotesize{CTTS} & \footnotesize{0} \\
53 & \footnotesize{$6^\mathrm{h}16^\mathrm{m}28.97^\mathrm{s}$} & \footnotesize{$22^\mathrm{d}32^\mathrm{m}39.40^\mathrm{s}$} & \footnotesize{15.688 $\pm$ 0.056} & \footnotesize{14.903 $\pm$ 0.064} & \footnotesize{14.459 $\pm$ 0.064} & \footnotesize{CTTS} & \footnotesize{0} \\
54 & \footnotesize{$6^\mathrm{h}16^\mathrm{m}25.67^\mathrm{s}$} & \footnotesize{$22^\mathrm{d}39^\mathrm{m}06.09^\mathrm{s}$} & \footnotesize{17.201 $\pm$ 0.219} & \footnotesize{16.343 $\pm$ 0.262} & \footnotesize{15.773 $\pm$ 0.198} & \footnotesize{CTTS} & \footnotesize{0} \\
55 & \footnotesize{$6^\mathrm{h}16^\mathrm{m}34.49^\mathrm{s}$} & \footnotesize{$22^\mathrm{d}35^\mathrm{m}22.21^\mathrm{s}$} & \footnotesize{16.21 $\pm$ 0.084} & \footnotesize{15.149 $\pm$ 0.084} & \footnotesize{14.537 $\pm$ 0.063} & \footnotesize{CTTS} & \footnotesize{0} \\
56 & \footnotesize{$6^\mathrm{h}16^\mathrm{m}30.92^\mathrm{s}$} & \footnotesize{$22^\mathrm{d}36^\mathrm{m}01.72^\mathrm{s}$} & \footnotesize{16.564 $\pm$ 0.128} & \footnotesize{15.298 $\pm$ 0.110} & \footnotesize{14.436 $\pm$ 0.062} & \footnotesize{CTTS} & \footnotesize{0} \\
57 & \footnotesize{$6^\mathrm{h}16^\mathrm{m}30.75^\mathrm{s}$} & \footnotesize{$22^\mathrm{d}39^\mathrm{m}28.69^\mathrm{s}$} & \footnotesize{16.132 $\pm$ 0.079} & \footnotesize{15.328 $\pm$ 0.099} & \footnotesize{14.737 $\pm$ 0.077} & \footnotesize{CTTS} & \footnotesize{0} \\
58 & \footnotesize{$6^\mathrm{h}16^\mathrm{m}32.27^\mathrm{s}$} & \footnotesize{$22^\mathrm{d}39^\mathrm{m}32.10^\mathrm{s}$} & \footnotesize{16.987 $\pm$ 0.166} & \footnotesize{15.842 $\pm$ 0.152} & \footnotesize{15.084 $\pm$ 0.107} & \footnotesize{CTTS} & \footnotesize{0} \\
59 & \footnotesize{$6^\mathrm{h}16^\mathrm{m}50.15^\mathrm{s}$} & \footnotesize{$22^\mathrm{d}33^\mathrm{m}04.16^\mathrm{s}$} & \footnotesize{16.416 $\pm$ 0.122} & \footnotesize{15.609 $\pm$ 0.138} & \footnotesize{15.122 $\pm$ 0.105} & \footnotesize{CTTS} & \footnotesize{0} \\
60 & \footnotesize{$6^\mathrm{h}16^\mathrm{m}49.96^\mathrm{s}$} & \footnotesize{$22^\mathrm{d}38^\mathrm{m}43.88^\mathrm{s}$} & \footnotesize{16.521 $\pm$ 0.136} & \footnotesize{15.645 $\pm$ 0.143} & \footnotesize{15.08 $\pm$ 0.113} & \footnotesize{CTTS} & \footnotesize{0} \\
61 & \footnotesize{$6^\mathrm{h}16^\mathrm{m}54.18^\mathrm{s}$} & \footnotesize{$22^\mathrm{d}34^\mathrm{m}56.75^\mathrm{s}$} & \footnotesize{16.184 $\pm$ 0.098} & \footnotesize{15.279 $\pm$ 0.107} & \footnotesize{14.72 $\pm$ 0.076} & \footnotesize{CTTS} & \footnotesize{0} \\
62 & \footnotesize{$6^\mathrm{h}16^\mathrm{m}48.71^\mathrm{s}$} & \footnotesize{$22^\mathrm{d}36^\mathrm{m}02.72^\mathrm{s}$} & \footnotesize{16.682 $\pm$ 0.152} & \footnotesize{15.854 $\pm$ 0.171} & \footnotesize{15.249 $\pm$ 0.126} & \footnotesize{CTTS} & \footnotesize{0} \\
\hline \\[-1.0em]
63 & \footnotesize{$6^\mathrm{h}16^\mathrm{m}20.50^\mathrm{s}$} & \footnotesize{$22^\mathrm{d}37^\mathrm{m}06.78^\mathrm{s}$} & \footnotesize{15.847 $\pm$ 0.074} & \footnotesize{15.298 $\pm$ 0.102} & \footnotesize{14.635 $\pm$ 0.083} & \footnotesize{HAeBe} & \footnotesize{0} \\
64 & \footnotesize{$6^\mathrm{h}16^\mathrm{m}20.35^\mathrm{s}$} & \footnotesize{$22^\mathrm{d}35^\mathrm{m}04.23^\mathrm{s}$} & \footnotesize{15.896 $\pm$ 0.073} & \footnotesize{15.477 $\pm$ 0.110} & \footnotesize{14.917 $\pm$ 0.100} & \footnotesize{HAeBe} & \footnotesize{0} \\
65 & \footnotesize{$6^\mathrm{h}16^\mathrm{m}37.47^\mathrm{s}$} & \footnotesize{$22^\mathrm{d}30^\mathrm{m}53.17^\mathrm{s}$} & \footnotesize{16.922 $\pm$ 0.187} & \footnotesize{16.207 $\pm$ 0.243} & \footnotesize{15.484 $\pm$ 0.150} & \footnotesize{HAeBe} & \footnotesize{0} \\
66 & \footnotesize{$6^\mathrm{h}16^\mathrm{m}25.69^\mathrm{s}$} & \footnotesize{$22^\mathrm{d}38^\mathrm{m}08.44^\mathrm{s}$} & \footnotesize{16.448 $\pm$ 0.140} & \footnotesize{15.86 $\pm$ 0.161} & \footnotesize{15.218 $\pm$ 0.132} & \footnotesize{HAeBe} & \footnotesize{0} \\
67 & \footnotesize{$6^\mathrm{h}16^\mathrm{m}31.69^\mathrm{s}$} & \footnotesize{$22^\mathrm{d}38^\mathrm{m}17.72^\mathrm{s}$} & \footnotesize{16.433 $\pm$ 0.108} & \footnotesize{16.133 $\pm$ 0.215} & \footnotesize{15.392 $\pm$ 0.138} & \footnotesize{HAeBe} & \footnotesize{0} \\
68 & \footnotesize{$6^\mathrm{h}16^\mathrm{m}57.74^\mathrm{s}$} & \footnotesize{$22^\mathrm{d}30^\mathrm{m}24.12^\mathrm{s}$} & \footnotesize{16.068 $\pm$ 0.092} & \footnotesize{15.669 $\pm$ 0.144} & \footnotesize{15.076 $\pm$ 0.101} & \footnotesize{HAeBe} & \footnotesize{0} \\
69 & \footnotesize{$6^\mathrm{h}16^\mathrm{m}57.50^\mathrm{s}$} & \footnotesize{$22^\mathrm{d}34^\mathrm{m}34.24^\mathrm{s}$} & \footnotesize{16.757 $\pm$ 0.159} & \footnotesize{16.112 $\pm$ 0.217} & \footnotesize{15.46 $\pm$ 0.157} & \footnotesize{HAeBe} & \footnotesize{0} \\
\hline \\[-1.0em]
70 & \footnotesize{$6^\mathrm{h}16^\mathrm{m}43.12^\mathrm{s}$} & \footnotesize{$22^\mathrm{d}33^\mathrm{m}01.40^\mathrm{s}$} & \footnotesize{14.73 $\pm$ 0.042} & \footnotesize{14.126 $\pm$ 0.053} & \footnotesize{13.625 $\pm$ 0.086} & \footnotesize{YSO} & \footnotesize{2} \\
71 & \footnotesize{$6^\mathrm{h}16^\mathrm{m}35.10^\mathrm{s}$} & \footnotesize{$22^\mathrm{d}31^\mathrm{m}33.98^\mathrm{s}$} & \footnotesize{15.086 $\pm$ 0.048} & \footnotesize{14.306 $\pm$ 0.051} & \footnotesize{13.628 $\pm$ 0.056} & \footnotesize{YSO} & \footnotesize{2} \\
72 & \footnotesize{$6^\mathrm{h}16^\mathrm{m}44.82^\mathrm{s}$} & \footnotesize{$22^\mathrm{d}34^\mathrm{m}42.82^\mathrm{s}$} & \footnotesize{15.207 $\pm$ 0.049} & \footnotesize{14.728 $\pm$ 0.067} & \footnotesize{14.395 $\pm$ 0.061} & \footnotesize{YSO} & \footnotesize{2} \\
\hline \\[-1.0em]
73 & \footnotesize{$6^\mathrm{h}16^\mathrm{m}40.21^\mathrm{s}$} & \footnotesize{$22^\mathrm{d}31^\mathrm{m}15.34^\mathrm{s}$} & \footnotesize{16.192 $\pm$ 0.104} & \footnotesize{15.358 $\pm$ 0.119} & \footnotesize{14.831 $\pm$ 0.119} & \footnotesize{CTTS} & \footnotesize{2} \\
74 & \footnotesize{$6^\mathrm{h}16^\mathrm{m}41.46^\mathrm{s}$} & \footnotesize{$22^\mathrm{d}31^\mathrm{m}28.03^\mathrm{s}$} & \footnotesize{15.451 $\pm$ 0.061} & \footnotesize{14.421 $\pm$ 0.058} & \footnotesize{13.757 $\pm$ 0.061} & \footnotesize{CTTS} & \footnotesize{2} \\
75 & \footnotesize{$6^\mathrm{h}16^\mathrm{m}43.47^\mathrm{s}$} & \footnotesize{$22^\mathrm{d}31^\mathrm{m}55.73^\mathrm{s}$} & \footnotesize{16.435 $\pm$ 0.121} & \footnotesize{15.65 $\pm$ 0.139} & \footnotesize{15.131 $\pm$ 0.135} & \footnotesize{CTTS} & \footnotesize{2} \\
76 & \footnotesize{$6^\mathrm{h}16^\mathrm{m}44.87^\mathrm{s}$} & \footnotesize{$22^\mathrm{d}34^\mathrm{m}10.32^\mathrm{s}$} & \footnotesize{14.743 $\pm$ 0.046} & \footnotesize{14.039 $\pm$ 0.057} & \footnotesize{13.62 $\pm$ 0.072} & \footnotesize{CTTS} & \footnotesize{1} \\
77 & \footnotesize{$6^\mathrm{h}16^\mathrm{m}45.52^\mathrm{s}$} & \footnotesize{$22^\mathrm{d}33^\mathrm{m}42.53^\mathrm{s}$} & \footnotesize{16.511 $\pm$ 0.132} & \footnotesize{15.626 $\pm$ 0.144} & \footnotesize{15.116 $\pm$ 0.162} & \footnotesize{CTTS} & \footnotesize{2} \\
78 & \footnotesize{$6^\mathrm{h}16^\mathrm{m}45.50^\mathrm{s}$} & \footnotesize{$22^\mathrm{d}34^\mathrm{m}58.63^\mathrm{s}$} & \footnotesize{16.432 $\pm$ 0.130} & \footnotesize{15.51 $\pm$ 0.140} & \footnotesize{14.975 $\pm$ 0.135} & \footnotesize{CTTS} & \footnotesize{1} \\
\hline \\[-1.0em]
79 & \footnotesize{$6^\mathrm{h}16^\mathrm{m}43.49^\mathrm{s}$} & \footnotesize{$22^\mathrm{d}33^\mathrm{m}37.44^\mathrm{s}$} & \footnotesize{16.867 $\pm$ 0.189} & \footnotesize{15.85 $\pm$ 0.178} & \footnotesize{14.899 $\pm$ 0.148} & \footnotesize{HAeBe} & \footnotesize{2} \\
\hline
\end{longtable}

\twocolumn

\end{appendix}

\end{document}